\documentclass[column,           % Format : preprint, twocolumn
               showpacs,            % Pacs : showpacs, noshowpacs
               nopreprintnumbers,     % Preprint: preprintnumbers,
               aps,                 % Society: ...
               prd,          	    % Journal Style : pra, prb, prc, prd, pre,
               letterpaper,             % Size : a4paper, ...
              groupeaddress,      % Affiliation (Title) : groupedaddress,
               nofootinbib,         % Footnote: footinbib, nofootinbib
               tightenlines,        % Remove additional spaces in a line
               floats,floatfix,      % Floating pictures and tables
               showkeys
               ]{revtex4-1}
               
\usepackage[toc,page]{appendix}
\usepackage{graphicx}% Include figure files
\usepackage{dcolumn}% Align table columns on decimal point
\usepackage{bm}% bold math
\usepackage{amsmath}
\usepackage{amsfonts,amssymb}
\usepackage{soul}
\usepackage{color}
\definecolor{v}{rgb}{0.6, 0.2, 0.8} %comentarios VM
\definecolor{MAGA}{rgb}{0.1, 0.43, 0.75}
\definecolor{jm}{rgb}{0.13, 0.48, 0.64}
\usepackage{xcolor}
\usepackage{enumerate}
\usepackage{float}
\usepackage{subfig}
\usepackage{ulem}

\begin{document}

\title{Taxonomy of Dark Energy Models}

\author{V. Motta$^1$}
\email{veronica.motta@uv.cl}

\author{Miguel A. Garc\'ia-Aspeitia$^{2,3}$}
\email{aspeitia@fisica.uaz.edu.mx}

\author{A. Hern\'andez-Almada$^4$}
\email{ahalmada@uaq.mx}

\author{Juan Maga\~na$^{5}$}
\email{jmagana@astro.puc.cl}

\author{Tom\'as Verdugo$^{6}$}
\email{tomasv@astro.unam.mx}

\affiliation{$^1$Instituto de F\'isica y Astronom\'ia, Facultad de Ciencias, Universidad de Valpara\'iso, Avda. Gran Breta\~na 1111, Valpara\'iso, Chile.}
\affiliation{$^2$Unidad Acad\'emica de F\'isica, Universidad Aut\'onoma de Zacatecas, Calzada Solidaridad esquina con Paseo a la Bufa S/N C.P. 98060, Zacatecas, M\'exico.}
\affiliation{$^3$Consejo Nacional de Ciencia y Tecnolog\'ia, \\ Av. Insurgentes Sur 1582. Colonia Cr\'edito Constructor, Del. Benito Ju\'arez C.P. 03940, Ciudad de M\'exico, M\'exico.}
\affiliation{$^4$Facultad de Ingenier\'ia, Universidad Aut\'onoma de Quer\'etaro, Centro Universitario Cerro de las Campanas, 76010, Santiago de Quer\'etaro, M\'exico.}
\affiliation{$^5$Instituto de Astrof\'isica \& Centro de Astro-Ingenier\'ia, Pontificia Universidad Cat\'olica de Chile, \\Av. Vicu\~na Mackenna, 4860, Santiago, Chile}
\affiliation{$^{6}$ Instituto de Astronom\'ia, Observatorio Astron\'omico Nacional, Universidad Nacional Aut\'onoma de M\'exico, Apartado postal 106, C.P. 22800,  Ensenada, B.C., M\'exico.}

\begin{abstract}
The accelerated expansion of the Universe is one of the main discoveries of the past decades, indicating the presence of an unknown component: the dark energy. Evidence of its presence is being gathered by a succession of observational experiments with increasing precision in its measurements. However, the most accepted model for explaining the dynamic of our Universe, the so-called Lambda cold dark matter, face several problems related to the nature of such energy component. This has lead to  a growing exploration of alternative models attempting to solve those drawbacks. In this review, we briefly summarize the characteristics of a (non-exhaustive) list of dark energy models as well as some of the most used cosmological samples. Next, we discuss how to constrain each model's parameters using observational data. Finally, we summarize the status of dark energy modeling.
\end{abstract}

\keywords{dark energy; modified gravity; cosmology}
%\draft
\pacs{}
\date{\today}
\maketitle

%%%%%%%%%%%%%%%%%%%%%%%%%%%%%%%%%%%%%%%%%%

%%%%%%%%%%%%%%%%%%%%%
\section{Introduction}
%%%%%%%%%%%%%%%%%%%%%

One of the most important discoveries in modern cosmology since the Universe expansion by George Lema{\^\i}tre \citep{Lemaitre:1927} and Edwin Hubble \citep{Hubble:1929}, and the detection of the cosmic microwave background (CMB) \citep{Penzias:1965,Dicke:1965} is the accelerated expansion at late times. This cosmic acceleration was detected at the end of the 90s, when two independent collaborations where observing high redshift type Ia supernovae  (SNIa) to measure the curvature and deceleration parameter of the Universe \citep{Riess:1998,Perlmutter:1999}; both groups established  a cosmological constant model of the Universe, with $\Omega_m \approx$ 0.3, and  $\Omega_{\Lambda} \approx$ 0.7.  Later, this result was confirmed by different cosmological probes, for example the CMB \citep{deBernardis:2000,Spergel:2003}, baryon acoustic oscillations \citep[e.g.,][]{Eisenstein:2005,Percival:2007}, and gravitational lensing, strong \citep[e.g,][]{Suyu:2010} and weak \citep[e.g.,][]{Schrabback:2010}.  

In the last decades, substantial effort has been made to understand the nature of DE \citep[e.g. see][for a recent review]{Huterer:2018} with the $\Lambda$CDM model being the cornerstone of modern cosmology and the simplest one, composed by two dominant components, known as cold dark matter (DM) and cosmological constant, and three subdominant components identified as baryons, photons and neutrinos. The $\Lambda$CDM model is not only effective at background level (isotropic and homogeneous Universe), but also robust at linear perturbations, having accurate predictions for the matter power spectrum and the small differences for photons temperature observed in the CMB \ \citep[][]{Bennett:2013,Planck:2018}. As mentioned previously, the $\Lambda$CDM model is composed by a $\sim32\%$ of cold dark matter which is essential for the structure formation, being the most suitable explanation for the rotation curves of galaxies \cite[see for instance][]{Feng:2010,Bertone:2018,Martino:2020}. It is assumed that dark matter is a non relativistic particle and the preferred  candidates are the particles  emerging from the Supersymmetric theory \cite{Martin:1997,Abazov:2007}. However, the lack of evidence of these particles strengthen the proposition of other candidates as Axions, Ultra light scalar field dark matter, among others \cite{Wilczek:1978,Lee:1996,Urena:2000}. Despite its remarkable achievements of the $\lambda$CDM model, it has important afflictions when describing the nature of the CC through the idea of a quantum vacuum fluctuations \cite{Zeldovich,Weinberg}. This idea generates, from the theoretical point of view, an error of $\sim10^{120}$ orders of magnitude and it is known as the {\it fine tuning} problem \cite{Carroll:2000fy}. In addition, the CC has the {\it coincidence problem} \cite{Carroll:2000fy}, i.e. why the Universe accelerate at late epochs and not before of after? On top of that, recent observations from Planck and Supernovae Type Ia differ in their values for $H_0$\footnote{See also \cite{Millon:2020,Birrer:2020} for recent values measured using gravitational lens systems.}, introducing tension between different observations at different redshifts,  \cite{Joudaki:2016mvz,Hildebrandt:2016iqg,Riess_2018,Riess:2019cxk,DiValentino:2021izs}. The community attribute the problem to the $\Lambda$CDM model and in particular to the CC, therefore new approaches are proposed to alleviate the discrepancy among observations \cite[see also][]{Lukovic:2016,Brax:2017,Kabath:2020}. 

In the context of these tensions and the problems with CC, a plethora of alternative dark energy models have been explored. Our aim is to review a set of those models consisting of fluids that can be described by a variety of formulations. Some of them can be described by different fluids and their particles that compose them, like scalar fields, while others that do not need extra fluids require modifications to general theory of relativity \cite[GTR,][]{PEDE:2019ApJ,PEDE:2020,Hernandez-Almada:2020uyr,m2000,MaartensCos,Garcia-Aspeitia:2016kak,Garcia-Aspeitia:2018fvw,Ellis,Gao:2014nia,Garcia-Aspeitia:2019yni,Glavan:2019inb,Garcia-Aspeitia:2020uwq,FREESE20021,Gondolo:2002fh,Jaime:2018ftn} (for recent overviews of GTR, see, e.g., \cite{Iorio:2015,Debono:2016,Vishwakarma:2016} and references therein). Specifically, the first category contains models avoiding the idea  of associating the Universe acceleration with quantum vacuum fluctuations (like in CC) and, thus assume a fluid expression manageable by the quantum field theory. The second one, consist of models introducing extensions to GTR, generating a  Universe acceleration without the addition of extra fluids. However, the current overabundance of models is overwhelming, posing difficulties to decide which is the best candidate to compete against the $\Lambda$CDM model.

Recent years have also witnessed the increase of observational surveys designed to obtain precise measurements aimed to probe DE nature. With this improvement in the instrument sensitivity came the mechanism to discriminate between models of DE, i.e. tensions between different probes could lead to rule out some of them. In this context, we consider widely used samples (such as SNIa \cite{Riess:1998}, CMB \cite{Planck_CP:2018}, baryon acoustic oscillations) as well as other recent compilations (strong gravitational lens systems \cite{Cao:2012,Cao:2015qja,Amante:2019xao}, starburst galaxies \cite{Chavez2012,Chavez2014,Terlevich2015,Chavez2016,GonzalezMoran2019,Cao:2020jgu}, and observational Hubble parameters \cite{Jimenez:2001gg,Amante:2019xao}. 

The outline of the review is as follows: Section \ref{basic} summarizes the basic equations, Section \ref{CS} presents the cosmological samples that we use to constraint the different theoretical models, in Section \ref{TDE} we describe the DE models, together with the constraints of the free parameters of each models. Finally, in  Section \ref{DC} we give a discussion and conclusion of the models mentioned, discussing the promising models and what is the contribution to the understanding of the Universe acceleration.

%%%%%%%%%%%%%%%%%%%%%%%%%%%%%%%%%%%%%%%%%%
\section{Basic Equations for the Background Cosmology} \label{basic}
%%%%%%%%%%%%%%%%%%%%%%%%%%%%%%%%%%%%%%%%%%

Modern cosmology is based on the general theory of relativity whose master equation is the field equation given by 
\begin{equation}
    G_{\mu\nu}=8\pi GT_{\mu\nu}, \label{EFEM}
\end{equation}
where $G_{\mu\nu}\equiv R_{\mu\nu}-g_{\mu\nu}R/2$ is known as the Einstein tensor, composed by the Ricci tensor ($R_{\mu\nu}$), the Ricci scalar ($R$) and the metric tensor ($g_{\mu\nu}$). The right side of Eq. \eqref{EFEM} shows the Newton gravitational constant ($G$) and the energy momentum tensor. Hereafter we will use natural units ($c=\hbar=1$), unless we explicitly mention the opposite.

Hereafter, we focus our attention in the background cosmology considering a homogeneous and isotropic Universe. In this vein, we use the standard line element of Friedmann-Lemaitre-Robertson-Walker (FLRW)  with flat geometry $k=0$, as 
\begin{equation}
ds^2=-dt^2+a(t)^2(dr^2+r^2d\Omega^2), \label{FLRW}
\end{equation}
where $d\Omega^2\equiv d\theta^2+\sin^2\theta d\varphi^2$ and $a(t)$ is the scale factor. The energy-momentum tensor is written as always as
\begin{equation}
    T_{\mu\nu}=pg_{\mu\nu}+(\rho+p)u_{\mu}u_{\nu}, \label{emt}
\end{equation}
being $p$, $\rho$, $u_{\mu}$ and $g_{\mu\nu}$ the pressure, density, four-velocity of the fluid and the metric tensor, respectively. The covariant derivative of the energy momentum tensor $\nabla^{\mu}T_{\mu\nu}=0$, generates the continuity equation, given by
\begin{equation}
    \sum_i\left[\dot{\rho}_i+3\frac{\dot{a}}{a}(\rho_i+p_i)\right]=0, \label{cont}
\end{equation}
where the sum is over all the species and dots stand for derivatives with respect to time.

Through the Einstein field equations, we arrive to the Friedmann equation, which is a first order non-linear differential equation composed by the scale factor and the densities of the species. Therefore, the equation takes the form 
\begin{equation}
H^2\equiv\left(\frac{\dot{a}}{a}\right)^2=\frac{8\pi G}{3}\sum_i\rho_i,
\label{eq:Hz_gen}
\end{equation}
where $H$ is known as the Hubble parameter, which indicates the Universe expansion rate. Hence, it is possible to define the dimensionless Friedmann equation in the form
\begin{equation}
E(z)^2\equiv\left(\frac{H}{H_0}\right)^2=\sum_i\Omega(z)_i,
\label{eq:Ez_gen}
\end{equation}
where $\Omega(z)\equiv8\pi G\rho(z)/3H_0^2$ is known as the density parameter, a function of the redshift ($z$), and the sum runs for the components of the Universe used in the model\footnote{$\Omega_b$, $\Omega_{dm}$, $\Omega_r$, $\Omega_{de}$ represents baryons, dark matter, radiation, and dark energy respectively.}, $H_0$ is the Hubble constant\footnote{This is the Hubble parameter evaluated at $z=0$.}. With the previous equations, it is possible to define the flat condition as 
\begin{equation}
    1=\sum_i\Omega_i, \label{FlatCond}
\end{equation}
notice that the density parameters are evaluated at $z=0$ and they are denoted with the label '0'.
The comoving distance from the observer to redshift $z$ is given by (units of $c$ are recovered in the following equations)
\begin{equation}
r(z)=\frac{c}{H_0}\int_0^z \frac{dz'}{E(z')}.
\label{eq:rz}
\end{equation}
Therefore, we define the luminosity distance, denoted as $d_L(z)$, as 
\begin{equation}\label{eq:dL}
    d_L(z)=\frac{c}{H_{0}}(1+z)\int_0^z\frac{dz^{\prime}}{E(z^{\prime})},
\end{equation}
where $c$ is the speed of light. The angular diameter distance is related to the comoving distance as
\begin{equation} \label{eq:dLDA}
D_A=d_L(z)/(1+z)^2,
\end{equation}
In addition, the angular diameter distance between two objects at redshifts $z_1$ and $z_2$ ($z_1<z_2$) is given by 
\begin{equation} \label{eq:D12}
         D_{12}(z)=\frac{c}{H_{0}(1+z)}\int_{z_1}^{z_2}\frac{dz^{\prime}}{E(z^{\prime})},
\end{equation}

%%%%%%%%%%%%%%%%%%%%%%%%%%%%%%%%%%%%%%%%%%
\section{Cosmological Samples} \label{CS}

Observational samples are required not only to investigate the consistency of any theoretical model prediction but also to discern  between different models.
In this section we review the most common samples used as ``standard candles'' to probe those models. Given a cosmological model, the astronomical distance to an object is generally measured using the redshift to the object and the distance-redshift relationship provided by the model (see equation \ref{eq:dL}). However, when the goal is to infer the cosmological model, an independent distance measurement is needed. The methods to obtain these independent measurements are based on empirical relationships and, thus, the astronomical objects following those relations are called standard candles.

Knowing the intrinsic flux of a standard candle\footnote{Recall the relation between the flux ($f$), intrinsic luminosity ($L$) and the luminosity distance ($d_L$): $f=L/(4 \pi d_L^2)$.}, we can obtain a relation between the apparent magnitude ($m$, observed) and the absolute magnitude ($M$, acquired from an empirical relation) 
\begin{equation} \label{dmodulus}
m=M+5 \log_{10} \left( \frac{d_L}{10pc} \right).
\end{equation}

In the following sections we include a general description of some of the most used cosmological samples with emphasis in the quantification of the goodness-of-fit (or figure-of-merit) by defining a function ($\chi^2$) associated to the sample errors or covariances, that is applied to investigate the cosmological models presented in Section \ref{TDE}.

%%%%%%%%%%%%%%%%%%%%%%%%%%%%%%%%%%%%%%
\subsection{Type Ia Supernovae}
%%%%%%%%%%%%%%%%%%%%%%%%%%%%%%%%%%%%%%

Type Ia supernova is believed to originate by a white dwarf accreating matter from a companion star. When the white dwarf exceeds the Chandrasekhar mass limit ($\sim 1.4 M_{\odot}$, where $M_{\odot}$ is 1 solar mass), there is a collapse and subsequent explosion. As all the SNIa have roughly the same luminosity\footnote{SNIa peak luminosities could have a scatter of $\sim 0.3$ mag but, after applying a correction related to the correlation between the peak luminosity and the light-curve decline time (the so-called ``stretch''), the scatter is reduced to $\leq 0.15$ mag. }, they can be used as standard candles.

Samples of SNIa (e.g. \cite{Scolnic:2018,Riess_2018,Hernandez-Almada:2020ulm,Magana:2017nfs,Garcia-Aspeitia:2020uwq}) provide distance modulus measurements (see Eq. \ref{dmodulus}). As the measurements in this kind of samples are correlated, it is convenient to build the chi square function as
\begin{equation}\label{eq:chi2SnIa}
    \chi_{\rm SNIa}^{2}=a +\log \left( \frac{e}{2\pi} \right)-\frac{b^{2}}{e},
\end{equation}
where
\begin{eqnarray}
    a &=& \Delta\boldsymbol{\tilde{\mu}}^{T}\cdot\mathbf{Cov_{P}^{-1}}\cdot\Delta\boldsymbol{\tilde{\mu}}, \nonumber\\
    b &=& \Delta\boldsymbol{\tilde{\mu}}^{T}\cdot\mathbf{Cov_{P}^{-1}}\cdot\Delta\mathbf{1}, \\
    e &=& \Delta\mathbf{1}^{T}\cdot\mathbf{Cov_{P}^{-1}}\cdot\Delta\mathbf{1}, \nonumber
\end{eqnarray}
and $\Delta\boldsymbol{\tilde{\mu}}$ is the vector of residuals between the theoretical distance modulus and the observed one, $\Delta\mathbf{1}=(1,1,\dots,1)^T$, $\mathbf{Cov_{P}}$ is the covariance matrix formed by adding the systematic and statistic uncertainties, i.e.   $\mathbf{Cov_{P}}=\mathbf{Cov_{P,sys}}+\mathbf{Cov_{P,stat}}$. The super-index $T$ on the above expressions denotes the transpose of the vectors.

The theoretical distance modulus is estimated by
\begin{equation}
    m_{th}=\mathcal{M}+5\log_{10}[d_L(z)/10\, pc],
\end{equation}
where $\mathcal{M}$ is a nuisance parameter which has been marginalized in \eqref{eq:chi2SnIa}.

%%%%%%%%%%%%%%%%%%%%%%%%%%%%%%%%%%%%%%
\subsection{Baryon Acoustic Oscillations}
%%%%%%%%%%%%%%%%%%%%%%%%%%%%%%%%%%%%%%

Another way to establish a constraint of model parameters is through the standard rules known as Baryon Acoustic Oscillations (BAO). These are primordial signatures in the matter power spectrum produced by the interaction between baryons and photons in a hot plasma in the pre-recombination epoch. 

The theoretical BAO angular scale ($\theta_{th}$) is estimated as
\begin{equation}
    \theta_{th}(z) = \frac{r_{drag}}{(1+z)D_A(z)}\,,
\end{equation}
where $d_L$ and $D_A$ are written in Eqs. \eqref{eq:dL} and \eqref{eq:dLDA}, respectively.
The parameter $r_{drag}$ indicates the sound horizon at baryon drag epoch. The comoving sound horizon, $r_s(z)$, is defined as
\begin{equation} \label{eq:rs}
 r_s(z) = \frac{c}{H_{0}} \int_z^\infty \frac{c_s(z')}{E(z')}dz',
 \end{equation}
where the sound speed $c_s(z) = 1/\sqrt{3\left(1+\bar{R}_b/\left(1+z\right)\right)}$, with
$\bar{R}_b = 31500\, \Omega_{b}h^2$($T_{CMB}$/2.7K)$^{-4}$, and
$T_{CMB}$ is the CMB temperature. 
The redshift $z_{drag}$ at the baryon drag epoch is well fitted with the
formula proposed by \citet{Eisenstein:1997ik}
\begin{equation}
z_{drag} =\frac{1291(\Omega_{m0}h^2)^{0.251}}{1+0.659\,(\Omega_{m0}h^2)^{0.828}}[1+b_1(\Omega_{b0}
h^2)^{b_2}],
\end{equation}
where
\begin{eqnarray}
b_1 &=& 0.313\left(\Omega_{m0}\,h^2\right)^{-0.419}\left[1+0.607\left(\Omega_{m0}\,h^2\right)^{0.674}\right], \\
b_2 &=& 0.238\left(\Omega_{m0}\,h^2\right)^{0.223}.
\end{eqnarray}
where $\Omega_{m0}$ and $\Omega_{b0}$ are the matter component (dark matter plus baryons) and baryon component  at $z=0$ respectively. For this work, we set the $r_{drag}=147.21 \pm 0.23$ obtained by Planck collaboration \cite{Planck:2018}. Notice that, as BAO data points are estimated using $r_{drag}$, which depends on the cosmological model, they could be considered as biased.

The most recent compilation of transversal BAO measurements $\theta_{BAO}(z)$ is presented in \cite{Nunes:2020hzy}. A total of 15 measurements \cite{Carvalho:2016, Alcaniz2017, CARVALHO:2020,de_Carvalho_2018, de_Carvalho:2020mnras} were obtained using the data realeases (DR), DR7, DR10, DR11, DR12, DR12Q (quasars), of Sloan Digital Sky Survey (SDSS) \cite{York_2000}. As transversal angular BAO points are considered uncorrelated, the chi square function is built as
\begin{equation}
\chi^2_{\rm BAO} = \sum_{i=1}^{N} \left( \frac{\theta_{\rm BAO}^i - \theta_{th}(z_i) }{\sigma_{\theta_{\rm BAO}^i}}\right)^2\,,
\end{equation}
where $\theta_{\rm BAO}^i \pm \sigma_{\theta_{\rm BAO}^i}$ is the BAO angular scale, $N$ is the number of data and its uncertainty at $68\%$ measured at $z_i$. 
It is worth to mention that there is a sample of 6 correlated data points, with their associated covariance matrix, collected in \cite{Giostri:2012} and measured by \cite{Percival:2010,Blake:2011,Beutler:2011hx}. In this case, the chi square function is
\begin{equation}
    \chi^2_{cBAO} = \Vec{X}^T  \rm{Cov}^{-1} \Vec{X}
\end{equation}
where $\Vec{X}$ is the difference between the theoretical and observational quantities of \\ $d_A(z_{drag})/D_V(z_i)$ measured at the redshift $z_i$, and  $\rm{Cov}^{-1}$ is the inverse covariance matrix (see \cite{Giostri:2012} for details), the dilation scale ($D_V$) is defined as \cite{Wigglez:Eisenstein2005}
\begin{equation} \label{eq:D_V}
    D_V = \left[ \frac{d_A^2(z)\,c\,z}{H_0 E(z)}\right]^{1/3}
\end{equation}
where $d_A(z)=(1+z)D_A(z)$ is the comoving angular-diameter distance.

%%%%%%%%%%%%%%%%%%%%%%%%%%%%%%%%%%%%%%
\subsection{Cosmic Microwave Background Radiation} \label{cmb}
%%%%%%%%%%%%%%%%%%%%%%%%%%%%%%%%%%%%%%

In the early Universe baryons and photons are coupled, leading to coherent oscillations that are observed in the power spectrum\footnote{The power spectrum is the statistical description of the temperature anisotropies observed in the CMB map.} of the CMB (e.g. WMAP \cite{Spergel:2003,Hinshaw:2013}, Planck \cite{Planck:2015XIII,Planck:2020VI}). This is a powerful probe due to its ability to estimate the cosmological parameters with high precision \cite{Hu:2002}. The information of the CMB acoustic peaks can be condensed in three quantities, their distance posteriors: the acoustic scale ($l_A$), the shift parameter ($R$), and the decoupling redshift ($z_*$). Several authors have proved that these quantities are almost independent of the DE model considered and, thus they can be used to test the parameters of alternative cosmologies \cite{Komatsu_2011,Wang:2006ts, Wang:2011sb}.
The acoustic scale is defined as
\begin{equation}
l_A = \frac{\pi r(z_*)}{r_s(z_*)},
\label{eq:lA}
\end{equation}
where $r_{s}$ is the sound horizon (Eq. \eqref{eq:rs}) at the redshift of decoupling $z_*$ given by \citet{Hu:1995en},
\begin{equation}
z_* = 1048[1+0.00124(\Omega_b0 h^2)^{-0.738}]
[1+g_1(\Omega_{m0}h^2)^{g_2}],
\end{equation}
where
\begin{equation}
g_1 = \frac{0.0783(\Omega_b h^2)^{-0.238}}{1+39.5(\Omega_b0 h^2)^{0.763}},\qquad
g_2 = \frac{0.560}{1+21.1(\Omega_b0 h^2)^{1.81}}.
\end{equation}
The shift parameter is defined as \citep{Bond:1997}
\begin{equation}
R = \frac{\sqrt{\Omega_{m0}H_{0}^2}}{c} r(z_{*}).
\end{equation}
where $\Omega_{m0}$ include the baryon and DM components.

Thus, the $\chi^2$ for the CMB data is constructed as
\begin{equation}\label{cmbchi}
 \chi^2_{\mathrm{CMB}} = X^T\,\mathrm{Cov}_{\mathrm{CMB}}^{-1}\,X,
\end{equation}
where $\mathrm{Cov}_{\mathrm{CMB}}^{-1}$ is the inverse covariance matrix of the distance posteriors and
\begin{equation}
 X =\left(
 \begin{array}{c}
 l_A^{th} - l_A^{obs} \\
 R^{th} -  R^{obs}\\
 z_*^{th} - z_{*}^{obs}
\end{array}\right),
\end{equation}
the superscripts $th$ and $obs$ refer to the theoretical and observational estimations respectively. 

To infer the parameters of the alternative cosmologies we employ the distance posteriors of WMAP \cite{Hinshaw:2013} and Planck \cite{Neveu:2016gxp}.

%%%%%%%%%%%%%%%%%%%%%%%%%%%%%%%%%%%%%
\subsection{Observational Hubble Parameter}
%%%%%%%%%%%%%%%%%%%%%%%%%%%%%%%%%%%%%%

The Hubble parameter is estimated mostly by using the differential age (DA, \cite{Jimenez:2001gg}) methodology and from BAO measurements. The former method consists of measuring the age between pairs of passive evolving galaxies (dubbed cosmic chronometers) with similar metallicity and separated by a small redshift interval\footnote{For example, \citep{Moresco:2012} measure $d z\sim0.04$ at $z<0.4$ and $d z\sim0.3$ at $z>0.4$.} with redshift $z\lesssim 2.0$.  Thus, the expansion rate is written as
\begin{equation}
    H(z) = - \frac{1}{1+z} \frac{dz}{dt}, \label{Loeb}
\end{equation}
where $dz$ is measured with high accuracy\footnote{Ref. \citep{Moresco:2012} indicates that spectroscopic redshifts of galaxies have typical uncertainties $\sigma_z < 0.001$.}. The OHD from the DA method are considered cosmological independent measurements. On the other hand, the OHD from BAO surveys are non-homogeneous since they depend on the cosmological model selected. By taking a unique value for $r_{drg}$ in these data, an OHD homogeneous sample can be obtained \cite[see][for further details]{Magana:2017nfs}. 

The observational Hubble parameter data (OHD) represents the most direct way to constrain the parameter space to mimic the observational expansion rate, the chi square function can be expressed as
\begin{equation} \label{eq:chiOHD}
    \chi^2_{{\rm OHD}}=\sum_{i=1}^{N}\left(\frac{H_{th}(z_i)-H_{obs}(z_i)}{\sigma^i_{obs}}\right)^2,
\end{equation}
where $H_{th}(z_i)$ is the theoretical estimate using \eqref{eq:Ez_gen} or a generalization, $H_{obs}(z_i)\pm \sigma_{obs}^i$ is the observational Hubble parameter (from DA,  (non)-homogeneous BAO points, or the joint of them) with its uncertainty at the redshift $z_i$, and $N$ is the number of points used.

%%%%%%%%%%%%%%%%%%%%%%%%%%%%%%%%%%%%%%
\subsection{Strong Gravitational Lens Systems}
%%%%%%%%%%%%%%%%%%%%%%%%%%%%%%%%%%%%%%

Strong gravitational lens systems (SLS) offer a unique opportunity to study the $\Omega_m-w$ plane because their confidence regions are almost orthogonal to those of standard rulers (like BAO and CMB). Different groups of SLS have been used to constrain cosmological parameters with different methods \citep{chae:2002,Biesiada, Cao_2012,Cao_2015,Maga_a_2015,Maga_a_2018,Amante:2019xao}. These systems have lenses in the region $0\lesssim z\lesssim1$ with their respective sources in the range $0.2\lesssim z\lesssim3.5$. The chi square function for SLS takes the form
\begin{equation}\label{eq:chiSLS}
    \chi^2_{\rm SLS}=\sum_i^{204}\frac{[D^{th}(z_L,z_S)-D^{obs}(\theta_E,\sigma^2)]^2}{(\delta D^{obs})^2}\,,
\end{equation}
where the observable to confront is $D^{obs}=c^2\theta_E/4\pi\sigma^2$, where $\theta_E$ is the Einstein radius of the lens obtained by assuming the gravitational lens potential is modeled by a Singular Isothermal Sphere (SIS) defined by
\begin{equation}
    \theta_E=4\pi\frac{\sigma_{SIS}^2D_{LS}}{c^2D_S}\,.
\end{equation}
In the above expression, $\sigma_{SIS}$ is the 3D velocity dispersion of the lens galaxy, $D_S$ is the angular diameter distance to the source, and $D_{LS}$ is the angular diameter distance from the lens to the source defined by Eq. \eqref{eq:D12}, where $1\to L$ and $2\to S$.  Notice that, as SLS data assumes a lens model for $\theta_E$ and $\sigma_{SIS}$ comes from spectroscopy, the sample is independent of $h$ and, as consequence, the parameter constraints do not depend on $h$.
The uncertainty of $D^{obs}$ is estimated by
\begin{equation} \label{eq:DLS}
    \delta D^{obs}=D^{obs}\left[\left(\frac{\delta\theta_E}{\theta_E}\right)^2+4\left(\frac{\delta\sigma}{\sigma}\right)^2\right]^{1/2}\,,
\end{equation}
where $\delta\theta_E$ and $\delta\sigma$ are the uncertainties of the Einstein radius and the observed line-of-sight (1D) velocity dispersion,  respectively.

The theoretical counterpart is estimated by the ratio
\begin{equation}
    D^{th}\equiv D_{LS}/D_S.
\end{equation}

A corrective parameter $f$ is often introduced in Eq. \eqref{eq:chiSLS} to take into account possible systematic differences among systems (e.g. elliptical instead of spherical profile for the lens halo, line-of-sight stellar velocity dispersion as opposed to the dark matter halo velocity dispersion, steeper mass distribution profile, see for example \cite{Ofek:2003,Cao:2011}). %However the inclusion of this parameter does not show considerable changes on the DE parameter estimation \citep{Treu:2006ApJ,Amante:2019xao}. 

%%%%%%%%%%%%%%%%%%%%%%%%%%%%%%%%%%%%%%
\subsection{Ionized Gas in Starburst Galaxies}
%%%%%%%%%%%%%%%%%%%%%%%%%%%%%%%%%%%%%%

Authors \citep[][and references therein]{Chavez2012,Chavez2014,Terlevich2015,Chavez2016,GonzalezMoran2019} argued that the correlation between the measured luminosity $L$ and the inferred velocity dispersion $\sigma$ of the ionized gas (e.g. $H\beta$, $H\alpha$, $[OIII]$ emission lines) in extreme starburst galaxies (i.e. containing a population of O and/or B stars) may be used as a cosmological tracer to constrain cosmological model parameters.  Compilations provide apparent magnitude, emission line luminosity and velocity dispersion (e.g. \cite{GonzalezMoran2019, Cao:2020jgu}) and the chi square function is estimated as
\begin{equation}\label{eq:chiHIIG}
    \chi^2_{{\rm HIIG}} = A - B^2/C\,,
\end{equation}
where
\begin{eqnarray}
    A &=& \sum_{i=1}^{153} \left( \frac{\mu_{th}(z_i)-\mu_{obs}^i}{\sigma_{\mu_{obs}^i}}\right)^2 \,,\\
    B &=& \sum_{i=1}^{153} \frac{\mu_{th}(z_i)-\mu_{obs}^i}{\sigma_{\mu_{obs}^i}} \,, \\
    C &=& \sum_{i=1}^{153} \frac{1}{(\sigma_{\mu_{obs}^i})^2}\,.
\end{eqnarray}
In the above expressions, $\mu_{obs}^i \pm \sigma_{obs}^i$ is the observed distance modulus with its uncertainty at redshift $z_i$. The theoretical estimate at the redshift $z$ is obtained by using \eqref{eq:dL} and
\begin{equation}
    \mu_{th}(z) = \mu_0 + 5 \log [\,d_L(z)\,] \,,
\end{equation}
where $\mu_0$ is a nuisance parameter which has been marginalized. 

%%%%%%%%%%%%%%%%%%%%%%%%%%%%%%%%%%%%%%
\subsection{Joint Analysis}
%%%%%%%%%%%%%%%%%%%%%%%%%%%%%%%%%%%%%%

Testing the consistency of a given cosmological model requires a range of observational samples with complementary sensitivity to the cosmological parameters. Constraint on those parameters is usually achieved by combining several samples also known as joint analysis. 

For instance, a Bayesian Markov Chain Monte Carlo (MCMC) analysis is able to constrain the phase-space parameter $\bf{\Theta}$  of a cosmological model given a number of cosmological samples. In general, the procedure consists of using \texttt{emcee} Python package \cite{Foreman:2013} for two phases: the burn-in and the MCMC. The first is performed  with a certain number of steps to achieve the convergence of the chains according to the Gelman-Rubin criterion \cite{Gelman:1992}.  The second phase is performed with an appropriate number of steps for sampling the confidence regions. Additionally, for each model, the  priors (flat or Gaussian) of the parameters are chosen according to values provided in the literature. 
For the joint analysis, the figure-of-merit to be optimized is given by
\begin{equation}\label{eq:chi2}
    \chi_{\rm Joint}^2 = \sum \chi_{\rm data},
\end{equation}
where the $\chi_{\rm data}$ represents the name of the different samples. In general, the joint analysis is calculated using the combination of at least three data samples, but ideally it should contain all of them.

%%%%%%%%%%%%%%%%%%%%%%%%%%%%%%%%%%%%%%%%%%
\section{Taxonomy of Dark Energy Models} \label{TDE}

This section is dedicated to describe the different constrictions through the samples mentioned previously for the different DE models studied in literature. We divide our study in DE models linked to a fluid with the capability of accelerating the Universe and models in which the Einstein field equations of General Theory of Relativity are modified. 

Among the featured models, the first category contains: constant DE equation of state, Parameterizations of DE, Chaplygin fluid, Viscous models,  and Phenomenological (Generalized) emergent DE (PEDE and GEDE) \cite{PEDE:2019ApJ,PEDE:2020,Hernandez-Almada:2020uyr}. The second category has: Brane models (with constant and variable tension), Unimodular Gravity, Einstein-Gauss-Bonet and Cardassian models \cite{m2000,MaartensCos,Garcia-Aspeitia:2016kak,Garcia-Aspeitia:2018fvw,Ellis,Gao:2014nia,Garcia-Aspeitia:2019yni,Garcia-Aspeitia:2019yod,Glavan:2019inb,Garcia-Aspeitia:2020uwq,FREESE20021,Gondolo:2002fh}.

%%%%%%%%%%%%%%%%%%%%%%%%%%%%%%%%%%%%%%%%%%%%%
\subsection{Accelerating Universe Fluids}

In this subsection we summarize all those models that involve a fluid enabling a late acceleration without modifications to GTR.

%%%%%%%%%%%%%%%%%%%%%%%%%%%%%%%%%%%%%%%%%%%
\subsubsection{The $\Lambda$CDM Model}
%%%%%%%%%%%%%%%%%%%%%%%%%%%%%%%%%%%%%%%%%%

The $\Lambda$CDM is the consensus model dominated by a cold dark matter and a cosmological constant component with subdominant species of baryons and relativistic particles (photons and neutrinos), being not only the most favoured by diverse observations but also the simplest. The dimensionless Friedmann function is given by Eq. \eqref{eq:Ez_gen}, which reads
\begin{equation}
    E(z)^2=\Omega_{m0}(z+1)^3+\Omega_{r0}(z+1)^4+\Omega_{\Lambda0}.
\end{equation}
The flatness condition is satisfied and written in the form
\begin{equation}
    1=\Omega_{m0}+\Omega_{r0}+\Omega_{\Lambda0}.
\end{equation}
Therefore, the CC density parameter can be written in terms of matter, while radiation takes the form
\begin{equation}
    \Omega_{r0}=2.469\times10^{-5}h^{-2}(1+0.2271 N_{eff}), \label{Rad0}
\end{equation}
where $N_{eff}=3.04$ is
the standard number of relativistic species \citep{Komatsu:2011} and $h=H_{0}/ 100\,$km s$^{-1}$Mpc$^{-1}$, where $H_0=67.66\pm0.42$km s$^{-1}$Mpc$^{-1}$ with Planck \cite{Planck_CP:2018}, while $H_0=73.2\pm1.3$km s$^{-1}$Mpc$^{-1}$ with Riess \cite{Riess:2020fzl}, presenting a tension between the observations and known as $H_0$ tension. Regarding the matter density parameter, the value is constrained as $\Omega_{m0}=0.3111\pm 0.0056$, using Planck satellite \cite{Planck_CP:2018}, which is a combination of baryonic and dark matter. Despite the model achievements, $\Lambda$CDM is afflicted with several problems, like the nature of CC \cite{Zeldovich,Weinberg}, the $\sigma_8$ tension \cite{Joudaki:2016mvz,Hildebrandt:2016iqg} and the $H_0$ tension \cite{Riess_2018,Riess:2019cxk,DiValentino:2021izs}.

%%%%%%%%%%%%%%%%%%%%%%%%%%%%%%%%%%%%%%%%%
\subsubsection{The $\omega$CDM Model}
%%%%%%%%%%%%%%%%%%%%%%%%%%%%%%%%%%%%%%%%%

This model is the simplest extension of the CC.  The dark energy has a constant equation of state (EoS) but it deviates from $w=-1$, and should satisfy $\omega<-1/3$ to obtain an accelerated Universe. The equation $E(z)$ can be written as: 
\begin{eqnarray}
E(z)_{\omega}^2=\Omega_{m0}(1+z)^3+\Omega_{r0}(1+z)^{4}+(1 -\Omega_{m0}-\Omega_{r0})(1+z)^{3(1+\omega)},
\end{eqnarray}

\begin{figure}[htb]
%\widefigure
\centering
\includegraphics[width=0.45\textwidth]{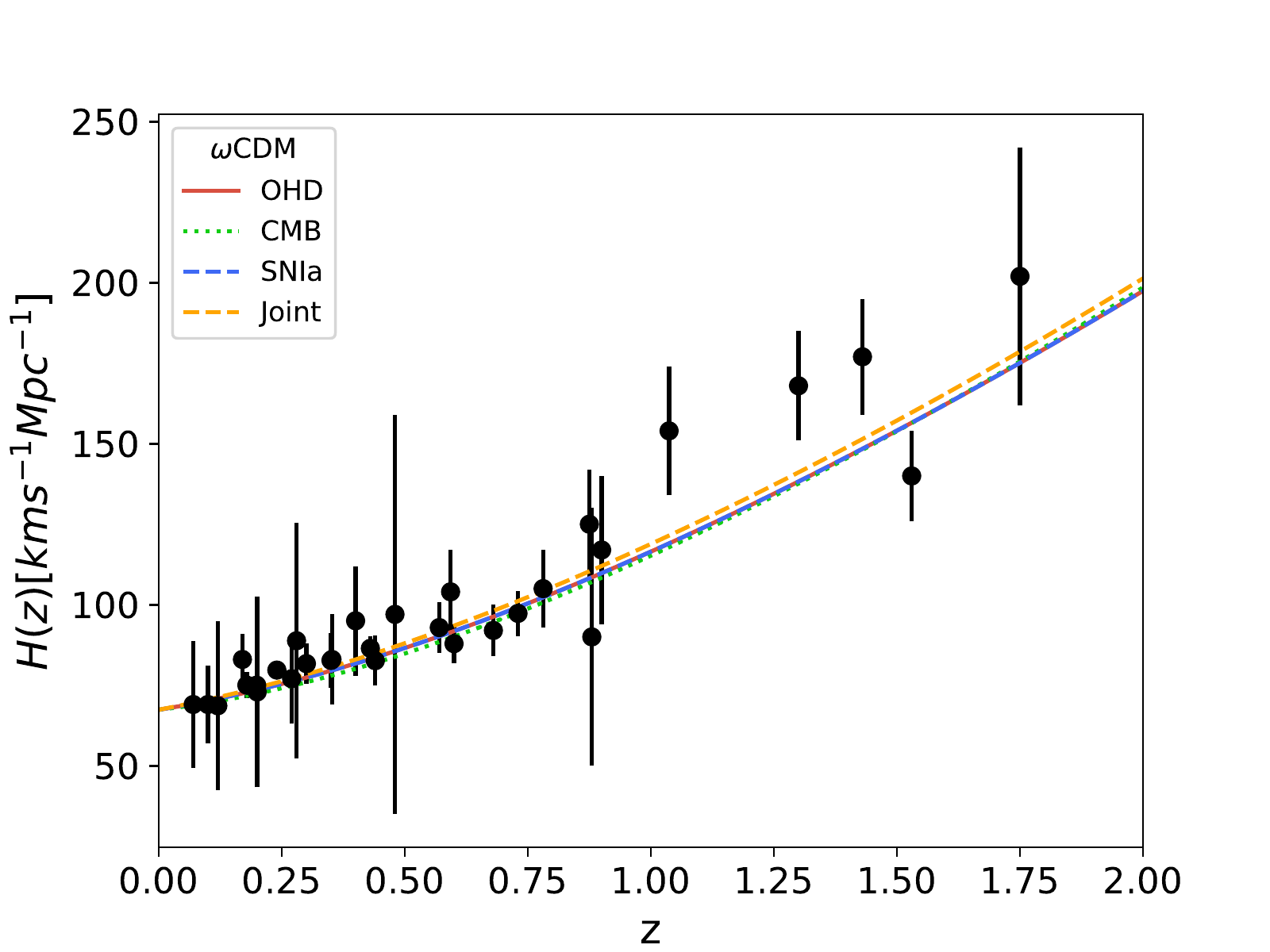}\\
\includegraphics[width=0.45\textwidth]{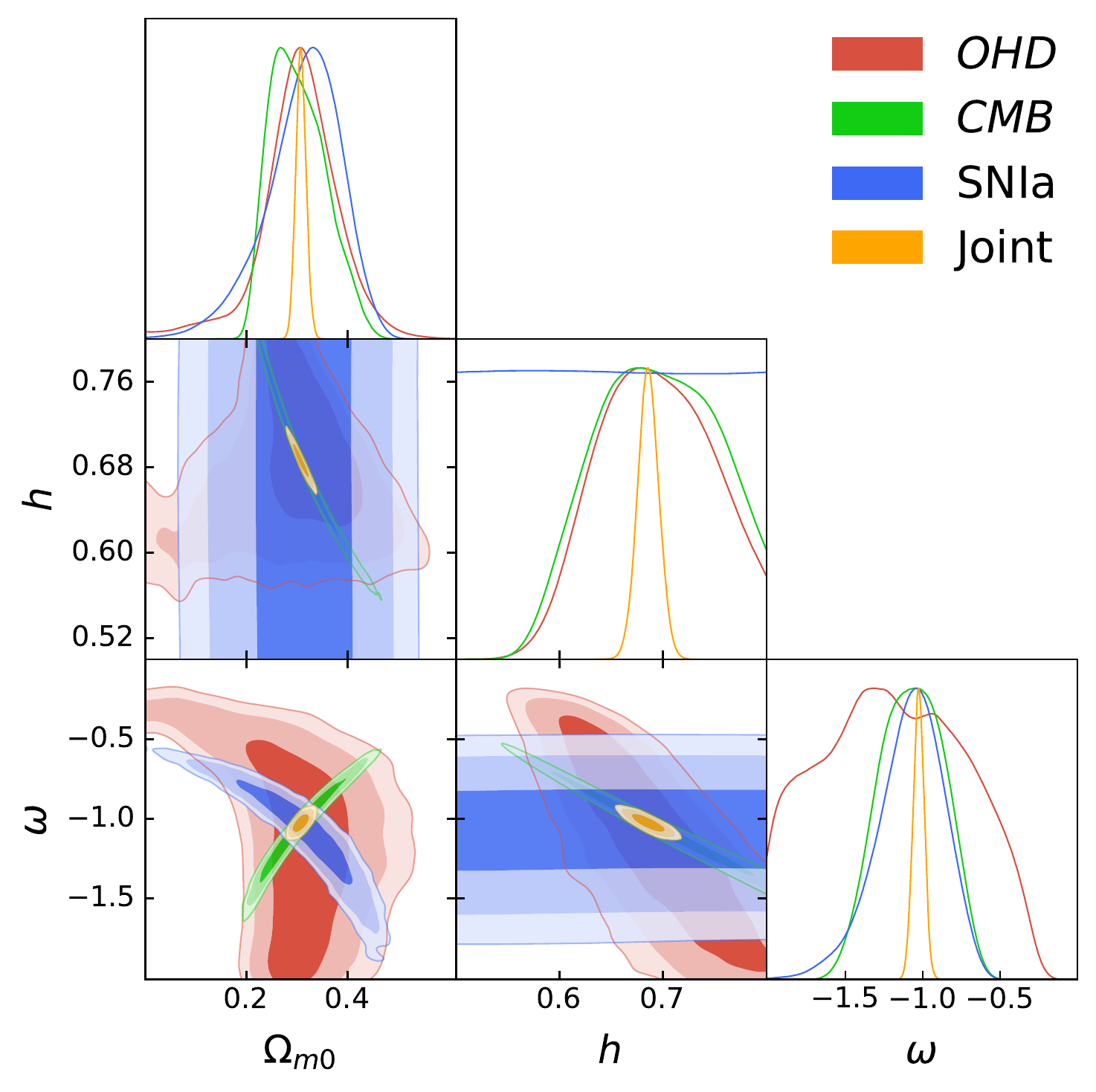}
\caption{Top panel: Best fit curve of the $H(z)$ function for the $w$CDM model using OHD (cosmic chronometers), CMB (Planck) and SNIa (Pantheon) data and the joint analysis of them. Bottom panel: 2D contours of the free model parameters at $1\sigma, 2\sigma$, and $3\sigma$ (from darker to lighter color bands) confidence levels, respectively.}
\label{fig:wcdm}
\end{figure}

The $\omega$CDM constrains are obtained assuming flat priors on the parameters. Table \ref{tab:wcdm} presents the mean values for the $\omega$CDM parameters using using independently OHD (31 data from cosmic chronometers), CMB (Planck) and SNIa (Pantheon) and the joint of them.
Fig. \ref{fig:wcdm} shows the mean value curve of the $H(z)$ function (top panel) for the $w$CDM model using these data. The bottom panel shows the constraint contours at $1\sigma, 2\sigma$, and $3\sigma$ confidence levels. Notice that, although SNIa data is not able to constrain the $h$ parameter, the three different samples provide consistent constraints on the $\Omega_{m0}-\omega$ space. Indeed, the joint analysis provides stringent constraints which are consistent with those of the $\Lambda$CDM model.

\begin{table}[htb]
\centering
\begin{tabular}{|lcccc|}
%\hline
\multicolumn{5}{c}{$\omega CDM$}\\
\hline
Data & $\chi^{2}_{min}$ & $\Omega_{m}$ & $w_{0}$&$h$\\
\hline
\multicolumn{5}{|c|}{}\\
OHD & $15.24$ & $0.30^{+0.06}_{-0.06}$ & $-1.19^{+0.51}_{-0.49}$ &$0.69^{+0.06}_{-0.05}$ \\
SNIa & $1035.92$ & $0.32^{+0.06}_{-0.08}$ & $-1.06^{+0.20}_{-0.23}$ &$0.60^{+0.26}_{-0.27}$\\
CMB & $0.05$ & $0.29^{+0.06}_{-0.05}$ & $-1.06^{+0.21}_{-0.22}$ &$0.69^{+0.06}_{-0.06}$\\
Joint & $1050.59$ & $0.30^{+0.01}_{-0.01}$ & $-1.02^{+0.03}_{-0.03}$ &$0.68^{+0.01}_{-0.01}$\\
\hline
\end{tabular}
\caption{Mean values for the $\omega$CDM parameters using the samples OHD, SNIa, CMB (Planck) and the joint of them.}
\label{tab:wcdm}
\end{table}

%%%%%%%%%%%%%%%%%%%%%%%%%%%%%%%%%%%%%%%%%%%
\subsubsection{Dark Energy Parameterizations}
%%%%%%%%%%%%%%%%%%%%%%%%%%%%%%%%%%%%%%%%%%

The natural alternatives to the $\omega$CDM is to consider DE varies with redshift through a parameterization $w(z)$. These functions are  proposed  phenomenologically to mimic the behaviour of the CC at late times. In the following we present some of these models for a Universe containing dark and baryonic matter, radiation, and dark energy. 
\begin{itemize}
\item The Chevallier-Polarski-Linder parametrization \citep[CPL,][]{Chevallier:2000qy, Linder:2003}.- An approach to study dynamical DE models is through a parametrization of its EoS.  The dimensionless Hubble parameter $E(z)$ for this Universe is given by
\begin{equation}
 E^{2}(z)=\Omega_{m0}(1+z)^{3} + \Omega_{r0}(1+z)^{4} +\Omega_{de}f_{de}(z),\quad
 \label{eq:Ez}
\end{equation}
We compute $\Omega_{r0}$ in the same form as is given by Eq. \eqref{Rad0}.

The density parameter for DE is written as $\Omega_{de}=1-\Omega_{m0}-\Omega_{r0}$, and the function $f_{de}(z)$ depends on $w(z)$ as
\begin{equation}
f_{de}(z)\equiv \frac{\rho_{de}(z)}{\rho_{de}(0)}=
\mathrm{exp}\left(3\int^{z}_{0}\frac{1+w(z)}{1+z}\mathrm{dz}\right),
\label{eq:fdez}
\end{equation}
where $\rho_{de}(z)$ is the energy density of DE at redshift $z$, and $\rho_{de}(0)$ is its present value. One of the most popular parameterization is proposed by \cite{Chevallier:2000qy,Linder:2002dt}, and reads as
\begin{equation}
\omega(z)=\omega_0+\omega_1\frac{z}{(1+z)},
\label{eq:eosCPL}
\end{equation}
where $\omega_{0}$ is the EoS at redshift $z=0$ and $\omega_{1}=\mathrm{d}w/\mathrm{d}z|_{z=0}$. Although this function is widely used it has a divergence problem when $z=-1$. The function $f_{de}(z)$ for the CPL parametrization is 
\begin{equation}
f_{de}(z)_{CPL}^2=(1+z)^{3(1+\omega_0+\omega_{1})}\exp\left(\frac{-3\omega_1z}{1+z}\right), \label{CPLE}
\end{equation}

The $h$, $\Omega_{m0}$, $\omega_{0}$, $\omega_{1}$ parameters are constrained using the OHD from cosmic chronometers \cite{Magana:2017}. Figure \ref{fig:cpl} shows the reconstruction of $H(z)$ using the best fit of the MCMC analysis: $h=0.73^{+0.10}_{-0.08}$, $\Omega_{m0}=0.29^{+0.09}_{-0.08}$, $\omega_{0}=-1.51^{+0.80}_{-0.91}$, and $\omega_{1}=-0.20^{+2.38}_{-2.53}$. 
The confidence contours of the parameters  at $1\sigma, 2\sigma$, and $3\sigma$ are also shown.

\begin{figure}[htb]
\centering
\includegraphics[width=0.45\textwidth]{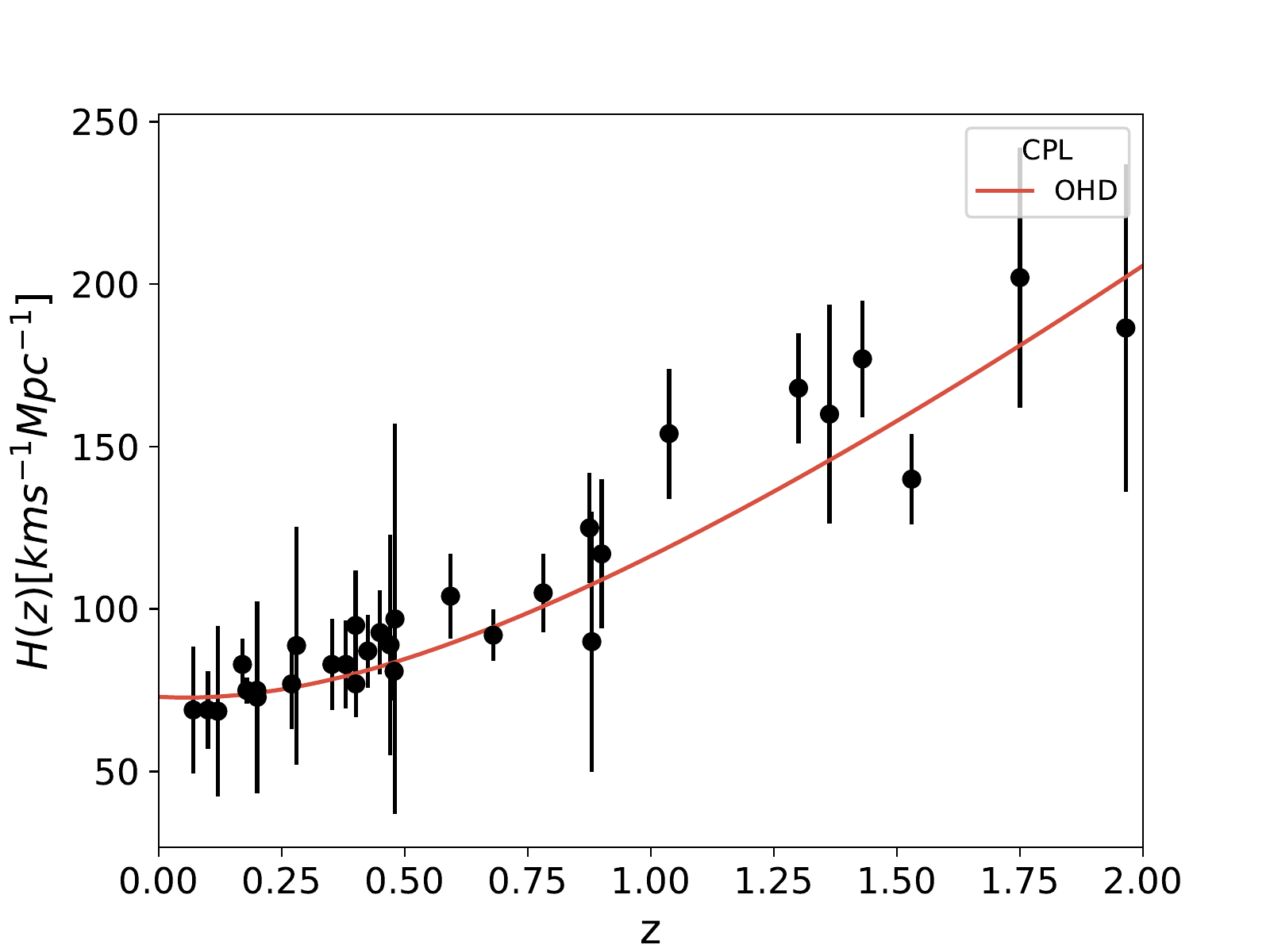}\\
\includegraphics[width=0.45\textwidth]{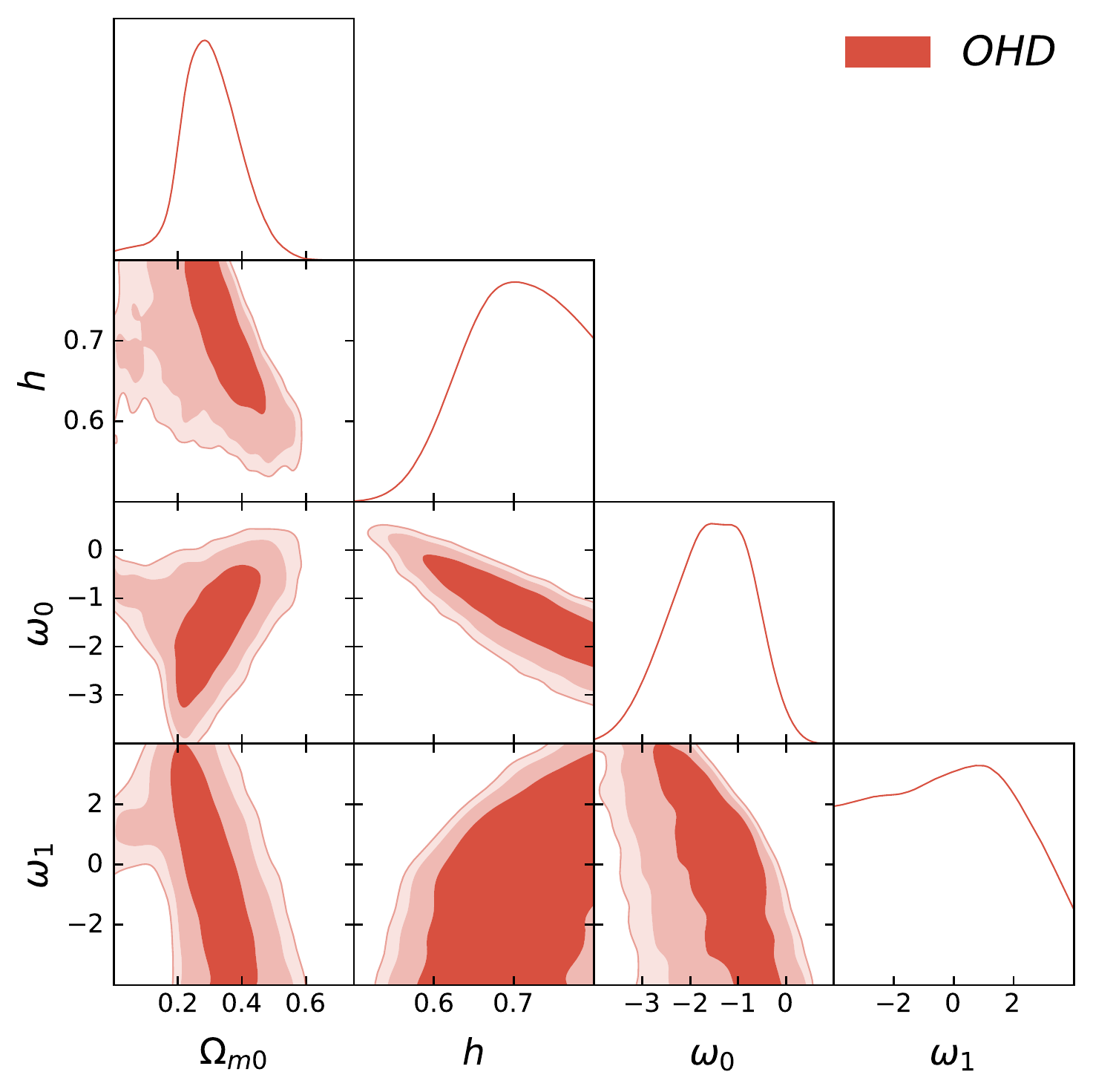}
\caption{Top panel: Best fit curve of the $H(z)$ function for the CPL parameterization using OHD (cosmic chronometers). Bottom panel: 2D contours of the free model parameters at $1\sigma, 2\sigma$, and $3\sigma$ (from darker to lighter color bands) confidence levels for OHD data.}
\label{fig:cpl}
\end{figure}

\item The Jassal-Bagla-Padmanabhan (JBP) parametrization.- \citet{Jassal} proposed the following \textit{ansatz} to parametrize the dark energy EoS
\begin{equation}
\omega(z)=\omega_0+\omega_1\frac{z}{(1+z)^2},
\label{eq:eosJBP}
\end{equation}
where $\omega_{0}$ is the EoS at redshift $z=0$ and $\omega_{1}=(\mathrm{d}w/\mathrm{d}z)|_{z=0}$. The function $f_{de}(z)$ is 
\begin{equation}
f_{de}(z)_{JBP}=(1+z)^{3(1+\omega_0)}\exp\left(\frac{3\omega_1z^2}{2(1+z)^2}\right), \label{JBPE}
\end{equation}

\item The Barbosa-Alcaniz (BA) parametrization.- \citet{Barboza:2008rh} considered a EoS given by:
\begin{equation}
w(z)=w_0 + w_1 \frac{z(1+z)}{1+z^2}.
\label{eq:wBA}
\end{equation}
This \textit{ansatz} behaves linearly at low redshifts as $w_{0}+w_{1}$, and $w\rightarrow w_{0}+w_{1}z$ when $z\rightarrow\infty$. In addition, $w(z)$ is well-behaved for  all epochs of the Universe.  For instance, the DE dynamics in the future, at $z=-1$, can be investigated without dealing with a divergence. Solving the integral in Eq. (\ref{eq:fdez}) and using Eq. (\ref{eq:wBA}) results in:
\begin{equation}
f_{de}(z)_{BA}=(1+z)^{3(1+w_0)}(1+z^2)^{\frac{3}{2}w_{1}}. 
\label{eq:fzBA}
\end{equation}

\item Feng-Shen-Li-Li \citep[FSLL,][]{Feng:2012gf} parametrizations.- The authors suggested two dark energy EoS given by:
\begin{equation}
w(z)=w_{0} + w_{1}\frac{z}{1+z^{2}}, \qquad \textrm{FSLLI}
\label{eq:wFSLLI}
\end{equation}
\begin{equation}
w(z)=w_{0} + w_{1}\frac{z^{2}}{1+z^{2}} \qquad \textrm{FSLLII}.
\label{eq:wFSLLII}
\end{equation}
\noindent
Both functions have the advantage of being divergence-free throughout the entire cosmic evolution, even at $z=-1$. At low redshifts,  $w(z)$ behaves as $w_{0}+w_{1}z$ and $w_{0}+w_{1}z^{2}$ for FSLLI and FSLLII, respectively. In addition, when $z\rightarrow \infty$, the EoS has the same value ($w_{0}$) as the present epoch for FSLLI and $w_{0}+w_{1}$ for FSLLII. Using Eqs. (\ref{eq:wFSLLI})-(\ref{eq:wFSLLII}) to solve Eq. (\ref{eq:fdez}) leads to:
\begin{equation}
f_{de\pm}(z)=(1+z)^{3(1+w_{0})}\mathrm{exp}\left[\pm\frac{3w_{1}}{2}\mathrm{arctan(z)}\right]
\left(1+z^{2}\right)^{\frac{3}{4}w_{1}}\left(1+z\right)^{\mp \frac{3}{2}w_{1}},
\label{eq:fzFSLL}
\end{equation}
\noindent
where $f_{+}$ and $f_{-}$ correspond to FSLLI and FSLLII, respectively.

\item Sendra-Lazkoz \citep[SL,][]{Sendra:2012} introduced new polynomial parameterizations to reduce the parameter correlation, so they can be better constrained by the observations at low redshifts. One of these parameterizations is given by:
\begin{equation}
w(z)=-1 + c_{1}\left(\frac{1+2z}{1+z}\right) + c_{2}\left(\frac{1+2z}{1+z}\right)^{2},\\ 
\label{eq:wSeLa}
\end{equation}
\noindent
where the constants are defined as $c_{1}=(16w_{0}-9w_{0.5}+7)/4$, and $c_{2}=-3w_{0}+ (9w_{0.5}-3)/4$, and $w_{0.5}$ is the value of the EoS at $z=0.5$. This $w(z)$ function is well-behaved at higher redshifts as $(-1-8w_{0}+9w_{0.5})/2$.  The substitution of Eq. (\ref{eq:wSeLa}) into Eq. (\ref{eq:fdez}) results:
\begin{equation}
f_{de}(z)_{SL}=(1+z)^{\frac{3}{2}(1-8w_{0}+9w_{0.5})}
\mathrm{exp}\left[\frac{3z\left\{w_{0}(52z+40)-9w_{0.5}(5z+4)+7z+4\right\}}{8(1+z)^2}\right]. 
\label{eq:fzSeLa}
\end{equation}
Notice that, although DE parameterizations are common and they could solve the coincidence problem, there is not a unique way to choose the form of the function. Furthermore, in many cases there are not strong arguments to justify the functional form by an association with a first-principles theory of quantum fields or gravity. A different approach, which is model-independent, consist of, for example investigating the cosmographic parameters that characterize the kinematics of the cosmic expansion  
\cite[e.g.,][]{Luongo:2011,Aviles:2012,padeaprox,Demianski:2012,Zhang:2016}. Some authors have used the Hubble parameter, the deceleration parameter 
($q(a)=-\ddot{a}a/\dot{a}^2$), or even higher order derivatives of the scale factor $a$, such as Jerk and Snap  \cite[e.g.,][]{2016IJMPD..2550032A,Lizardo:2020wxw}. By estimating these cosmographic parameters using cosmological data, it is possible to associate its features to a given DE model 
\cite[see][]{Santos:2011,delCampo:2012,Nair:2012,2016IJMPD..2550032A,Mamon:2017,RomanGarza:2019}.

The cosmological constrains for the aforementioned models are obtained assuming flat priors on the DE parameters and a Gaussian prior on $h$. 
Table \ref{tab:parameterizations} provides the mean values for the $\Omega_{m}$, $w_{0}$, and $w_{1}$ ($w_{0.5}$) parameters of the JBP, BA, FSLLI, FSLLII, and SL DE parameterizations using the joint of the OHD sample (34 data points from DA and BAO measurements) in the redshift range $0.07 < z < 2.3$ \cite{Sharov:2014voa}, distance posteriors from Planck \cite{Planck:2015XIV}, and different BAO measurements \cite[see details in][]{Magana:2017usz}. Fig. \ref{fig:Eparams} shows the reconstruction of $H(z)$ for these parametrizations using the parameter mean values (top panel) and the $1 \sigma$ and $3 \sigma$ confidence contour of the cosmological constrains (bottom panel). Notice that the DE parameterizations are consistent for $\Omega_{m0}$ and $\omega_{0}$.

\begin{table}[htb]
\centering
\begin{tabular}{|lccccc|}
%\hline
\multicolumn{6}{c}{DE parametrizations}\\
\hline
Model & $\chi^{2}_{min}$&$\Omega_{m}$& $w_{0}$&$w_{1} (w_{0.5})$&$h$\\
\hline
\multicolumn{6}{|c|}{}\\
JBP & $67.22$ & $0.29^{+0.01}_{-0.01}$ & $-1.22^{+0.21}_{-0.16}$ & $0.55^{+0.91}_{-1.18}$&$0.71^{+0.014}_{-0.014}$ \\
BA & $67.46$ & $0.29^{+0.01}_{-0.01}$ & $-1.12^{+0.13}_{-0.13}$ & $0.007^{+0.22}_{-0.24}$&$0.71^{+0.015}_{-0.015}$\\
FSLLI & $67.01$ & $0.29^{+0.01}_{-0.01}$ & $-1.22^{+0.18}_{-0.17}$ & $0.32^{+0.54}_{-0.57}$&$0.71^{+0.015}_{-0.015}$\\
FSLLII & $67.61$ & $0.29^{+0.01}_{-0.01}$ & $-1.09^{+0.10}_{-0.10}$ & $-0.13^{+0.39}_{-0.44}$&$0.70^{+0.014}_{-0.014}$\\
SL & $68.52$ & $0.29^{+0.01}_{-0.01}$ & $-1.10^{+0.13}_{-0.13}$ & $-1.13^{+0.05}_{-0.05}$&$0.70^{+0.015}_{-0.015}$ \\
\hline
\end{tabular}
\caption{Mean values for the $\Omega_{m}$, $w_{0}$, and $w_{1} (w_{0.5})$ parameters using 
the joint analysis of OHD, CMB, and BAO data for the JBP, BA, FSLLI, FSLLII and SL DE parameterizations \cite[see][]{Magana:2017usz}.}
\label{tab:parameterizations}
\end{table}

\begin{figure}[thb]
\centering
\includegraphics[width=0.45\textwidth]{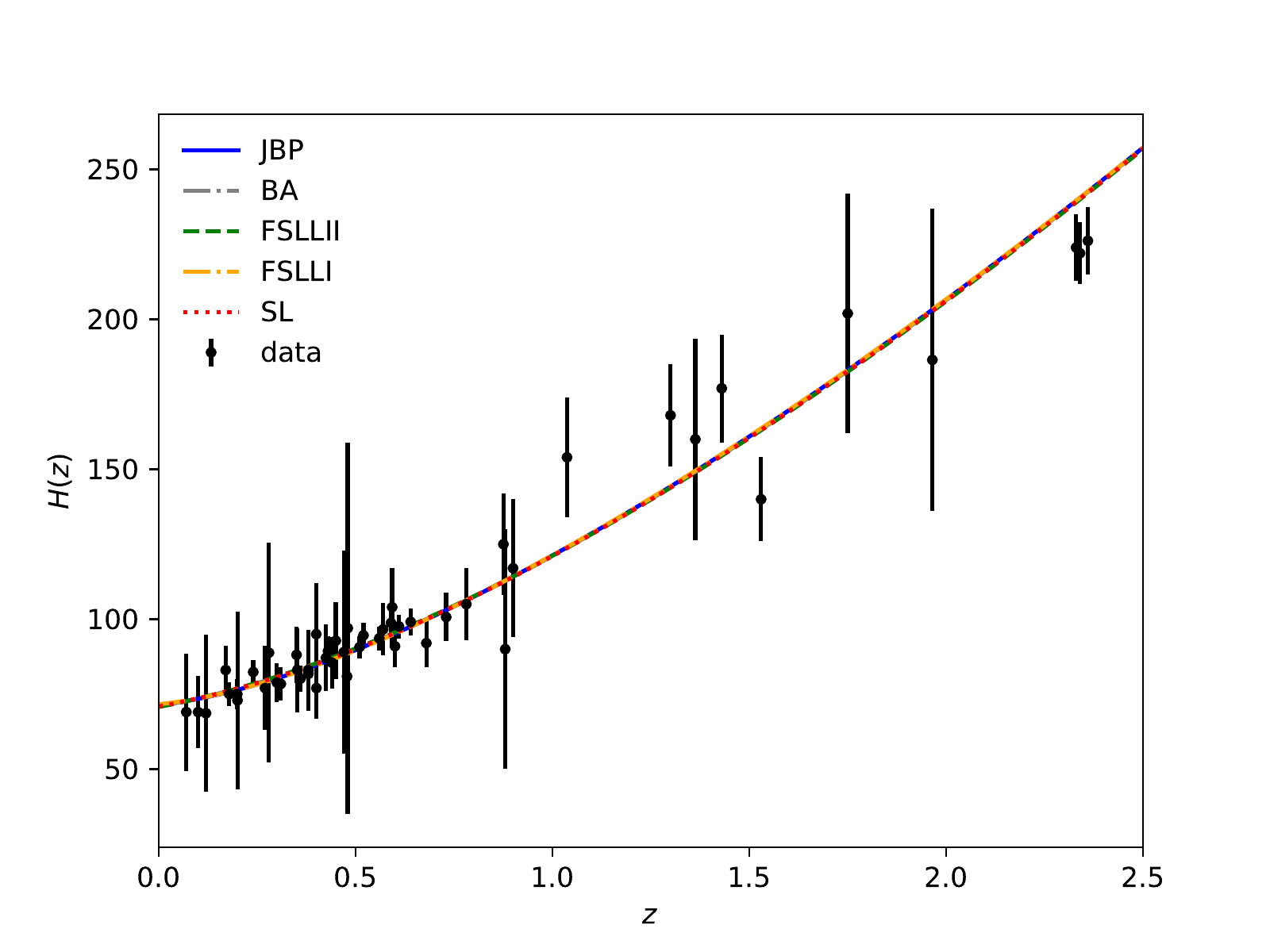}\\
\includegraphics[width=0.45\textwidth]{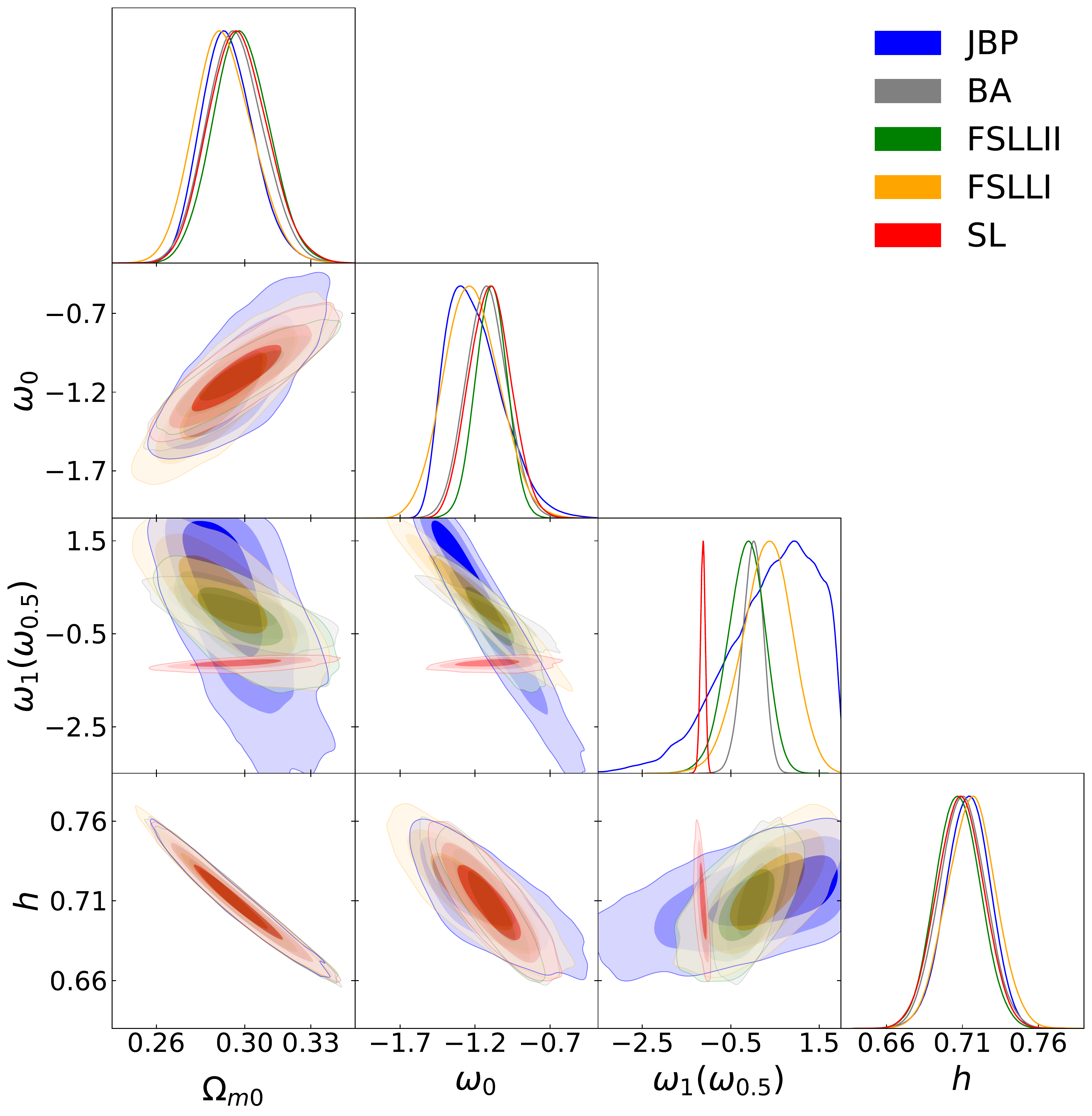}
\caption{Top panel: Best fit curve of $H(z)$ for the DE parameterization functions from the joint analysis of OHD, CMB, and BAO measurements. Bottom panel: 2D contours of the free model parameters at $1\sigma$, and $3\sigma$ (from darker to lighter color bands) confidence levels using this joint analysis.}
\label{fig:Eparams}
\end{figure}

\end{itemize}

%%%%%%%%%%%%%%%%%%%%%%%%%%%%%%%%%%%%%%%%%%%%%
\subsubsection{Chaplygin-Like Fluid}

One point of view for studying the DE and DM problems is through the unified dark fluids approach, which is known as Chaplygin gas  \cite[see, for instance,][]{Shenavar:2020,Dymnikova:2017,Deng:2011,Bhadra:2012,Pourhassan:2013}. An example of this is the well-known generalized Chaplygin gas \cite{Bento2003, Bento:2004} described by the EoS $p=-A\rho^{-\alpha}$ where $A$ and $\alpha$ are constants (the case $\alpha=1$ is the original model proposed by S. Chaplygin \cite{Chaplygin}). This fluid behaves as DM at early epoch and DE at late times and may have its origen from the Nambu-Goto $d$-brane action. Although this interesting formulation reproduces the accelerated expansion of the Universe, it presents flaws to describe the CMB anisotropies \cite{Amendola_2003}. In this context, an alternative to Chaplygin gas was proposed by \citet{Hova2017}, dubbed generalized Chaplygin gas-like, with EoS 
\begin{equation}
p_{df}=-\rho_{df}+ \rho_{df} \, \text{sinc}(\mu\pi\rho_{df0}/\rho_{df})\,,
\end{equation}
being $\text{sinc}(x)\equiv\sin(x)/x$ and $\rho_{df}$ the dark fluid density, which plays the role of the mixture of DE and DM densities. In this case, $\mu$ is a dimensionless parameter constrained as $\mu\gtrsim0.688$  to be consistent with the stellar age bound\footnote{ Hova and Yang \citet{Hova2017} adopt $\mu\approx0.876$ to obtain an Universe age of $t\approx13.7$Gyrs.} and $\rho_{df0}$ is the present energy density of this fluid, constrained in terms of the density parameter as $\Omega_{df0}\sim0.96$ in \citet{Hova2017}. It behaves as a CC in the late stage of the universe and as DM at the matter domination epoch. The evolution of the EoS of the dark fluid is given by
\begin{equation}
\omega_{df}(z)\equiv-1+\frac{(z+1)^3\tan(\lambda)}{[(z+1)^6+\tan^2\lambda]\xi(z)},
\label{eq:eosdf}
\end{equation}
where $\xi(z)\equiv\text{arctan}[(z+1)^{-3}\tan\lambda]$ and $\lambda\equiv\mu\pi/2$. To explore the universe dynamics in this context, we consider a general FLRW metric including baryonic and radiation components, hence we write the Friedmann and acceleration equations as
\begin{eqnarray}
H^2&=&\frac{8\pi G}{3}\left(\rho_{df}+\sum_i\rho_i\right)-\frac{k}{a^2}, \label{1}\\
\frac{\ddot{a}}{a}&=&-\frac{4\pi G}{3}\left\lbrace\left[3\,\text{sinc}\left(\frac{2\lambda\rho_{df0}}{\rho_{df}}\right)-2\right]\rho_{df}+\sum_i(1+3\omega_i)\rho_i\right\rbrace,
\end{eqnarray}
From Eq. \eqref{eq:Ez_gen} we have \cite{Hova2017, Hernandez-Almada:2018osh}
\begin{equation}
E(z)^2=\frac{\lambda\Omega_{df0}}{\xi(z)}+\sum_i\Omega_{i0}(z+1)^{3(1+\omega_i)}+\Omega_{k}(z+1)^2, \label{Ez}
\end{equation}
where $\Omega_{df0}\equiv 8\pi G\rho_{df0}/3H_0^2$ is the density parameter associated with the Chaplygin gas-like fluid, $\Omega_{i0}$ and $\omega_{i}$ are the density parameters and the EoS for baryonic matter and radiation (according to Eq. \ref{Rad0}), $\Omega_{k}\equiv-k/H_0^2$ is the curvature density parameter and $H_{0}=h\times 100\, \mathrm{km\,s^{-1}Mpc^{-1}}$. In addition, from \eqref{FlatCond} we have the constraint $\Omega_{df0}+\Omega_{b0}+\Omega_{r0}=1-\Omega_{k}$. 

\begin{figure}[thb]
\centering
\includegraphics[width=0.45\textwidth]{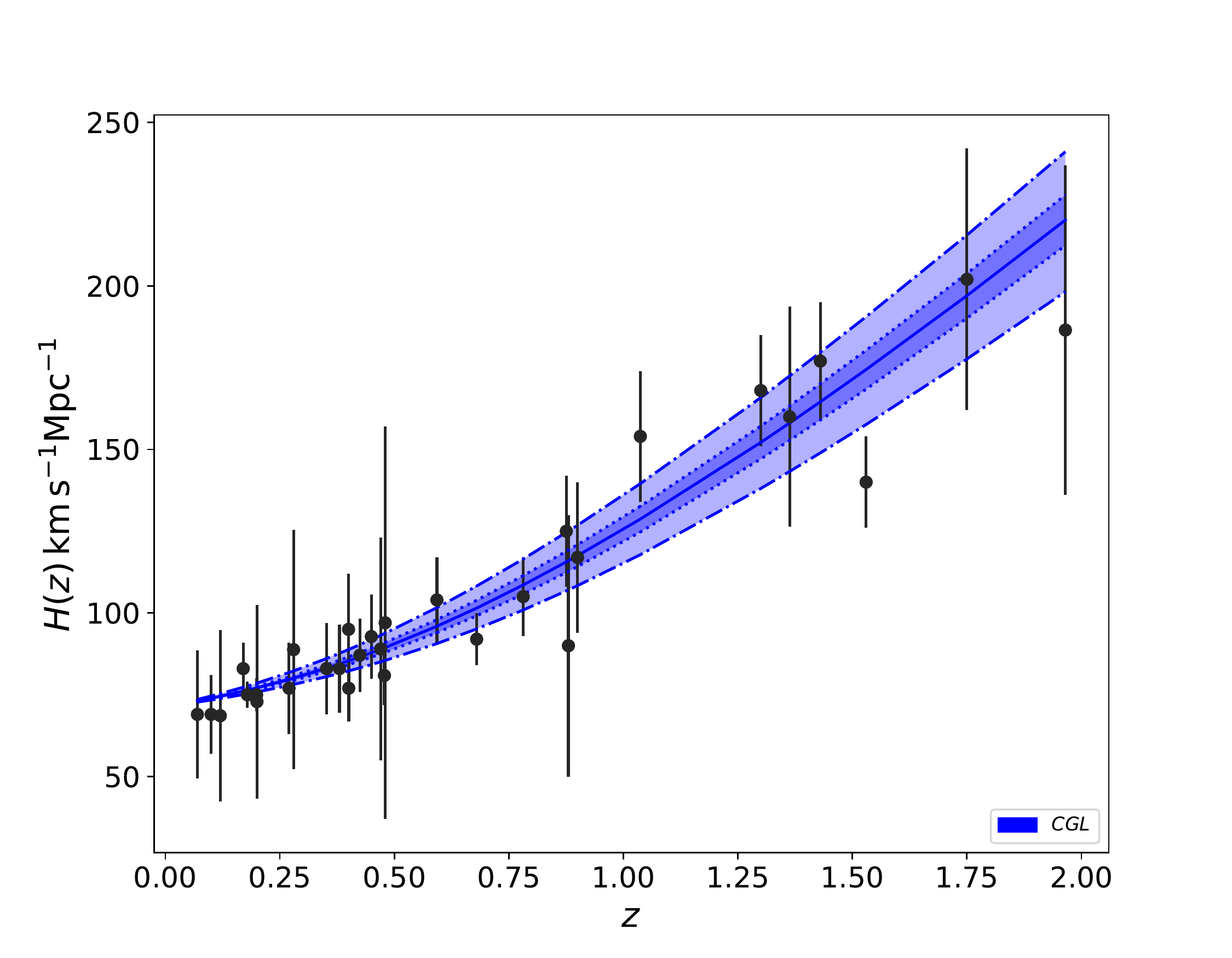}\vspace{0.2ex}\\
\includegraphics[width=0.45\textwidth]{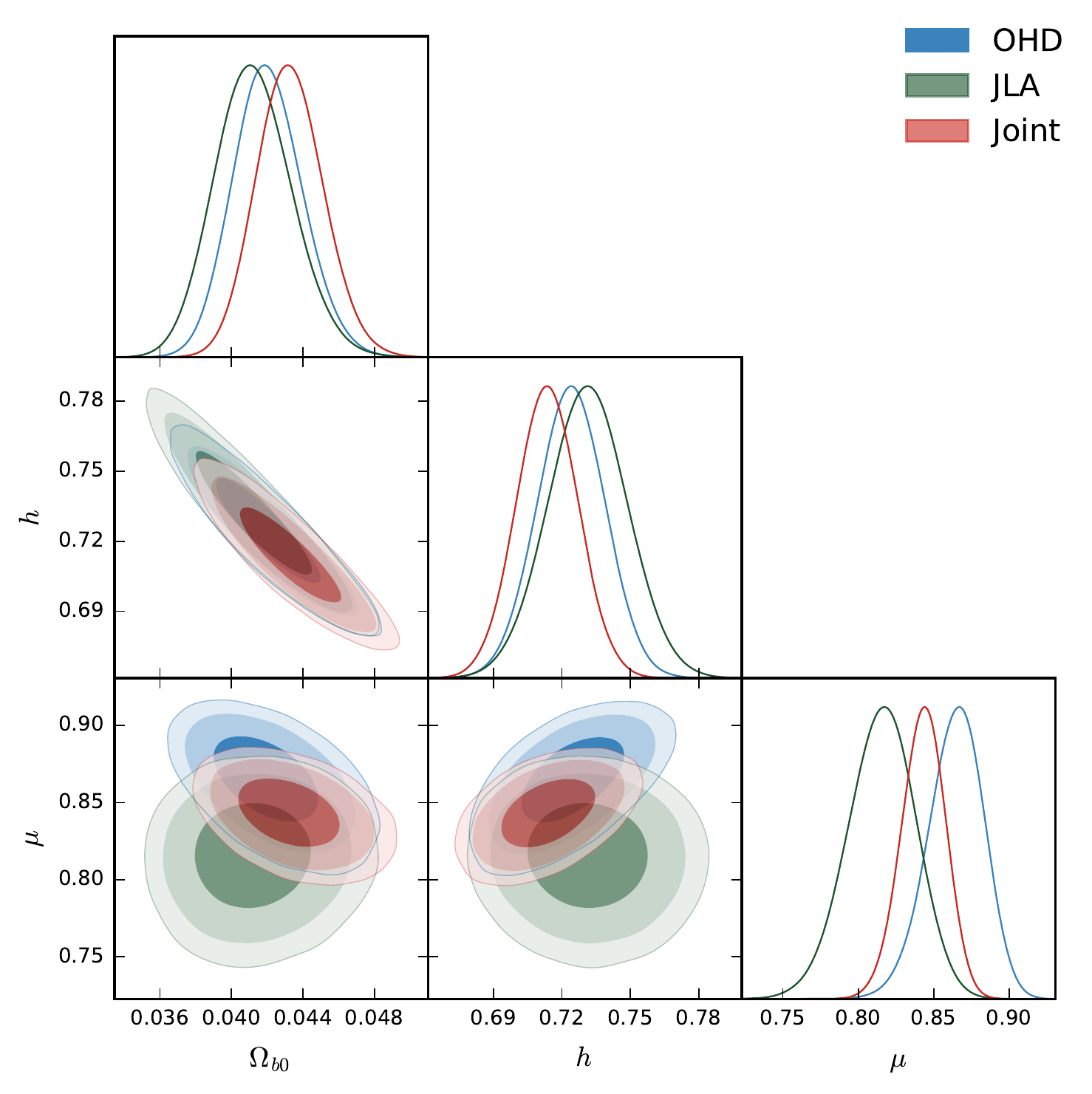}
\caption{Top panel: Best fit curve of Chaplygin-like gas and its uncertainty at $1\sigma$. Bottom panel: 2D contours of the free model parameters at $1\sigma, 2\sigma$, and $3\sigma$ (from darker to lighter color bands) CL using DA OHD, SNIa (JLA), and OHD+SNIa (Joint) data. Adapted from \cite{Hernandez-Almada:2018osh}.}
\label{fig:chaplyginGas}
\end{figure}

Figure \ref{fig:chaplyginGas} shows the best fit curve (top panel) for the Chaplygin-like gas when the curvature term is neglected using the OHD, SNIa and OHD+SNIa (Joint) samples. Additionally, 2D contours at $1\sigma$, $2\sigma$, and $3\sigma$ confidence level (CL) and 1D posterior distribution of the free parameters are displayed for each sample. Table \ref{tab:bf_Chaplygin} presents the best fit values and their uncertainties at $1\sigma$ for the model free parameters. 

\begin{table}[htb]
\caption{Mean values for the model parameters ($\Omega_{b0}$, $h$, $\mu$) derived from OHD and SNIa measurements for a flat universe \cite[see][]{Hernandez-Almada:2018osh}.}
\centering
\begin{tabular}{|l ccc |}
\hline
Data set &        OHD    &    JLA &     Joint   \\
\hline
$\chi^2_{min}$ & $14.9$ & $690.8$ & $706.7$  \\ [0.5ex]
$\Omega_{b0}$  & $0.042^{+0.002}_{-0.002}$ & $0.041^{+0.002}_{-0.002}$ & $0.043^{+0.002}_{-0.002}$ \\ [0.7ex]
%$\Omega_{k}$   & - & - & - & $0.128^{+0.086}_{-0.090}$ & $0.392^{+0.187}_{-0.369}$ & $0.183^{+0.073}_{-0.079}$ \\ [0.7ex]
$h$            & $0.724^{+0.015}_{-0.015}$ & $0.724^{+0.018}_{-0.017}$ & $0.714^{+0.014}_{-0.014}$  \\ [0.7ex]
$\mu$          & $0.865^{+0.018}_{-0.019}$ & $0.816^{+0.021}_{-0.023}$ & $0.843^{+0.014}_{-0.015}$ \\ [0.7ex]
$a$            & -& $0.141^{+0.007}_{-0.007}$& $0.142^{+0.007}_{-0.007}$  \\ [0.7ex]
$b$            & -& $3.11^{+0.08}_{-0.08}$ & $3.12^{+0.08}_{-0.08}$  \\ [0.7ex]
$M_b^1$        & -& $-19.00^{+0.06}_{-0.06}$ & $-19.01^{+0.04}_{-0.04}$  \\ [0.7ex]
$\delta_M$     & -& $0.07^{+0.02}_{-0.02}$ & $0.07^{+0.02}_{-0.02}$  \\ [0.7ex]
\hline
\end{tabular}
\label{tab:bf_Chaplygin}
\end{table}

%%%%%%%%%%%%%%%%%%%%%%%%%%%%%%%%%%%%%%%%%%%%%
\subsubsection{Viscous Model}

\begin{figure}[H]
\centering
\includegraphics[width=0.4\textwidth]{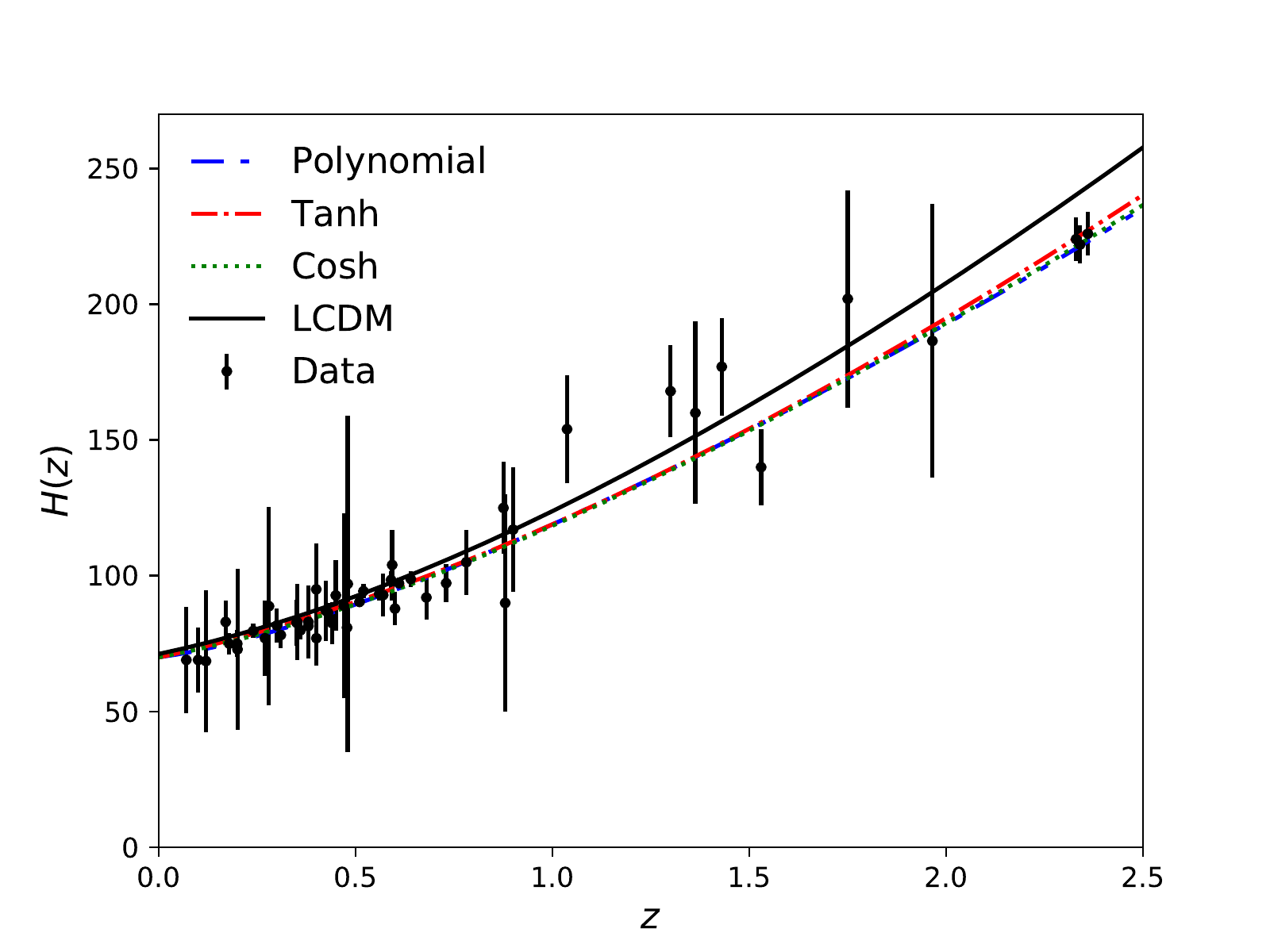}\\
\includegraphics[width=0.35\textwidth]{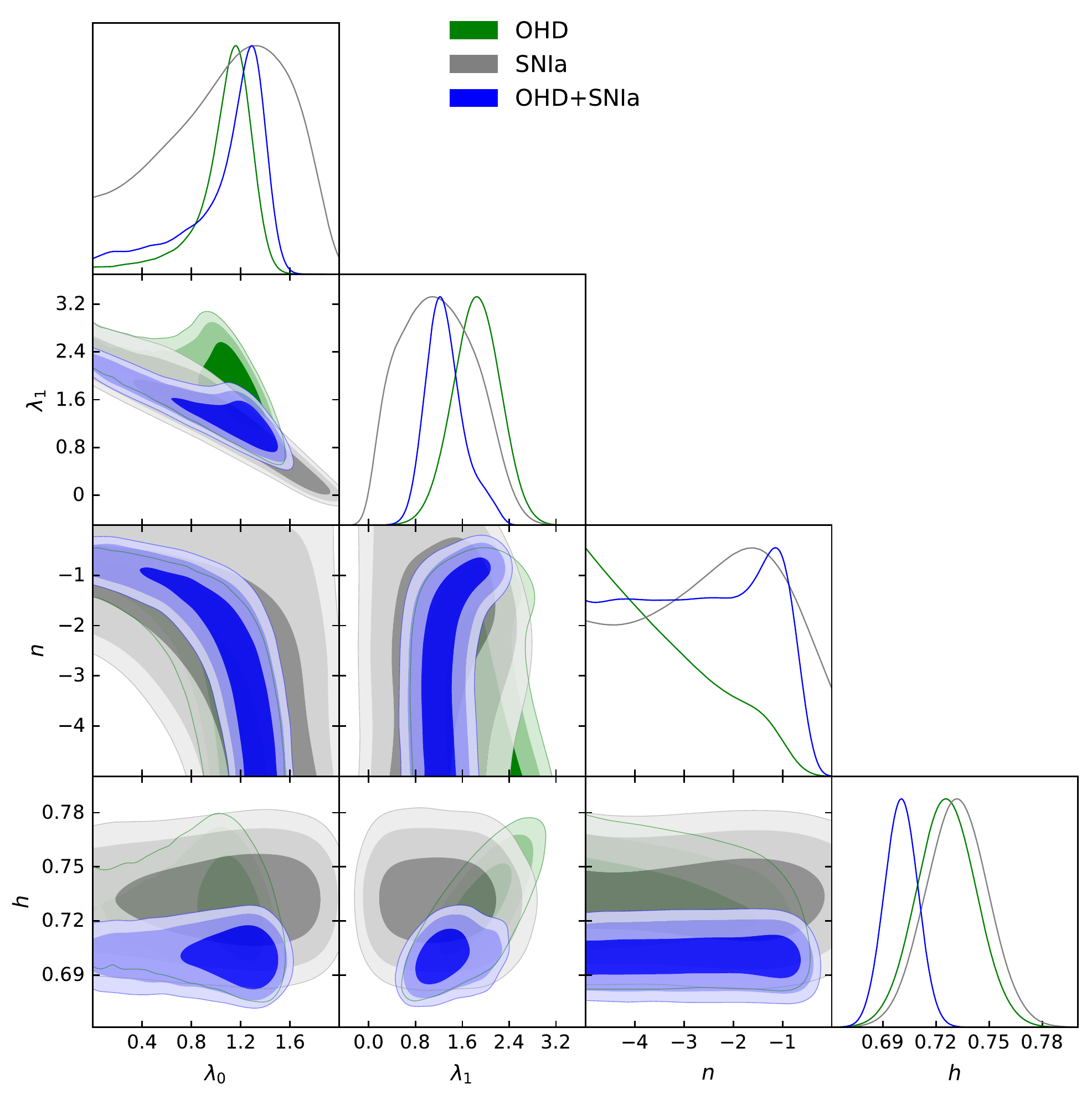}
\includegraphics[width=0.35\textwidth]{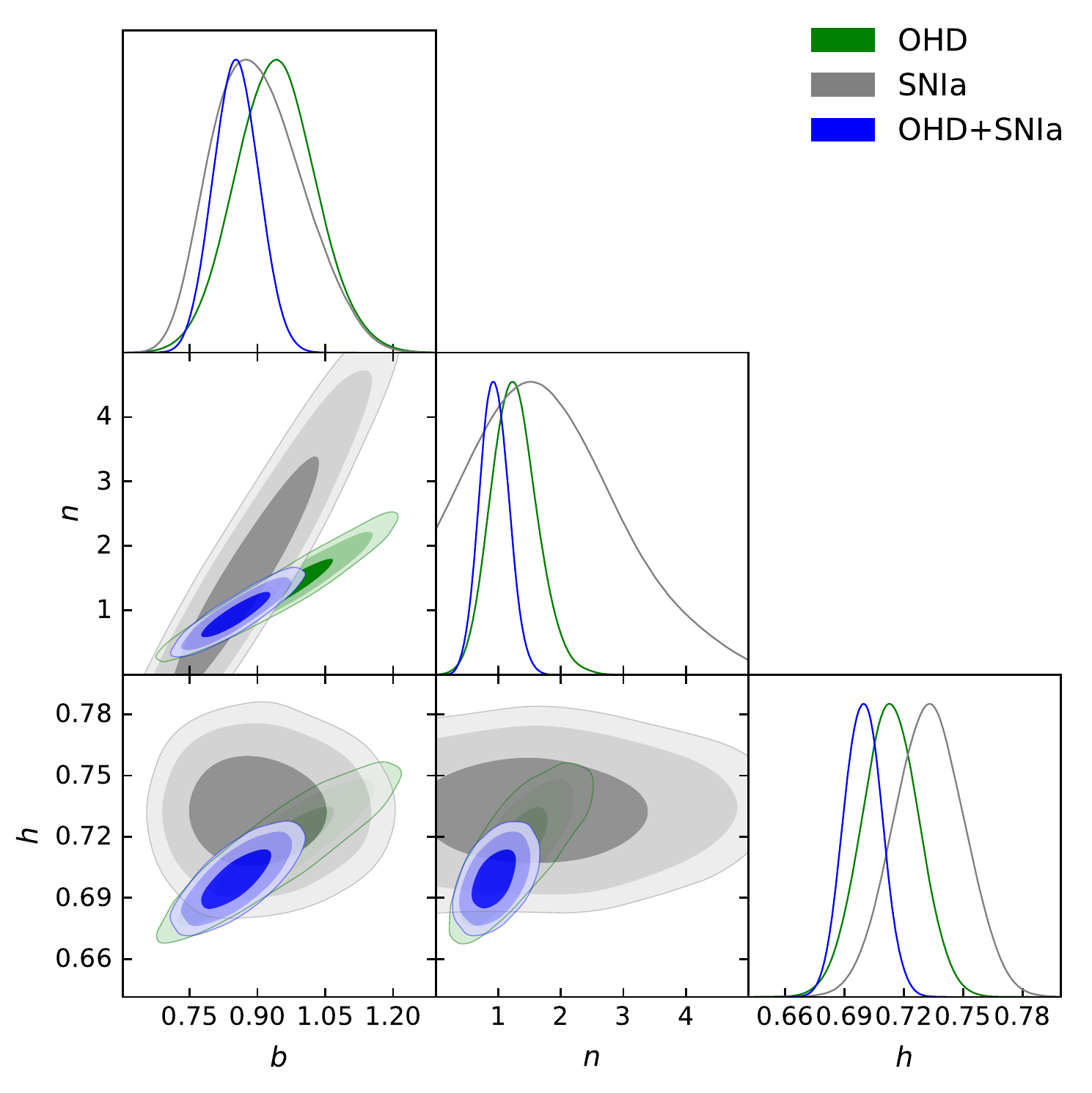}
\includegraphics[width=0.35\textwidth]{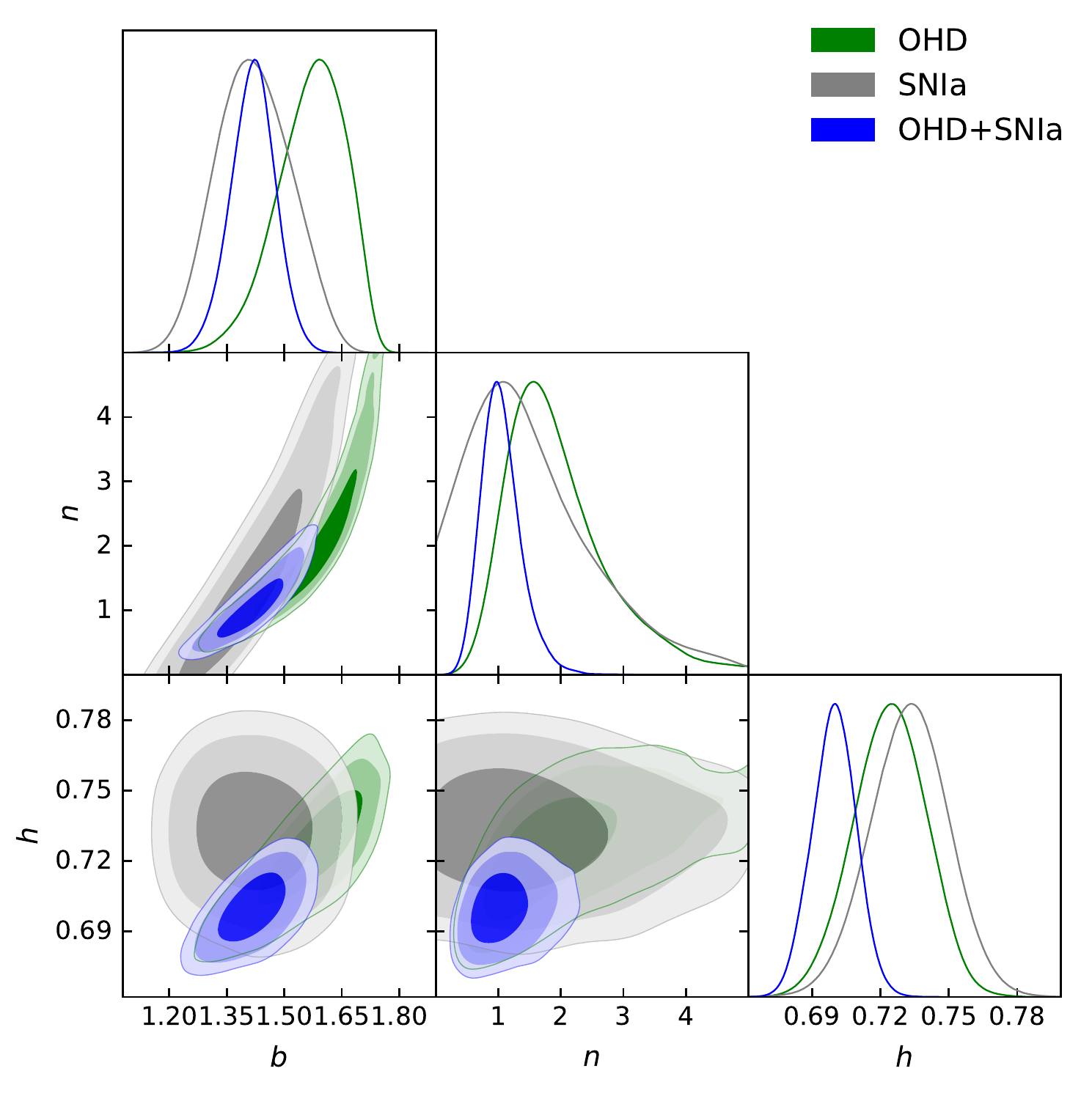}
\caption{Top panel: Best fit curves of $H(z)$ in the single viscous fluid model for different bulk viscosity coefficient  and its uncertainty at $1\sigma$ and $3\sigma$. Bottom panel: 2D contours of the free model parameters for the polynomial (left), tanh (middle), cosh (right) models respectively at $1\sigma, 2\sigma$, and $3\sigma$ (from darker to lighter color bands) CL  and 1D posterior distributions of the model parameters using nonhomogeneous OHD, SNIa, OHD+SNIa data. Figure adapted from \cite{Almada:2019}.}
\label{fig:disipative1}
\end{figure}

The accelerated expansion of the Universe may be also described by considering dissipative effects in the Universe components, mainly in the matter component. The bulk viscosity coefficient, which satisfies the cosmological principle, is introduced in the energy-momentum tensor, Eq. \ref{emt}, as an effective pressure as $p\to \tilde{p} = p + \Pi$ where $\Pi = -3\xi H$ based on the Eckart formalism. Under this argument, several models for $\xi$ have been addressed such as:
\begin{itemize}
    \item $\xi = \xi_0 \rho^s_m$. Probably this model, where $\rho_m$ is the energy density of dust matter and $\xi_0,s$ are constants, is the simplest one that successfully reproduce the late accelerated stage of the Universe. Some studies that consider a single fluid in the Universe are presented in \cite{Brevik:2005bj} (see for example \cite{Cruz:2017bcv} for a case in a causal theory). Additionally, there are other works that include several components such as radiation and DE \cite{Hernandez-Almada:2020ulm}.
    \item $\xi=\xi(z)$. In spite of the success of the previous model at late epochs of the Universe, it has problems in early epochs because $\xi$ diverges. This motivates the use of alternative viscosity models such as those proposed by \cite{Almada:2019}, in particular polynomial forms of the redshift.
    \item $\xi = A \cosh (bE^{-n})$ and $\xi = A \tanh (bE^{-n})$. Alternatively, more complex models are investigated in \cite{Almada:2019} by proposing the viscosity as \textcolor{v}{a} hyperbolic function of the dimensionless Hubble parameter $E$.
\end{itemize}
When a single dust matter fluid is contained in the Universe, the dimensionless Hubble parameter is obtained using Eq. \eqref{eq:Ez_gen}. Then, we obtain the following system to be solved
\begin{equation}\label{eq:diffE}
    -2(1+z) \frac{dE(z)}{dz} + 3E(z) = 9\lambda(z)\,,
\end{equation}
where $\lambda(z) = \xi(z) \kappa^2 / 3H_0^2$. For $\lambda(z)=\lambda_0 + \lambda_1(1+z)^n$, we obtain,
\begin{equation} \label{eq:Epol}
    E(z) = \lambda_2 (1+z)^{3/2} - \frac{\lambda_1}{2n-3}(1+z)^n + \frac{\lambda_0}{3}\,,
\end{equation}
where
\begin{equation}
   \lambda_2 =  1  + \frac{\lambda_1}{2n-3} - \frac{\lambda_0}{3} \,.
\end{equation}
Fig. \ref{fig:disipative1} shows the best fit curve of $H(z)$ (top panel) using non-homogeneous OHD+SNIa data. Two-dimensional contours at at $1\sigma$, $2\sigma$, $3\sigma$ CL of the free model parameters are also displayed at the bottom panel for OHD, SNIa, OHD+SNIa data. Additionally, we include the best fit curves and 2D contours when $\lambda(z)=1/3 \tanh(bE^{-n})$ and $\lambda(z)=1/3 \cosh(bE^{-n})$. Tables \ref{tab:bf_model1} and \ref{tab:bf_model23} show the best fit values and their uncertainties at $1\sigma$ of the single fluid models.

\begin{table}[htb]
\caption{Best fitting parameters of the polynomial model \cite[see][for details]{Almada:2019}.}
\centering
\begin{tabular}{| ccccccc |}
\hline
%%%\multicolumn{7}{|c|}{Polynomial model} \\
Data     & $\chi^2$ &    $\lambda_0$            & $\lambda_1$               & $n$                        & $h$ & $\mathcal{M}$       \\
\hline
OHD      & $15.1$ & $1.112^{+0.154}_{-0.256}$ & $1.844^{+0.399}_{-0.408}$ & $-3.628^{+1.534}_{-0.990}$ & $0.726^{+0.017}_{-0.017}$ &  -  \\ [0.7ex]
SNIa     & $1027.9$ & $1.129^{+0.459}_{-0.618}$ & $1.159^{+0.702}_{-0.668}$ & $-2.351^{+1.374}_{-1.712}$ & $0.732^{+0.017}_{-0.017}$  & $5.741^{+0.053}_{-0.055}$ \\ [0.7ex]
OHD+SNIa & $1053.2$ & $1.183^{+0.177}_{-0.487}$ & $1.273^{+0.363}_{-0.281}$ & $-2.656^{+1.494}_{-1.596}$ & $0.700^{+0.009}_{-0.009}$  &  $5.634^{+0.023}_{-0.023}$ \\ [0.7ex]
\hline
\end{tabular}
\label{tab:bf_model1}
\end{table}

\begin{table}[htb]
\caption{Best fitting parameters of the hyperbolic models \cite[see][for details]{Almada:2019}.}
\centering
\begin{tabular}{| cccccc |}
\hline
 \multicolumn{6}{|c|}{tanh model} \\
Data     &$\chi^2$  &    $b$                      & $n$                         & $h$                       & $\mathcal{M}$   \\
\hline
 OHD     & $28.8$    & $0.937^{+0.088}_{-0.087}$   & $1.230^{+0.376}_{-0.350}$   & $0.713^{+0.014}_{-0.015}$  & -   \\ [0.7ex]
SNIa     & $1026.3$    & $0.894^{+0.109}_{-0.094}$   & $1.727^{+1.271}_{-1.005}$   & $0.733^{+0.018}_{-0.018}$  &  $5.747^{+0.054}_{- 0.056}$ \\ [0.7ex]
OHD+SNIa & $1055.7$    & $0.853^{+0.050}_{-0.050}$   & $0.933^{+0.236}_{-0.222}$   & $0.699^{+0.009}_{-0.009}$  & $5.644^{+0.023}_{-0.023}$  \\ [0.7ex]
\hline
 \multicolumn{6}{|c|}{cosh model} \\
Data     &$\chi^2$  &    $b$                      & $n$                         & $h$        & $\mathcal{M}$                   \\
 \hline
 OHD     & $26.7$   & $1.580^{+0.084}_{-0.098}$   & $1.790^{+0.939}_{-0.605}$   & $0.724^{+0.016}_{-0.016}$ &  -   \\ [0.7ex]
SNIa     & $1041.5$ & $1.417^{+0.106}_{-0.096}$   & $1.348^{+1.225}_{-0.782}$   & $0.733^{+0.017}_{-0.017}$ &  $5.747^{+0.051}_{-0.053}$   \\ [0.7ex]
OHD+SNIa & $1054.7$ & $1.420^{+0.056}_{-0.059}$   & $1.014^{+0.339}_{-0.273}$   & $0.700^{+0.009}_{-0.010}$ &  $5.640^{+0.023}_{-0.024}$   \\ [0.7ex]
\hline
\end{tabular}
\label{tab:bf_model23}
\end{table}

\begin{figure}[htb]
\centering
\includegraphics[width=0.3\textwidth]{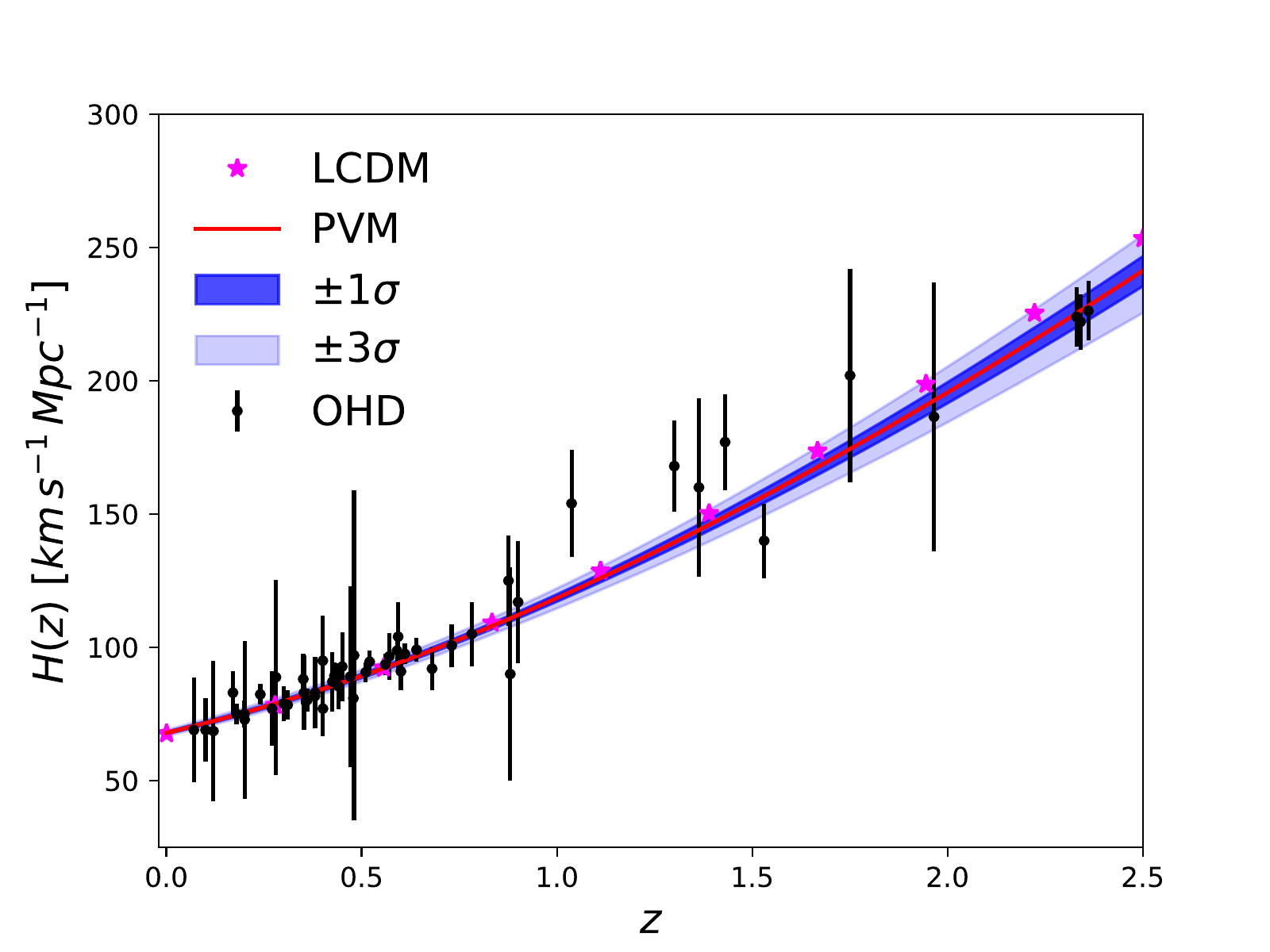}
\includegraphics[width=0.3\textwidth]{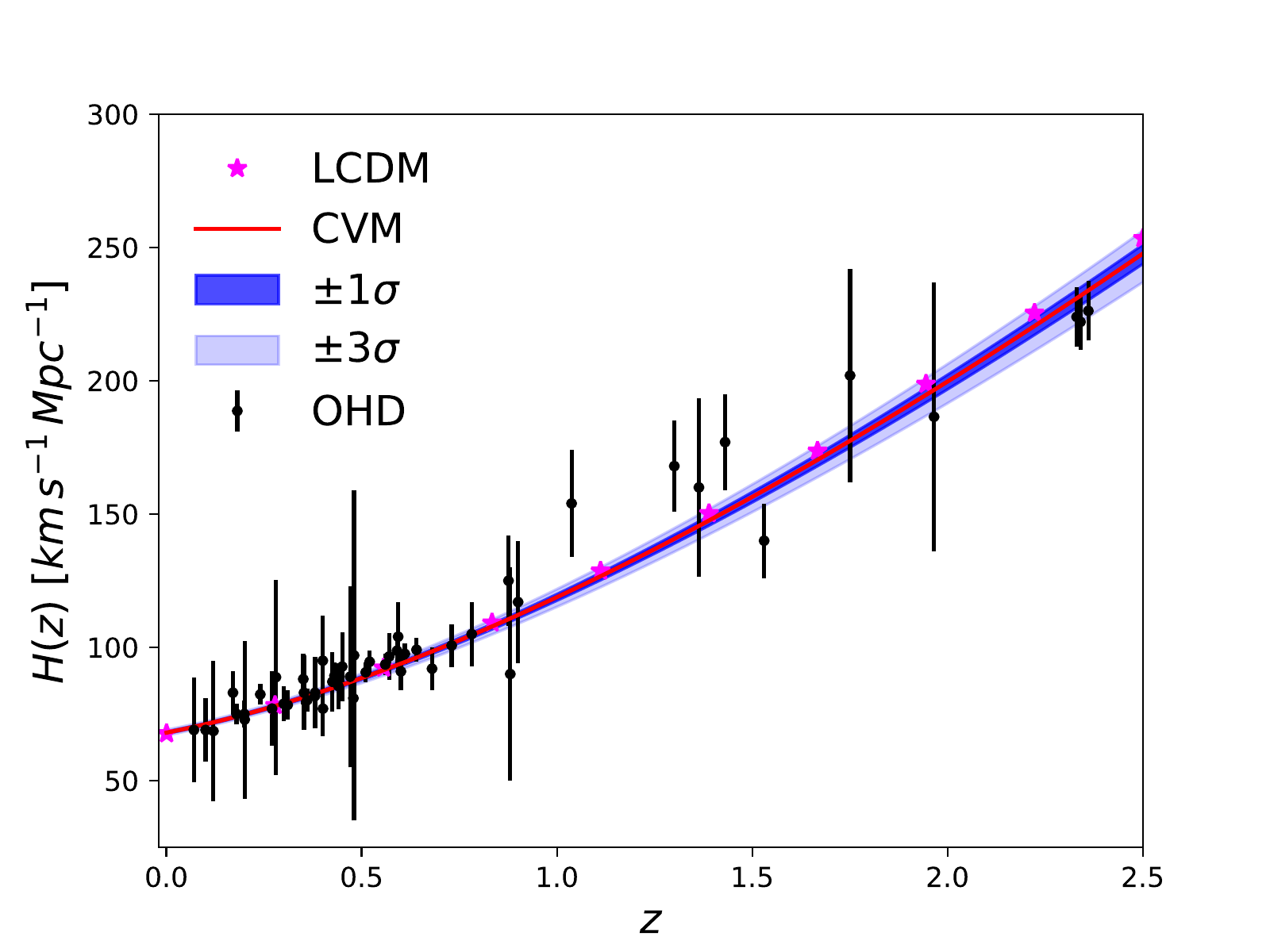}\\
\includegraphics[width=0.3\textwidth]{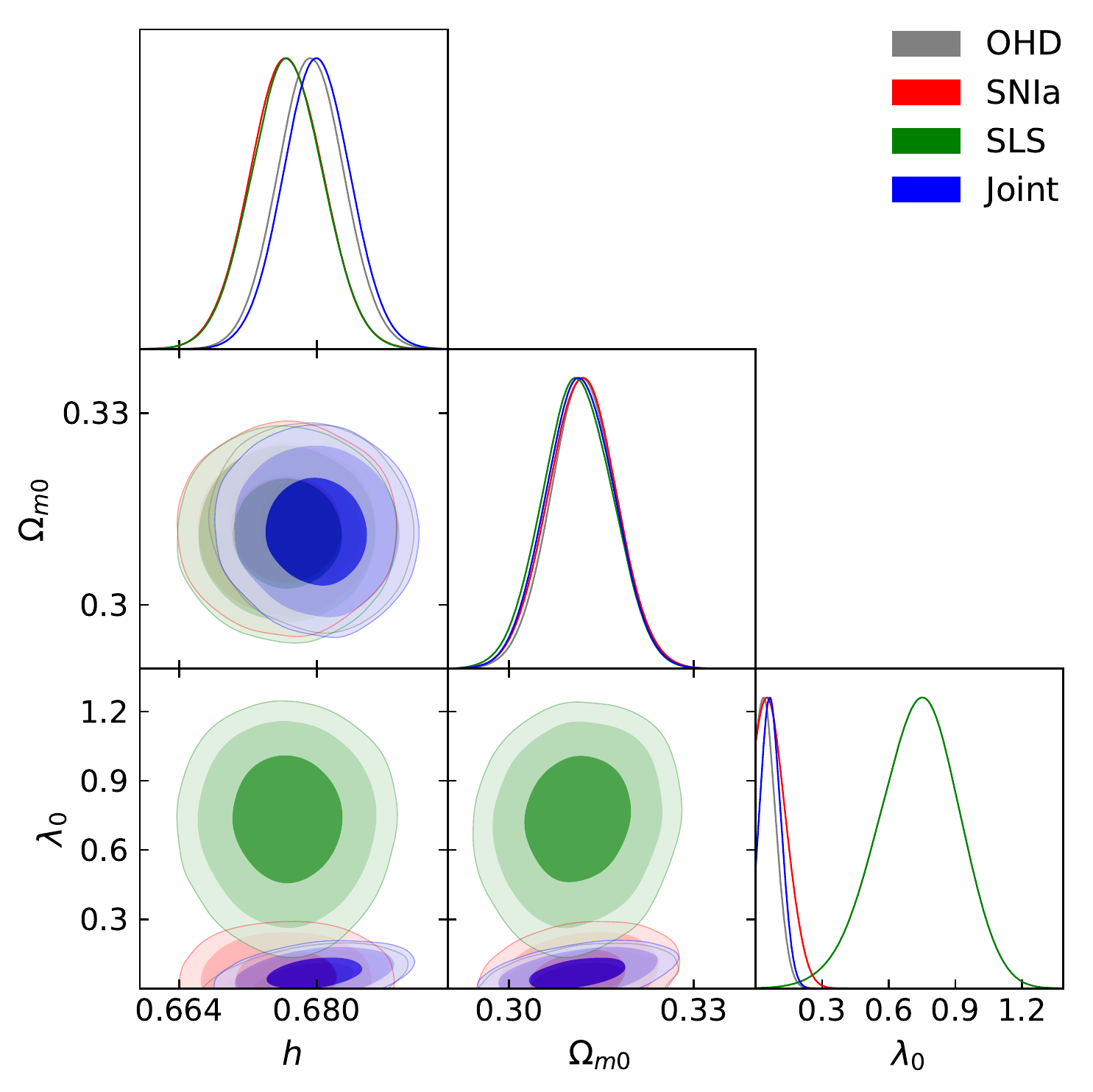}
\includegraphics[width=0.3\textwidth]{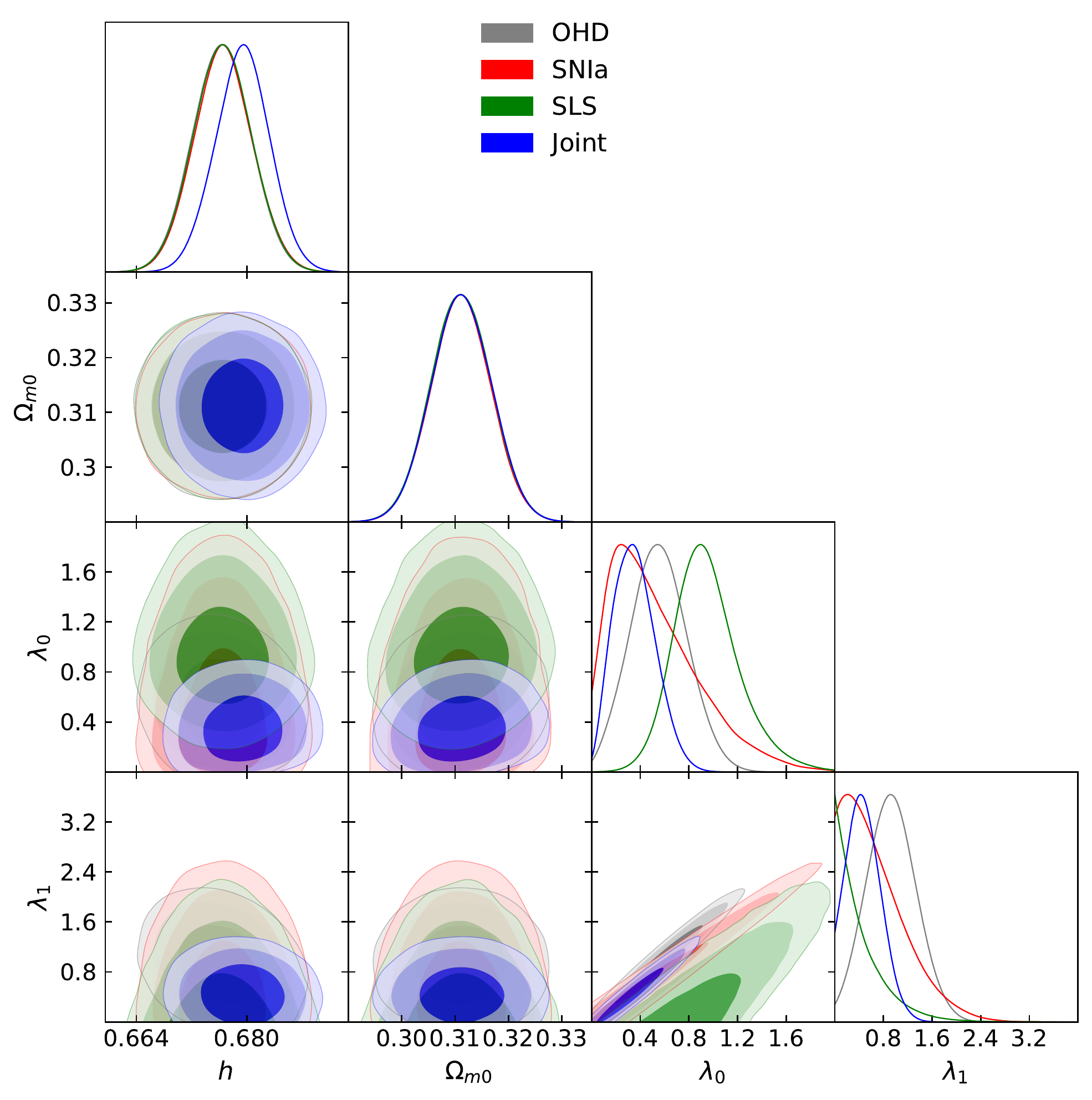}
\caption{Top panel: Best fit curve of viscous fluid model and its uncertainty at $1\sigma$ and $3\sigma$ and $\Lambda$CDM. Bottom panel: 2D contours of the free model parameters at $1\sigma, 2\sigma$, and $3\sigma$ (from darker to lighter color bands) CL and 1D posterior distribution of the model parameters using nonhomogeneous OHD, SNIa, SLS and OHD+SNIa+SLS (Joint) data. Figure adapted from \cite{Herrera:2020}.}
\label{fig:disipative2}
\end{figure}

A generalized form of the previous model (Eq. \eqref{eq:Epol}) considers one additional fluid to the dust matter component. Authors in \cite{Herrera:2020} analyze the case that includes a DE component with EoS $w=-1$ to describe the late time stage of the Universe. Notice that the radiation component at this time can be considered negligible. In this case, the Hubble parameter is given by
\begin{eqnarray} \label{eq:E2pol}
    E(z) &=& \sqrt{\Omega(z)}\left[1 + \frac{\lambda_0}{3\sqrt{\Omega_{de0}}}\operatorname{sinh}^{-1}{\left (\sqrt{\frac{\Omega_{de0}}{ \Omega_{m0}(1+z)^3}} \right )} \right. \nonumber \\
    && \left. - \frac{\lambda_0}{3\sqrt{\Omega_{de0}}} \operatorname{sinh}^{-1}{\left (\sqrt{\frac{\Omega_{de0}}{\Omega_{m0}}} \right )} \right] \nonumber \\
         & & +\sqrt{\Omega(z)} \left[  \frac{\lambda_1}{2n\sqrt{\Omega_{de0}}}(1+z)^n \times \right. \nonumber \\ 
         & & \left. \,{{}_{2}F_{1}}\left( \frac{1}{2}, \frac{n}{3}, 1+\frac{n}{3}, -\frac{\Omega_{m0}(1+z)^3}{\Omega_{de0}} \right. \right) \nonumber \\
         & & \left. - \frac{\lambda_1}{2n\sqrt{\Omega_{de0}}}\,{{}_{2}F_{1}}\left( \frac{1}{2}, \frac{n}{3}, 1+\frac{n}{3}, -\frac{\Omega_{m0}}{\Omega_{de0}}  \right)  \right]\,,
\end{eqnarray}
where $E(0)=\Omega(0)=1$, $\Omega(z) = \Omega_{m0}(1+z)^3+\Omega_{de0}$, and ${}_{2}F_{1}$ is the hypergeometric function. Based on the results obtained in \cite{Almada:2019}, Eq. \eqref{eq:E2pol} is obtained assuming $n=-2$. It is straightforward that the case for a constant viscosity coefficient is obtained when $\lambda_1=0$. Figure \ref{fig:disipative2} displays the best fit curves (top panel) of Eq. \eqref{eq:E2pol} over OHD data, obtained by confronting to OHD+SNIa+SLS data, and 2D contours at $1\sigma$, $2\sigma$, $3\sigma$ CL of the free model parameters are presented at the bottom panel for OHD, SNIa, SLS, and OHD+SNIa+SLS data. Table \ref{tab:bestfits_VLCDM} shows the best fit values and their uncertainties at $1\sigma$ of the free model parameters.

\begin{table}[htb]
\caption{Best fitting values of the free model parameters \cite[see][for details]{Herrera:2020}.}
\centering
\begin{tabular}{|ccccccc|}
\hline
Sample     &    $\chi^2$     &  $h$ & $\Omega_{m0}$ & $\lambda_0$ & $\lambda_1$ &  $\mathcal{M}$             \\
\hline
\multicolumn{7}{|c|}{$\xi_0={\rm Constant}$} \\ 
 OHD   & $25.9$ & $0.679^{+0.004}_{-0.004}$ & $0.312^{+0.005}_{-0.005}$ & $0.053^{+0.047}_{-0.035}$ & -- & -- \\ [0.7ex]
SNIa  & $1027.1$ & $0.676^{+0.004}_{-0.004}$ & $0.312^{+0.005}_{-0.005}$ & $0.080^{+0.071}_{-0.072}$ & -- & $-19.400^{+0.016}_{- 0.016}$  \\ [0.7ex]
SLS  & $602.3$ & $0.677^{+0.004}_{-0.004}$ & $0.311^{+0.006}_{-0.006}$ & $0.737^{+0.175}_{-0.188}$ & -- & --   \\ [0.7ex]
 Joint & $1684.1$ & $0.680^{+0.004}_{-0.004}$ & $0.311^{+0.006}_{-0.005}$ & $0.071^{+0.047}_{-0.040}$ & -- & $-19.400^{+0.012}_{- 0.012}$  \\ [0.7ex]
\hline
\multicolumn{7}{|c|}{ $\xi_0={\rm Polynomial}$ } \\ [0.7ex]
OHD  & $20.9$   & $0.676^{+0.004}_{-0.004}$ & $0.311^{+0.006}_{-0.006}$ & $0.551^{+0.237}_{-0.228}$ & $0.929^{+0.412}_{-0.401}$ & --   \\ [0.7ex]
SNIa & $1044.5$ & $0.676^{+0.004}_{-0.004}$ & $0.311^{+0.006}_{-0.006}$ & $0.461^{+0.441}_{-0.280}$ & $0.580^{+0.620}_{-0.395}$ & $-19.400^{+0.019}_{- 0.019}$  \\ [0.7ex]
SLS  & $603.6$  & $0.676^{+0.004}_{-0.004}$ & $0.311^{+0.006}_{-0.006}$ & $0.927^{+0.284}_{-0.235}$ & $0.312^{+0.484}_{-0.231}$ & --   \\ [0.7ex]
Joint & $1668.8$ & $0.679^{+0.004}_{-0.004}$ & $0.311^{+0.006}_{-0.006}$ & $0.347^{+0.183}_{-0.164}$ & $0.465^{+0.301}_{-0.263}$& $-19.400^{+0.014}_{- 0.014}$   \\ [0.7ex]
\hline
\end{tabular}
\label{tab:bestfits_VLCDM}
\end{table}

%%%%%%%%%%%%%%%%%%%%%%%%%%%%%%%%%%%%%%%%%%%%%
\subsubsection{Interacting Viscous Models}

\begin{figure}[ht]
\centering
\includegraphics[width=0.45\textwidth]{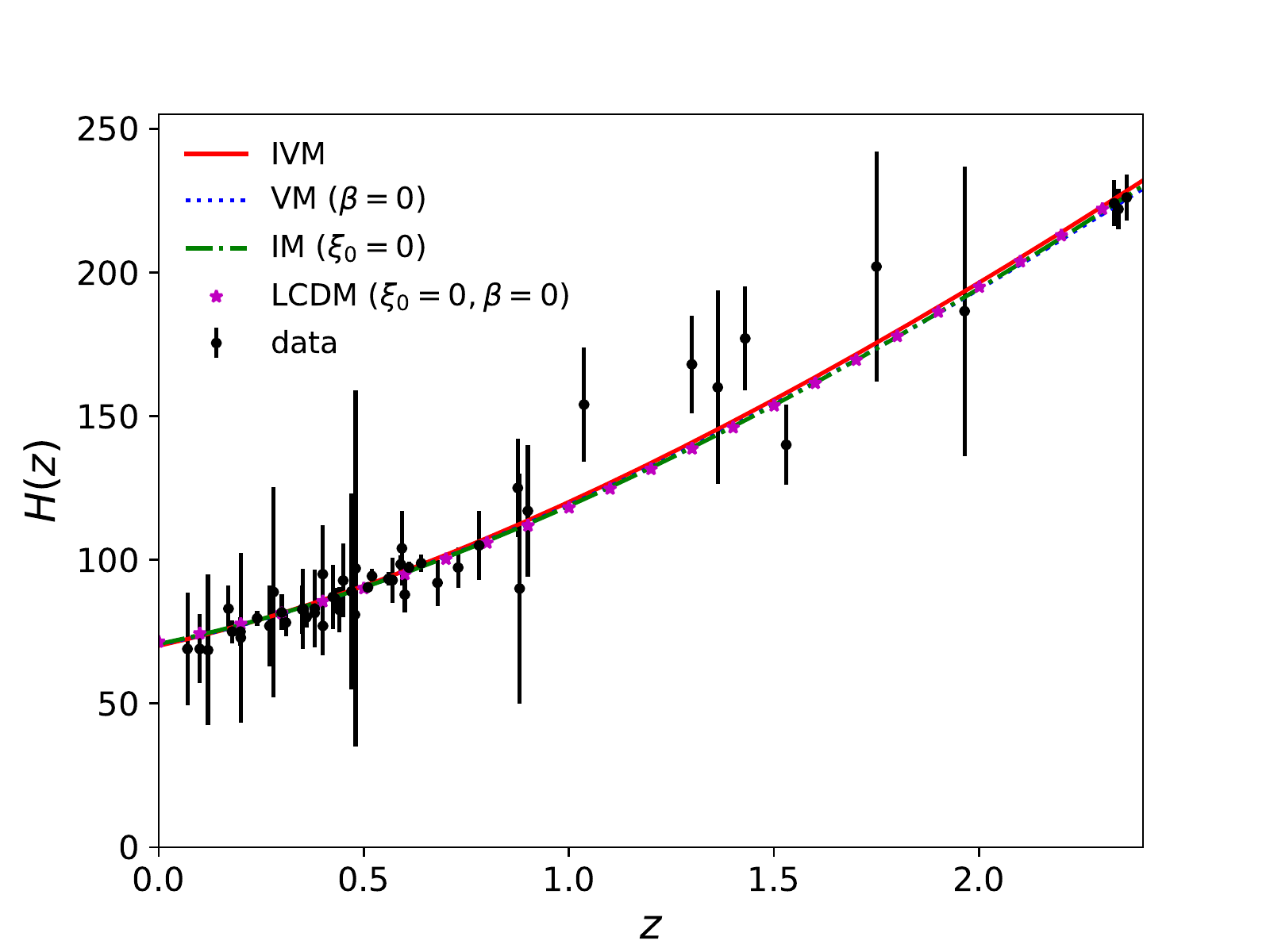}\\
\includegraphics[width=0.45\textwidth]{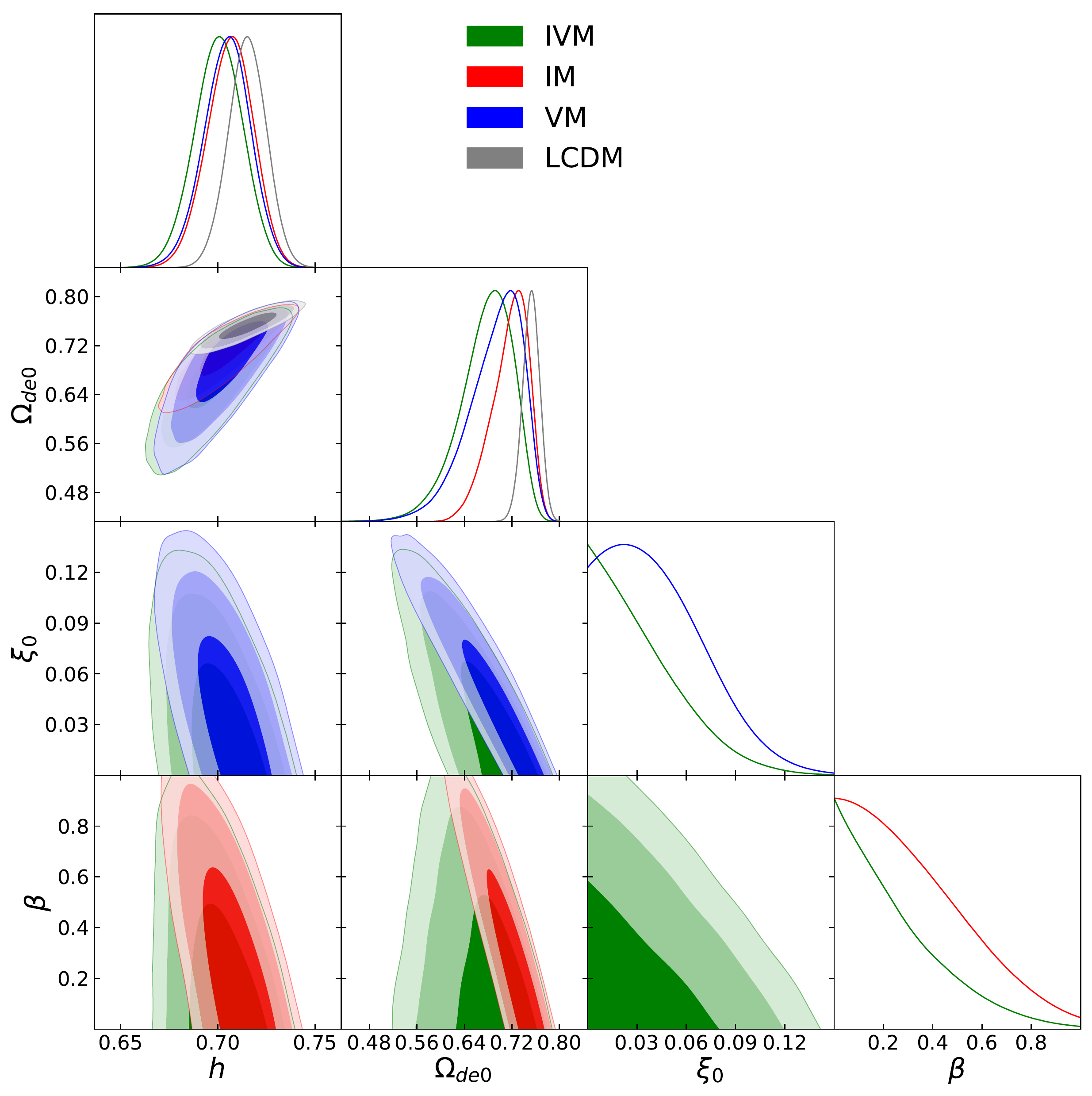}
\caption{Top panel: Best fit curve of IVM and $\Lambda$CDM. Bottom panel: 2D contours of the free model parameters at $1\sigma, 2\sigma$, and $3\sigma$ (from darker to lighter color bands) CL using nonhomogeneous OHD data. Figure adapted from \cite{Hernandez-Almada:2020ulm}.}
\label{fig:IVM}
\end{figure}

A generalized case of the viscous models presented in the previous section is to consider a flat FLRW Universe which contains a non-perfect fluid as dust matter (dm) component that interacts with a perfect fluid as the DE component, together with the radiation fluid. Similarly, through the energy-momentum tensor, Eq. \eqref{emt}, the viscous term is included in the field equations by changing $p\to\tilde{p}=p+\Pi$ as the sum of the total barotropic pressure of the fluids ($p$) and the bulk viscosity coefficient ($\Pi$), where $\rho$ is the energy density of the fluid and $u_{\mu}$ is the associated four-velocity. Inspired by the viscosity behavior in fluid mechanics that is proportional to the speed, we assume $\Pi=-3 \zeta H$. Furthermore, the matter component and DE interacts through an energy exchange term $Q$, and a viscosity effect encoded in the terms containing the bulk viscosity coefficient $\zeta$. In this approach, the Friedmann, continuity and acceleration equations are \cite{Hernandez-Almada:2020ulm}
\begin{eqnarray}
    &&H^2=\frac{8\pi G}{3}\left( \rho_r + \rho_{dm} + \rho_{de} \right), \label{eq:VFEa} \\
    &&\dot{\rho}_{r} + 4 H \rho_r = 0 \,,\label{eq:VFEb} \\
    &&\dot{\rho}_{dm} + 3 H \rho_{dm} = 9 H^2\zeta + Q \,, \label{eq:VFEc}\\
    &&\dot{\rho}_{de} + 3 \gamma_{de} H \rho_{de} = - Q \,, \label{eq:VFEd}\\
    &&2\dot{H} - 24\pi G H \zeta = -8\pi G \left( \rho_{dm} + \frac{4}{3}\rho_{r} + \gamma_{de}\rho_{de} \right) \,, \label{eq:VFEe}
\end{eqnarray}
where $\rho_r$, $\rho_{dm}$, and $\rho_{de}$ are the relativistic species, dust matter and dark energy  densities, respectively. Notice that the DE component behaves as CC when $\gamma_{de}=0$. In particular, the typical \textit{ansatz} for the viscosity coefficient is considered and given by
\begin{equation}
    \zeta = \frac{\xi}{\kappa^2} \left( \frac{\rho_{dm}}{\rho_{dm0}} \right)^{1/2} \,,
\end{equation}
where $\rho_{dm0}$ is the dm density at present epoch and $\xi$ is a free parameter with units of $[\xi]=$[eV]. It is convenient to use the dimensionless parameter of $\xi$ defined as $\xi_0=\sqrt{3}\xi/\kappa \rho_{dm0}$. Additionally, the interacting term $Q$ is consider to be \cite{Zimdahl:2003}
\begin{equation}
    Q = \beta H \frac{\rho_{de}\rho_{dm}}{\rho_{de}+\rho_{dm}},
\end{equation}
where $\beta$ is a free parameter. It is straightforward that an Universe with only viscosity effects is obtained when $\beta=0$.
Figure \ref{fig:IVM} shows the best fit curve (top panel) to OHD data for interacting viscous model ($\xi_0 \neq 0, \beta \neq 0$), viscous model ($\beta=0$), interacting model ($\xi_0=0$) and $\Lambda$CDM, respectively. 2D contours at $1\sigma$, $2\sigma$, $3\sigma$ CL and 1D posterior distributions of the free model parameters are presented at the bottom panel. Authors in \cite{Hernandez-Almada:2020ulm} found that the energy density dynamics of the mentioned models are similar to the evolution of $\Lambda$CDM. Table \ref{tab:bestfit_IVM} reports the best fit values and their uncertainties at $1\sigma$ for the free parameters of IVM, IM, VM and LCDM models.

\begin{table}[htb]
\caption{Best fit values for the free parameters of IVM, IM, VM and LCDM models using the OHD sample. The uncertainties are at $1\sigma$ \cite[see][for details]{Hernandez-Almada:2020ulm}.}
\centering
\begin{tabular}{| lccccc|}
\hline
Model & $\chi^2$ &  $h$ & $\Omega_{de0}$ & $\xi_0$ & $\beta$               \\
\hline 
 & & & & &  \\
IVM & $30.5$ & $0.701^{+0.012}_{-0.013}$ &  $0.682^{+0.040}_{-0.040}$ & $0.028^{+0.033}_{-0.020}$ & $0.200^{+0.260}_{-0.145}$    \\ [1.1ex]
IM & $29.2$ & $0.707^{+0.011}_{-0.012}$ & $0.721^{+0.026}_{-0.037}$ & $0$ & $0.283^{+0.290}_{-0.197}$   \\ [1.1ex]
VM & $29.1$ & $0.705^{+0.011}_{-0.012}$ & $0.698^{+0.038}_{-0.054}$ & $0.040^{+0.035}_{-0.026}$ & $0$  \\ [1.1ex]
LCDM & $28.9$ & $0.715^{+0.010}_{-0.010}$ & $0.753^{+0.014}_{-0.015}$ & $0$ & $0$     \\ [1.1ex]
\hline
\end{tabular}
\label{tab:bestfit_IVM}
\end{table}

%%%%%%%%%%%%%%%%%%%%%%%%%%%%%%%%%%%%%%%%%%%%
\subsubsection{ Phenomenological Emergent Dark Energy Model}

The phenomenological emergent dark energy model (PEDE) was proposed by \cite{PEDE:2019ApJ} and assume that the DE is negligible at early times, emerging at late times. These kind of models are known as emergent and contribute to elucidate a solution to the $H_0$ tension. The idea consist in proposing a function that mimics the evolution of DE density parameter from a phenomenological point of view. The PEDE model has the same degrees of freedom as $\Lambda$CDM model.

\begin{figure}
\centering
\includegraphics[width=0.45\textwidth]{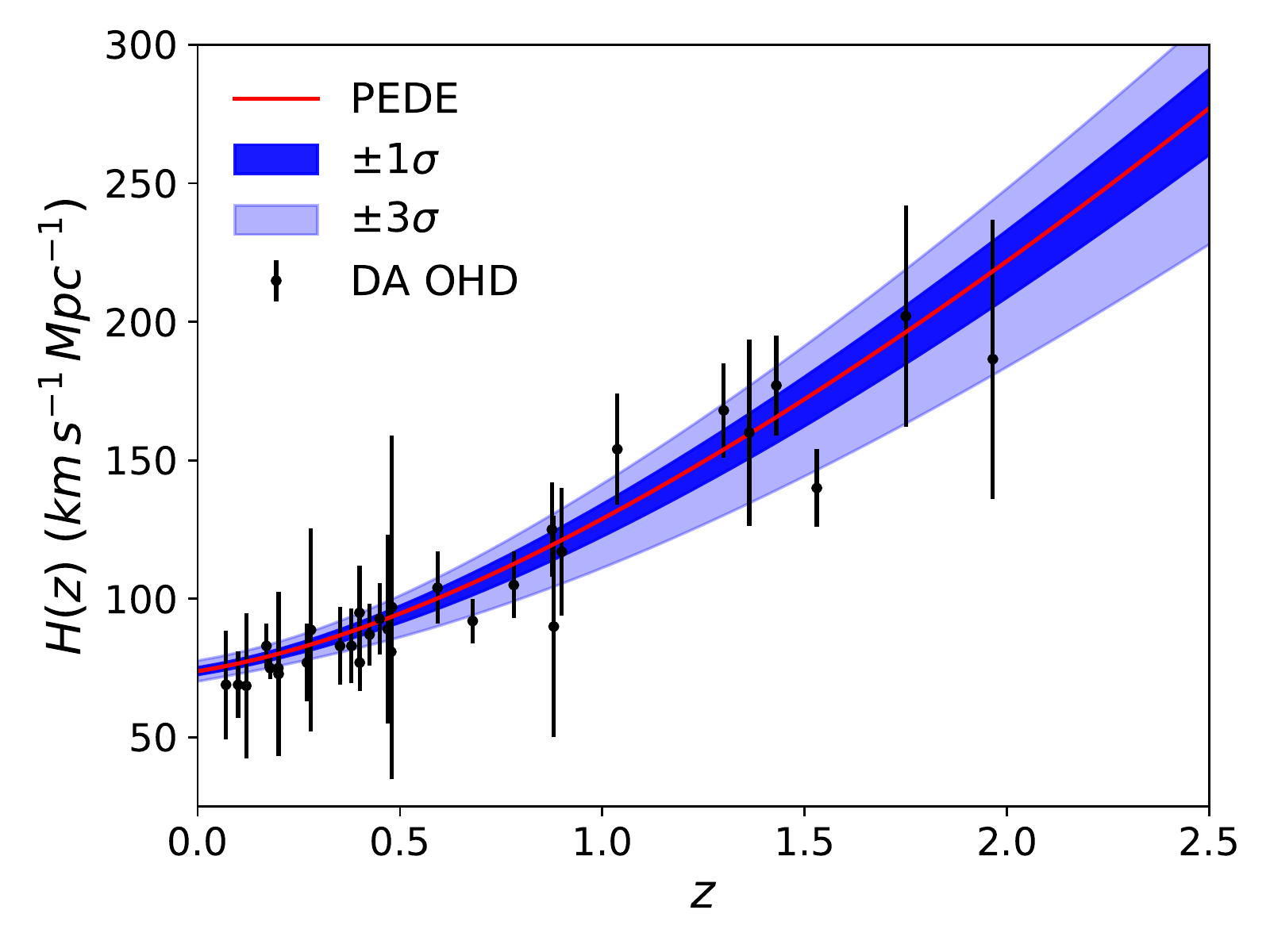}\\
\includegraphics[width=0.45\textwidth]{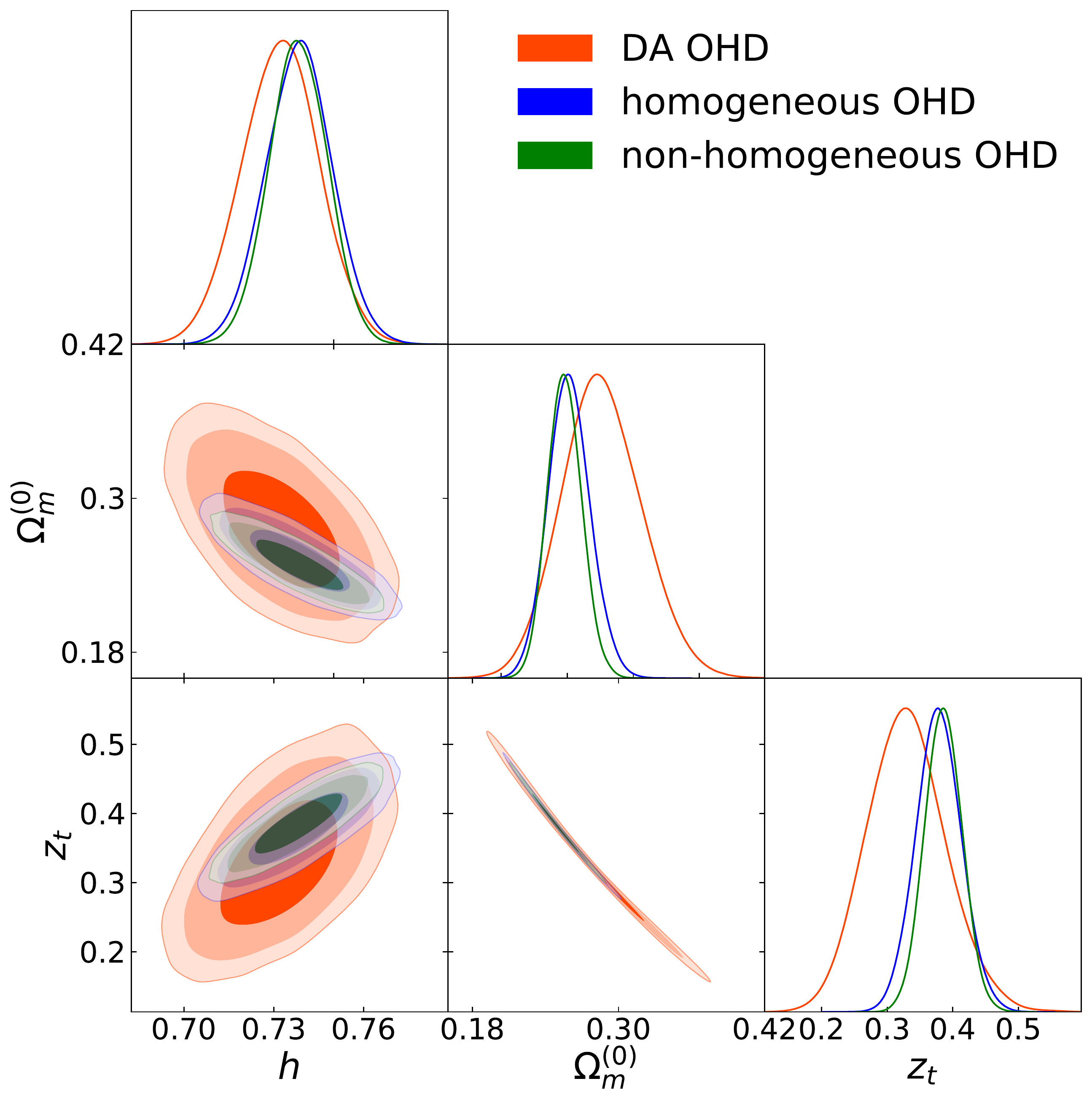}
\caption{Top panel: Best fit curve of PEDE model and its uncertainty at $1\sigma$ and $3\sigma$. Bottom panel: 2D contours of the free model parameters at $1\sigma, 2\sigma$, and $3\sigma$ (from darker to lighter color bands) CL using DA, homogeneous and non-homogeneous OHD data. Figure adapted from \cite{Hernandez-Almada:2020uyr}.}
\label{fig:pede}
\end{figure}

We consider a FLRW metric which contains matter ($m$, dark matter plus baryons), radiation ($r$), and PEDE. The dynamics of this Universe is described by the Friedmann equation \eqref{eq:Hz_gen} and the continuity equation for each component as: 
\begin{subequations}
\label{non_min}
	\begin{eqnarray}
	%&&H^2\equiv \left(\frac{\dot{a}}{a}\right)^2=\frac{8 \pi G}{3}(\rho_{DE}+\rho_{\rm{m}}+\rho_{r}),\label{Fried_rho}\\
	&&\dot{\rho}_{DE}+3H (1+ w_{DE})\rho_{DE}=0,\label{ce_de}\\
	&&\dot{\rho}_{\rm{m}}+3H(1+w_{m})\rho_{\rm{m}}=0,\label{ce_m}\\
	&&\dot{\rho}_{r}+3H(1+w_{r})\rho_{r}=0, \label{ce_rad}
	\end{eqnarray}
\end{subequations}
By solving Eqs. \eqref{ce_de}, \eqref{ce_m}, \eqref{ce_rad} we can rewrite the Eq. \eqref{eq:Ez_gen} in terms of the density parameters and redshift, as 
\begin{equation}
    H(z)^{2}=H_{0}^{2}\left[\Omega_{m0}\left(1+z\right)^{3}+\Omega_{r0}\left(1+z\right)^{4}+ \widetilde{\Omega}_{\rm{DE}}(z) \right],
\end{equation}
where $\widetilde{\Omega}_{\rm{DE}}(z)=\Omega_{\rm{DE}0}f(z)$, where ${\Omega}_{\rm{DE}}$ follows Eq. \eqref{FlatCond}. 
%\begin{equation}
 %   {\Omega}_{\rm{DE}}=1-\Omega_{m}-{\Omega}_{\rm{r}}. 
%\label{eq:flatness}
%\end{equation}
Notice that \cite{PEDE:2019ApJ} propose a phenomenological functional form for $f(z)$, described by Eq. \eqref{eq:fdez}
%\begin{equation}
%f(z)\equiv %\frac{\rho_{de}(z)}{\rho_{de}(0)}=
%\mathrm{exp}\left(3\int^{z}_{0}\frac{1+w_{DE}(z)}{1+z}\mathrm{d}z\right).
%\label{eq:fz}
%\end{equation}
and hence $\widetilde{\Omega}_{\rm{DE}}(z)$ as\footnote{Where it is defined  $\widetilde{\Omega}_{\rm{DE}}(z)\equiv \rho_{DE}/ \rho_{c}^{(0)}$.}
\begin{equation}
\widetilde{\Omega}_{\rm{DE}}(z)\,=\,\Omega_{\rm{DE}}^{(0)}\left[ 1 - {\rm{tanh}}\left( {\log}_{10}(1+z) \right) \right],
\label{eq:omegatilde_pede}
\end{equation}
\noindent
where $\widetilde{\Omega}_{\rm{DE}}\rightarrow 0$ at $z \rightarrow\infty$ and $\widetilde{\Omega}_{\rm{DE}}\rightarrow 1.4$ at $z \rightarrow -1$. Notice that
\begin{eqnarray}
    & \Omega_{DE}(z)= \frac{H_0^2}{H(z)^2}\widetilde{\Omega}_{\rm{DE}}(z)=\frac{H_0^2}{H(z)^2}\,\Omega_{\rm{DE}}^{(0)}\left[ 1 - {\rm{tanh}}\left( {\log}_{10}(1+z) \right) \right],  
\end{eqnarray}
Therefore, the dimensionless Friedmann equation results as
\begin{eqnarray}
    E(z)^2=\Omega_{m0}(1+z)^3+\Omega_{r0}(1+z)^4+\Omega_{\rm{DE}0} [ 1 - \rm{tanh}(\log_{10}(1+z))],
\end{eqnarray}
where the radiation density parameter at current epoch is calculated with Eq. \eqref{Rad0}. To constraint the PEDE parameters different OHD are employed: those from the DA technique (i.e. cosmic chronometers), and a full sample (homogeneous and non-homogeneous) of BAO measurements. Results of the constrictions are presented in Figs. \ref{fig:pede}; the top panel illustrates the $H(z)$ reconstruction and the bottom panel the confidence contours for the case $\Omega_m(z_t)=\Omega_{de}(z_t)$. Table \ref{tab:PEDE} presents the constraints for the model free parameters together with the associated $\chi^2$ \cite[see][for details]{Hernandez-Almada:2020uyr}.

\begin{table*}
\caption{Mean values of the free parameters for PEDE model using homogeneous, non-homogeneous and DA OHD and a Gaussian prior on $h=0.7403\pm 0.0142$ \citep{Riess_2019}. The last column shows the estimated redsfhit $z_t$ using the condition $\Omega_m(z_t) = \Omega_{DE}(z_t)$  \cite[see][for details]{Hernandez-Almada:2020uyr}. The uncertainties reported correspond to $1\sigma$ confidence level. In parenthesis are the best fit
values when a flat prior on $h$ is considered in the region $[0,1]$.} 
\centering
\resizebox{\textwidth}{!}{%
\begin{tabular}{|lccccc|}
\hline
Sample     &    $\chi^2$     &  $h$ & $\Omega_m^{(0)}$ & $\Delta$ & $z_t$ \\
\hline
\multicolumn{6}{|c|}{PEDE} \\ [0.9ex]
homogeneous OHD  & $24.5$ ($24.5$) & $0.740^{+0.011}_{-0.011}$ ($0.738^{+0.018}_{-0.018}$)& $0.252^{+0.016}_{-0.015}$ ($0.254^{+0.024}_{-0.022}$) & $1.0$  & $0$ \\ [0.9ex]
non-homogeneous OHD    & $32.1$ ($32.1$) & $0.740^{+0.010}_{-0.010}$ ($0.740^{+0.014}_{-0.014}$) & $0.249^{+0.013}_{-0.013}$ ($0.249^{+0.018}_{-0.016}$) & $1.0$  & $0$\\ [0.9ex]
DA OHD    & $14.7$ ($14.6$) & $0.739^{+0.014}_{-0.014}$ ($0.723^{+0.049}_{-0.044}$) & $0.319^{+0.035}_{-0.039}$ ($0.329^{+0.057}_{-0.045}$) & $1.0$  & $0$\\ [0.9ex]
%homogeneous OHD  & $24.2$ ($24.2$) & $0.739^{+0.011}_{-0.011}$ ($0.735^{+0.018}_{-0.018}$) & $0.251^{+0.016}_{-0.015}$ ($0.255^{+0.024}_{-0.022}$) & $1.0$  & $0.378^{+0.035}_{-0.034}$ ($0.371^{+0.049}_{-0.049}$) \\ [0.9ex]
%non-homogeneous OHD    & $31.6$ ($31.6$) & $0.738^{+0.010}_{-0.010}$ ($0.736^{+0.013}_{-0.013}$)& $0.248^{+0.013}_{-0.013}$ ($0.250^{+0.017}_{-0.016}$) & $1.0$  & $0.386^{+0.028}_{-0.028}$ ($0.381^{+0.037}_{-0.037}$)\\ [0.9ex]
%DA OHD    & $16.1$ ($14.4$)& $0.732^{+0.013}_{-0.013}$ ($0.691^{+0.032}_{-0.032}$) & $0.275^{+0.031}_{-0.029}$ ($0.333^{+0.064}_{-0.054}$)& $1.0$  & $0.328^{+0.060}_{-0.058}$ ($0.226^{+0.096}_{-0.096}$)\\ [0.9ex]
%\hline
%\multicolumn{6}{|c|}{GEDE} \\ [0.9ex]
%homogeneous OHD  & $23.7$ ($23.0$) & $0.735^{+0.012}_{-0.012}$ ($0.725^{+0.023}_{-0.020}$) & $0.247^{+0.018}_{-0.017}$ ($0.256^{+0.025}_{-0.022}$) & $0.690^{+0.624}_{-0.457}$ ($0.533^{+0.712}_{-0.390}$) & $0.403^{+0.058}_{-0.057}$ ($0.385^{+0.058}_{-0.056}$) \\ [0.9ex]
%non-homogeneous OHD    & $30.2$ ($28.6$)& $0.731^{+0.012}_{-0.011}$ ($0.718^{+0.017}_{-0.015}$) & $0.245^{+0.014}_{-0.013}$ ($0.255^{+0.018}_{-0.017}$) & $0.539^{+0.470}_{-0.352}$ ($0.332^{+0.472}_{-0.244}$) & $0.417^{+0.044}_{-0.043}$ ($0.403^{+0.043}_{-0.043}$)\\ [0.9ex]
%DA OHD  & $14.7$ ($14.6$) & $0.739^{+0.014}_{-0.014}$ ($0.723^{+0.048}_{-0.044}$) & $0.319^{+0.036}_{-0.039}$ ($0.329^{+0.057}_{-0.046}$) & $3.930^{+2.304}_{-2.083}$ ($3.264^{+3.258}_{-2.230}$) & $0.183^{+0.094}_{-0.057}$ ($0.174^{+0.083}_{-0.064}$) \\ [0.9ex]
\hline
\end{tabular}}
\label{tab:PEDE}
\end{table*}

%%%%%%%%%%%%%%%%%%%%%%%%%%%%%%%
\subsubsection{Generalized Emergent Dark Energy} \label{Gen}

\begin{figure}
\centering
\includegraphics[width=0.45\textwidth]{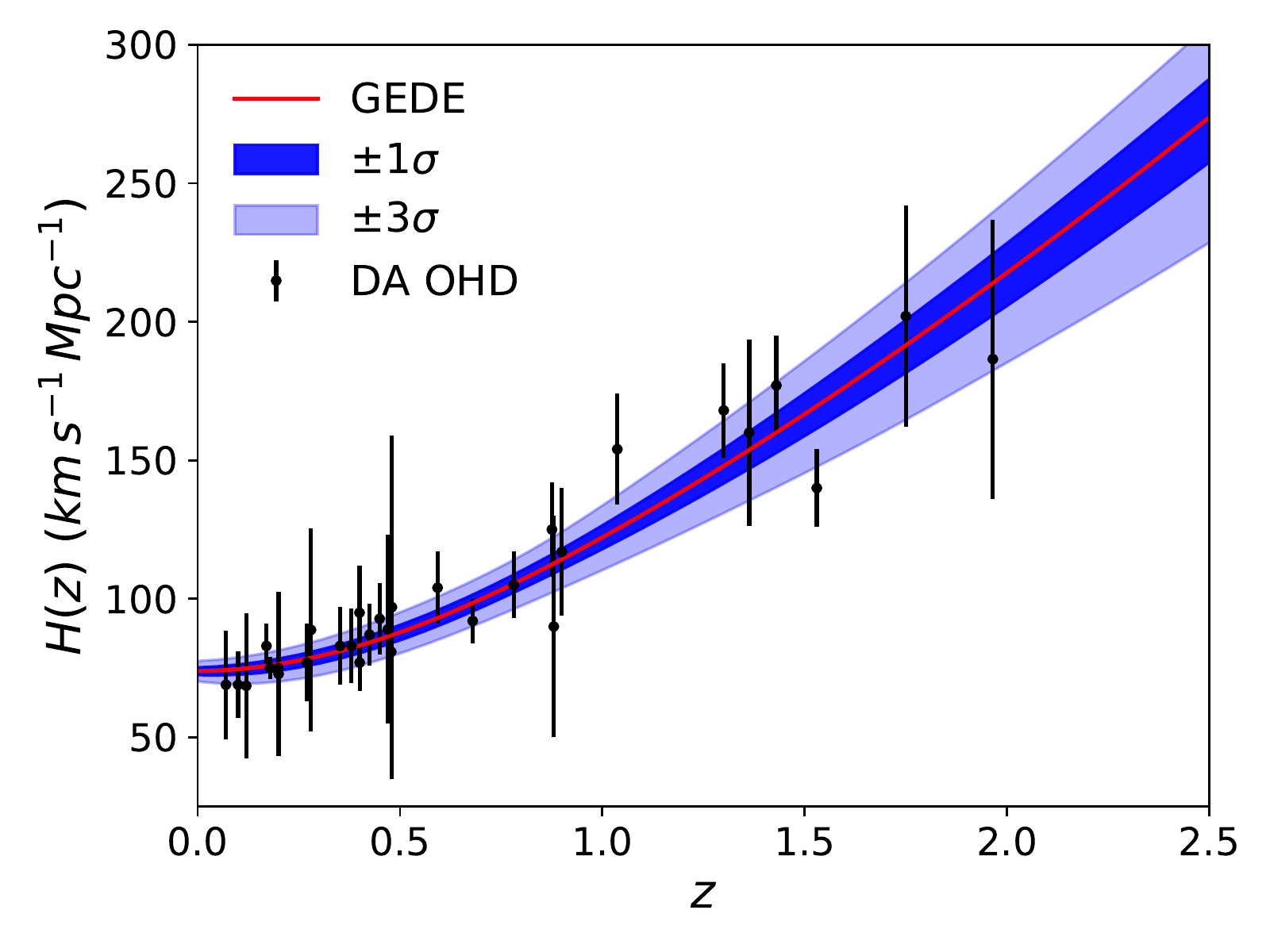}\\
\includegraphics[width=0.45\textwidth]{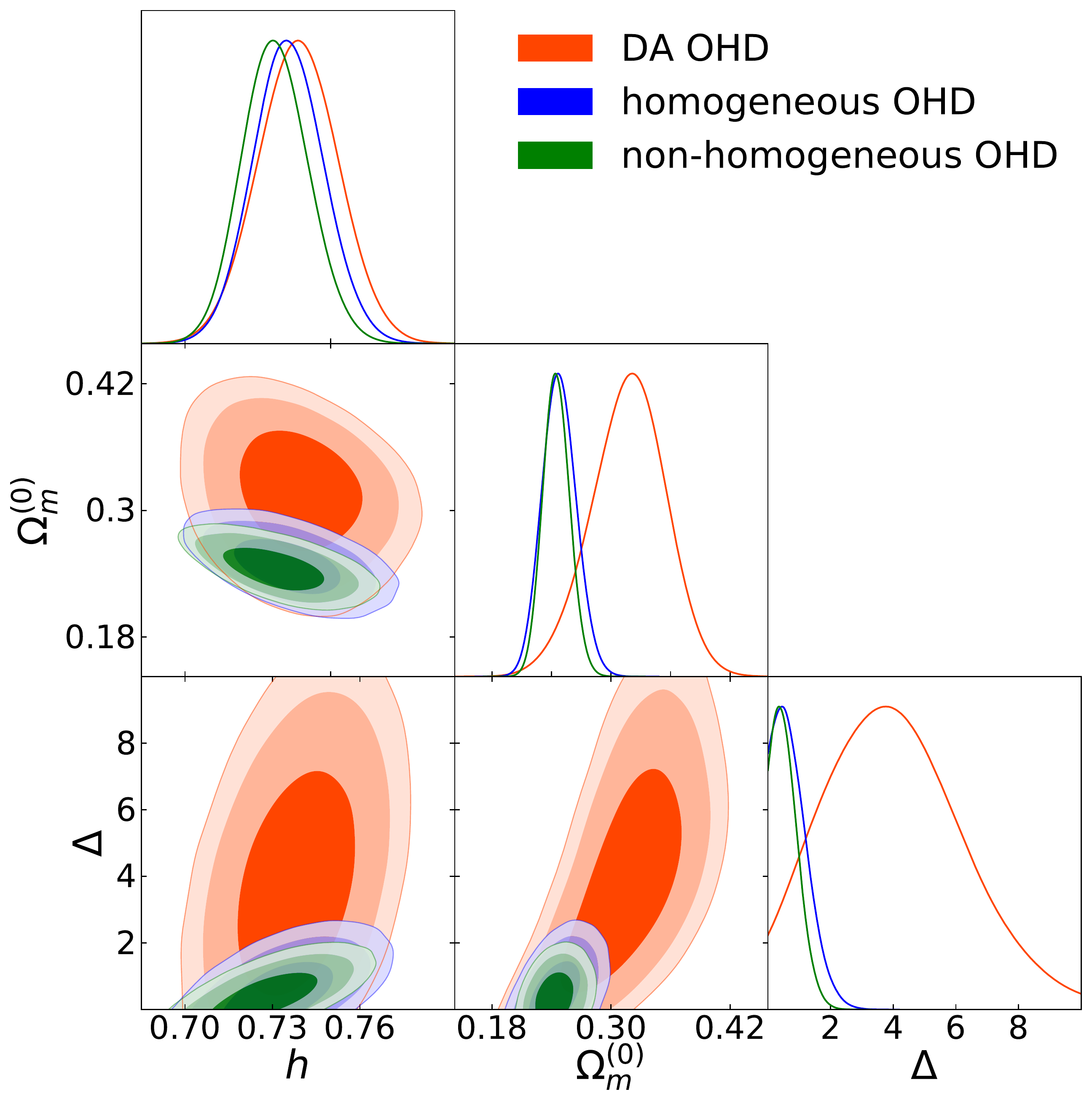}
\caption{Top panel: Best fit curve of GEDE model and its uncertainty at $1\sigma$ and $3\sigma$. Bottom panel: 2D contours of the free model parameters at $1\sigma, 2\sigma$, and $3\sigma$ (from darker to lighter color bands) CL using DA, homogeneous and non-homogeneous OHD data. Figure adapted from \cite{Hernandez-Almada:2020uyr}.}
\label{fig:gede}
\end{figure}

Recently, \cite{PEDE:2020} proposed a generalisation for the PEDE model, also known as Generalized Emergent Dark Energy Model (GEDE) model, by introducing
\begin{equation}
\widetilde{\Omega}_{\rm{DE}}(z)\,=\,\Omega_{\rm{DE}}^{(0)} \frac{ 1 - {\rm{tanh}}\left( {\Delta \log}_{10}(\frac{1+z}{1+z_{t}}) \right)}{ 1 + {\rm{tanh}}\left(\Delta {\log}_{10}(1+z_{t}) \right)},
\label{eq:omega_pede}
\end{equation}
where $z_t$ is a transition redshift, $\Omega_{DE}(z_t)=\Omega_{m0}(1+z_t)^3$, $\Delta$ is an appropriate dimensionless non-negative free parameter with the characteristic that if $\Delta=0$ the $\Lambda$CDM model is recovered, and when $\Delta=1$ and $z_t=0$ the previously PEDE model is obtained. As $z_{t}$ can be related to $\Omega_{m0}$ and $\Delta$, then $z_{t}$ is not a free parameter. Notice that the DE density parameter is given by
\begin{eqnarray}
&{\Omega}_{\rm{DE}}\, =\frac{H_0^2}{H^2} (1-\Omega_{m0}-\Omega_{r0})\frac{ 1 - {\rm{tanh}}\left( {\Delta \log}_{10}(\frac{1+z}{1+z_{t}}) \right)}{ 1 + {\rm{tanh}}\left(\Delta {\log}_{10}(1+z_{t}) \right)}.
\label{eq:omegade_gene}
\end{eqnarray}
The GEDE Friedmann equation is written as
\begin{eqnarray}
    &&E(z)=\left[\Omega_{m0}\left(1+z\right)^{3}+\Omega_{r0}\left(1+z\right)^{4}+\Omega_{\rm{DE}}^{(0)} \frac{ 1 - {\rm{tanh}}\left( {\Delta \log}_{10}(\frac{1+z}{1+z_{t}}) \right)}{ 1 + {\rm{tanh}}\left(\Delta \,{\log}_{10}(1+z_{t}) \right)} \right]^{1/2}.\nonumber\\
\end{eqnarray}
The results obtained from the MCMC analysis using the same data as PEDE model are shown in Fig. \ref{fig:gede}, presenting the best fit curve for $H(z)$ confronting with the OHD data and the constraints for $\Omega_m^{(0)}$ and $\Delta$ which is the free parameter for GEDE. Table \ref{tab:GEDE}  presents the constraints for all the free parameters together with their respective $\chi^2$ \cite[see][for details]{Hernandez-Almada:2020uyr}.

\begin{table*}
\caption{Mean values of the free parameters for GEDE model using homogeneous, non-homogeneous and DA OHD and a Gaussian prior on $h=0.7403\pm 0.0142$ \citep{Riess_2019}. The last column shows the estimated redsfhit $z_t$ using the condition $\Omega_m(z_t) = \Omega_{DE}(z_t)$  \cite[see][for details]{Hernandez-Almada:2020uyr}. The uncertainties reported correspond to $1\sigma$ confidence level. In parenthesis are the best fit
values when a flat prior on $h$ is considered in the region $[0,1]$.}
\centering
\resizebox{\textwidth}{!}{%
\begin{tabular}{|lccccc|}
\hline
Sample     &    $\chi^2$     &  $h$ & $\Omega_m^{(0)}$ & $\Delta$ & $z_t$ \\
\hline
%\multicolumn{6}{|c|}{PEDE} \\ [0.9ex]
%homogeneous OHD  & $24.5$ ($24.5$) & $0.740^{+0.011}_{-0.011}$ ($0.738^{+0.018}_{-0.018}$)& $0.252^{+0.016}_{-0.015}$ ($0.254^{+0.024}_{-0.022}$) & $1.0$  & $0$ \\ [0.9ex]
%non-homogeneous OHD    & $32.1$ ($32.1$) & $0.740^{+0.010}_{-0.010}$ ($0.740^{+0.014}_{-0.014}$) & $0.249^{+0.013}_{-0.013}$ ($0.249^{+0.018}_{-0.016}$) & $1.0$  & $0$\\ [0.9ex]
%DA OHD    & $14.7$ ($14.6$) & $0.739^{+0.014}_{-0.014}$ ($0.723^{+0.049}_{-0.044}$) & $0.319^{+0.035}_{-0.039}$ ($0.329^{+0.057}_{-0.045}$) & $1.0$  & $0$\\ [0.9ex]
%homogeneous OHD  & $24.2$ ($24.2$) & $0.739^{+0.011}_{-0.011}$ ($0.735^{+0.018}_{-0.018}$) & $0.251^{+0.016}_{-0.015}$ ($0.255^{+0.024}_{-0.022}$) & $1.0$  & $0.378^{+0.035}_{-0.034}$ ($0.371^{+0.049}_{-0.049}$) \\ [0.9ex]
%non-homogeneous OHD    & $31.6$ ($31.6$) & $0.738^{+0.010}_{-0.010}$ ($0.736^{+0.013}_{-0.013}$)& $0.248^{+0.013}_{-0.013}$ ($0.250^{+0.017}_{-0.016}$) & $1.0$  & $0.386^{+0.028}_{-0.028}$ ($0.381^{+0.037}_{-0.037}$)\\ [0.9ex]
%DA OHD    & $16.1$ ($14.4$)& $0.732^{+0.013}_{-0.013}$ ($0.691^{+0.032}_{-0.032}$) & $0.275^{+0.031}_{-0.029}$ ($0.333^{+0.064}_{-0.054}$)& $1.0$  & $0.328^{+0.060}_{-0.058}$ ($0.226^{+0.096}_{-0.096}$)\\ [0.9ex]
\hline
\multicolumn{6}{|c|}{GEDE} \\ [0.9ex]
homogeneous OHD  & $23.7$ ($23.0$) & $0.735^{+0.012}_{-0.012}$ ($0.725^{+0.023}_{-0.020}$) & $0.247^{+0.018}_{-0.017}$ ($0.256^{+0.025}_{-0.022}$) & $0.690^{+0.624}_{-0.457}$ ($0.533^{+0.712}_{-0.390}$) & $0.403^{+0.058}_{-0.057}$ ($0.385^{+0.058}_{-0.056}$) \\ [0.9ex]
non-homogeneous OHD    & $30.2$ ($28.6$)& $0.731^{+0.012}_{-0.011}$ ($0.718^{+0.017}_{-0.015}$) & $0.245^{+0.014}_{-0.013}$ ($0.255^{+0.018}_{-0.017}$) & $0.539^{+0.470}_{-0.352}$ ($0.332^{+0.472}_{-0.244}$) & $0.417^{+0.044}_{-0.043}$ ($0.403^{+0.043}_{-0.043}$)\\ [0.9ex]
DA OHD  & $14.7$ ($14.6$) & $0.739^{+0.014}_{-0.014}$ ($0.723^{+0.048}_{-0.044}$) & $0.319^{+0.036}_{-0.039}$ ($0.329^{+0.057}_{-0.046}$) & $3.930^{+2.304}_{-2.083}$ ($3.264^{+3.258}_{-2.230}$) & $0.183^{+0.094}_{-0.057}$ ($0.174^{+0.083}_{-0.064}$) \\ [0.9ex]

\hline
\end{tabular}}
\label{tab:GEDE}
\end{table*}

%%%%%%%%%%%%%%%%%%%%%%%%%%%%%%%%%%%%%%%%%%%%
\subsection{Modifications to General Theory of Relativity}

In this subsection, we present models that modify the GTR in order to obtain a late Universe acceleration.

%%%%%%%%%%%%%%%%%%%%%%%%%%%%%%%%%%%%%%%%%%%%%
\subsubsection{Constant Brane Tension} \label{brane}

Brane world models are inspired by the seminal papers of \cite{Randall-I,Randall-II} in which they assume a four dimensional manifold called the brane immersed in a five dimensional Anti-d'Sitter space time called the bulk. The mentioned configuration is a via to understand the hierarchy problem but could be also extended to describe the cosmology. The main parameter of the theory is called the brane tension, which differentiate between the high and low energy physics involved and becomes a free parameter that needs to be constrained by different cosmological samples, for this model in particular, the brane tension is constant, hence, we call it a Constant Brane Tension (CBT) model.

First of all, we introduce the Einstein's field equation projected onto the brane
\begin{equation}
G_{\mu\nu}+\xi_{\mu\nu}=\kappa^2_{(4)}T_{\mu\nu} + \kappa^4_{(5)}\Pi_{\mu\nu} + \kappa^2_{(5)}F_{\mu\nu}, \label{1}
\end{equation}
where $T_{\mu\nu}$ is defined in Eq.\eqref{emt} of the matter trapped in the brane, $G_{\mu\nu}$ is the classical Einstein's tensor described by \eqref{EFEM} and the rest of terms in the right and left sides of this equation are explicitly given by:
\begin{subequations}
\begin{eqnarray}
\kappa^2_{(4)}&=&8\pi G_{N}=\frac{\kappa^4_{(5)}}{6}\lambda, \\
\Pi_{\mu\nu}&=&-\frac{1}{4}T_{\mu\alpha}T_{\nu}^{\alpha}+\frac{TT_{\mu\nu}}{12}+\frac{g_{\mu\nu}}{24}(3T_{\alpha\beta}T^{\alpha\beta}-T^2), \\
F_{\mu\nu}&=&\frac{2T_{AB}g_{\mu}^{A}g_{\nu}^{B}}{3}+\frac{2g_{\mu\nu}}{3}\left(T_{AB}n^An^B-\frac{^{(5)}T}{4}\right), \\
\xi_{\mu\nu}&=&^{(5)}C^E_{AFB}n_En^Fg^{A}_{\mu}g^{B}_{\nu}.
\end{eqnarray}
\end{subequations}
Here $G_N$ is the Newton's gravitational constant, $\lambda$ is the previously mentioned brane tension, $\kappa_{(4)}$ and $\kappa_{(5)}$ are the four- and five-dimensional coupling constants of gravity, respectively. The tensor $\Pi_{\mu\nu}$ represents the quadratic corrections on the brane generated by the energy-momentum tensor,  $F_{\mu\nu}$ gives the contributions of the energy-momentum tensor in the bulk, which is projected onto the brane through the unit normal vector $n_A$. The tensor $\xi_{\mu\nu}$ provides the contribution of the five-dimensional Weyl's tensor projected onto the brane manifold \cite{sms} \footnote{Notice that the latin letters take the values $0,1,2,3,4$.}. It is worth to note that non-local corrections are negligible in cosmological cases \cite{m2000}, under the assumption of a AdS$_{(5)}$ bulk. 

To derive the Friedmann equations under the modified field equations, we consider an homogeneous and isotropic  Universe in which a line element is given by Eq. \eqref{FLRW}. We consider radiation and dark matter components as perfect fluids in the brane. We assume that the bulk has no matter component. Using Eqs. \eqref{1}, we obtain the modified Friedmann equation:
\begin{equation}
H^2=\frac{8\pi G}{3}\sum_i\rho_i\left(1+\frac{ \rho_i }{2\lambda}\right). \label{FLRWM}
\end{equation}
Notice that $\rho_i$ is the energy density for the radiation, dark matter and DE.
It is worth to notice that the low energy regime, i.e. the canonical Friedmann equation, is recovered when $\rho_i/2\lambda\to0$. Crossed terms were not used in the Friedmann equation, i.e. there is not interaction between different species. In addition, if we consider, for instance, that the bulk black hole mass vanishes, the bulk geometry reduces to $\rm AdS_{5}$ and $\rho_{\epsilon }=0$ \cite{m2000,MaartensCos}. Thus, the Friedmann equation can be written as:
\begin{eqnarray}
H^2=\frac{8\pi G}{3}\left[\frac{\rho_{0m}}{a^3}\left(1+\frac{\rho_{0m}}{2\lambda a^3}\right)+
\frac{\rho_{0r}}{a^4}\left(1+\frac{\rho_{0r}}{2\lambda a^4}\right)
+\frac{\rho_{0de}}{a^{3(1+\omega_{de})}}\left(1+\frac{\rho_{0de} }{2\lambda a^{3(1+\omega_{de})}}\right) \right]. \label{eq1}
\end{eqnarray}
The above equation can be expressed in terms of the density parameters through the dimensionless Friedmann equation
\begin{eqnarray}
&&E(z)^{2}=\Omega_{0m}(1+z)^{3}+
\Omega_{0r}(1+z)^{4}+\Omega_{0de}(1+z)^{3(1+\omega_{de})}\nonumber\\
&&+\mathcal{M}\left[\Omega_{0m}^2(1+z)^{6}+\Omega_{0r}^2(1+z)^{8}+\Omega_{0de}^2(1+z)^{6(1+\omega_{de})}\right], \qquad
\label{eq:H}
\end{eqnarray}
where 
\begin{equation}
\mathcal{M}\equiv \frac{H_0^2}{2\kappa^2\lambda}, \label{const}
\end{equation}
being $\rho_{crit}$ the Universe critical density. Notice that when $\mathcal{M}\to0$, the canonical Friedmann equation with $w_{de}$ is recovered. 
If $w_{de}=\omega_{\Lambda}=-1$, i.e. the DE is the CC, we obtain the traditional $\Lambda$CDM dynamics.

At early times the brane dynamics dominate over other terms in the Universe, but is negligible at late time. Indeed, given a value for the brane tension, we can infer  the limits of high and low energies in terms of the redshift: $z+1\gg\sum_i(\lambda/\rho_{0i})^{1/3(1+\omega_i)}$ and $z+1\ll\sum_i(\lambda/\rho_{0i})^{1/3(1+\omega_i)}$ respectively. For example, in matter domination epoch, the previous expressions can be rewritten as: $z\gg(\lambda/\rho_{0m})^{1/3}-1$ and $z\ll(\lambda/\rho_{0m})^{1/3}-1$, for high and low energy limits, respectively.

\begin{figure}
\centering
\includegraphics[width=0.45\textwidth]{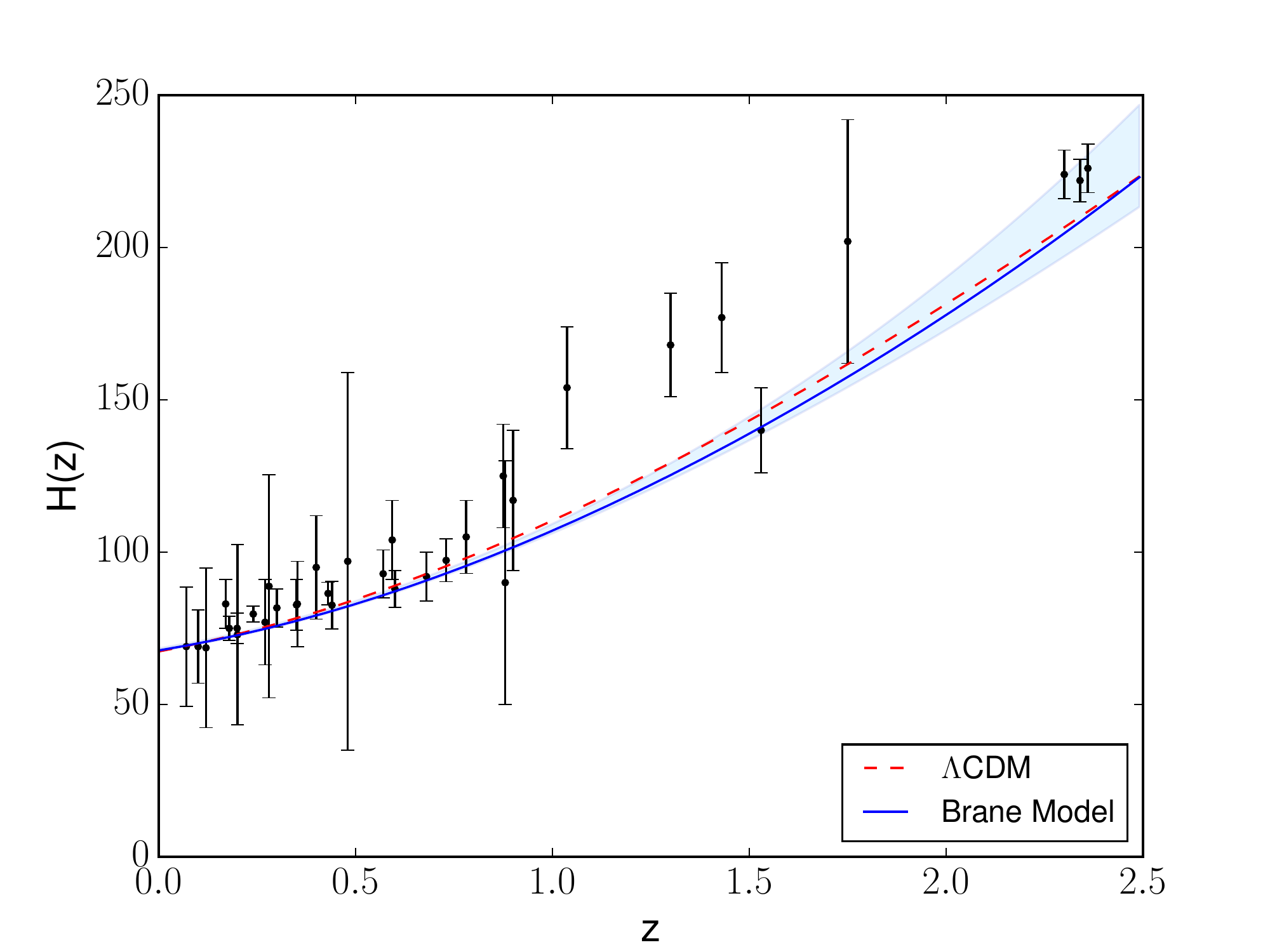}\\
\includegraphics[width=0.45\textwidth]{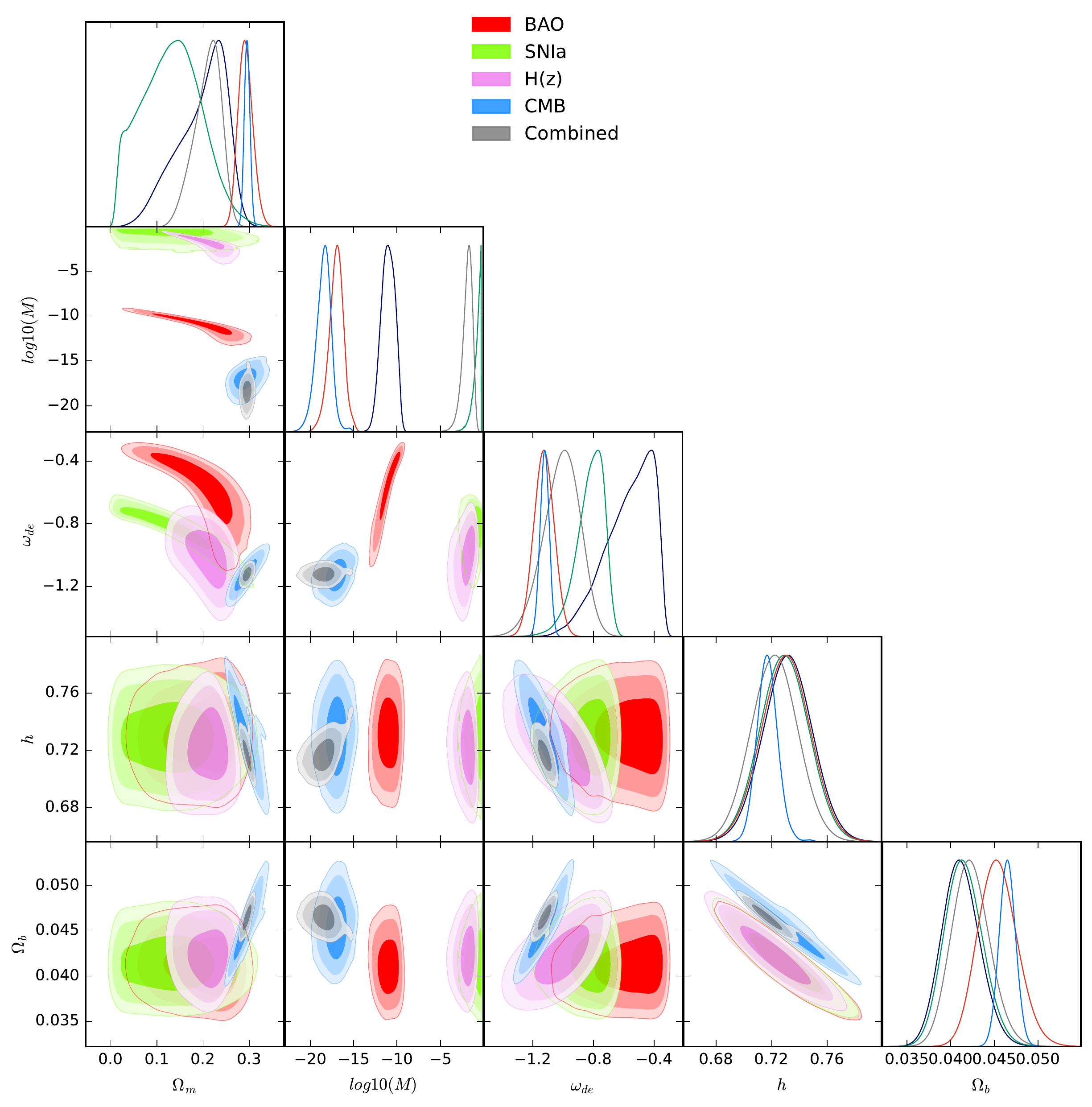}
\caption{Top panel: Best fit curve of constant brane tension model and its uncertainty at $1\sigma$. Bottom panel: 2D contours of the free model parameters at $1\sigma, 2\sigma$, and $3\sigma$ (from darker to lighter color bands) CL using OHD, SNIa (Pantheon), SLS, HIIG, BAO and Joint data.}
\label{fig:braneCte}
\end{figure}

Table \ref{tab:BRANE} shows the best-fit for the different free parameters together with the estimated $\chi^2$.
Fig. \ref{fig:braneCte} shows the severe tension between the different cosmological samples (OHD, SNIa, SLS, HIIG, BAO), hence concluding that the model is not viable to replace the $\Lambda$CDM model unless an additional DE component is included, which defeat the purpose of choosing this model.

\begin{table}
\caption{Best fitting values of the free parameters for the constant brane tension model with the different samples used in this paper.}
\centering
\begin{tabular}{|cccccc|}
\hline
Sample     &    $\chi^2$     &  $h$ & $\Omega_{m0}$ & $w_{de}$ &  $\log_{10}{\mathcal{M}}$          \\
\hline
 OHD   & $18.19$ & $0.72^{+0.01}_{-0.01}$ & $0.21^{+0.02}_{-0.03}$ & $-1.00^{+0.11}_{-0.12}$ & $<-0.88$  \\ [0.7ex]
 BAO    & $5.46$ & $0.73^{+0.01}_{-0.01}$ & $0.20^{+0.04}_{-0.07}$ & $-0.53^{+0.13}_{-0.19}$ & $<-9.52$ \\ [0.7ex]
SNIa    & $574.73$ & $0.72^{+0.01}_{-0.01}$ & $0.13^{+0.06}_{-0.07}$ & $-0.81^{+0.07}_{-0.10}$ &  $<-0.31$   \\ [0.7ex]
CMB    & $10.87$ & $0.73^{+0.01}_{-0.01}$ & $0.29^{+0.01}_{-0.01}$ & $-1.12^{+0.06}_{-0.06}$ & $<-15.0$    \\ [0.7ex]
 Joint  & $636.70$ & $0.71^{+0.01}_{-0.01}$ & $0.30^{+0.01}_{-0.01}$ & $-1.12^{+0.03}_{-0.03}$ & $<-16.2$ \\ [0.7ex]
\hline
\end{tabular}
\label{tab:BRANE}
\end{table}

%%%%%%%%%%%%%%%%%%%%%%%%%%%%%%%%%%%%%%%%%%%%%
\subsubsection{Variable Brane Tension}

A natural extension to the previous model is the one called variable brane tension (BVT). The framework is the same as the on in Section \ref{brane}, but now  an extra degree of freedom is assumed, a brane tension emerge as a function of the redshift. Naturally, the model resolve the problems associated with the presence of the brane tension at early epochs but also generates a CC with five dimensional origins.  We briefly discuss the theoretical framework of a BVT model which was previously studied in \cite{Garcia-Aspeitia:2018fvw}. We start from the BVT field equation as
\begin{equation}
    G_{\mu\nu}-8\pi GT_{\mu\nu}=\frac{1}{\lambda}\left[48\pi G\Pi_{\mu\nu}+\frac{3}{4\pi G}\xi_{\mu\nu}\right], \label{fe}
\end{equation}
where
\begin{eqnarray}
&&\xi_{\mu\nu}=\mathcal{U}\left(u_{\mu}u_{\nu}+\frac{1}{3}\epsilon_{\mu\nu}\right)+\mathcal{P}_{\mu\nu}, \\
&&\Pi_{\mu\nu}=-\frac{1}{4}T_{\mu\alpha}T^{\alpha}_{\nu}+\frac{1}{12}T^{\alpha}_{\alpha}T_{\mu\nu}+\frac{1}{24}g_{\mu\nu}[3T_{\alpha\beta}T^{\alpha\beta}-(T^{\alpha}_{\alpha})^2].
\end{eqnarray}
where $G_{\mu\nu}$ is described in \eqref{EFEM}, $\xi_{\mu\nu}$ is a non-local Weyl tensor decomposed in its irreducibility form which also contains $\mathcal{U}$ is the non-local energy density, $\mathcal{P}_{\mu\nu}$ is the non-local anisotropic stress tensor, $u_{\alpha}$ is the four-velocity and $\epsilon_{\mu\nu}\equiv g_{\mu\nu}+u_{\mu}u_{\nu}$. In addition $T_{\mu\nu}$ is the standard energy-momentum tensor and $\Pi_{\mu\nu}$ contains a quadratic form of the energy-momentum tensor. Notice that the corrective terms that comes from brane world is contingent to the brane tension defined by $\lambda$, which in this model is not a constant. Therefore, the low energy limit is considered when $\lambda\to\infty$ recovering the traditional field equation of GR, while in the other limit $\lambda\to0$ extra terms play a preponderant role. Finally, notice that in this case we do not consider extra fields onto the bulk, neglecting the terms that come from $F_{\mu\nu}$ and only considering those fields living in the brane.

Therefore, if we introduce the previously line element in Eq. \eqref{fe} together with the perfect fluid energy-momentum tensor (Eq. \eqref{emt}) we have the following Friedmann equation \cite{Garcia-Aspeitia:2018fvw}
\begin{equation}
E(z)^2=\Omega_{0m}(z+1)^3+\Omega_{0r}(z+1)^4+\frac{\mathcal{M}}{\hat{\lambda}(z)}[\Omega_{0m}^2(z+1)^6+\Omega_{0r}^2(z+1)^8], \label{FriedmannBranesX}
\end{equation}
here we have already considered  matter and radiation components, where their evolution come from the conservation of the energy-momentum tensor ($\nabla^{\mu}T_{\mu\nu}=0$) and their separability from the quadratic part of the field equation. The brane tension evolves homogeneous and isotropically because it only on the temporal function and can be chosen using other physical assumptions. In addition, the brane tension is not directly coupled with the continuity equation of the fluids and it is defined as $\hat{\lambda}(z)\equiv\lambda_0\lambda(z)$, being $\hat{\lambda}(z)$ a dimensionless function that can be selected appropriately. Moreover, $\mathcal{M}\equiv3H_0^2/16\pi G\lambda_0$ and, under the flatness condition, we have the constriction
\begin{equation}
\mathcal{M}=\frac{1-\Omega_{0m}-\Omega_{0r}}{\Omega_{0m}^2+\Omega_{0r}^2}\hat{\lambda}(z=0).
\end{equation}
Regarding the choice of the $\hat{\lambda}(z)$ function, we pick a polynomial form as $\hat{\lambda}(z)=(z+1)^n$, where $n$ is a free parameter and $n\in\mathbb{R}$. \citet{Garcia-Aspeitia:2018fvw} discuss the inspiration for this function, arguing that it could be a generalization of the E\"{o}tv\"{o}s law, similar functions can be found in tracker behavior for scalar fields.

\begin{figure}
\centering
\includegraphics[width=0.45\textwidth]{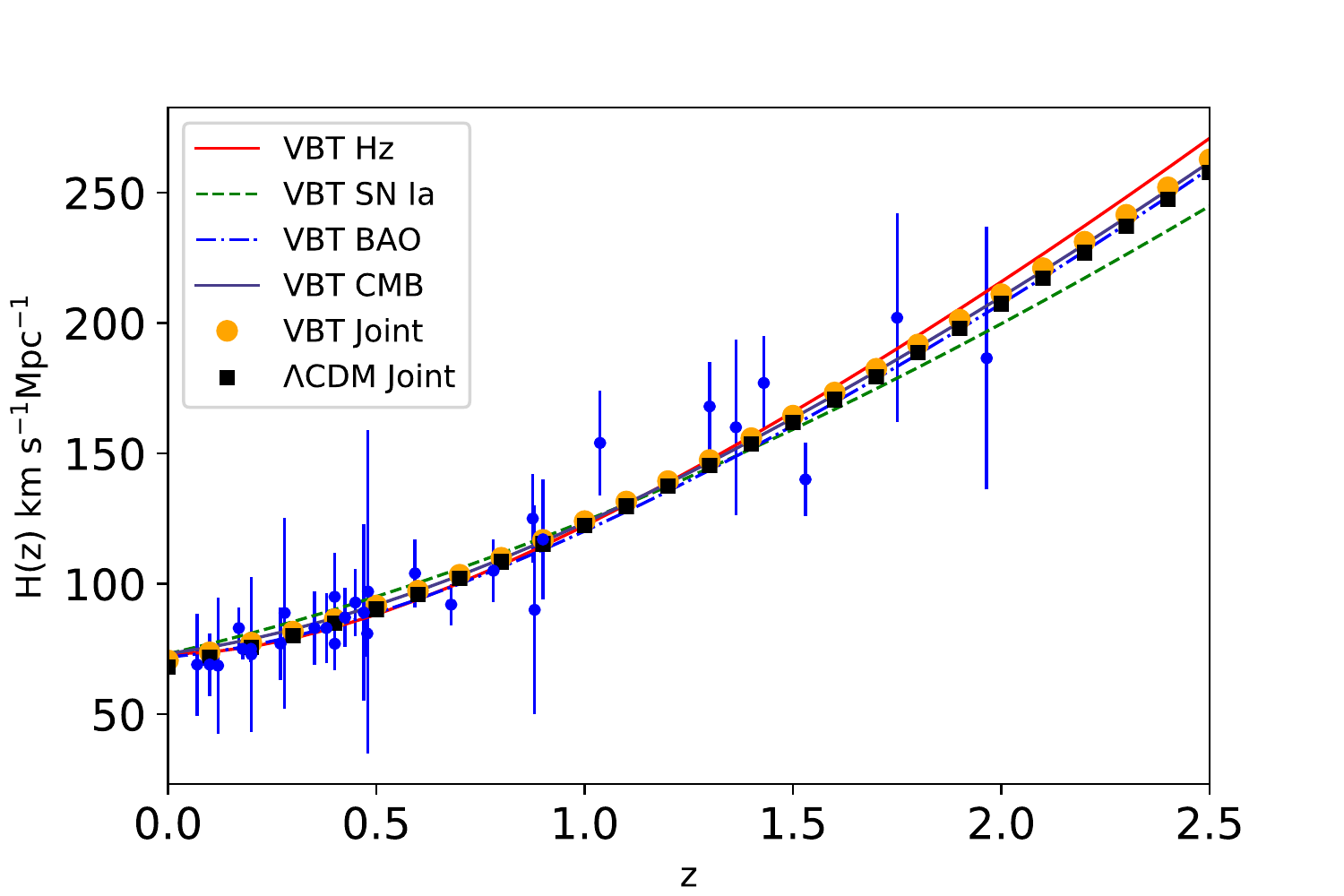}\\
\includegraphics[width=0.45\textwidth]{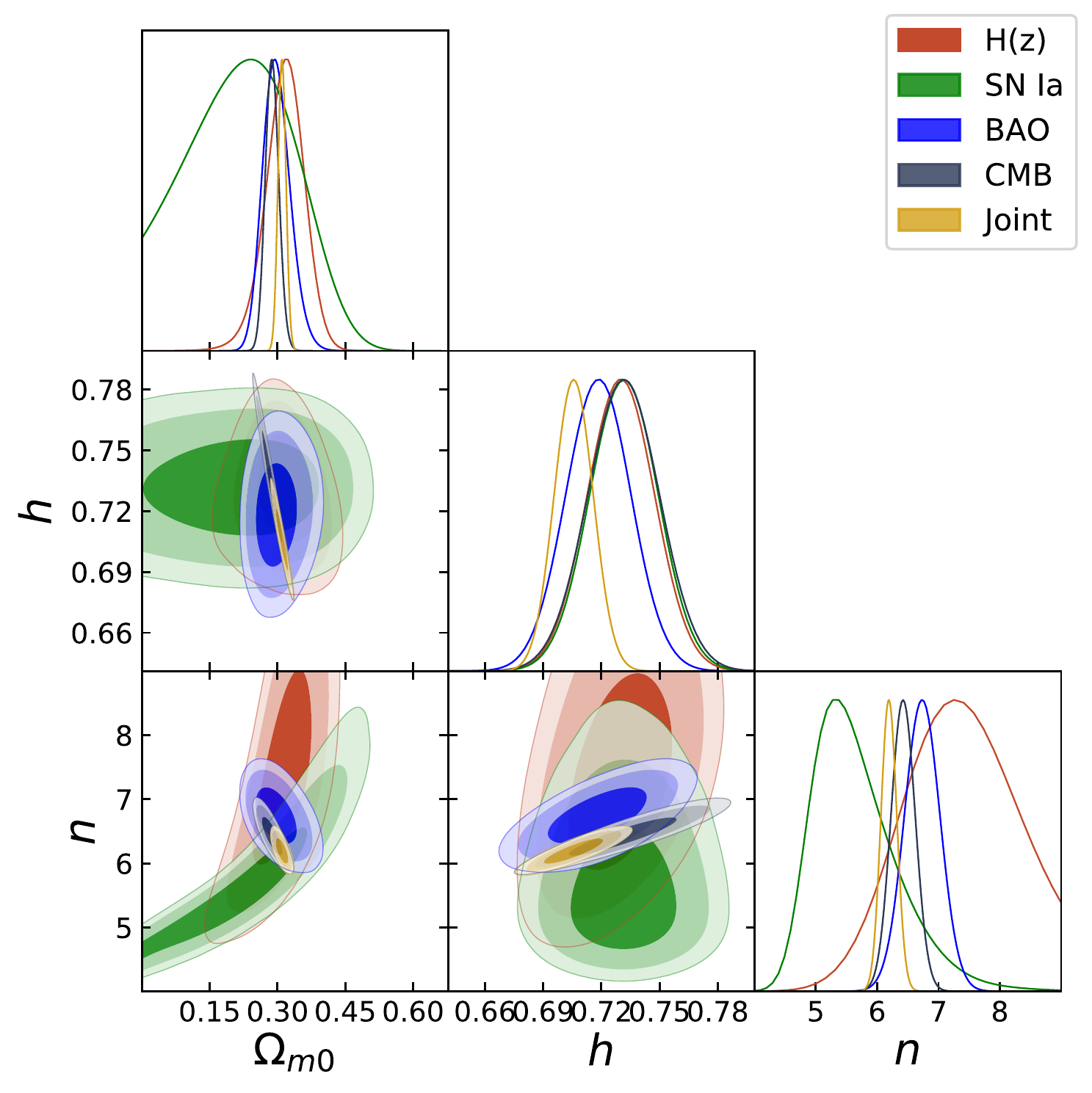}
\caption{Top panel: Best fit curve for the variable brane tension model and its uncertainty at $1\sigma$. Bottom panel: 2D contours of the free model parameters at $1\sigma, 2\sigma$, and $3\sigma$ (from darker to lighter color bands) CL using OHD, SNIa, BAO, CMB and Joint data.}
\label{fig:braneVar}
\end{figure}

Using a joint analysis that contains OHD, CMB, BAO and SNIa observations (see Fig. \ref{fig:braneVar} and Table \ref{tab:BVT}), \cite{Garcia-Aspeitia:2018fvw} found out $n=6.19.\pm0.12$. Notice that the result is consistent with predictions because if we use Eq. \eqref{FriedmannBranesX}, neglecting $(z+1)^8$, the term $\mathcal{M}$ will behave as a CC at late times but with extra dimensions origin.

\clearpage
\begin{table}
\caption{Best fitting values of the free parameters for the Variable Brane Tension model with the different samples used in this paper.}
\centering
\begin{tabular}{|cccccc|}
\hline
Sample     &    $\chi^2_{min}$     &  $h$ & $\Omega_{m0}$ & $n$ &  $\lambda_0(10^{-12}$eV$^4)$          \\
\hline
 OHD   & $14.46$ & $0.730^{+0.017}_{-0.017}$ & $0.318^{+0.039}_{-0.042}$ & $7.400^{+1.100}_{-0.926}$ & $3.20^{+1.05}_{-0.95}$  \\ [0.7ex]
 BAO    & $9.49$ & $0.718^{+0.016}_{-0.016}$ & $0.297^{+0.031}_{-0.028}$ & $6.730^{+0.287}_{-0.289}$ & $2.62^{+0.77}_{-0.57}$ \\ [0.7ex]
SNIa    & $691.10$ & $0.731^{+0.017}_{-0.017}$ & $0.231^{+0.114}_{-0.120}$ & $5.580^{+0.815}_{-0.568}$ &  $1.48^{+2.40}_{-1.16}$   \\ [0.7ex]
CMB    & $3.64$ & $0.732^{+0.017}_{-0.017}$ & $0.288^{+0.014}_{-0.013}$ & $6.420^{+0.185}_{-0.185}$ & $2.52^{+0.19}_{-0.17}$    \\ [0.7ex]
 Joint  & $716.43$ & $0.706^{+0.009}_{-0.009}$ & $0.31^{+0.008}_{-0.008}$ & $6.190^{+0.121}_{-0.120}$ & $2.81^{+0.12}_{-0.11}$ \\ [0.7ex]
\hline
\end{tabular}
\label{tab:BVT}
\end{table}

%%%%%%%%%%%%%%%%%%%%%%%%%%%%%%%%%%%%%%%%%%%%%
\subsubsection{Unimodular Gravity}

Unimodular Gravity (UG) is a remarkable proposition to tackle the problem of the CC by limiting the metric in the following way $\sqrt{-g}=\xi$, where $\xi$ is a constant, restricting the field equations at only nine linear independent equations and the field equation is trace-free \cite{Ellis}. The possibility to integrate the line element of FLRW give us the opportunity to obtain clues about the nature of CC, tracing its presence at epochs of reionization \cite{Garcia-Aspeitia:2019yni}.

UG can be described by the following field equation
\begin{equation}
R_{\mu\nu}-\frac{1}{4}g_{\mu\nu}R=8\pi G\left(T_{\mu\nu}-\frac{1}{4}g_{\mu\nu}T\right), \label{UGfield}
\end{equation}
where all the tensors are the standards of GR and $G$ is the Newton's gravitational constant.

In order to study the background cosmology, we consider an isotropic, homogeneous FLRW metric \eqref{FLRW}, the perfect fluid energy momentum tensor is written as show Eq. \eqref{emt}. Hence, we have \cite{Ellis,Gao:2014nia}
\begin{equation}
\dot{H}=\frac{\ddot{a}}{a}-H^2=-4\pi G\sum_i(\rho_i+p_i), \label{Friedmann_acc}
\end{equation}
where the dots stands for time derivative. In addition, a general conservation for UG theory is now written in the form
\begin{equation}
\nabla^{\mu}[32\pi GT_{\mu\nu}-(R+8\pi GT)g_{\mu\nu}]=0. \label{Eq:noncons}
\end{equation}
Without independently assuming the energy momentum conservation ($\nabla^{\mu}T_{\mu\nu}=0$), the Eq. \eqref{Eq:noncons} introduces new Friedmann, acceleration and fluid equations coupled with third order derivatives in the scale factor. Hence, in the case of non traditional conservation of the energy-momentum tensor, Eq. \eqref{Eq:noncons} must be solved to obtain the characteristic fluid equation. Solving for \eqref{Eq:noncons} under a FLRW metric and perfect fluid we have
\begin{equation}
\sum_i\left[\frac{d}{dt}(\rho_i+p_i)+3H(\rho_i+p_i)\right]=\frac{H^3}{4\pi G}(1-j), \label{chida2}
\end{equation}
where the sum is over all the species in the Universe and $j\equiv\dddot{a}/aH^3$ is the Jerk Parameter (JP) \cite{Zhang:2016,Mamon:2018dxf}, well known in cosmography and proposed by \cite{Garcia-Aspeitia:2019yni} for the study of UG.

On the other hand, the integral-transcendent-Friedmann equation can be computed with the help of Eq. \eqref{eq:Hz_gen} and \eqref{chida2}, obtaining the Friedmann equation as 
\begin{equation}
H^2=\frac{8\pi G}{3}\sum_i\rho_i+H^2_{corr}. \label{Frie}
\end{equation}
where the non-canonical extra term in Eq. \eqref{Frie}, i.e. the UG correction to the Friedmann and acceleration equations, is defined in the form
\begin{equation}
H_{corr}^2\equiv\frac{8\pi G}{3}\sum_ip_i+\frac{2}{3}\int_{a_{ini}}^{a(t)} H(a^{\prime})^2[j(a^{\prime})-1]\frac{da^{\prime}}{a^{\prime}}, \label{HUG}
\end{equation}
where the sum runs over the different species in the Universe and $a_{ini}$ is some constant initial value. 

According to \cite{Garcia-Aspeitia:2019yni}, it is plausible to consider an \textit{ansatz} for the JP in terms of the redshift with the following characteristics
\begin{equation}
j(z)=\frac{9(1+w)w}{2E(z)^2}\Omega_{0i}(z+1)^{3(w+1)}+1, \label{JX}
\end{equation}
where $w$ is the EoS for any fluid. If we choose $\Omega_{0i}\to\Omega_{0r}$ as the radiation density parameter\footnote{Matter emerges naturally from Eq. \eqref{chida2}, therefore the other expected fluid should be radiation to avoid introducing an exotic fluid.} and $w\to w_r=1/3$ as the EoS of radiation, then the functional form reproduce the $\Lambda$CDM jerk parameter in all eras \cite{Garcia-Aspeitia:2019yni}. 

From the previous equation and Eq. \eqref{Frie}  it is possible to deduce
\begin{eqnarray}
E(z)^2=\Omega_{0m}(z+1)^3+\Omega_{0r}(z+1)^4+\Omega_{0exs}(z_{ini}+1)^4, \label{HX} 
\end{eqnarray}
where $\Omega_{0exs}\equiv w_r\Omega_{0r}$. 

Notice that the source of the Universe acceleration is the constant term in the previous equation \eqref{HX}, where we can naturally relate $\Omega_{0\Lambda}\to\Omega_{0exs}(z_{ini}+1)^4$. Since our choice in Eq. \eqref{JX} depends on the EoS and the energy density parameter of the radiation, the constant inherits those terms. 

\begin{figure}
\centering
\includegraphics[width=0.45\textwidth]{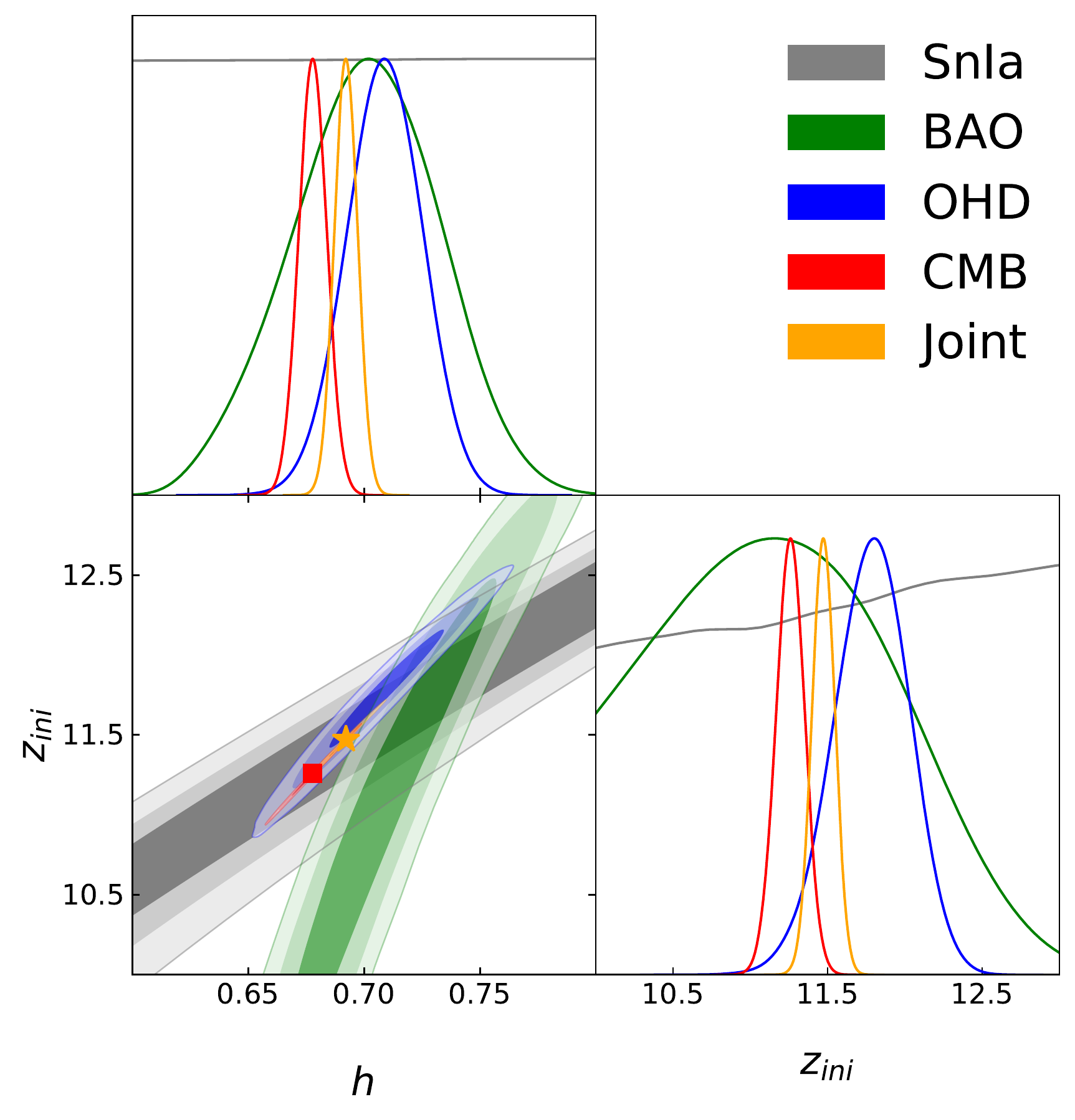}
\caption{1D marginalized posterior distributions and the 2D $68\%$, $95\%$, $99.7\%$ of CL for the $h$, and $z_{ini}$ parameters of the UG model. The star (square) marker represents the best fit value of Joint (CMB) data.}
\label{fig:UG}
\end{figure}

Fig. \ref{fig:UG} shows the constraints for $z_{ini}$ and $h$  considering the SNIa, BAO, OHD, CMB and a joint data, with the best-fit values in Table \ref{tab:UG} \cite[see][for details]{Garcia-Aspeitia:2019yni,Garcia-Aspeitia:2019yod}. The results suggest a $z_{ini}=11.47$, which is in the reionization era. The interpretation of the result is that UG is an emergent DE theory, with DE arising during the reionization period. 

\begin{table}
\caption{Mean values for the UG
model parameters ($h$, $z_{ini}$) and $\chi^2_{min}$, derived from each data set and the joint analysis \cite[see][]{Garcia-Aspeitia:2019yod}.} 
\centering
\resizebox{0.4\textwidth}{!}{
\begin{tabular}{|cccc|}
%\hline
\multicolumn{4}{c}{}\\
\hline
Sample & $\chi^{2}_{min}$& $h$ & $z_{ini}$ \\
\hline
OHD  &   $22.0$ & $0.709^{+0.016}_{-0.016}$  & $11.788^{+0.237}_{-0.250}$ \\
SnIa & $1036.0$ & $0.602^{+0.270}_{-0.272}$  & $10.623^{+2.366}_{-3.021}$ \\
CMB  & $0.0001$ & $0.678^{+0.006}_{-0.006}$ & $11.259^{+0.091}_{-0.092}$  \\
BAO  & $12.9$ & $0.701^{+0.031}_{-0.033}$ &  $10.847^{+0.979}_{-1.383}$  \\
Joint & $1097.6$ & $0.692^{+0.005}_{-0.005}$ & $11.473^{+0.074}_{-0.073}$ \\
\hline
\end{tabular}}
\label{tab:UG}
\end{table}

%%%%%%%%%%%%%%%%%%%%%%%%%%%%%%%%%%%%%%%%%%%%%
\subsubsection{Einstein-Gauss-Bonet}

Einstein-Gauss-Bonet (EGB) is a recent proposition that modifies the geometrical part of the field equations \cite{Glavan:2019inb}, maintaining the continuity equation in its original form. In \cite{Garcia-Aspeitia:2020uwq} they constrained the free parameter through diverse observations finding results compatibles with the cosmological standard model. However, the authors also found that specific values of the free parameter could generate an eternal acceleration, even in epochs in which is not expected (reionization, nucleosynthesis, etc). Other mathematical flaws of the EGB model have just been found  \cite{Ai:2020peo,Gurses:2020ofy,Lu:2020iav, Fernandes:2020nbq,Mahapatra:2020rds}.

The action of the EGB gravity can be written in the form \cite{Glavan:2019inb}
\begin{eqnarray}
    S_{EGB}[g_{\mu\nu}]=\int d^{d+1}x\sqrt{-g}\Big[\frac{1}{16\pi G}(R-2\Lambda)+\mathcal{L}_m+\frac{\alpha}{d-3}\mathcal{G}\Big],
\end{eqnarray}
where $\Lambda$ is an effective cosmological constant, $R$ is the Ricci scalar, $\mathcal{L}_m$  is the matter Lagrangian, $\alpha$ is an appropriate free parameter, $\mathcal{G}=6R^{\mu\nu}_{\;\;\;\;[\mu\nu}R^{\rho\sigma}_{\;\;\;\;\rho\sigma]}$ is the Gauss-Bonnet contribution to the Einstein-Hilbert action and $d+1$ is considered in the limit when $ \lim_{d\to3}d+1$ as presented in \cite{Glavan:2019inb}. Minimizing the action, the field equation can be written as
\begin{eqnarray}
    &&G_{\mu\nu}+\Lambda g_{\mu\nu}+\frac{\alpha}{(d-3)}(4RR_{\mu\nu}-8R_{\mu\alpha}R^{\alpha}_{\nu}-8R_{\mu\alpha\nu\beta}R^{\alpha\beta}+4R_{\mu\alpha\beta\sigma}R^{\alpha\beta\sigma}_{\nu}\nonumber\\&&-g_{\mu\nu}\mathcal{G})=8\pi GT_{\mu\nu}.
\end{eqnarray}
Notice that when $\alpha=0$, the standard Einstein field equation with a CC is recovered.

In order to study the background cosmology, we assume a flat FLRW given by Eq. \eqref{FLRW}, the energy-momentum tensor is the usual given by \eqref{emt}. After some manipulations of the previous expressions, the Friedmann equation for EGB reads
\begin{equation}
H^2+3\alpha H^4=\frac{8\pi G}{3}\sum_i\rho_i+\frac{\Lambda}{3}. \label{Friedmann}
\end{equation}
Moreover, the continuity equation takes its traditional form as \eqref{cont}. In terms of the dimensionless variables, Eq. \eqref{Friedmann} is re-written as
\begin{equation}
    E(z)^2+\bar{\alpha}E(z)^4=\Omega_{m0}(z+1)^3+\Omega_{r0}(z+1)^4+\Omega_{\Lambda 0}, \label{Friedmannnon}
\end{equation}
where $\bar{\alpha}\equiv3\alpha H_0^2$, $\Omega_{i0}\equiv\kappa^2\rho_i/3H_0^2$ and $\Omega_{\Lambda0}\equiv\Lambda/3H_0^2$, composed by  matter (baryons and dark matter) and relativistic particles (photons and neutrinos). Another important consideration is that $\bar{\alpha}$ is a positive value as inflation demands (see Ref. \cite{Clifton:2020xhc} for details).

In order to constrain the $\bar{\alpha}$ parameter, we divide the problem in two branches through Eq. \eqref{Friedmannnon}. Therefore, if we only consider the branch where we have a real value of $E(z)$, then we have \cite{Garcia-Aspeitia:2020uwq}
\begin{eqnarray}
E(z)^2=\frac{1}{2\bar{\alpha}}\left[\sqrt{1+4\bar{\alpha}\Omega(z)_{std}}-1\right], \label{FriedmannAd}
\end{eqnarray}
where
\begin{equation}
\Omega(z)_{std}\equiv\Omega_{m0}(z+1)^3+\Omega_{r0}(z+1)^4+\Omega_{\Lambda 0}\,,
\end{equation}
is the standard cosmological model. Eq. \eqref{FriedmannAd} is constrained to the condition $E(0)=1$, having the following relation \cite{Garcia-Aspeitia:2020uwq}
\begin{equation}
    \Omega_{\Lambda 0} = \frac{(2\bar{\alpha}+1)^2-1}{4\bar{\alpha}} - \Omega_{m0} - \Omega_{r0}\,. \label{constriction}
\end{equation}
Notice that when $\bar{\alpha}\to 0$, in \eqref{FriedmannAd} the standard Friedmann equation is recovered.

\begin{figure}
\centering
\includegraphics[width=0.45\textwidth]{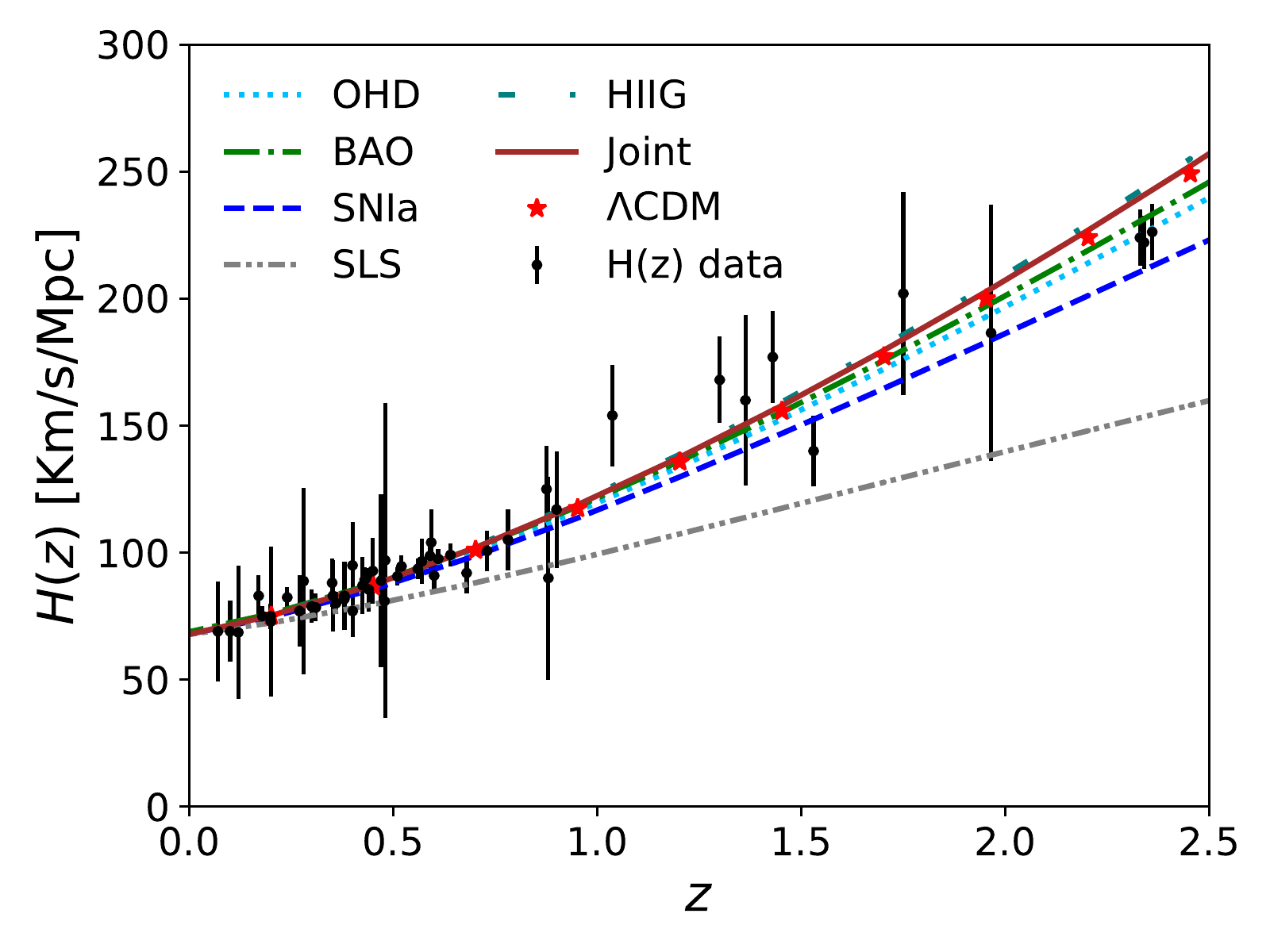}\\
\includegraphics[width=0.45\textwidth]{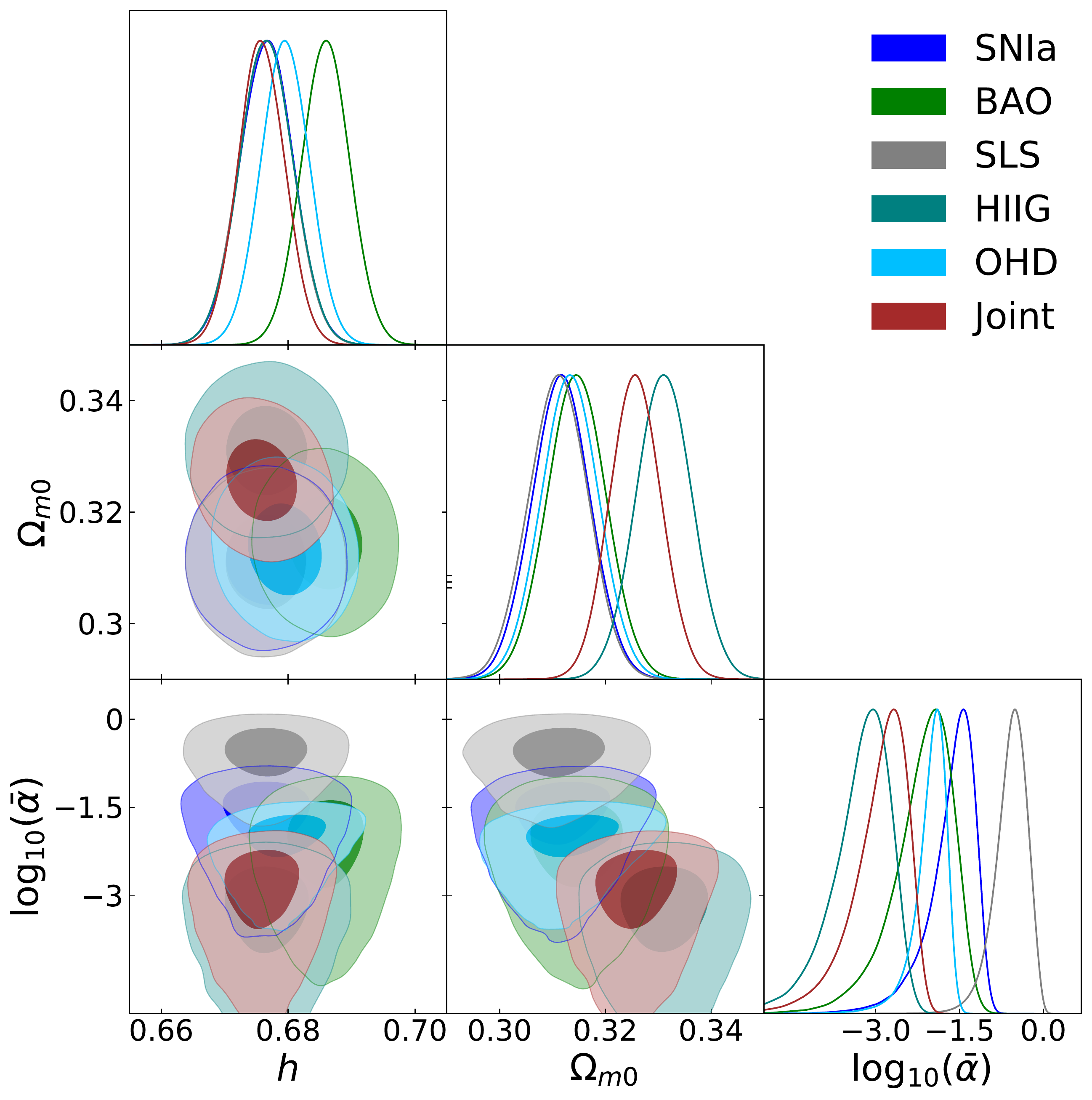}
\caption{Top panel: Best fit curve of Einstein-Gauss-Bonnet model and its uncertainty at $1\sigma$. Bottom panel: 2D contours of the free model parameters at $1\sigma, 2\sigma$, and $3\sigma$ (from darker to lighter color bands) CL using OHD, SNIa (Pantheon), SLS, HIIG, BAO and Joint data. Figure adapted from \cite{Garcia-Aspeitia:2020uwq}.}
\label{fig:egb}
\end{figure}

Fig. \ref{fig:egb} show the MCMC analysis implemented using OHD, SNIa, SLS, HIIG and BAO samples. The upper panel shows the reconstruction of $H(z)$ and the bottom panel presents the CL contours for the free parameter of the theory $\bar{\alpha}$. In addition, best-fits for the free parameters if the model are presented in Table \ref{tab:EGB} in conjunction with the $\chi^2$ parameter (see details in \cite{Garcia-Aspeitia:2020uwq}).

\begin{table}
\caption{Best fitting values of the free parameters for the EGB model with the different samples used in this paper. \cite[see][for details]{Garcia-Aspeitia:2020uwq}.}
\centering
\begin{tabular}{|cccccc|}
\hline
Sample     &    $\chi^2$     &  $h$ & $\Omega_{m0}$ & $\bar{\alpha}$ &  $\mathcal{M}$          \\
\hline
 OHD   & $25.8$ & $0.677^{+0.004}_{-0.004}$ & $0.312^{+0.005}_{-0.005}$ & $0.011^{+0.007}_{-0.005}$ & --  \\ [0.7ex]
 BAO    & $40.7$ & $0.686^{+0.004}_{-0.004}$ & $0.315^{+0.006}_{-0.006}$ & $0.008^{+0.014}_{-0.006}$ & -- \\ [0.7ex]
SNIa    & $39.8$ & $0.677^{+0.004}_{-0.004}$ & $0.312^{+0.005}_{-0.005}$ & $0.028^{+0.028}_{-0.018}$ &  $-19.400^{+0.016}_{- 0.016}$   \\ [0.7ex]
SLS     & $577.8$  & $0.676^{+0.004}_{-0.004}$ & $0.311^{+0.006}_{-0.006}$ & $0.281^{+0.218}_{-0.143}$  & --    \\ [0.7ex]
HIIG    & $2269.5$ & $0.677^{+0.004}_{-0.004}$ & $0.331^{+0.005}_{-0.005}$ & $0.0006^{+0.0011}_{-0.0005}$ & --    \\ [0.7ex]
 Joint  & $6181.9$ & $0.676^{+0.004}_{-0.004}$ & $0.326^{+0.005}_{-0.005}$ & $0.001^{+0.002}_{-0.001}$ & $-19.400^{+0.012}_{- 0.012}$ \\ [0.7ex]
\hline
\end{tabular}
\label{tab:EGB}
\end{table}

%%%%%%%%%%%%%%%%%%%%%%%%%%%%%%%%%%%%%%%%%%%%%
\subsubsection{Cardassian Models}

Cardassian\footnote{The name Cardassian comes from  alien creatures shown in the television series Star Trek.} models are a phenomenological form to describe the late time acceleration of the Universe that could be theoretically sustained under the assumption of extra dimensions. The models assume an extra function in the Friedmann equations whose form is related to those of the polytropic fluids. Therefore, just the presence of matter and radiation in this functional form  automatically produce a late time Universe acceleration without the need for a CC. The main disadvantage is the apparition of additional terms that must be constrained with observations and the difficulties to describe the model from theoretical arguments. It is important to mention that the Cardassian models are divided into the Original Cardassian (OC) and the Modified Polytropic Cardassian (MPC). 

The theoretical details of OC model, introduced by Ref. \cite{FREESE20021}, is given by the following Friedmann equation
\begin{eqnarray}
   && E(z)^2=\Omega_{m0}(1+z)^3+\Omega_{r0}(1+z)^4+(1-\Omega_{m0}-\Omega_{r0})\nonumber\\&&\times\left(\frac{\Omega_{m0}(1+z)^3+\Omega_{r0}(1+z)^4}{\Omega_{m0}+\Omega_{r0}}\right)^n,
\end{eqnarray}
where $n$ is a free parameter. Despite OC model rise from phenomenological assumption in order to obtain an accelerated Universe, it is important to notice that the mathematical structure can be strictly obtained from brane world models with n-branes immersed in a five dimensional bulk.

\begin{figure}
\centering
\includegraphics[width=0.36\textwidth]{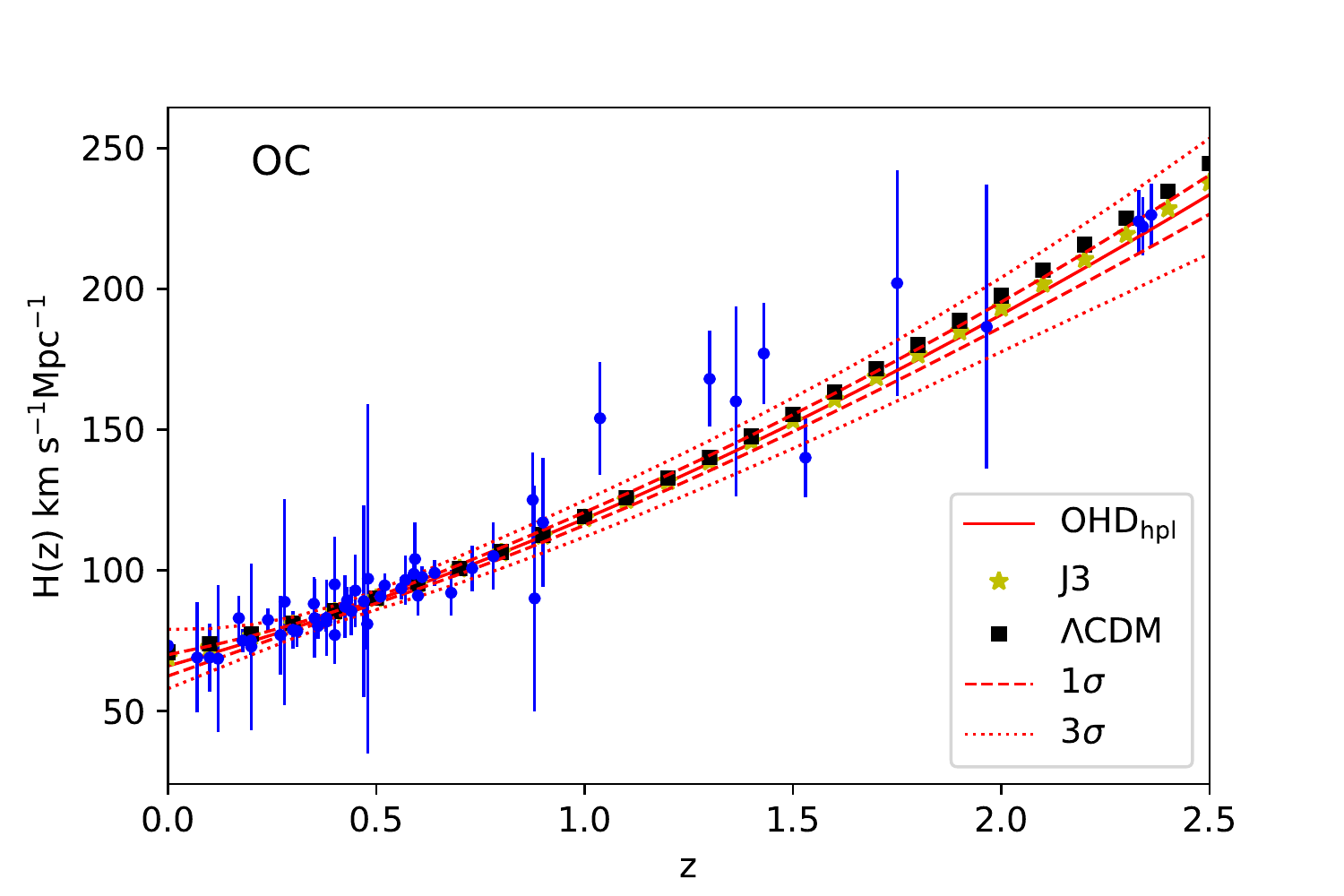}
\includegraphics[width=0.36\textwidth]{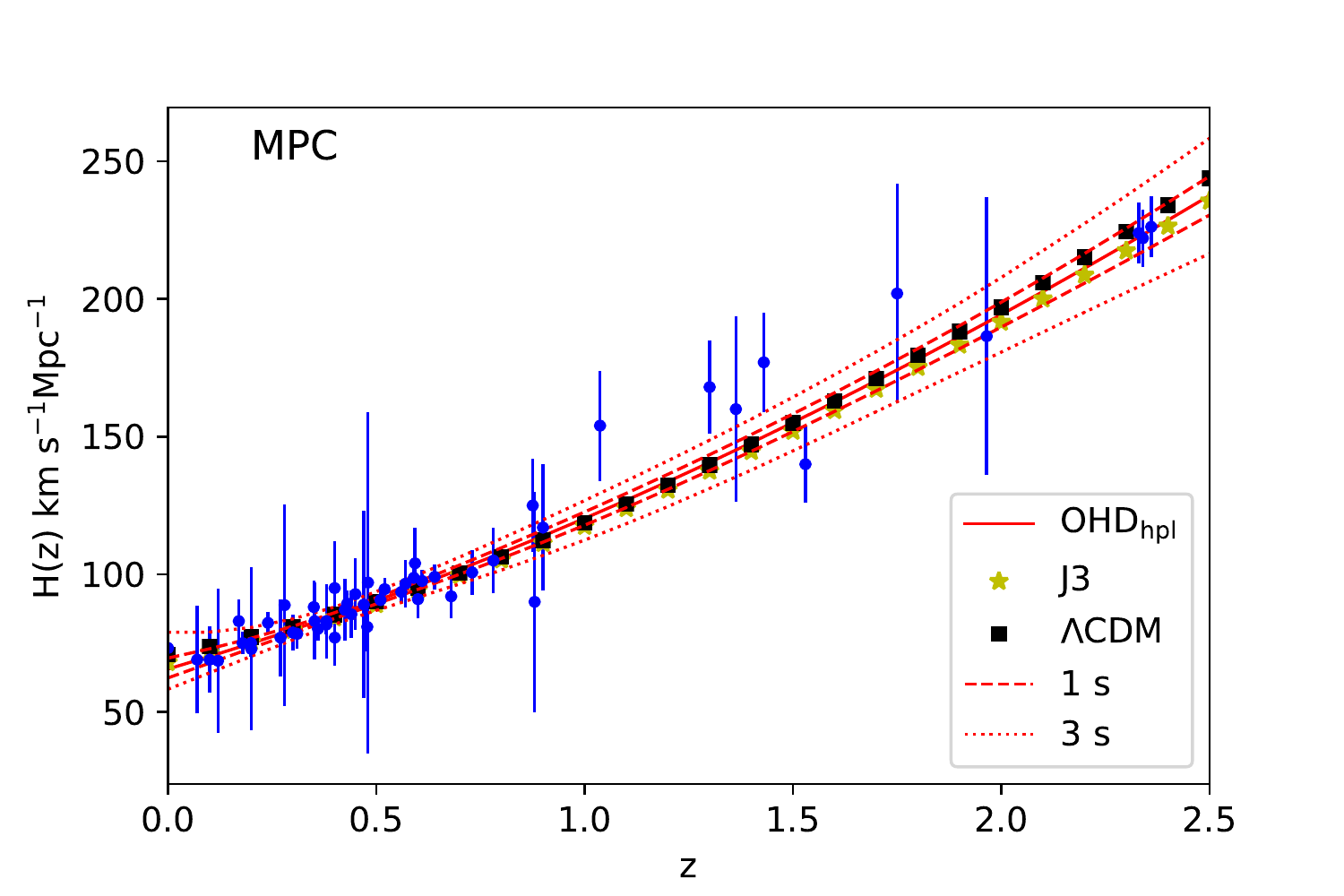}\\
\includegraphics[width=0.36\textwidth]{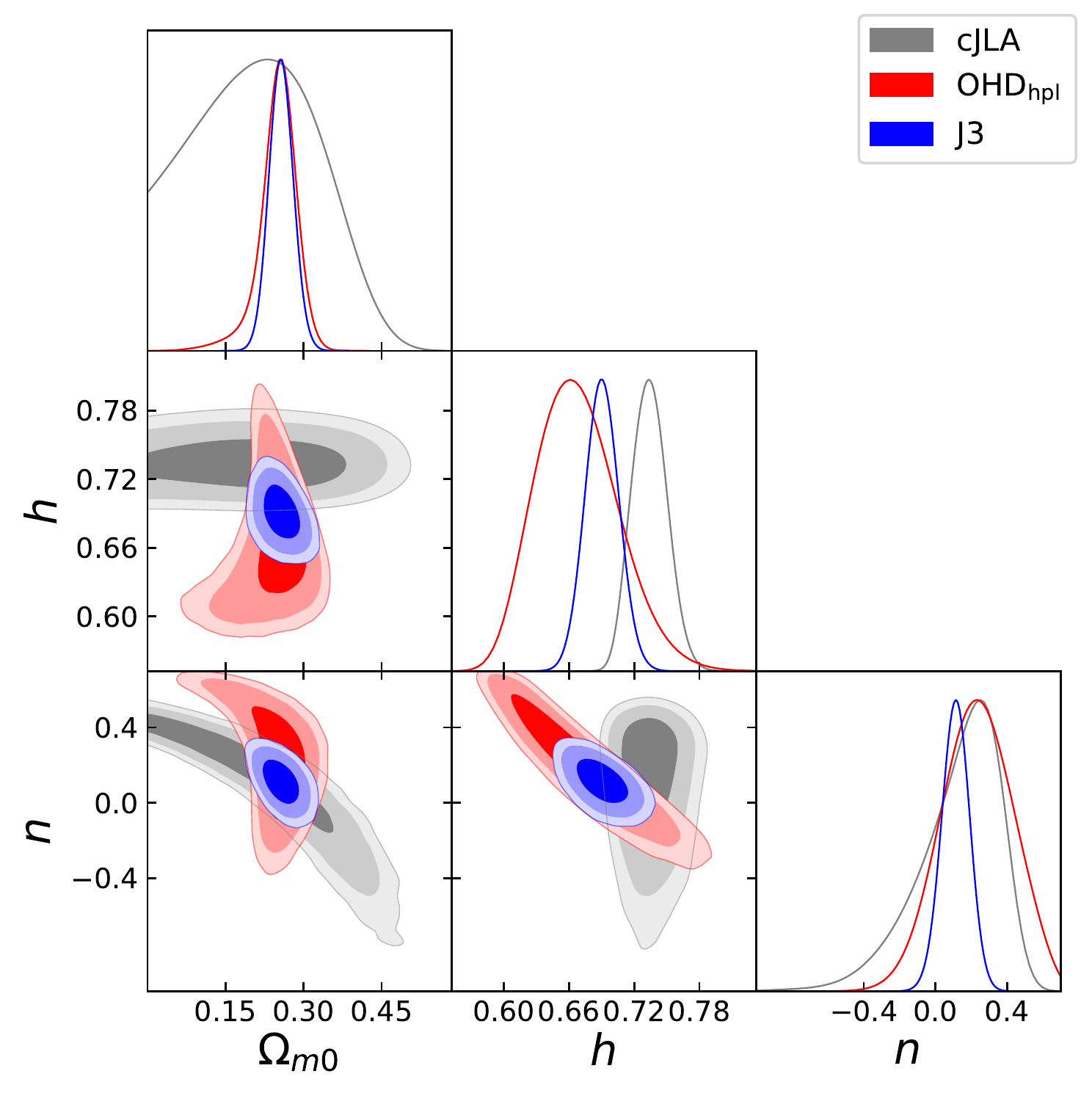}
\includegraphics[width=0.36\textwidth]{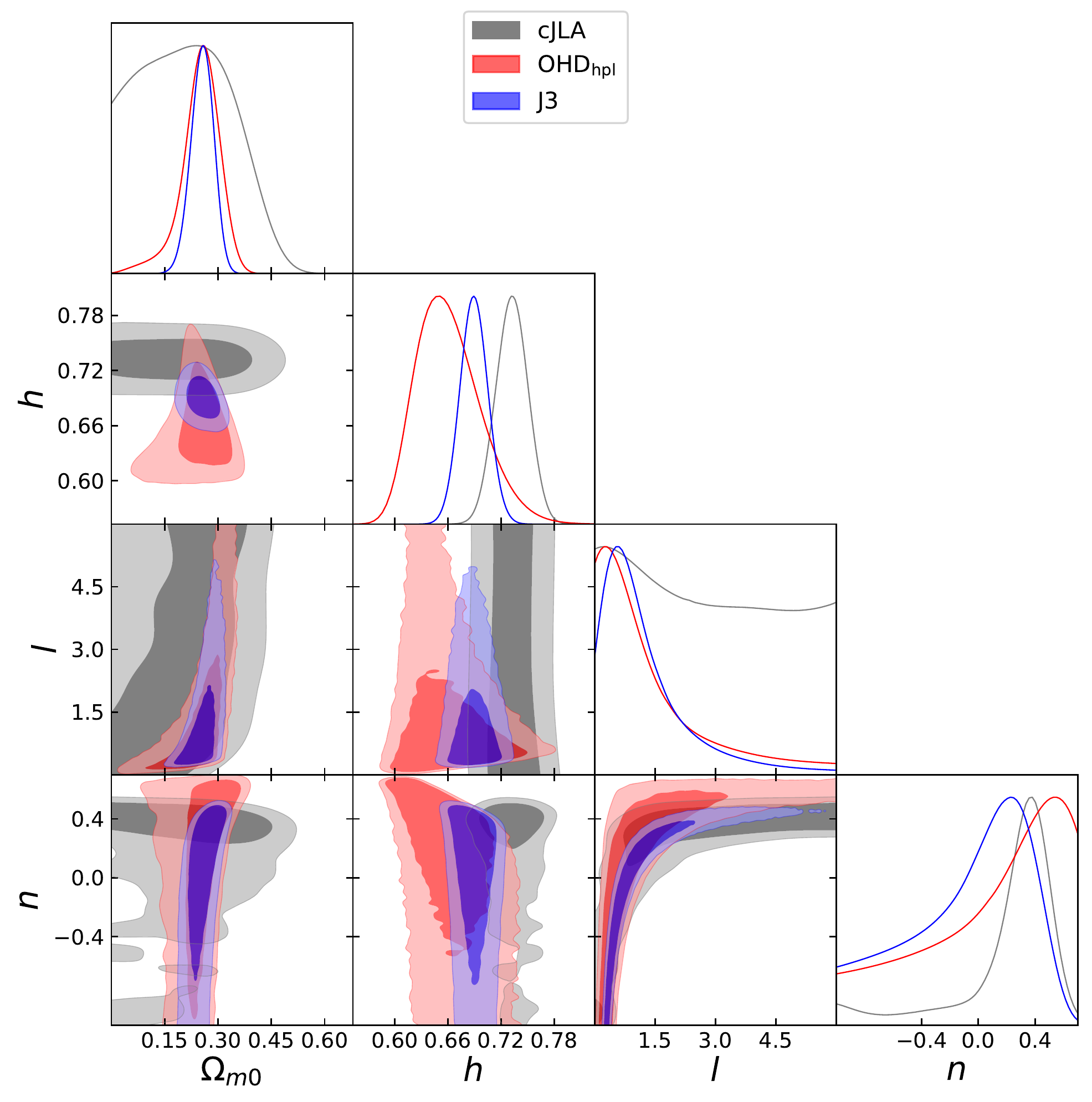}
\caption{Top panel: Best fit curve of $H(z)$ for the original and modified polytropic Cardassian models and uncertainties at $1\sigma$ and $1\sigma$. Bottom panel: 2D contours of the free model parameters at $1\sigma, 2\sigma$, and $3\sigma$ (from darker to lighter color bands) CL using SNIa (cJLA, compressed JLA), OHD, and the Joint analysis of both data (denoted J3).}
\label{fig:Cardassian}
\end{figure}

In addition, the MPC model \cite{Gondolo:2002fh}, takes the form
\begin{eqnarray}
E(z)^2=\Omega_{r0}(1+z)^4+\Omega_{m0}(1+z)^3\beta(z)^{1/l},
\end{eqnarray}
where
\begin{eqnarray}
\beta(z)\equiv1+\left[\left(\frac{1-\Omega_{r0}}{\Omega_{m0}}\right)^l-1\right](1+z)^{3/(n-1)},
\end{eqnarray}
here $l$ and $n$ are free parameters.

Table \ref{tab:Cardassians} gives the minimum chi-square and mean values for the OC and MPC parameters using the samples OHD (DA), SNIa (compressed JLA), and the joint analysis of these data sets.
Fig. \ref{fig:Cardassian}  shows the best fit of $H(z)$ for the original (left panel) and modified polytropic Cardassian models (right), respectively. The figures also show the confidence contours (bottom panels) for both Cardassian models using SNIa (cJLA, compressed JLA), OHD uniformed with the sound horizon estimation from Planck data, and the Joint analysis of both data (denoted J3). 
It is worth to note that both models can reproduce the dynamics of the Hubble measurements.
\begin{table}[htb]
\caption{Mean values for the OC and MPC parameters using the samples OHD (DA), SNIa (compressed JLA), and the joint analysis of these data sets. \cite[see][for details]{Magana:2017nfs}.}
\centering
\begin{tabular}{|lccccc|}
\hline
\multicolumn{6}{|c|}{Orignal Cardassian}\\
\hline
Data & $\chi^{2}_{min}$&$\Omega_{m}$& n&$l$&$h$\\
\hline
\multicolumn{6}{|c|}{}\\
OHD (DA) & $15.22$ & $0.30^{+0.06}_{-0.06}$ & $-0.19^{+0.51}_{-0.50}$ & $--$&$0.69^{+0.06}_{-0.05}$ \\
SNIa & $32.95$ & $0.22^{+0.11}_{-0.12}$ & $0.16^{+0.17}_{-0.26}$ & $--$&$0.72^{+0.19}_{-0.19}$\\
Joint & $54.28$ & $0.25^{+0.02}_{-0.02}$ & $0.11^{+0.07}_{-0.07}$ &$--$&$0.69^{+0.01}_{-0.01}$\\
\hline
\multicolumn{6}{|c|}{Modified polytropic Cardassian}\\
\hline
\multicolumn{6}{|c|}{}\\
OHD (DA) & $17.95$ & $0.32^{+0.06}_{-0.07}$ & $0.10^{+0.38}_{-0.60}$ & $2.13^{+2.34}_{-1.33}$&$0.68^{+0.07}_{-0.05}$ \\
SNIa & $33.76$ & $0.22^{+0.12}_{-0.13}$ & $0.36^{+0.07}_{-0.33}$ & $2.61^{+2.27}_{-1.83}$&$0.72^{+0.18}_{-0.19}$\\
Joint & $54.23$ & $0.25^{+0.03}_{-0.03}$ & $0.06^{+0.29}_{-0.58}$ &$0.89^{+1.29}_{-0.47}$&$0.68^{+0.01}_{-0.01}$\\
\hline
\end{tabular}
\label{tab:Cardassians}
\end{table}

%%%%%%%%%%%%%%%%%%%%%%%%%%%%%%%%%%%%%%%%%%
\section{Discussion and Conclusions} \label{DC}

In this review we presented a brief and non-exhaustive review on DE models that however summarizes some important scientific results,  settling the ground for the ideas exposed in this work.

Our motivation to explore alternatives to $\Lambda$CDM lies on the problems that afflicts the cosmological constant and the recently observed tension in the $\sigma_8$ and $H_0$ parameters estimated from different samples. Notice how the CC is directly related to the problem of the Universe acceleration and, therefore a competitive model should be implemented. However, regarding the $H_0$ tension, it is not clear whether it is related to dark energy or not \cite[see for example][]{Efstathiou:2021}, although some authors assume that this is the case \cite[][]{DiValentino:2021izs}.

For each dark energy presented in this review,  we have summarized its main characteristics, its theoretical structure and in its capabilities to fit the data provided by modern observations. Parameterizations are always the standard form to tackle the problem of Universe acceleration but there is not a unique way of choosing the form of the function. Furthermore, in many cases there are not solid arguments to justify the chosen form and it is anfractuous to associate the parameterization with some quantum field or with a model that modifies the GR. Models like Chaplygin and Viscous encompass the dark energy and dark matter contribution in just one theoretical framework through a diffusion function in the continuity equations. The theoretical background is robust, with its equations deduced from a quantum field theory in which a scalar field is involved. In addition, the diverse data samples tend to prefer the mentioned models over others with extra complexities. 

We also described models with a late apparition of the dark energy (CC always exist in the Universe evolution). The PEDE, GEDE and UG models allow us to estimate the birth of DE at reionization epoch. These kind of models are valuable because their mathematical expression is able to resolve the tension between Planck and SNIa data. Observational constraints also favour them over other models. Furthermore, extra dimensional models contain a theoretical complexity that undermines its recognition, for instance, the RS model has severe faults when it is contrasted with observations. However, the addition of extra degrees of freedom to obtain a variable brane tension reduce the disagreement with observations and shed light on the nature of CC, i.e. it comes from the existence of extra dimensions. The disadvantage is that the CC problem is now dragged to the extra dimensions scenario, introducing difficulties to calculate the expected value in our four-dimensional Universe. EGB model surge as a curiosity in recent literature but, after a more rigorous mathematical exam, there are several notorious flaws in its theoretical arguments. In addition, from a cosmological point of view, it is based on an spurious early acceleration that could cause severe problems with the well-known characteristics of the Universe in epochs such as the nucleosynthesis. Constraints on EGB obtained with the SLS data sample and applied to dynamical systems point out that the early acceleration never ends. Finally, Cardassian models are phenomenological models that could be justified by the assumption of extra dimensions. Although this complicate the equations, they have the advantage of avoiding tensions when contrasted with observations. Notice that, as expected, Cardassian models tend to reproduce the CC, not showing a dynamical DE.

As a final remark, we would like to reflect on the $H_0$ tension considering the  different models discussed in this review. In Table \ref{tab:H0allModels} we present the compilation of the dimensionless Hubble parameter ($h$) for all the models  with their respective selected samples and priors. We have used a Gaussian prior in most of the cases using Riess or Planck data \cite{Riess_2018,Planck_CP:2018}, while only in four models we consider a flat prior. For the purpose of the following discussion, we will compare only those models with flat priors which are CPL, UG and both Cardassian models. Only those models using joint samples that include CMB are able to constraint $H_0$ and, thus the best-fit value is in agreement with Planck's result. 
However this is not the final word about the $H_0$ tension, \citep{Efstathiou:2021} has recently suggested this might be related to a misunderstanding of the distance ladder measurements (i.e. a need for a better agreement between the SNIa absolute magnitude and the Cepheid-based distance ladder) instead of an 'exotic late-time physics'.

Finally, we summarize our results in the following way. The Brane model with constant tension deduced from the RS paradigm and the EGB have several failures which may call into question the viability of both models. Variable brane tension, Cardassian and viscous models can be  considered as promising having problems with the complexity of their theoretical background. Parameterizations, Chaplygin, UG, PEDE, and GEDE are excellent competitors against the standard paradigm, with a strong possibility of resolving the $H_0$ tension and contributing to elucidate the nature of the dark energy  component, the Universe acceleration, and its possible consequences in the fate of our Universe.

\begin{table*}[thb]
\caption{Best fitting values of the $h$ parameter for all the models presented.}
\centering
\begin{tabular}{|cccc|}
\hline
Model     &    $h$  & Sample   &  Prior \\
\hline
$w$CDM   &  $0.68^{+0.01}_{-0.01}$ & OHD, SNIa, CMB & Gaussian ($0.6766^{+0.0042}_{-0.0042})$ \\ [0.7ex]
CPL    & $0.73^{+0.10}_{-0.08}$ & OHD  & Flat [0.2,1.0]\\ [0.7ex]
JBP    & $0.71^{+0.014}_{-0.014}$ & OHD, CMB, BAO & Gaussian ($0.73^{+0.0175}_{-0.0175}$) \\ [0.7ex]
BA     & $0.71^{+0.015}_{-0.015}$  & OHD, CMB, BAO & Gaussian ($0.73^{+0.0175}_{-0.0175}$)\\ [0.7ex]
FSLLI    & $0.71^{+0.015}_{-0.015}$ & OHD, CMB, BAO & Gaussian ($0.73^{+0.0175}_{-0.0175}$)\\ [0.7ex]
FSLLII    & $0.70^{+0.014}_{-0.014}$ & OHD, CMB, BAO & Gaussian ($0.73^{+0.0175}_{-0.0175}$)\\ [0.7ex]
SL  & $0.70^{+0.015}_{-0.015}$ & OHD, CMB, BAO & Gaussian ($0.73^{+0.0175}_{-0.0175}$)\\ [0.7ex]
Chaplygin-Like Fluid & $0.71^{+0.014}_{-0.014}$ & OHD, SNIa & Gaussian ($0.723^{+0.017}_{-0.017}$)\\ [0.7ex]
Viscous (Polynomial)    & $0.70^{+0.009}_{-0.009}$ & OHD, SNIa & Gaussian ($0.7324^{+0.0174}_{-0.0174}$)\\ [0.7ex]
Viscous (Hyperbolic tanh)    & $0.69^{+0.009}_{-0.009}$ &  OHD, SNIa & Gaussian ($0.7324^{+0.0174}_{-0.0174}$) \\ [0.7ex]
Viscous (Hyperbolic cosh)    &  $0.70^{+0.009}_{-0.010}$ &  OHD, SNIa & Gaussian ($0.7324^{+0.0174}_{-0.0174}$) \\ [0.7ex]
Viscous ($\xi_0=$Constant)    & $0.68^{+0.004}_{-0.004}$ & OHD, SNIa, SLS & Gaussian ($0.6766^{+0.0042}_{-0.0042}$)\\ [0.7ex]
Viscous ($\xi_0=$Polynomial)    & $0.67^{+0.004}_{-0.004}$ & OHD, SNIa, SLS & Gaussian ($0.6766^{+0.0042}_{-0.0042}$)\\ [0.7ex]
Interacting Viscous    & $0.70^{+0.012}_{-0.013}$ & OHD & Gaussian ($0.7324^{+0.0174}_{-0.0174}$)\\ [0.7ex]
PEDE   & $0.74^{+0.011}_{-0.011}$ & Homogeneous OHD & Gaussian ($0.7403^{+0.0142}_{-0.0142}$) \\ [0.7ex]
GEDE    & $0.73^{+0.012}_{-0.012}$ & Homogeneous OHD & Gaussian ($0.7403^{+0.0142}_{-0.0142}$)\\ [0.7ex]
CBT    & $0.71^{+0.01}_{-0.01}$ & BAO, SNIa, OHD, CMB & Gaussian ($0.7324^{+0.0174}_{-0.0174}$) \\ [0.7ex]
VBT   & $0.70^{+0.009}_{-0.009}$ & OHD, SNIa, BAO, BAO & Gaussian ($0.7324^{+0.0174}_{-0.0174}$)\\ [0.7ex]
UG   & $0.69^{+0.005}_{-0.005}$ &  OHD, SNIa, CMB, BAO & Flat [0.2,1.0]\\ [0.7ex]
EGB   & $0.67^{+0.004}_{-0.004}$ &  SNIa, BAO, OHD, SLS, HIIG & Gaussian ($0.6766^{+0.0042}_{-0.0042}$) \\ [0.7ex]
Original Cardassian   & $0.69^{+0.01}_{-0.01}$ & SNIa, OHD & Flat [0,1]\\ [0.7ex]
Modified Polytropic Cardassian   & $0.68^{+0.01}_{-0.01}$ & SNIa, OHD & Flat [0,1]\\ [0.7ex]
\hline
\end{tabular}
\label{tab:H0allModels}
\end{table*}

\begin{acknowledgments}
We thank the anonymous referee for thoughtful remarks and suggestions. The authors are grateful to Mario H. Amante for the contribution to graphic material. V.M. acknowledges support from Centro de Astrof\'{\i}sica de Valpara\'{\i}so (CAV) and ANID REDES (190147). M.A.G.-A. acknowledges support from CONACYT research fellow, Sistema Nacional de Investigadores (SNI), COZCYT and Instituto Avanzado de Cosmolog\'ia (IAC) collaborations. AHA thanks support from SNI-M\'exico. J.M. acknowledges support from CONICYT project Basal AFB-170002 and CONICYT/FONDECYT 3160674.
\end{acknowledgments}

\bibliography{librero2.bib}

\begin{thebibliography}{156}%
\makeatletter
\providecommand \@ifxundefined [1]{%
 \@ifx{#1\undefined}
}%
\providecommand \@ifnum [1]{%
 \ifnum #1\expandafter \@firstoftwo
 \else \expandafter \@secondoftwo
 \fi
}%
\providecommand \@ifx [1]{%
 \ifx #1\expandafter \@firstoftwo
 \else \expandafter \@secondoftwo
 \fi
}%
\providecommand \natexlab [1]{#1}%
\providecommand \enquote  [1]{``#1''}%
\providecommand \bibnamefont  [1]{#1}%
\providecommand \bibfnamefont [1]{#1}%
\providecommand \citenamefont [1]{#1}%
\providecommand \href@noop [0]{\@secondoftwo}%
\providecommand \href [0]{\begingroup \@sanitize@url \@href}%
\providecommand \@href[1]{\@@startlink{#1}\@@href}%
\providecommand \@@href[1]{\endgroup#1\@@endlink}%
\providecommand \@sanitize@url [0]{\catcode `\\12\catcode `\$12\catcode
  `\&12\catcode `\#12\catcode `\^12\catcode `\_12\catcode `\%12\relax}%
\providecommand \@@startlink[1]{}%
\providecommand \@@endlink[0]{}%
\providecommand \url  [0]{\begingroup\@sanitize@url \@url }%
\providecommand \@url [1]{\endgroup\@href {#1}{\urlprefix }}%
\providecommand \urlprefix  [0]{URL }%
\providecommand \Eprint [0]{\href }%
\providecommand \doibase [0]{http://dx.doi.org/}%
\providecommand \selectlanguage [0]{\@gobble}%
\providecommand \bibinfo  [0]{\@secondoftwo}%
\providecommand \bibfield  [0]{\@secondoftwo}%
\providecommand \translation [1]{[#1]}%
\providecommand \BibitemOpen [0]{}%
\providecommand \bibitemStop [0]{}%
\providecommand \bibitemNoStop [0]{.\EOS\space}%
\providecommand \EOS [0]{\spacefactor3000\relax}%
\providecommand \BibitemShut  [1]{\csname bibitem#1\endcsname}%
\let\auto@bib@innerbib\@empty
%</preamble>
\bibitem [{\citenamefont {{Lema{\^\i}tre}}(1927)}]{Lemaitre:1927}%
  \BibitemOpen
  \bibfield  {author} {\bibinfo {author} {\bibfnamefont {G.}~\bibnamefont
  {{Lema{\^\i}tre}}},\ }\href@noop {} {\bibfield  {journal} {\bibinfo
  {journal} {Annales de la Soci\&eacute;t\&eacute; Scientifique de Bruxelles}\
  }\textbf {\bibinfo {volume} {47}},\ \bibinfo {pages} {49} (\bibinfo {year}
  {1927})}\BibitemShut {NoStop}%
\bibitem [{\citenamefont {{Hubble}}(1929)}]{Hubble:1929}%
  \BibitemOpen
  \bibfield  {author} {\bibinfo {author} {\bibfnamefont {E.}~\bibnamefont
  {{Hubble}}},\ }\href {\doibase 10.1073/pnas.15.3.168} {\bibfield  {journal}
  {\bibinfo  {journal} {Proceedings of the National Academy of Science}\
  }\textbf {\bibinfo {volume} {15}},\ \bibinfo {pages} {168} (\bibinfo {year}
  {1929})}\BibitemShut {NoStop}%
\bibitem [{\citenamefont {{Penzias}}\ and\ \citenamefont
  {{Wilson}}(1965)}]{Penzias:1965}%
  \BibitemOpen
  \bibfield  {author} {\bibinfo {author} {\bibfnamefont {A.~A.}\ \bibnamefont
  {{Penzias}}}\ and\ \bibinfo {author} {\bibfnamefont {R.~W.}\ \bibnamefont
  {{Wilson}}},\ }\href {\doibase 10.1086/148307} {\bibfield  {journal}
  {\bibinfo  {journal} {ApJ}\ }\textbf {\bibinfo {volume} {142}},\ \bibinfo
  {pages} {419} (\bibinfo {year} {1965})}\BibitemShut {NoStop}%
\bibitem [{\citenamefont {{Dicke}}\ \emph {et~al.}(1965)\citenamefont
  {{Dicke}}, \citenamefont {{Peebles}}, \citenamefont {{Roll}},\ and\
  \citenamefont {{Wilkinson}}}]{Dicke:1965}%
  \BibitemOpen
  \bibfield  {author} {\bibinfo {author} {\bibfnamefont {R.~H.}\ \bibnamefont
  {{Dicke}}}, \bibinfo {author} {\bibfnamefont {P.~J.~E.}\ \bibnamefont
  {{Peebles}}}, \bibinfo {author} {\bibfnamefont {P.~G.}\ \bibnamefont
  {{Roll}}}, \ and\ \bibinfo {author} {\bibfnamefont {D.~T.}\ \bibnamefont
  {{Wilkinson}}},\ }\href {\doibase 10.1086/148306} {\bibfield  {journal}
  {\bibinfo  {journal} {ApJ}\ }\textbf {\bibinfo {volume} {142}},\ \bibinfo
  {pages} {414} (\bibinfo {year} {1965})}\BibitemShut {NoStop}%
\bibitem [{\citenamefont {Riess}\ \emph {et~al.}(1998)\citenamefont {Riess},
  \citenamefont {Filippenko}, \citenamefont {Challis}, \citenamefont
  {Clocchiatti}, \citenamefont {Diercks} \emph {et~al.}}]{Riess:1998}%
  \BibitemOpen
  \bibfield  {author} {\bibinfo {author} {\bibfnamefont {A.~G.}\ \bibnamefont
  {Riess}}, \bibinfo {author} {\bibfnamefont {A.~V.}\ \bibnamefont
  {Filippenko}}, \bibinfo {author} {\bibfnamefont {P.}~\bibnamefont {Challis}},
  \bibinfo {author} {\bibfnamefont {A.}~\bibnamefont {Clocchiatti}}, \bibinfo
  {author} {\bibfnamefont {A.}~\bibnamefont {Diercks}},  \emph {et~al.},\
  }\href {http://stacks.iop.org/1538-3881/116/i=3/a=1009} {\bibfield  {journal}
  {\bibinfo  {journal} {The Astronomical Journal}\ }\textbf {\bibinfo {volume}
  {116}},\ \bibinfo {pages} {1009} (\bibinfo {year} {1998})}\BibitemShut
  {NoStop}%
\bibitem [{\citenamefont {Perlmutter}\ \emph {et~al.}(1999)\citenamefont
  {Perlmutter}, \citenamefont {Aldering}, \citenamefont {Goldhaber},
  \citenamefont {Knop}, \citenamefont {Nugent}, \citenamefont {others},\ and\
  \citenamefont {Project}}]{Perlmutter:1999}%
  \BibitemOpen
  \bibfield  {author} {\bibinfo {author} {\bibfnamefont {S.}~\bibnamefont
  {Perlmutter}}, \bibinfo {author} {\bibfnamefont {G.}~\bibnamefont
  {Aldering}}, \bibinfo {author} {\bibfnamefont {G.}~\bibnamefont {Goldhaber}},
  \bibinfo {author} {\bibfnamefont {R.~A.}\ \bibnamefont {Knop}}, \bibinfo
  {author} {\bibfnamefont {P.}~\bibnamefont {Nugent}}, \bibinfo {author}
  {\bibnamefont {others}}, \ and\ \bibinfo {author} {\bibfnamefont {T.~S.~C.}\
  \bibnamefont {Project}},\ }\href
  {http://stacks.iop.org/0004-637X/517/i=2/a=565} {\bibfield  {journal}
  {\bibinfo  {journal} {The Astrophysical Journal}\ }\textbf {\bibinfo {volume}
  {517}},\ \bibinfo {pages} {565} (\bibinfo {year} {1999})}\BibitemShut
  {NoStop}%
\bibitem [{\citenamefont {{de Bernardis}}\ \emph {et~al.}(2000)\citenamefont
  {{de Bernardis}}, \citenamefont {{Ade}}, \citenamefont {{Bock}},
  \citenamefont {{Bond}}, \citenamefont {{Borrill}}, \citenamefont
  {{Boscaleri}}, \citenamefont {{Coble}}, \citenamefont {{Crill}},
  \citenamefont {{De Gasperis}}, \citenamefont {{Farese}}, \citenamefont
  {{Ferreira}}, \citenamefont {{Ganga}}, \citenamefont {{Giacometti}},
  \citenamefont {{Hivon}}, \citenamefont {{Hristov}}, \citenamefont
  {{Iacoangeli}}, \citenamefont {{Jaffe}}, \citenamefont {{Lange}},
  \citenamefont {{Martinis}}, \citenamefont {{Masi}}, \citenamefont {{Mason}},
  \citenamefont {{Mauskopf}}, \citenamefont {{Melchiorri}}, \citenamefont
  {{Miglio}}, \citenamefont {{Montroy}}, \citenamefont {{Netterfield}},
  \citenamefont {{Pascale}}, \citenamefont {{Piacentini}}, \citenamefont
  {{Pogosyan}}, \citenamefont {{Prunet}}, \citenamefont {{Rao}}, \citenamefont
  {{Romeo}}, \citenamefont {{Ruhl}}, \citenamefont {{Scaramuzzi}},
  \citenamefont {{Sforna}},\ and\ \citenamefont
  {{Vittorio}}}]{deBernardis:2000}%
  \BibitemOpen
  \bibfield  {author} {\bibinfo {author} {\bibfnamefont {P.}~\bibnamefont {{de
  Bernardis}}}, \bibinfo {author} {\bibfnamefont {P.~A.~R.}\ \bibnamefont
  {{Ade}}}, \bibinfo {author} {\bibfnamefont {J.~J.}\ \bibnamefont {{Bock}}},
  \bibinfo {author} {\bibfnamefont {J.~R.}\ \bibnamefont {{Bond}}}, \bibinfo
  {author} {\bibfnamefont {J.}~\bibnamefont {{Borrill}}}, \bibinfo {author}
  {\bibfnamefont {A.}~\bibnamefont {{Boscaleri}}}, \bibinfo {author}
  {\bibfnamefont {K.}~\bibnamefont {{Coble}}}, \bibinfo {author} {\bibfnamefont
  {B.~P.}\ \bibnamefont {{Crill}}}, \bibinfo {author} {\bibfnamefont
  {G.}~\bibnamefont {{De Gasperis}}}, \bibinfo {author} {\bibfnamefont {P.~C.}\
  \bibnamefont {{Farese}}}, \bibinfo {author} {\bibfnamefont {P.~G.}\
  \bibnamefont {{Ferreira}}}, \bibinfo {author} {\bibfnamefont
  {K.}~\bibnamefont {{Ganga}}}, \bibinfo {author} {\bibfnamefont
  {M.}~\bibnamefont {{Giacometti}}}, \bibinfo {author} {\bibfnamefont
  {E.}~\bibnamefont {{Hivon}}}, \bibinfo {author} {\bibfnamefont {V.~V.}\
  \bibnamefont {{Hristov}}}, \bibinfo {author} {\bibfnamefont {A.}~\bibnamefont
  {{Iacoangeli}}}, \bibinfo {author} {\bibfnamefont {A.~H.}\ \bibnamefont
  {{Jaffe}}}, \bibinfo {author} {\bibfnamefont {A.~E.}\ \bibnamefont
  {{Lange}}}, \bibinfo {author} {\bibfnamefont {L.}~\bibnamefont {{Martinis}}},
  \bibinfo {author} {\bibfnamefont {S.}~\bibnamefont {{Masi}}}, \bibinfo
  {author} {\bibfnamefont {P.~V.}\ \bibnamefont {{Mason}}}, \bibinfo {author}
  {\bibfnamefont {P.~D.}\ \bibnamefont {{Mauskopf}}}, \bibinfo {author}
  {\bibfnamefont {A.}~\bibnamefont {{Melchiorri}}}, \bibinfo {author}
  {\bibfnamefont {L.}~\bibnamefont {{Miglio}}}, \bibinfo {author}
  {\bibfnamefont {T.}~\bibnamefont {{Montroy}}}, \bibinfo {author}
  {\bibfnamefont {C.~B.}\ \bibnamefont {{Netterfield}}}, \bibinfo {author}
  {\bibfnamefont {E.}~\bibnamefont {{Pascale}}}, \bibinfo {author}
  {\bibfnamefont {F.}~\bibnamefont {{Piacentini}}}, \bibinfo {author}
  {\bibfnamefont {D.}~\bibnamefont {{Pogosyan}}}, \bibinfo {author}
  {\bibfnamefont {S.}~\bibnamefont {{Prunet}}}, \bibinfo {author}
  {\bibfnamefont {S.}~\bibnamefont {{Rao}}}, \bibinfo {author} {\bibfnamefont
  {G.}~\bibnamefont {{Romeo}}}, \bibinfo {author} {\bibfnamefont {J.~E.}\
  \bibnamefont {{Ruhl}}}, \bibinfo {author} {\bibfnamefont {F.}~\bibnamefont
  {{Scaramuzzi}}}, \bibinfo {author} {\bibfnamefont {D.}~\bibnamefont
  {{Sforna}}}, \ and\ \bibinfo {author} {\bibfnamefont {N.}~\bibnamefont
  {{Vittorio}}},\ }\href {\doibase 10.1038/35010035} {\bibfield  {journal}
  {\bibinfo  {journal} {Nat.}\ }\textbf {\bibinfo {volume} {404}},\ \bibinfo
  {pages} {955} (\bibinfo {year} {2000})},\ \Eprint
  {http://arxiv.org/abs/astro-ph/0004404} {arXiv:astro-ph/0004404 [astro-ph]}
  \BibitemShut {NoStop}%
\bibitem [{\citenamefont {{Spergel}}\ \emph {et~al.}(2003)\citenamefont
  {{Spergel}}, \citenamefont {{Verde}}, \citenamefont {{Peiris}}, \citenamefont
  {{Komatsu}}, \citenamefont {{Nolta}}, \citenamefont {{Bennett}},
  \citenamefont {{Halpern}}, \citenamefont {{Hinshaw}}, \citenamefont
  {{Jarosik}}, \citenamefont {{Kogut}}, \citenamefont {{Limon}}, \citenamefont
  {{Meyer}}, \citenamefont {{Page}}, \citenamefont {{Tucker}}, \citenamefont
  {{Weiland}}, \citenamefont {{Wollack}},\ and\ \citenamefont
  {{Wright}}}]{Spergel:2003}%
  \BibitemOpen
  \bibfield  {author} {\bibinfo {author} {\bibfnamefont {D.~N.}\ \bibnamefont
  {{Spergel}}}, \bibinfo {author} {\bibfnamefont {L.}~\bibnamefont {{Verde}}},
  \bibinfo {author} {\bibfnamefont {H.~V.}\ \bibnamefont {{Peiris}}}, \bibinfo
  {author} {\bibfnamefont {E.}~\bibnamefont {{Komatsu}}}, \bibinfo {author}
  {\bibfnamefont {M.~R.}\ \bibnamefont {{Nolta}}}, \bibinfo {author}
  {\bibfnamefont {C.~L.}\ \bibnamefont {{Bennett}}}, \bibinfo {author}
  {\bibfnamefont {M.}~\bibnamefont {{Halpern}}}, \bibinfo {author}
  {\bibfnamefont {G.}~\bibnamefont {{Hinshaw}}}, \bibinfo {author}
  {\bibfnamefont {N.}~\bibnamefont {{Jarosik}}}, \bibinfo {author}
  {\bibfnamefont {A.}~\bibnamefont {{Kogut}}}, \bibinfo {author} {\bibfnamefont
  {M.}~\bibnamefont {{Limon}}}, \bibinfo {author} {\bibfnamefont {S.~S.}\
  \bibnamefont {{Meyer}}}, \bibinfo {author} {\bibfnamefont {L.}~\bibnamefont
  {{Page}}}, \bibinfo {author} {\bibfnamefont {G.~S.}\ \bibnamefont
  {{Tucker}}}, \bibinfo {author} {\bibfnamefont {J.~L.}\ \bibnamefont
  {{Weiland}}}, \bibinfo {author} {\bibfnamefont {E.}~\bibnamefont
  {{Wollack}}}, \ and\ \bibinfo {author} {\bibfnamefont {E.~L.}\ \bibnamefont
  {{Wright}}},\ }\href {\doibase 10.1086/377226} {\bibfield  {journal}
  {\bibinfo  {journal} {ApJS}\ }\textbf {\bibinfo {volume} {148}},\ \bibinfo
  {pages} {175} (\bibinfo {year} {2003})},\ \Eprint
  {http://arxiv.org/abs/astro-ph/0302209} {arXiv:astro-ph/0302209 [astro-ph]}
  \BibitemShut {NoStop}%
\bibitem [{\citenamefont {Eisenstein}(2005)}]{Eisenstein:2005}%
  \BibitemOpen
  \bibfield  {author} {\bibinfo {author} {\bibfnamefont {D.}~\bibnamefont
  {Eisenstein}},\ }\href {\doibase https://doi.org/10.1016/j.newar.2005.08.005}
  {\bibfield  {journal} {\bibinfo  {journal} {New Astronomy Reviews}\ }\textbf
  {\bibinfo {volume} {49}},\ \bibinfo {pages} {360 } (\bibinfo {year}
  {2005})},\ \bibinfo {note} {wide-Field Imaging from Space}\BibitemShut
  {NoStop}%
\bibitem [{\citenamefont {{Percival}}\ \emph {et~al.}(2007)\citenamefont
  {{Percival}}, \citenamefont {{Cole}}, \citenamefont {{Eisenstein}},
  \citenamefont {{Nichol}}, \citenamefont {{Peacock}}, \citenamefont {{Pope}},\
  and\ \citenamefont {{Szalay}}}]{Percival:2007}%
  \BibitemOpen
  \bibfield  {author} {\bibinfo {author} {\bibfnamefont {W.~J.}\ \bibnamefont
  {{Percival}}}, \bibinfo {author} {\bibfnamefont {S.}~\bibnamefont {{Cole}}},
  \bibinfo {author} {\bibfnamefont {D.~J.}\ \bibnamefont {{Eisenstein}}},
  \bibinfo {author} {\bibfnamefont {R.~C.}\ \bibnamefont {{Nichol}}}, \bibinfo
  {author} {\bibfnamefont {J.~A.}\ \bibnamefont {{Peacock}}}, \bibinfo {author}
  {\bibfnamefont {A.~C.}\ \bibnamefont {{Pope}}}, \ and\ \bibinfo {author}
  {\bibfnamefont {A.~S.}\ \bibnamefont {{Szalay}}},\ }\href {\doibase
  10.1111/j.1365-2966.2007.12268.x} {\bibfield  {journal} {\bibinfo  {journal}
  {Monthly Notices of the Royal Astronomical Society}\ }\textbf {\bibinfo
  {volume} {381}},\ \bibinfo {pages} {1053} (\bibinfo {year} {2007})},\ \Eprint
  {http://arxiv.org/abs/0705.3323} {arXiv:0705.3323 [astro-ph]} \BibitemShut
  {NoStop}%
\bibitem [{\citenamefont {{Suyu}}\ \emph {et~al.}(2010)\citenamefont {{Suyu}},
  \citenamefont {{Marshall}}, \citenamefont {{Auger}}, \citenamefont
  {{Hilbert}}, \citenamefont {{Blandford}}, \citenamefont {{Koopmans}},
  \citenamefont {{Fassnacht}},\ and\ \citenamefont {{Treu}}}]{Suyu:2010}%
  \BibitemOpen
  \bibfield  {author} {\bibinfo {author} {\bibfnamefont {S.~H.}\ \bibnamefont
  {{Suyu}}}, \bibinfo {author} {\bibfnamefont {P.~J.}\ \bibnamefont
  {{Marshall}}}, \bibinfo {author} {\bibfnamefont {M.~W.}\ \bibnamefont
  {{Auger}}}, \bibinfo {author} {\bibfnamefont {S.}~\bibnamefont {{Hilbert}}},
  \bibinfo {author} {\bibfnamefont {R.~D.}\ \bibnamefont {{Blandford}}},
  \bibinfo {author} {\bibfnamefont {L.~V.~E.}\ \bibnamefont {{Koopmans}}},
  \bibinfo {author} {\bibfnamefont {C.~D.}\ \bibnamefont {{Fassnacht}}}, \ and\
  \bibinfo {author} {\bibfnamefont {T.}~\bibnamefont {{Treu}}},\ }\href
  {\doibase 10.1088/0004-637X/711/1/201} {\bibfield  {journal} {\bibinfo
  {journal} {ApJ}\ }\textbf {\bibinfo {volume} {711}},\ \bibinfo {pages} {201}
  (\bibinfo {year} {2010})},\ \Eprint {http://arxiv.org/abs/0910.2773}
  {arXiv:0910.2773 [astro-ph.CO]} \BibitemShut {NoStop}%
\bibitem [{\citenamefont {{Schrabback}}\ \emph {et~al.}(2010)\citenamefont
  {{Schrabback}}, \citenamefont {{Hartlap}}, \citenamefont {{Joachimi}},
  \citenamefont {{Kilbinger}}, \citenamefont {{Simon}}, \citenamefont
  {{Benabed}}, \citenamefont {{Brada{\v{c}}}}, \citenamefont {{Eifler}},
  \citenamefont {{Erben}}, \citenamefont {{Fassnacht}}, \citenamefont {{High}},
  \citenamefont {{Hilbert}}, \citenamefont {{Hildebrandt}}, \citenamefont
  {{Hoekstra}}, \citenamefont {{Kuijken}}, \citenamefont {{Marshall}},
  \citenamefont {{Mellier}}, \citenamefont {{Morganson}}, \citenamefont
  {{Schneider}}, \citenamefont {{Semboloni}}, \citenamefont {{van Waerbeke}},\
  and\ \citenamefont {{Velander}}}]{Schrabback:2010}%
  \BibitemOpen
  \bibfield  {author} {\bibinfo {author} {\bibfnamefont {T.}~\bibnamefont
  {{Schrabback}}}, \bibinfo {author} {\bibfnamefont {J.}~\bibnamefont
  {{Hartlap}}}, \bibinfo {author} {\bibfnamefont {B.}~\bibnamefont
  {{Joachimi}}}, \bibinfo {author} {\bibfnamefont {M.}~\bibnamefont
  {{Kilbinger}}}, \bibinfo {author} {\bibfnamefont {P.}~\bibnamefont
  {{Simon}}}, \bibinfo {author} {\bibfnamefont {K.}~\bibnamefont {{Benabed}}},
  \bibinfo {author} {\bibfnamefont {M.}~\bibnamefont {{Brada{\v{c}}}}},
  \bibinfo {author} {\bibfnamefont {T.}~\bibnamefont {{Eifler}}}, \bibinfo
  {author} {\bibfnamefont {T.}~\bibnamefont {{Erben}}}, \bibinfo {author}
  {\bibfnamefont {C.~D.}\ \bibnamefont {{Fassnacht}}}, \bibinfo {author}
  {\bibfnamefont {F.~W.}\ \bibnamefont {{High}}}, \bibinfo {author}
  {\bibfnamefont {S.}~\bibnamefont {{Hilbert}}}, \bibinfo {author}
  {\bibfnamefont {H.}~\bibnamefont {{Hildebrandt}}}, \bibinfo {author}
  {\bibfnamefont {H.}~\bibnamefont {{Hoekstra}}}, \bibinfo {author}
  {\bibfnamefont {K.}~\bibnamefont {{Kuijken}}}, \bibinfo {author}
  {\bibfnamefont {P.~J.}\ \bibnamefont {{Marshall}}}, \bibinfo {author}
  {\bibfnamefont {Y.}~\bibnamefont {{Mellier}}}, \bibinfo {author}
  {\bibfnamefont {E.}~\bibnamefont {{Morganson}}}, \bibinfo {author}
  {\bibfnamefont {P.}~\bibnamefont {{Schneider}}}, \bibinfo {author}
  {\bibfnamefont {E.}~\bibnamefont {{Semboloni}}}, \bibinfo {author}
  {\bibfnamefont {L.}~\bibnamefont {{van Waerbeke}}}, \ and\ \bibinfo {author}
  {\bibfnamefont {M.}~\bibnamefont {{Velander}}},\ }\href {\doibase
  10.1051/0004-6361/200913577} {\bibfield  {journal} {\bibinfo  {journal}
  {A\&A}\ }\textbf {\bibinfo {volume} {516}},\ \bibinfo {eid} {A63} (\bibinfo
  {year} {2010})},\ \Eprint {http://arxiv.org/abs/0911.0053} {arXiv:0911.0053
  [astro-ph.CO]} \BibitemShut {NoStop}%
\bibitem [{\citenamefont {{Huterer}}\ and\ \citenamefont
  {{Shafer}}(2018)}]{Huterer:2018}%
  \BibitemOpen
  \bibfield  {author} {\bibinfo {author} {\bibfnamefont {D.}~\bibnamefont
  {{Huterer}}}\ and\ \bibinfo {author} {\bibfnamefont {D.~L.}\ \bibnamefont
  {{Shafer}}},\ }\href {\doibase 10.1088/1361-6633/aa997e} {\bibfield
  {journal} {\bibinfo  {journal} {Reports on Progress in Physics}\ }\textbf
  {\bibinfo {volume} {81}},\ \bibinfo {eid} {016901} (\bibinfo {year}
  {2018})},\ \Eprint {http://arxiv.org/abs/1709.01091} {arXiv:1709.01091
  [astro-ph.CO]} \BibitemShut {NoStop}%
\bibitem [{\citenamefont {{Bennett}}\ \emph {et~al.}(2013)\citenamefont
  {{Bennett}}, \citenamefont {{Larson}}, \citenamefont {{Weiland}},
  \citenamefont {{Jarosik}}, \citenamefont {{Hinshaw}}, \citenamefont
  {{Odegard}}, \citenamefont {{Smith}}, \citenamefont {{Hill}}, \citenamefont
  {{Gold}}, \citenamefont {{Halpern}}, \citenamefont {{Komatsu}}, \citenamefont
  {{Nolta}}, \citenamefont {{Page}}, \citenamefont {{Spergel}}, \citenamefont
  {{Wollack}}, \citenamefont {{Dunkley}}, \citenamefont {{Kogut}},
  \citenamefont {{Limon}}, \citenamefont {{Meyer}}, \citenamefont {{Tucker}},\
  and\ \citenamefont {{Wright}}}]{Bennett:2013}%
  \BibitemOpen
  \bibfield  {author} {\bibinfo {author} {\bibfnamefont {C.~L.}\ \bibnamefont
  {{Bennett}}}, \bibinfo {author} {\bibfnamefont {D.}~\bibnamefont {{Larson}}},
  \bibinfo {author} {\bibfnamefont {J.~L.}\ \bibnamefont {{Weiland}}}, \bibinfo
  {author} {\bibfnamefont {N.}~\bibnamefont {{Jarosik}}}, \bibinfo {author}
  {\bibfnamefont {G.}~\bibnamefont {{Hinshaw}}}, \bibinfo {author}
  {\bibfnamefont {N.}~\bibnamefont {{Odegard}}}, \bibinfo {author}
  {\bibfnamefont {K.~M.}\ \bibnamefont {{Smith}}}, \bibinfo {author}
  {\bibfnamefont {R.~S.}\ \bibnamefont {{Hill}}}, \bibinfo {author}
  {\bibfnamefont {B.}~\bibnamefont {{Gold}}}, \bibinfo {author} {\bibfnamefont
  {M.}~\bibnamefont {{Halpern}}}, \bibinfo {author} {\bibfnamefont
  {E.}~\bibnamefont {{Komatsu}}}, \bibinfo {author} {\bibfnamefont {M.~R.}\
  \bibnamefont {{Nolta}}}, \bibinfo {author} {\bibfnamefont {L.}~\bibnamefont
  {{Page}}}, \bibinfo {author} {\bibfnamefont {D.~N.}\ \bibnamefont
  {{Spergel}}}, \bibinfo {author} {\bibfnamefont {E.}~\bibnamefont
  {{Wollack}}}, \bibinfo {author} {\bibfnamefont {J.}~\bibnamefont
  {{Dunkley}}}, \bibinfo {author} {\bibfnamefont {A.}~\bibnamefont {{Kogut}}},
  \bibinfo {author} {\bibfnamefont {M.}~\bibnamefont {{Limon}}}, \bibinfo
  {author} {\bibfnamefont {S.~S.}\ \bibnamefont {{Meyer}}}, \bibinfo {author}
  {\bibfnamefont {G.~S.}\ \bibnamefont {{Tucker}}}, \ and\ \bibinfo {author}
  {\bibfnamefont {E.~L.}\ \bibnamefont {{Wright}}},\ }\href {\doibase
  10.1088/0067-0049/208/2/20} {\bibfield  {journal} {\bibinfo  {journal}
  {ApJS}\ }\textbf {\bibinfo {volume} {208}},\ \bibinfo {eid} {20} (\bibinfo
  {year} {2013})},\ \Eprint {http://arxiv.org/abs/1212.5225} {arXiv:1212.5225
  [astro-ph.CO]} \BibitemShut {NoStop}%
\bibitem [{\citenamefont {{Planck Collaboration}}\ \emph
  {et~al.}(2020)\citenamefont {{Planck Collaboration}}, \citenamefont
  {{Aghanim}}, \citenamefont {{Akrami}}, \citenamefont {{Ashdown}},
  \citenamefont {{Aumont}}, \citenamefont {{Baccigalupi}}, \citenamefont
  {{Ballardini}}, \citenamefont {{Banday}}, \citenamefont {{Barreiro}},
  \citenamefont {{Bartolo}}, \citenamefont {{Basak}}, \citenamefont {{Battye}},
  \citenamefont {{Benabed}}, \citenamefont {{Bernard}}, \citenamefont
  {{Bersanelli}}, \citenamefont {{Bielewicz}}, \citenamefont {{Bock}},
  \citenamefont {{Bond}}, \citenamefont {{Borrill}}, \citenamefont {{Bouchet}},
  \citenamefont {{Boulanger}}, \citenamefont {{Bucher}}, \citenamefont
  {{Burigana}}, \citenamefont {{Butler}}, \citenamefont {{Calabrese}},
  \citenamefont {{Cardoso}}, \citenamefont {{Carron}}, \citenamefont
  {{Challinor}}, \citenamefont {{Chiang}}, \citenamefont {{Chluba}},
  \citenamefont {{Colombo}}, \citenamefont {{Combet}}, \citenamefont
  {{Contreras}}, \citenamefont {{Crill}}, \citenamefont {{Cuttaia}},
  \citenamefont {{de Bernardis}}, \citenamefont {{de Zotti}}, \citenamefont
  {{Delabrouille}}, \citenamefont {{Delouis}}, \citenamefont {{Di Valentino}},
  \citenamefont {{Diego}}, \citenamefont {{Dor{\'e}}}, \citenamefont
  {{Douspis}}, \citenamefont {{Ducout}}, \citenamefont {{Dupac}}, \citenamefont
  {{Dusini}}, \citenamefont {{Efstathiou}}, \citenamefont {{Elsner}},
  \citenamefont {{En{\ss}lin}}, \citenamefont {{Eriksen}}, \citenamefont
  {{Fantaye}}, \citenamefont {{Farhang}}, \citenamefont {{Fergusson}},
  \citenamefont {{Fernandez-Cobos}}, \citenamefont {{Finelli}}, \citenamefont
  {{Forastieri}}, \citenamefont {{Frailis}}, \citenamefont {{Fraisse}},
  \citenamefont {{Franceschi}}, \citenamefont {{Frolov}}, \citenamefont
  {{Galeotta}}, \citenamefont {{Galli}}, \citenamefont {{Ganga}}, \citenamefont
  {{G{\'e}nova-Santos}}, \citenamefont {{Gerbino}}, \citenamefont {{Ghosh}},
  \citenamefont {{Gonz{\'a}lez-Nuevo}}, \citenamefont {{G{\'o}rski}},
  \citenamefont {{Gratton}}, \citenamefont {{Gruppuso}}, \citenamefont
  {{Gudmundsson}}, \citenamefont {{Hamann}}, \citenamefont {{Handley}},
  \citenamefont {{Hansen}}, \citenamefont {{Herranz}}, \citenamefont
  {{Hildebrandt}}, \citenamefont {{Hivon}}, \citenamefont {{Huang}},
  \citenamefont {{Jaffe}}, \citenamefont {{Jones}}, \citenamefont {{Karakci}},
  \citenamefont {{Keih{\"a}nen}}, \citenamefont {{Keskitalo}}, \citenamefont
  {{Kiiveri}}, \citenamefont {{Kim}}, \citenamefont {{Kisner}}, \citenamefont
  {{Knox}}, \citenamefont {{Krachmalnicoff}}, \citenamefont {{Kunz}},
  \citenamefont {{Kurki-Suonio}}, \citenamefont {{Lagache}}, \citenamefont
  {{Lamarre}}, \citenamefont {{Lasenby}}, \citenamefont {{Lattanzi}},
  \citenamefont {{Lawrence}}, \citenamefont {{Le Jeune}}, \citenamefont
  {{Lemos}}, \citenamefont {{Lesgourgues}}, \citenamefont {{Levrier}},
  \citenamefont {{Lewis}}, \citenamefont {{Liguori}}, \citenamefont {{Lilje}},
  \citenamefont {{Lilley}}, \citenamefont {{Lindholm}}, \citenamefont
  {{L{\'o}pez-Caniego}}, \citenamefont {{Lubin}}, \citenamefont {{Ma}},
  \citenamefont {{Mac{\'\i}as-P{\'e}rez}}, \citenamefont {{Maggio}},
  \citenamefont {{Maino}}, \citenamefont {{Mandolesi}}, \citenamefont
  {{Mangilli}}, \citenamefont {{Marcos-Caballero}}, \citenamefont {{Maris}},
  \citenamefont {{Martin}}, \citenamefont {{Martinelli}}, \citenamefont
  {{Mart{\'\i}nez-Gonz{\'a}lez}}, \citenamefont {{Matarrese}}, \citenamefont
  {{Mauri}}, \citenamefont {{McEwen}}, \citenamefont {{Meinhold}},
  \citenamefont {{Melchiorri}}, \citenamefont {{Mennella}}, \citenamefont
  {{Migliaccio}}, \citenamefont {{Millea}}, \citenamefont {{Mitra}},
  \citenamefont {{Miville-Desch{\^e}nes}}, \citenamefont {{Molinari}},
  \citenamefont {{Montier}}, \citenamefont {{Morgante}}, \citenamefont
  {{Moss}}, \citenamefont {{Natoli}}, \citenamefont {{N{\o}rgaard-Nielsen}},
  \citenamefont {{Pagano}}, \citenamefont {{Paoletti}}, \citenamefont
  {{Partridge}}, \citenamefont {{Patanchon}}, \citenamefont {{Peiris}},
  \citenamefont {{Perrotta}}, \citenamefont {{Pettorino}}, \citenamefont
  {{Piacentini}}, \citenamefont {{Polastri}}, \citenamefont {{Polenta}},
  \citenamefont {{Puget}}, \citenamefont {{Rachen}}, \citenamefont
  {{Reinecke}}, \citenamefont {{Remazeilles}}, \citenamefont {{Renzi}},
  \citenamefont {{Rocha}}, \citenamefont {{Rosset}}, \citenamefont {{Roudier}},
  \citenamefont {{Rubi{\~n}o-Mart{\'\i}n}}, \citenamefont {{Ruiz-Granados}},
  \citenamefont {{Salvati}}, \citenamefont {{Sandri}}, \citenamefont
  {{Savelainen}}, \citenamefont {{Scott}}, \citenamefont {{Shellard}},
  \citenamefont {{Sirignano}}, \citenamefont {{Sirri}}, \citenamefont
  {{Spencer}}, \citenamefont {{Sunyaev}}, \citenamefont {{Suur-Uski}},
  \citenamefont {{Tauber}}, \citenamefont {{Tavagnacco}}, \citenamefont
  {{Tenti}}, \citenamefont {{Toffolatti}}, \citenamefont {{Tomasi}},
  \citenamefont {{Trombetti}}, \citenamefont {{Valenziano}}, \citenamefont
  {{Valiviita}}, \citenamefont {{Van Tent}}, \citenamefont {{Vibert}},
  \citenamefont {{Vielva}}, \citenamefont {{Villa}}, \citenamefont
  {{Vittorio}}, \citenamefont {{Wandelt}}, \citenamefont {{Wehus}},
  \citenamefont {{White}}, \citenamefont {{White}}, \citenamefont {{Zacchei}},\
  and\ \citenamefont {{Zonca}}}]{Planck:2018}%
  \BibitemOpen
  \bibfield  {author} {\bibinfo {author} {\bibnamefont {{Planck
  Collaboration}}}, \bibinfo {author} {\bibfnamefont {N.}~\bibnamefont
  {{Aghanim}}}, \bibinfo {author} {\bibfnamefont {Y.}~\bibnamefont {{Akrami}}},
  \bibinfo {author} {\bibfnamefont {M.}~\bibnamefont {{Ashdown}}}, \bibinfo
  {author} {\bibfnamefont {J.}~\bibnamefont {{Aumont}}}, \bibinfo {author}
  {\bibfnamefont {C.}~\bibnamefont {{Baccigalupi}}}, \bibinfo {author}
  {\bibfnamefont {M.}~\bibnamefont {{Ballardini}}}, \bibinfo {author}
  {\bibfnamefont {A.~J.}\ \bibnamefont {{Banday}}}, \bibinfo {author}
  {\bibfnamefont {R.~B.}\ \bibnamefont {{Barreiro}}}, \bibinfo {author}
  {\bibfnamefont {N.}~\bibnamefont {{Bartolo}}}, \bibinfo {author}
  {\bibfnamefont {S.}~\bibnamefont {{Basak}}}, \bibinfo {author} {\bibfnamefont
  {R.}~\bibnamefont {{Battye}}}, \bibinfo {author} {\bibfnamefont
  {K.}~\bibnamefont {{Benabed}}}, \bibinfo {author} {\bibfnamefont {J.~P.}\
  \bibnamefont {{Bernard}}}, \bibinfo {author} {\bibfnamefont {M.}~\bibnamefont
  {{Bersanelli}}}, \bibinfo {author} {\bibfnamefont {P.}~\bibnamefont
  {{Bielewicz}}}, \bibinfo {author} {\bibfnamefont {J.~J.}\ \bibnamefont
  {{Bock}}}, \bibinfo {author} {\bibfnamefont {J.~R.}\ \bibnamefont {{Bond}}},
  \bibinfo {author} {\bibfnamefont {J.}~\bibnamefont {{Borrill}}}, \bibinfo
  {author} {\bibfnamefont {F.~R.}\ \bibnamefont {{Bouchet}}}, \bibinfo {author}
  {\bibfnamefont {F.}~\bibnamefont {{Boulanger}}}, \bibinfo {author}
  {\bibfnamefont {M.}~\bibnamefont {{Bucher}}}, \bibinfo {author}
  {\bibfnamefont {C.}~\bibnamefont {{Burigana}}}, \bibinfo {author}
  {\bibfnamefont {R.~C.}\ \bibnamefont {{Butler}}}, \bibinfo {author}
  {\bibfnamefont {E.}~\bibnamefont {{Calabrese}}}, \bibinfo {author}
  {\bibfnamefont {J.~F.}\ \bibnamefont {{Cardoso}}}, \bibinfo {author}
  {\bibfnamefont {J.}~\bibnamefont {{Carron}}}, \bibinfo {author}
  {\bibfnamefont {A.}~\bibnamefont {{Challinor}}}, \bibinfo {author}
  {\bibfnamefont {H.~C.}\ \bibnamefont {{Chiang}}}, \bibinfo {author}
  {\bibfnamefont {J.}~\bibnamefont {{Chluba}}}, \bibinfo {author}
  {\bibfnamefont {L.~P.~L.}\ \bibnamefont {{Colombo}}}, \bibinfo {author}
  {\bibfnamefont {C.}~\bibnamefont {{Combet}}}, \bibinfo {author}
  {\bibfnamefont {D.}~\bibnamefont {{Contreras}}}, \bibinfo {author}
  {\bibfnamefont {B.~P.}\ \bibnamefont {{Crill}}}, \bibinfo {author}
  {\bibfnamefont {F.}~\bibnamefont {{Cuttaia}}}, \bibinfo {author}
  {\bibfnamefont {P.}~\bibnamefont {{de Bernardis}}}, \bibinfo {author}
  {\bibfnamefont {G.}~\bibnamefont {{de Zotti}}}, \bibinfo {author}
  {\bibfnamefont {J.}~\bibnamefont {{Delabrouille}}}, \bibinfo {author}
  {\bibfnamefont {J.~M.}\ \bibnamefont {{Delouis}}}, \bibinfo {author}
  {\bibfnamefont {E.}~\bibnamefont {{Di Valentino}}}, \bibinfo {author}
  {\bibfnamefont {J.~M.}\ \bibnamefont {{Diego}}}, \bibinfo {author}
  {\bibfnamefont {O.}~\bibnamefont {{Dor{\'e}}}}, \bibinfo {author}
  {\bibfnamefont {M.}~\bibnamefont {{Douspis}}}, \bibinfo {author}
  {\bibfnamefont {A.}~\bibnamefont {{Ducout}}}, \bibinfo {author}
  {\bibfnamefont {X.}~\bibnamefont {{Dupac}}}, \bibinfo {author} {\bibfnamefont
  {S.}~\bibnamefont {{Dusini}}}, \bibinfo {author} {\bibfnamefont
  {G.}~\bibnamefont {{Efstathiou}}}, \bibinfo {author} {\bibfnamefont
  {F.}~\bibnamefont {{Elsner}}}, \bibinfo {author} {\bibfnamefont {T.~A.}\
  \bibnamefont {{En{\ss}lin}}}, \bibinfo {author} {\bibfnamefont {H.~K.}\
  \bibnamefont {{Eriksen}}}, \bibinfo {author} {\bibfnamefont {Y.}~\bibnamefont
  {{Fantaye}}}, \bibinfo {author} {\bibfnamefont {M.}~\bibnamefont
  {{Farhang}}}, \bibinfo {author} {\bibfnamefont {J.}~\bibnamefont
  {{Fergusson}}}, \bibinfo {author} {\bibfnamefont {R.}~\bibnamefont
  {{Fernandez-Cobos}}}, \bibinfo {author} {\bibfnamefont {F.}~\bibnamefont
  {{Finelli}}}, \bibinfo {author} {\bibfnamefont {F.}~\bibnamefont
  {{Forastieri}}}, \bibinfo {author} {\bibfnamefont {M.}~\bibnamefont
  {{Frailis}}}, \bibinfo {author} {\bibfnamefont {A.~A.}\ \bibnamefont
  {{Fraisse}}}, \bibinfo {author} {\bibfnamefont {E.}~\bibnamefont
  {{Franceschi}}}, \bibinfo {author} {\bibfnamefont {A.}~\bibnamefont
  {{Frolov}}}, \bibinfo {author} {\bibfnamefont {S.}~\bibnamefont
  {{Galeotta}}}, \bibinfo {author} {\bibfnamefont {S.}~\bibnamefont {{Galli}}},
  \bibinfo {author} {\bibfnamefont {K.}~\bibnamefont {{Ganga}}}, \bibinfo
  {author} {\bibfnamefont {R.~T.}\ \bibnamefont {{G{\'e}nova-Santos}}},
  \bibinfo {author} {\bibfnamefont {M.}~\bibnamefont {{Gerbino}}}, \bibinfo
  {author} {\bibfnamefont {T.}~\bibnamefont {{Ghosh}}}, \bibinfo {author}
  {\bibfnamefont {J.}~\bibnamefont {{Gonz{\'a}lez-Nuevo}}}, \bibinfo {author}
  {\bibfnamefont {K.~M.}\ \bibnamefont {{G{\'o}rski}}}, \bibinfo {author}
  {\bibfnamefont {S.}~\bibnamefont {{Gratton}}}, \bibinfo {author}
  {\bibfnamefont {A.}~\bibnamefont {{Gruppuso}}}, \bibinfo {author}
  {\bibfnamefont {J.~E.}\ \bibnamefont {{Gudmundsson}}}, \bibinfo {author}
  {\bibfnamefont {J.}~\bibnamefont {{Hamann}}}, \bibinfo {author}
  {\bibfnamefont {W.}~\bibnamefont {{Handley}}}, \bibinfo {author}
  {\bibfnamefont {F.~K.}\ \bibnamefont {{Hansen}}}, \bibinfo {author}
  {\bibfnamefont {D.}~\bibnamefont {{Herranz}}}, \bibinfo {author}
  {\bibfnamefont {S.~R.}\ \bibnamefont {{Hildebrandt}}}, \bibinfo {author}
  {\bibfnamefont {E.}~\bibnamefont {{Hivon}}}, \bibinfo {author} {\bibfnamefont
  {Z.}~\bibnamefont {{Huang}}}, \bibinfo {author} {\bibfnamefont {A.~H.}\
  \bibnamefont {{Jaffe}}}, \bibinfo {author} {\bibfnamefont {W.~C.}\
  \bibnamefont {{Jones}}}, \bibinfo {author} {\bibfnamefont {A.}~\bibnamefont
  {{Karakci}}}, \bibinfo {author} {\bibfnamefont {E.}~\bibnamefont
  {{Keih{\"a}nen}}}, \bibinfo {author} {\bibfnamefont {R.}~\bibnamefont
  {{Keskitalo}}}, \bibinfo {author} {\bibfnamefont {K.}~\bibnamefont
  {{Kiiveri}}}, \bibinfo {author} {\bibfnamefont {J.}~\bibnamefont {{Kim}}},
  \bibinfo {author} {\bibfnamefont {T.~S.}\ \bibnamefont {{Kisner}}}, \bibinfo
  {author} {\bibfnamefont {L.}~\bibnamefont {{Knox}}}, \bibinfo {author}
  {\bibfnamefont {N.}~\bibnamefont {{Krachmalnicoff}}}, \bibinfo {author}
  {\bibfnamefont {M.}~\bibnamefont {{Kunz}}}, \bibinfo {author} {\bibfnamefont
  {H.}~\bibnamefont {{Kurki-Suonio}}}, \bibinfo {author} {\bibfnamefont
  {G.}~\bibnamefont {{Lagache}}}, \bibinfo {author} {\bibfnamefont {J.~M.}\
  \bibnamefont {{Lamarre}}}, \bibinfo {author} {\bibfnamefont {A.}~\bibnamefont
  {{Lasenby}}}, \bibinfo {author} {\bibfnamefont {M.}~\bibnamefont
  {{Lattanzi}}}, \bibinfo {author} {\bibfnamefont {C.~R.}\ \bibnamefont
  {{Lawrence}}}, \bibinfo {author} {\bibfnamefont {M.}~\bibnamefont {{Le
  Jeune}}}, \bibinfo {author} {\bibfnamefont {P.}~\bibnamefont {{Lemos}}},
  \bibinfo {author} {\bibfnamefont {J.}~\bibnamefont {{Lesgourgues}}}, \bibinfo
  {author} {\bibfnamefont {F.}~\bibnamefont {{Levrier}}}, \bibinfo {author}
  {\bibfnamefont {A.}~\bibnamefont {{Lewis}}}, \bibinfo {author} {\bibfnamefont
  {M.}~\bibnamefont {{Liguori}}}, \bibinfo {author} {\bibfnamefont {P.~B.}\
  \bibnamefont {{Lilje}}}, \bibinfo {author} {\bibfnamefont {M.}~\bibnamefont
  {{Lilley}}}, \bibinfo {author} {\bibfnamefont {V.}~\bibnamefont
  {{Lindholm}}}, \bibinfo {author} {\bibfnamefont {M.}~\bibnamefont
  {{L{\'o}pez-Caniego}}}, \bibinfo {author} {\bibfnamefont {P.~M.}\
  \bibnamefont {{Lubin}}}, \bibinfo {author} {\bibfnamefont {Y.~Z.}\
  \bibnamefont {{Ma}}}, \bibinfo {author} {\bibfnamefont {J.~F.}\ \bibnamefont
  {{Mac{\'\i}as-P{\'e}rez}}}, \bibinfo {author} {\bibfnamefont
  {G.}~\bibnamefont {{Maggio}}}, \bibinfo {author} {\bibfnamefont
  {D.}~\bibnamefont {{Maino}}}, \bibinfo {author} {\bibfnamefont
  {N.}~\bibnamefont {{Mandolesi}}}, \bibinfo {author} {\bibfnamefont
  {A.}~\bibnamefont {{Mangilli}}}, \bibinfo {author} {\bibfnamefont
  {A.}~\bibnamefont {{Marcos-Caballero}}}, \bibinfo {author} {\bibfnamefont
  {M.}~\bibnamefont {{Maris}}}, \bibinfo {author} {\bibfnamefont {P.~G.}\
  \bibnamefont {{Martin}}}, \bibinfo {author} {\bibfnamefont {M.}~\bibnamefont
  {{Martinelli}}}, \bibinfo {author} {\bibfnamefont {E.}~\bibnamefont
  {{Mart{\'\i}nez-Gonz{\'a}lez}}}, \bibinfo {author} {\bibfnamefont
  {S.}~\bibnamefont {{Matarrese}}}, \bibinfo {author} {\bibfnamefont
  {N.}~\bibnamefont {{Mauri}}}, \bibinfo {author} {\bibfnamefont {J.~D.}\
  \bibnamefont {{McEwen}}}, \bibinfo {author} {\bibfnamefont {P.~R.}\
  \bibnamefont {{Meinhold}}}, \bibinfo {author} {\bibfnamefont
  {A.}~\bibnamefont {{Melchiorri}}}, \bibinfo {author} {\bibfnamefont
  {A.}~\bibnamefont {{Mennella}}}, \bibinfo {author} {\bibfnamefont
  {M.}~\bibnamefont {{Migliaccio}}}, \bibinfo {author} {\bibfnamefont
  {M.}~\bibnamefont {{Millea}}}, \bibinfo {author} {\bibfnamefont
  {S.}~\bibnamefont {{Mitra}}}, \bibinfo {author} {\bibfnamefont {M.~A.}\
  \bibnamefont {{Miville-Desch{\^e}nes}}}, \bibinfo {author} {\bibfnamefont
  {D.}~\bibnamefont {{Molinari}}}, \bibinfo {author} {\bibfnamefont
  {L.}~\bibnamefont {{Montier}}}, \bibinfo {author} {\bibfnamefont
  {G.}~\bibnamefont {{Morgante}}}, \bibinfo {author} {\bibfnamefont
  {A.}~\bibnamefont {{Moss}}}, \bibinfo {author} {\bibfnamefont
  {P.}~\bibnamefont {{Natoli}}}, \bibinfo {author} {\bibfnamefont {H.~U.}\
  \bibnamefont {{N{\o}rgaard-Nielsen}}}, \bibinfo {author} {\bibfnamefont
  {L.}~\bibnamefont {{Pagano}}}, \bibinfo {author} {\bibfnamefont
  {D.}~\bibnamefont {{Paoletti}}}, \bibinfo {author} {\bibfnamefont
  {B.}~\bibnamefont {{Partridge}}}, \bibinfo {author} {\bibfnamefont
  {G.}~\bibnamefont {{Patanchon}}}, \bibinfo {author} {\bibfnamefont {H.~V.}\
  \bibnamefont {{Peiris}}}, \bibinfo {author} {\bibfnamefont {F.}~\bibnamefont
  {{Perrotta}}}, \bibinfo {author} {\bibfnamefont {V.}~\bibnamefont
  {{Pettorino}}}, \bibinfo {author} {\bibfnamefont {F.}~\bibnamefont
  {{Piacentini}}}, \bibinfo {author} {\bibfnamefont {L.}~\bibnamefont
  {{Polastri}}}, \bibinfo {author} {\bibfnamefont {G.}~\bibnamefont
  {{Polenta}}}, \bibinfo {author} {\bibfnamefont {J.~L.}\ \bibnamefont
  {{Puget}}}, \bibinfo {author} {\bibfnamefont {J.~P.}\ \bibnamefont
  {{Rachen}}}, \bibinfo {author} {\bibfnamefont {M.}~\bibnamefont
  {{Reinecke}}}, \bibinfo {author} {\bibfnamefont {M.}~\bibnamefont
  {{Remazeilles}}}, \bibinfo {author} {\bibfnamefont {A.}~\bibnamefont
  {{Renzi}}}, \bibinfo {author} {\bibfnamefont {G.}~\bibnamefont {{Rocha}}},
  \bibinfo {author} {\bibfnamefont {C.}~\bibnamefont {{Rosset}}}, \bibinfo
  {author} {\bibfnamefont {G.}~\bibnamefont {{Roudier}}}, \bibinfo {author}
  {\bibfnamefont {J.~A.}\ \bibnamefont {{Rubi{\~n}o-Mart{\'\i}n}}}, \bibinfo
  {author} {\bibfnamefont {B.}~\bibnamefont {{Ruiz-Granados}}}, \bibinfo
  {author} {\bibfnamefont {L.}~\bibnamefont {{Salvati}}}, \bibinfo {author}
  {\bibfnamefont {M.}~\bibnamefont {{Sandri}}}, \bibinfo {author}
  {\bibfnamefont {M.}~\bibnamefont {{Savelainen}}}, \bibinfo {author}
  {\bibfnamefont {D.}~\bibnamefont {{Scott}}}, \bibinfo {author} {\bibfnamefont
  {E.~P.~S.}\ \bibnamefont {{Shellard}}}, \bibinfo {author} {\bibfnamefont
  {C.}~\bibnamefont {{Sirignano}}}, \bibinfo {author} {\bibfnamefont
  {G.}~\bibnamefont {{Sirri}}}, \bibinfo {author} {\bibfnamefont {L.~D.}\
  \bibnamefont {{Spencer}}}, \bibinfo {author} {\bibfnamefont {R.}~\bibnamefont
  {{Sunyaev}}}, \bibinfo {author} {\bibfnamefont {A.~S.}\ \bibnamefont
  {{Suur-Uski}}}, \bibinfo {author} {\bibfnamefont {J.~A.}\ \bibnamefont
  {{Tauber}}}, \bibinfo {author} {\bibfnamefont {D.}~\bibnamefont
  {{Tavagnacco}}}, \bibinfo {author} {\bibfnamefont {M.}~\bibnamefont
  {{Tenti}}}, \bibinfo {author} {\bibfnamefont {L.}~\bibnamefont
  {{Toffolatti}}}, \bibinfo {author} {\bibfnamefont {M.}~\bibnamefont
  {{Tomasi}}}, \bibinfo {author} {\bibfnamefont {T.}~\bibnamefont
  {{Trombetti}}}, \bibinfo {author} {\bibfnamefont {L.}~\bibnamefont
  {{Valenziano}}}, \bibinfo {author} {\bibfnamefont {J.}~\bibnamefont
  {{Valiviita}}}, \bibinfo {author} {\bibfnamefont {B.}~\bibnamefont {{Van
  Tent}}}, \bibinfo {author} {\bibfnamefont {L.}~\bibnamefont {{Vibert}}},
  \bibinfo {author} {\bibfnamefont {P.}~\bibnamefont {{Vielva}}}, \bibinfo
  {author} {\bibfnamefont {F.}~\bibnamefont {{Villa}}}, \bibinfo {author}
  {\bibfnamefont {N.}~\bibnamefont {{Vittorio}}}, \bibinfo {author}
  {\bibfnamefont {B.~D.}\ \bibnamefont {{Wandelt}}}, \bibinfo {author}
  {\bibfnamefont {I.~K.}\ \bibnamefont {{Wehus}}}, \bibinfo {author}
  {\bibfnamefont {M.}~\bibnamefont {{White}}}, \bibinfo {author} {\bibfnamefont
  {S.~D.~M.}\ \bibnamefont {{White}}}, \bibinfo {author} {\bibfnamefont
  {A.}~\bibnamefont {{Zacchei}}}, \ and\ \bibinfo {author} {\bibfnamefont
  {A.}~\bibnamefont {{Zonca}}},\ }\href {\doibase 10.1051/0004-6361/201833910}
  {\bibfield  {journal} {\bibinfo  {journal} {A\&A}\ }\textbf {\bibinfo
  {volume} {641}},\ \bibinfo {eid} {A6} (\bibinfo {year} {2020})},\ \Eprint
  {http://arxiv.org/abs/1807.06209} {arXiv:1807.06209 [astro-ph.CO]}
  \BibitemShut {NoStop}%
\bibitem [{\citenamefont {Feng}(2010)}]{Feng:2010}%
  \BibitemOpen
  \bibfield  {author} {\bibinfo {author} {\bibfnamefont {J.~L.}\ \bibnamefont
  {Feng}},\ }\href {\doibase 10.1146/annurev-astro-082708-101659} {\bibfield
  {journal} {\bibinfo  {journal} {Ann. Rev. Astron. Astrophys.}\ }\textbf
  {\bibinfo {volume} {48}},\ \bibinfo {pages} {495} (\bibinfo {year} {2010})},\
  \Eprint {http://arxiv.org/abs/1003.0904} {arXiv:1003.0904 [astro-ph.CO]}
  \BibitemShut {NoStop}%
\bibitem [{\citenamefont {Bertone}\ and\ \citenamefont
  {Hooper}(2018)}]{Bertone:2018}%
  \BibitemOpen
  \bibfield  {author} {\bibinfo {author} {\bibfnamefont {G.}~\bibnamefont
  {Bertone}}\ and\ \bibinfo {author} {\bibfnamefont {D.}~\bibnamefont
  {Hooper}},\ }\href {\doibase 10.1103/RevModPhys.90.045002} {\bibfield
  {journal} {\bibinfo  {journal} {Rev. Mod. Phys.}\ }\textbf {\bibinfo {volume}
  {90}},\ \bibinfo {pages} {045002} (\bibinfo {year} {2018})}\BibitemShut
  {NoStop}%
\bibitem [{\citenamefont {de~Martino}\ \emph {et~al.}(2020)\citenamefont
  {de~Martino}, \citenamefont {Chakrabarty}, \citenamefont {Cesare},
  \citenamefont {Gallo}, \citenamefont {Ostorero},\ and\ \citenamefont
  {Diaferio}}]{Martino:2020}%
  \BibitemOpen
  \bibfield  {author} {\bibinfo {author} {\bibfnamefont {I.}~\bibnamefont
  {de~Martino}}, \bibinfo {author} {\bibfnamefont {S.~S.}\ \bibnamefont
  {Chakrabarty}}, \bibinfo {author} {\bibfnamefont {V.}~\bibnamefont {Cesare}},
  \bibinfo {author} {\bibfnamefont {A.}~\bibnamefont {Gallo}}, \bibinfo
  {author} {\bibfnamefont {L.}~\bibnamefont {Ostorero}}, \ and\ \bibinfo
  {author} {\bibfnamefont {A.}~\bibnamefont {Diaferio}},\ }\href {\doibase
  10.3390/universe6080107} {\bibfield  {journal} {\bibinfo  {journal}
  {Universe}\ }\textbf {\bibinfo {volume} {6}} (\bibinfo {year} {2020}),\
  10.3390/universe6080107}\BibitemShut {NoStop}%
\bibitem [{\citenamefont {Martin}(1998)}]{Martin:1997}%
  \BibitemOpen
  \bibfield  {author} {\bibinfo {author} {\bibfnamefont {S.~P.}\ \bibnamefont
  {Martin}},\ }\href {\doibase 10.1142/9789812839657_0001} {\bibfield
  {journal} {\bibinfo  {journal} {Adv. Ser. Direct. High Energy Phys.}\
  }\textbf {\bibinfo {volume} {18}},\ \bibinfo {pages} {1} (\bibinfo {year}
  {1998})},\ \Eprint {http://arxiv.org/abs/hep-ph/9709356}
  {arXiv:hep-ph/9709356} \BibitemShut {NoStop}%
\bibitem [{\citenamefont {Abazov}\ \emph {et~al.}(2008)\citenamefont {Abazov}
  \emph {et~al.}}]{Abazov:2007}%
  \BibitemOpen
  \bibfield  {author} {\bibinfo {author} {\bibfnamefont {V.~M.}\ \bibnamefont
  {Abazov}} \emph {et~al.} (\bibinfo {collaboration} {D0}),\ }\href {\doibase
  10.1016/j.physletb.2008.01.042} {\bibfield  {journal} {\bibinfo  {journal}
  {Phys. Lett. B}\ }\textbf {\bibinfo {volume} {660}},\ \bibinfo {pages} {449}
  (\bibinfo {year} {2008})},\ \Eprint {http://arxiv.org/abs/0712.3805}
  {arXiv:0712.3805 [hep-ex]} \BibitemShut {NoStop}%
\bibitem [{\citenamefont {Wilczek}(1978)}]{Wilczek:1978}%
  \BibitemOpen
  \bibfield  {author} {\bibinfo {author} {\bibfnamefont {F.}~\bibnamefont
  {Wilczek}},\ }\href {\doibase 10.1103/PhysRevLett.40.279} {\bibfield
  {journal} {\bibinfo  {journal} {Phys. Rev. Lett.}\ }\textbf {\bibinfo
  {volume} {40}},\ \bibinfo {pages} {279} (\bibinfo {year} {1978})}\BibitemShut
  {NoStop}%
\bibitem [{\citenamefont {Lee}\ and\ \citenamefont {Koh}(1996)}]{Lee:1996}%
  \BibitemOpen
  \bibfield  {author} {\bibinfo {author} {\bibfnamefont {J.-w.}\ \bibnamefont
  {Lee}}\ and\ \bibinfo {author} {\bibfnamefont {I.-g.}\ \bibnamefont {Koh}},\
  }\href {\doibase 10.1103/PhysRevD.53.2236} {\bibfield  {journal} {\bibinfo
  {journal} {Phys. Rev. D}\ }\textbf {\bibinfo {volume} {53}},\ \bibinfo
  {pages} {2236} (\bibinfo {year} {1996})}\BibitemShut {NoStop}%
\bibitem [{\citenamefont {Ure\~na L\'opez}\ and\ \citenamefont
  {Matos}(2000)}]{Urena:2000}%
  \BibitemOpen
  \bibfield  {author} {\bibinfo {author} {\bibfnamefont {L.~A.}\ \bibnamefont
  {Ure\~na L\'opez}}\ and\ \bibinfo {author} {\bibfnamefont {T.}~\bibnamefont
  {Matos}},\ }\href {\doibase 10.1103/PhysRevD.62.081302} {\bibfield  {journal}
  {\bibinfo  {journal} {Phys. Rev. D}\ }\textbf {\bibinfo {volume} {62}},\
  \bibinfo {pages} {081302} (\bibinfo {year} {2000})}\BibitemShut {NoStop}%
\bibitem [{\citenamefont {Zeldovich}(1968)}]{Zeldovich}%
  \BibitemOpen
  \bibfield  {author} {\bibinfo {author} {\bibfnamefont {Y.~B.}\ \bibnamefont
  {Zeldovich}},\ }\href@noop {} {\bibfield  {journal} {\bibinfo  {journal}
  {Soviet Physics Uspekhi}\ }\textbf {\bibinfo {volume} {11}} (\bibinfo {year}
  {1968})}\BibitemShut {NoStop}%
%%CITATION = HEP-PH/0210296;%%
\bibitem [{\citenamefont {Weinberg}(1989)}]{Weinberg}%
  \BibitemOpen
  \bibfield  {author} {\bibinfo {author} {\bibfnamefont {S.}~\bibnamefont
  {Weinberg}},\ }\href@noop {} {\bibfield  {journal} {\bibinfo  {journal}
  {Reviews of Modern Physics}\ }\textbf {\bibinfo {volume} {61}} (\bibinfo
  {year} {1989})}\BibitemShut {NoStop}%
%%CITATION = HEP-PH/0210296;%%
\bibitem [{\citenamefont {Carroll}(2001)}]{Carroll:2000fy}%
  \BibitemOpen
  \bibfield  {author} {\bibinfo {author} {\bibfnamefont {S.~M.}\ \bibnamefont
  {Carroll}},\ }\href {\doibase 10.12942/lrr-2001-1} {\bibfield  {journal}
  {\bibinfo  {journal} {Living Rev. Rel.}\ }\textbf {\bibinfo {volume} {4}},\
  \bibinfo {pages} {1} (\bibinfo {year} {2001})},\ \Eprint
  {http://arxiv.org/abs/astro-ph/0004075} {arXiv:astro-ph/0004075} \BibitemShut
  {NoStop}%
\bibitem [{\citenamefont {{Millon}}\ \emph {et~al.}(2020)\citenamefont
  {{Millon}}, \citenamefont {{Galan}}, \citenamefont {{Courbin}}, \citenamefont
  {{Treu}}, \citenamefont {{Suyu}}, \citenamefont {{Ding}}, \citenamefont
  {{Birrer}}, \citenamefont {{Chen}}, \citenamefont {{Shajib}}, \citenamefont
  {{Sluse}}, \citenamefont {{Wong}}, \citenamefont {{Agnello}}, \citenamefont
  {{Auger}}, \citenamefont {{Buckley-Geer}}, \citenamefont {{Chan}},
  \citenamefont {{Collett}}, \citenamefont {{Fassnacht}}, \citenamefont
  {{Hilbert}}, \citenamefont {{Koopmans}}, \citenamefont {{Motta}},
  \citenamefont {{Mukherjee}}, \citenamefont {{Rusu}}, \citenamefont
  {{Sonnenfeld}}, \citenamefont {{Spiniello}},\ and\ \citenamefont {{Van de
  Vyvere}}}]{Millon:2020}%
  \BibitemOpen
  \bibfield  {author} {\bibinfo {author} {\bibfnamefont {M.}~\bibnamefont
  {{Millon}}}, \bibinfo {author} {\bibfnamefont {A.}~\bibnamefont {{Galan}}},
  \bibinfo {author} {\bibfnamefont {F.}~\bibnamefont {{Courbin}}}, \bibinfo
  {author} {\bibfnamefont {T.}~\bibnamefont {{Treu}}}, \bibinfo {author}
  {\bibfnamefont {S.~H.}\ \bibnamefont {{Suyu}}}, \bibinfo {author}
  {\bibfnamefont {X.}~\bibnamefont {{Ding}}}, \bibinfo {author} {\bibfnamefont
  {S.}~\bibnamefont {{Birrer}}}, \bibinfo {author} {\bibfnamefont {G.~C.~F.}\
  \bibnamefont {{Chen}}}, \bibinfo {author} {\bibfnamefont {A.~J.}\
  \bibnamefont {{Shajib}}}, \bibinfo {author} {\bibfnamefont {D.}~\bibnamefont
  {{Sluse}}}, \bibinfo {author} {\bibfnamefont {K.~C.}\ \bibnamefont {{Wong}}},
  \bibinfo {author} {\bibfnamefont {A.}~\bibnamefont {{Agnello}}}, \bibinfo
  {author} {\bibfnamefont {M.~W.}\ \bibnamefont {{Auger}}}, \bibinfo {author}
  {\bibfnamefont {E.~J.}\ \bibnamefont {{Buckley-Geer}}}, \bibinfo {author}
  {\bibfnamefont {J.~H.~H.}\ \bibnamefont {{Chan}}}, \bibinfo {author}
  {\bibfnamefont {T.}~\bibnamefont {{Collett}}}, \bibinfo {author}
  {\bibfnamefont {C.~D.}\ \bibnamefont {{Fassnacht}}}, \bibinfo {author}
  {\bibfnamefont {S.}~\bibnamefont {{Hilbert}}}, \bibinfo {author}
  {\bibfnamefont {L.~V.~E.}\ \bibnamefont {{Koopmans}}}, \bibinfo {author}
  {\bibfnamefont {V.}~\bibnamefont {{Motta}}}, \bibinfo {author} {\bibfnamefont
  {S.}~\bibnamefont {{Mukherjee}}}, \bibinfo {author} {\bibfnamefont {C.~E.}\
  \bibnamefont {{Rusu}}}, \bibinfo {author} {\bibfnamefont {A.}~\bibnamefont
  {{Sonnenfeld}}}, \bibinfo {author} {\bibfnamefont {C.}~\bibnamefont
  {{Spiniello}}}, \ and\ \bibinfo {author} {\bibfnamefont {L.}~\bibnamefont
  {{Van de Vyvere}}},\ }\href {\doibase 10.1051/0004-6361/201937351} {\bibfield
   {journal} {\bibinfo  {journal} {A\&A}\ }\textbf {\bibinfo {volume} {639}},\
  \bibinfo {eid} {A101} (\bibinfo {year} {2020})},\ \Eprint
  {http://arxiv.org/abs/1912.08027} {arXiv:1912.08027 [astro-ph.CO]}
  \BibitemShut {NoStop}%
\bibitem [{\citenamefont {{Birrer}}\ \emph {et~al.}(2020)\citenamefont
  {{Birrer}}, \citenamefont {{Shajib}}, \citenamefont {{Galan}}, \citenamefont
  {{Millon}}, \citenamefont {{Treu}}, \citenamefont {{Agnello}}, \citenamefont
  {{Auger}}, \citenamefont {{Chen}}, \citenamefont {{Christensen}},
  \citenamefont {{Collett}}, \citenamefont {{Courbin}}, \citenamefont
  {{Fassnacht}}, \citenamefont {{Koopmans}}, \citenamefont {{Marshall}},
  \citenamefont {{Park}}, \citenamefont {{Rusu}}, \citenamefont {{Sluse}},
  \citenamefont {{Spiniello}}, \citenamefont {{Suyu}}, \citenamefont
  {{Wagner-Carena}}, \citenamefont {{Wong}}, \citenamefont {{Barnab{\`e}}},
  \citenamefont {{Bolton}}, \citenamefont {{Czoske}}, \citenamefont {{Ding}},
  \citenamefont {{Frieman}},\ and\ \citenamefont {{Van de
  Vyvere}}}]{Birrer:2020}%
  \BibitemOpen
  \bibfield  {author} {\bibinfo {author} {\bibfnamefont {S.}~\bibnamefont
  {{Birrer}}}, \bibinfo {author} {\bibfnamefont {A.~J.}\ \bibnamefont
  {{Shajib}}}, \bibinfo {author} {\bibfnamefont {A.}~\bibnamefont {{Galan}}},
  \bibinfo {author} {\bibfnamefont {M.}~\bibnamefont {{Millon}}}, \bibinfo
  {author} {\bibfnamefont {T.}~\bibnamefont {{Treu}}}, \bibinfo {author}
  {\bibfnamefont {A.}~\bibnamefont {{Agnello}}}, \bibinfo {author}
  {\bibfnamefont {M.}~\bibnamefont {{Auger}}}, \bibinfo {author} {\bibfnamefont
  {G.~C.~F.}\ \bibnamefont {{Chen}}}, \bibinfo {author} {\bibfnamefont
  {L.}~\bibnamefont {{Christensen}}}, \bibinfo {author} {\bibfnamefont
  {T.}~\bibnamefont {{Collett}}}, \bibinfo {author} {\bibfnamefont
  {F.}~\bibnamefont {{Courbin}}}, \bibinfo {author} {\bibfnamefont {C.~D.}\
  \bibnamefont {{Fassnacht}}}, \bibinfo {author} {\bibfnamefont {L.~V.~E.}\
  \bibnamefont {{Koopmans}}}, \bibinfo {author} {\bibfnamefont {P.~J.}\
  \bibnamefont {{Marshall}}}, \bibinfo {author} {\bibfnamefont {J.~W.}\
  \bibnamefont {{Park}}}, \bibinfo {author} {\bibfnamefont {C.~E.}\
  \bibnamefont {{Rusu}}}, \bibinfo {author} {\bibfnamefont {D.}~\bibnamefont
  {{Sluse}}}, \bibinfo {author} {\bibfnamefont {C.}~\bibnamefont
  {{Spiniello}}}, \bibinfo {author} {\bibfnamefont {S.~H.}\ \bibnamefont
  {{Suyu}}}, \bibinfo {author} {\bibfnamefont {S.}~\bibnamefont
  {{Wagner-Carena}}}, \bibinfo {author} {\bibfnamefont {K.~C.}\ \bibnamefont
  {{Wong}}}, \bibinfo {author} {\bibfnamefont {M.}~\bibnamefont
  {{Barnab{\`e}}}}, \bibinfo {author} {\bibfnamefont {A.~S.}\ \bibnamefont
  {{Bolton}}}, \bibinfo {author} {\bibfnamefont {O.}~\bibnamefont {{Czoske}}},
  \bibinfo {author} {\bibfnamefont {X.}~\bibnamefont {{Ding}}}, \bibinfo
  {author} {\bibfnamefont {J.~A.}\ \bibnamefont {{Frieman}}}, \ and\ \bibinfo
  {author} {\bibfnamefont {L.}~\bibnamefont {{Van de Vyvere}}},\ }\href
  {\doibase 10.1051/0004-6361/202038861} {\bibfield  {journal} {\bibinfo
  {journal} {A\&A}\ }\textbf {\bibinfo {volume} {643}},\ \bibinfo {eid} {A165}
  (\bibinfo {year} {2020})},\ \Eprint {http://arxiv.org/abs/2007.02941}
  {arXiv:2007.02941 [astro-ph.CO]} \BibitemShut {NoStop}%
\bibitem [{\citenamefont {Joudaki}\ \emph {et~al.}(2017)\citenamefont {Joudaki}
  \emph {et~al.}}]{Joudaki:2016mvz}%
  \BibitemOpen
  \bibfield  {author} {\bibinfo {author} {\bibfnamefont {S.}~\bibnamefont
  {Joudaki}} \emph {et~al.},\ }\href {\doibase 10.1093/mnras/stw2665}
  {\bibfield  {journal} {\bibinfo  {journal} {Mon. Not. Roy. Astron. Soc.}\
  }\textbf {\bibinfo {volume} {465}},\ \bibinfo {pages} {2033} (\bibinfo {year}
  {2017})},\ \Eprint {http://arxiv.org/abs/1601.05786} {arXiv:1601.05786
  [astro-ph.CO]} \BibitemShut {NoStop}%
\bibitem [{\citenamefont {Hildebrandt}\ \emph {et~al.}(2017)\citenamefont
  {Hildebrandt} \emph {et~al.}}]{Hildebrandt:2016iqg}%
  \BibitemOpen
  \bibfield  {author} {\bibinfo {author} {\bibfnamefont {H.}~\bibnamefont
  {Hildebrandt}} \emph {et~al.},\ }\href {\doibase 10.1093/mnras/stw2805}
  {\bibfield  {journal} {\bibinfo  {journal} {Mon. Not. Roy. Astron. Soc.}\
  }\textbf {\bibinfo {volume} {465}},\ \bibinfo {pages} {1454} (\bibinfo {year}
  {2017})},\ \Eprint {http://arxiv.org/abs/1606.05338} {arXiv:1606.05338
  [astro-ph.CO]} \BibitemShut {NoStop}%
\bibitem [{\citenamefont {Riess}\ \emph {et~al.}(2018)\citenamefont {Riess},
  \citenamefont {Casertano}, \citenamefont {Yuan}, \citenamefont {Macri},
  \citenamefont {Anderson}, \citenamefont {MacKenty}, \citenamefont {Bowers},
  \citenamefont {Clubb}, \citenamefont {Filippenko}, \citenamefont {Jones},\
  and\ \citenamefont {Tucker}}]{Riess_2018}%
  \BibitemOpen
  \bibfield  {author} {\bibinfo {author} {\bibfnamefont {A.~G.}\ \bibnamefont
  {Riess}}, \bibinfo {author} {\bibfnamefont {S.}~\bibnamefont {Casertano}},
  \bibinfo {author} {\bibfnamefont {W.}~\bibnamefont {Yuan}}, \bibinfo {author}
  {\bibfnamefont {L.}~\bibnamefont {Macri}}, \bibinfo {author} {\bibfnamefont
  {J.}~\bibnamefont {Anderson}}, \bibinfo {author} {\bibfnamefont {J.~W.}\
  \bibnamefont {MacKenty}}, \bibinfo {author} {\bibfnamefont {J.~B.}\
  \bibnamefont {Bowers}}, \bibinfo {author} {\bibfnamefont {K.~I.}\
  \bibnamefont {Clubb}}, \bibinfo {author} {\bibfnamefont {A.~V.}\ \bibnamefont
  {Filippenko}}, \bibinfo {author} {\bibfnamefont {D.~O.}\ \bibnamefont
  {Jones}}, \ and\ \bibinfo {author} {\bibfnamefont {B.~E.}\ \bibnamefont
  {Tucker}},\ }\href {\doibase 10.3847/1538-4357/aaadb7} {\bibfield  {journal}
  {\bibinfo  {journal} {The Astrophysical Journal}\ }\textbf {\bibinfo {volume}
  {855}},\ \bibinfo {pages} {136} (\bibinfo {year} {2018})}\BibitemShut
  {NoStop}%
\bibitem [{\citenamefont {Riess}\ \emph
  {et~al.}(2019{\natexlab{a}})\citenamefont {Riess}, \citenamefont {Casertano},
  \citenamefont {Yuan}, \citenamefont {Macri},\ and\ \citenamefont
  {Scolnic}}]{Riess:2019cxk}%
  \BibitemOpen
  \bibfield  {author} {\bibinfo {author} {\bibfnamefont {A.~G.}\ \bibnamefont
  {Riess}}, \bibinfo {author} {\bibfnamefont {S.}~\bibnamefont {Casertano}},
  \bibinfo {author} {\bibfnamefont {W.}~\bibnamefont {Yuan}}, \bibinfo {author}
  {\bibfnamefont {L.~M.}\ \bibnamefont {Macri}}, \ and\ \bibinfo {author}
  {\bibfnamefont {D.}~\bibnamefont {Scolnic}},\ }\href {\doibase
  10.3847/1538-4357/ab1422} {\bibfield  {journal} {\bibinfo  {journal}
  {Astrophys. J.}\ }\textbf {\bibinfo {volume} {876}},\ \bibinfo {pages} {85}
  (\bibinfo {year} {2019}{\natexlab{a}})},\ \Eprint
  {http://arxiv.org/abs/1903.07603} {arXiv:1903.07603 [astro-ph.CO]}
  \BibitemShut {NoStop}%
\bibitem [{\citenamefont {Di~Valentino}\ \emph {et~al.}(2021)\citenamefont
  {Di~Valentino}, \citenamefont {Mena}, \citenamefont {Pan}, \citenamefont
  {Visinelli}, \citenamefont {Yang}, \citenamefont {Melchiorri}, \citenamefont
  {Mota}, \citenamefont {Riess},\ and\ \citenamefont
  {Silk}}]{DiValentino:2021izs}%
  \BibitemOpen
  \bibfield  {author} {\bibinfo {author} {\bibfnamefont {E.}~\bibnamefont
  {Di~Valentino}}, \bibinfo {author} {\bibfnamefont {O.}~\bibnamefont {Mena}},
  \bibinfo {author} {\bibfnamefont {S.}~\bibnamefont {Pan}}, \bibinfo {author}
  {\bibfnamefont {L.}~\bibnamefont {Visinelli}}, \bibinfo {author}
  {\bibfnamefont {W.}~\bibnamefont {Yang}}, \bibinfo {author} {\bibfnamefont
  {A.}~\bibnamefont {Melchiorri}}, \bibinfo {author} {\bibfnamefont {D.~F.}\
  \bibnamefont {Mota}}, \bibinfo {author} {\bibfnamefont {A.~G.}\ \bibnamefont
  {Riess}}, \ and\ \bibinfo {author} {\bibfnamefont {J.}~\bibnamefont {Silk}},\
  }\href@noop {} {\  (\bibinfo {year} {2021})},\ \Eprint
  {http://arxiv.org/abs/2103.01183} {arXiv:2103.01183 [astro-ph.CO]}
  \BibitemShut {NoStop}%
\bibitem [{\citenamefont {Lukovi\'c}\ \emph {et~al.}(2016)\citenamefont
  {Lukovi\'c}, \citenamefont {D'Agostino},\ and\ \citenamefont
  {Vittorio}}]{Lukovic:2016}%
  \BibitemOpen
  \bibfield  {author} {\bibinfo {author} {\bibfnamefont {V.~V.}\ \bibnamefont
  {Lukovi\'c}}, \bibinfo {author} {\bibfnamefont {R.}~\bibnamefont
  {D'Agostino}}, \ and\ \bibinfo {author} {\bibfnamefont {N.}~\bibnamefont
  {Vittorio}},\ }\href {\doibase 10.1051/0004-6361/201628217} {\bibfield
  {journal} {\bibinfo  {journal} {Astron. Astrophys.}\ }\textbf {\bibinfo
  {volume} {595}},\ \bibinfo {pages} {A109} (\bibinfo {year} {2016})},\ \Eprint
  {http://arxiv.org/abs/1607.05677} {arXiv:1607.05677 [astro-ph.CO]}
  \BibitemShut {NoStop}%
\bibitem [{\citenamefont {Brax}(2017)}]{Brax:2017}%
  \BibitemOpen
  \bibfield  {author} {\bibinfo {author} {\bibfnamefont {P.}~\bibnamefont
  {Brax}},\ }\href {\doibase 10.1088/1361-6633/aa8e64} {\bibfield  {journal}
  {\bibinfo  {journal} {Reports on Progress in Physics}\ }\textbf {\bibinfo
  {volume} {81}},\ \bibinfo {pages} {016902} (\bibinfo {year}
  {2017})}\BibitemShut {NoStop}%
\bibitem [{\citenamefont {{Kab{\'a}th}}\ \emph {et~al.}(2020)\citenamefont
  {{Kab{\'a}th}}, \citenamefont {{Jones}},\ and\ \citenamefont
  {{Skarka}}}]{Kabath:2020}%
  \BibitemOpen
  \bibfield  {author} {\bibinfo {author} {\bibfnamefont {P.}~\bibnamefont
  {{Kab{\'a}th}}}, \bibinfo {author} {\bibfnamefont {D.}~\bibnamefont
  {{Jones}}}, \ and\ \bibinfo {author} {\bibfnamefont {M.}~\bibnamefont
  {{Skarka}}},\ }\href {\doibase 10.1007/978-3-030-38509-5} {\emph {\bibinfo
  {title} {{Reviews in Frontiers of Modern Astrophysics; From Space Debris to
  Cosmology}}}}\ (\bibinfo {year} {2020})\BibitemShut {NoStop}%
\bibitem [{\citenamefont {{Li}}\ and\ \citenamefont
  {{Shafieloo}}(2019)}]{PEDE:2019ApJ}%
  \BibitemOpen
  \bibfield  {author} {\bibinfo {author} {\bibfnamefont {X.}~\bibnamefont
  {{Li}}}\ and\ \bibinfo {author} {\bibfnamefont {A.}~\bibnamefont
  {{Shafieloo}}},\ }\href {\doibase 10.3847/2041-8213/ab3e09} {\bibfield
  {journal} {\bibinfo  {journal} {ApJL}\ }\textbf {\bibinfo {volume} {883}},\
  \bibinfo {eid} {L3} (\bibinfo {year} {2019})},\ \Eprint
  {http://arxiv.org/abs/1906.08275} {arXiv:1906.08275 [astro-ph.CO]}
  \BibitemShut {NoStop}%
\bibitem [{\citenamefont {{Li}}\ and\ \citenamefont
  {{Shafieloo}}(2020)}]{PEDE:2020}%
  \BibitemOpen
  \bibfield  {author} {\bibinfo {author} {\bibfnamefont {X.}~\bibnamefont
  {{Li}}}\ and\ \bibinfo {author} {\bibfnamefont {A.}~\bibnamefont
  {{Shafieloo}}},\ }\href@noop {} {\bibfield  {journal} {\bibinfo  {journal}
  {arXiv e-prints}\ ,\ \bibinfo {eid} {arXiv:2001.05103}} (\bibinfo {year}
  {2020})},\ \Eprint {http://arxiv.org/abs/2001.05103} {arXiv:2001.05103
  [astro-ph.CO]} \BibitemShut {NoStop}%
\bibitem [{\citenamefont {Hern\'andez-Almada}\ \emph
  {et~al.}(2020{\natexlab{a}})\citenamefont {Hern\'andez-Almada}, \citenamefont
  {Leon}, \citenamefont {Maga\~na}, \citenamefont {Garc\'\i{}a-Aspeitia},\ and\
  \citenamefont {Motta}}]{Hernandez-Almada:2020uyr}%
  \BibitemOpen
  \bibfield  {author} {\bibinfo {author} {\bibfnamefont {A.}~\bibnamefont
  {Hern\'andez-Almada}}, \bibinfo {author} {\bibfnamefont {G.}~\bibnamefont
  {Leon}}, \bibinfo {author} {\bibfnamefont {J.}~\bibnamefont {Maga\~na}},
  \bibinfo {author} {\bibfnamefont {M.~A.}\ \bibnamefont
  {Garc\'\i{}a-Aspeitia}}, \ and\ \bibinfo {author} {\bibfnamefont
  {V.}~\bibnamefont {Motta}},\ }\href {\doibase 10.1093/mnras/staa2052}
  {\bibfield  {journal} {\bibinfo  {journal} {Mon. Not. Roy. Astron. Soc.}\
  }\textbf {\bibinfo {volume} {497}},\ \bibinfo {pages} {1590} (\bibinfo {year}
  {2020}{\natexlab{a}})},\ \Eprint {http://arxiv.org/abs/2002.12881}
  {arXiv:2002.12881 [astro-ph.CO]} \BibitemShut {NoStop}%
\bibitem [{\citenamefont {Maartens}(2000{\natexlab{a}})}]{m2000}%
  \BibitemOpen
  \bibfield  {author} {\bibinfo {author} {\bibfnamefont {R.}~\bibnamefont
  {Maartens}},\ }\href@noop {} {\bibfield  {journal} {\bibinfo  {journal}
  {Phys. Rev. D}\ }\textbf {\bibinfo {volume} {62}},\ \bibinfo {pages} {084023}
  (\bibinfo {year} {2000}{\natexlab{a}})}\BibitemShut {NoStop}%
\bibitem [{\citenamefont {Maartens}(2000{\natexlab{b}})}]{MaartensCos}%
  \BibitemOpen
  \bibfield  {author} {\bibinfo {author} {\bibfnamefont {R.}~\bibnamefont
  {Maartens}},\ }\href@noop {} {\bibfield  {journal} {\bibinfo  {journal}
  {Phys. Rev. D}\ }\textbf {\bibinfo {volume} {62}},\ \bibinfo {pages} {084023}
  (\bibinfo {year} {2000}{\natexlab{b}})}\BibitemShut {NoStop}%
%%CITATION = ARXIV:1001.4538;%%
\bibitem [{\citenamefont {Garc\'\i{}a-Aspeitia}\ \emph
  {et~al.}(2017)\citenamefont {Garc\'\i{}a-Aspeitia}, \citenamefont {Maga\~na},
  \citenamefont {Hern\'andez-Almada},\ and\ \citenamefont
  {Motta}}]{Garcia-Aspeitia:2016kak}%
  \BibitemOpen
  \bibfield  {author} {\bibinfo {author} {\bibfnamefont {M.~A.}\ \bibnamefont
  {Garc\'\i{}a-Aspeitia}}, \bibinfo {author} {\bibfnamefont {J.}~\bibnamefont
  {Maga\~na}}, \bibinfo {author} {\bibfnamefont {A.}~\bibnamefont
  {Hern\'andez-Almada}}, \ and\ \bibinfo {author} {\bibfnamefont
  {V.}~\bibnamefont {Motta}},\ }\href {\doibase 10.1142/S0218271818500062}
  {\bibfield  {journal} {\bibinfo  {journal} {Int. J. Mod. Phys. D}\ }\textbf
  {\bibinfo {volume} {27}},\ \bibinfo {pages} {1850006} (\bibinfo {year}
  {2017})},\ \Eprint {http://arxiv.org/abs/1609.08220} {arXiv:1609.08220
  [astro-ph.CO]} \BibitemShut {NoStop}%
\bibitem [{\citenamefont {Garcia-Aspeitia}\ \emph {et~al.}(2018)\citenamefont
  {Garcia-Aspeitia}, \citenamefont {Hernandez-Almada}, \citenamefont
  {Maga\~na}, \citenamefont {Amante}, \citenamefont {Motta},\ and\
  \citenamefont {Mart\'\i{}nez-Robles}}]{Garcia-Aspeitia:2018fvw}%
  \BibitemOpen
  \bibfield  {author} {\bibinfo {author} {\bibfnamefont {M.~A.}\ \bibnamefont
  {Garcia-Aspeitia}}, \bibinfo {author} {\bibfnamefont {A.}~\bibnamefont
  {Hernandez-Almada}}, \bibinfo {author} {\bibfnamefont {J.}~\bibnamefont
  {Maga\~na}}, \bibinfo {author} {\bibfnamefont {M.~H.}\ \bibnamefont
  {Amante}}, \bibinfo {author} {\bibfnamefont {V.}~\bibnamefont {Motta}}, \
  and\ \bibinfo {author} {\bibfnamefont {C.}~\bibnamefont
  {Mart\'\i{}nez-Robles}},\ }\href {\doibase 10.1103/PhysRevD.97.101301}
  {\bibfield  {journal} {\bibinfo  {journal} {Phys. Rev. D}\ }\textbf {\bibinfo
  {volume} {97}},\ \bibinfo {pages} {101301} (\bibinfo {year} {2018})},\
  \Eprint {http://arxiv.org/abs/1804.05085} {arXiv:1804.05085 [gr-qc]}
  \BibitemShut {NoStop}%
\bibitem [{\citenamefont {Wainwright}\ and\ \citenamefont
  {Ellis}(1997)}]{Ellis}%
  \BibitemOpen
  \bibfield  {author} {\bibinfo {author} {\bibfnamefont {J.}~\bibnamefont
  {Wainwright}}\ and\ \bibinfo {author} {\bibfnamefont {G.~F.~R.}\ \bibnamefont
  {Ellis}},\ }\href@noop {} {\emph {\bibinfo {title} {{Dynamical Systems in
  Cosmology}}}}\ (\bibinfo  {publisher} {Cambridge University Press},\ \bibinfo
  {year} {1997})\BibitemShut {NoStop}%
\bibitem [{\citenamefont {Gao}\ \emph {et~al.}(2014)\citenamefont {Gao},
  \citenamefont {Brandenberger}, \citenamefont {Cai},\ and\ \citenamefont
  {Chen}}]{Gao:2014nia}%
  \BibitemOpen
  \bibfield  {author} {\bibinfo {author} {\bibfnamefont {C.}~\bibnamefont
  {Gao}}, \bibinfo {author} {\bibfnamefont {R.~H.}\ \bibnamefont
  {Brandenberger}}, \bibinfo {author} {\bibfnamefont {Y.}~\bibnamefont {Cai}},
  \ and\ \bibinfo {author} {\bibfnamefont {P.}~\bibnamefont {Chen}},\ }\href
  {\doibase 10.1088/1475-7516/2014/09/021} {\bibfield  {journal} {\bibinfo
  {journal} {JCAP}\ }\textbf {\bibinfo {volume} {1409}},\ \bibinfo {pages}
  {021} (\bibinfo {year} {2014})},\ \Eprint {http://arxiv.org/abs/1405.1644}
  {arXiv:1405.1644 [gr-qc]} \BibitemShut {NoStop}%
%%CITATION = ARXIV:1405.1644;%%
\bibitem [{\citenamefont {Garc\'ia-Aspeitia}\ \emph {et~al.}(2019)\citenamefont
  {Garc\'ia-Aspeitia}, \citenamefont {Mart\'inez-Robles}, \citenamefont
  {Hern\'andez-Almada}, \citenamefont {Maga\~na},\ and\ \citenamefont
  {Motta}}]{Garcia-Aspeitia:2019yni}%
  \BibitemOpen
  \bibfield  {author} {\bibinfo {author} {\bibfnamefont {M.~A.}\ \bibnamefont
  {Garc\'ia-Aspeitia}}, \bibinfo {author} {\bibfnamefont {C.}~\bibnamefont
  {Mart\'inez-Robles}}, \bibinfo {author} {\bibfnamefont {A.}~\bibnamefont
  {Hern\'andez-Almada}}, \bibinfo {author} {\bibfnamefont {J.}~\bibnamefont
  {Maga\~na}}, \ and\ \bibinfo {author} {\bibfnamefont {V.}~\bibnamefont
  {Motta}},\ }\href {\doibase 10.1103/PhysRevD.99.123525} {\bibfield  {journal}
  {\bibinfo  {journal} {Phys. Rev.}\ }\textbf {\bibinfo {volume} {D99}},\
  \bibinfo {pages} {123525} (\bibinfo {year} {2019})},\ \Eprint
  {http://arxiv.org/abs/1903.06344} {arXiv:1903.06344 [gr-qc]} \BibitemShut
  {NoStop}%
%%CITATION = ARXIV:1903.06344;%%
\bibitem [{\citenamefont {Glavan}\ and\ \citenamefont
  {Lin}(2020)}]{Glavan:2019inb}%
  \BibitemOpen
  \bibfield  {author} {\bibinfo {author} {\bibfnamefont {D.~z.}\ \bibnamefont
  {Glavan}}\ and\ \bibinfo {author} {\bibfnamefont {C.}~\bibnamefont {Lin}},\
  }\href {\doibase 10.1103/PhysRevLett.124.081301} {\bibfield  {journal}
  {\bibinfo  {journal} {Phys. Rev. Lett.}\ }\textbf {\bibinfo {volume} {124}},\
  \bibinfo {pages} {081301} (\bibinfo {year} {2020})},\ \Eprint
  {http://arxiv.org/abs/1905.03601} {arXiv:1905.03601 [gr-qc]} \BibitemShut
  {NoStop}%
\bibitem [{\citenamefont {Garc\'ia-Aspeitia}\ and\ \citenamefont
  {Hern\'andez-Almada}(2021)}]{Garcia-Aspeitia:2020uwq}%
  \BibitemOpen
  \bibfield  {author} {\bibinfo {author} {\bibfnamefont {M.~A.}\ \bibnamefont
  {Garc\'ia-Aspeitia}}\ and\ \bibinfo {author} {\bibfnamefont {A.}~\bibnamefont
  {Hern\'andez-Almada}},\ }\href {\doibase
  https://doi.org/10.1016/j.dark.2021.100799} {\bibfield  {journal} {\bibinfo
  {journal} {Physics of the Dark Universe}\ }\textbf {\bibinfo {volume} {32}},\
  \bibinfo {pages} {100799} (\bibinfo {year} {2021})},\ \Eprint
  {http://arxiv.org/abs/2007.06730} {arXiv:2007.06730 [astro-ph.CO]}
  \BibitemShut {NoStop}%
\bibitem [{\citenamefont {Freese}\ and\ \citenamefont
  {Lewis}(2002)}]{FREESE20021}%
  \BibitemOpen
  \bibfield  {author} {\bibinfo {author} {\bibfnamefont {K.}~\bibnamefont
  {Freese}}\ and\ \bibinfo {author} {\bibfnamefont {M.}~\bibnamefont {Lewis}},\
  }\href {\doibase https://doi.org/10.1016/S0370-2693(02)02122-6} {\bibfield
  {journal} {\bibinfo  {journal} {Physics Letters B}\ }\textbf {\bibinfo
  {volume} {540}},\ \bibinfo {pages} {1} (\bibinfo {year} {2002})}\BibitemShut
  {NoStop}%
\bibitem [{\citenamefont {Gondolo}\ and\ \citenamefont
  {Freese}(2003)}]{Gondolo:2002fh}%
  \BibitemOpen
  \bibfield  {author} {\bibinfo {author} {\bibfnamefont {P.}~\bibnamefont
  {Gondolo}}\ and\ \bibinfo {author} {\bibfnamefont {K.}~\bibnamefont
  {Freese}},\ }\href {\doibase 10.1103/PhysRevD.68.063509} {\bibfield
  {journal} {\bibinfo  {journal} {Phys. Rev. D}\ }\textbf {\bibinfo {volume}
  {68}},\ \bibinfo {pages} {063509} (\bibinfo {year} {2003})},\ \Eprint
  {http://arxiv.org/abs/hep-ph/0209322} {arXiv:hep-ph/0209322} \BibitemShut
  {NoStop}%
\bibitem [{\citenamefont {Jaime}\ \emph {et~al.}(2018)\citenamefont {Jaime},
  \citenamefont {Jaber},\ and\ \citenamefont
  {Escamilla-Rivera}}]{Jaime:2018ftn}%
  \BibitemOpen
  \bibfield  {author} {\bibinfo {author} {\bibfnamefont {L.~G.}\ \bibnamefont
  {Jaime}}, \bibinfo {author} {\bibfnamefont {M.}~\bibnamefont {Jaber}}, \ and\
  \bibinfo {author} {\bibfnamefont {C.}~\bibnamefont {Escamilla-Rivera}},\
  }\href {\doibase 10.1103/PhysRevD.98.083530} {\bibfield  {journal} {\bibinfo
  {journal} {Phys. Rev. D}\ }\textbf {\bibinfo {volume} {98}},\ \bibinfo
  {pages} {083530} (\bibinfo {year} {2018})},\ \Eprint
  {http://arxiv.org/abs/1804.04284} {arXiv:1804.04284 [astro-ph.CO]}
  \BibitemShut {NoStop}%
\bibitem [{\citenamefont {Iorio}(2015)}]{Iorio:2015}%
  \BibitemOpen
  \bibfield  {author} {\bibinfo {author} {\bibfnamefont {L.}~\bibnamefont
  {Iorio}},\ }\href {\doibase 10.3390/universe1010038} {\bibfield  {journal}
  {\bibinfo  {journal} {Universe}\ }\textbf {\bibinfo {volume} {1}},\ \bibinfo
  {pages} {38} (\bibinfo {year} {2015})}\BibitemShut {NoStop}%
\bibitem [{\citenamefont {Debono}\ and\ \citenamefont
  {Smoot}(2016)}]{Debono:2016}%
  \BibitemOpen
  \bibfield  {author} {\bibinfo {author} {\bibfnamefont {I.}~\bibnamefont
  {Debono}}\ and\ \bibinfo {author} {\bibfnamefont {G.~F.}\ \bibnamefont
  {Smoot}},\ }\href {\doibase 10.3390/universe2040023} {\bibfield  {journal}
  {\bibinfo  {journal} {Universe}\ }\textbf {\bibinfo {volume} {2}} (\bibinfo
  {year} {2016}),\ 10.3390/universe2040023}\BibitemShut {NoStop}%
\bibitem [{\citenamefont {Vishwakarma}(2016)}]{Vishwakarma:2016}%
  \BibitemOpen
  \bibfield  {author} {\bibinfo {author} {\bibfnamefont {R.~G.}\ \bibnamefont
  {Vishwakarma}},\ }\href {\doibase 10.3390/universe2020011} {\bibfield
  {journal} {\bibinfo  {journal} {Universe}\ }\textbf {\bibinfo {volume} {2}}
  (\bibinfo {year} {2016}),\ 10.3390/universe2020011}\BibitemShut {NoStop}%
\bibitem [{\citenamefont {Aghanim}\ \emph {et~al.}(2018)\citenamefont {Aghanim}
  \emph {et~al.}}]{Planck_CP:2018}%
  \BibitemOpen
  \bibfield  {author} {\bibinfo {author} {\bibfnamefont {N.}~\bibnamefont
  {Aghanim}} \emph {et~al.} (\bibinfo {collaboration} {Planck}),\ }\href@noop
  {} {\  (\bibinfo {year} {2018})},\ \Eprint {http://arxiv.org/abs/1807.06209}
  {arXiv:1807.06209 [astro-ph.CO]} \BibitemShut {NoStop}%
%%CITATION = ARXIV:1807.06209;%%
\bibitem [{\citenamefont {{Cao}}\ \emph {et~al.}(2012)\citenamefont {{Cao}},
  \citenamefont {{Pan}}, \citenamefont {{Biesiada}}, \citenamefont
  {{Godlowski}},\ and\ \citenamefont {{Zhu}}}]{Cao:2012}%
  \BibitemOpen
  \bibfield  {author} {\bibinfo {author} {\bibfnamefont {S.}~\bibnamefont
  {{Cao}}}, \bibinfo {author} {\bibfnamefont {Y.}~\bibnamefont {{Pan}}},
  \bibinfo {author} {\bibfnamefont {M.}~\bibnamefont {{Biesiada}}}, \bibinfo
  {author} {\bibfnamefont {W.}~\bibnamefont {{Godlowski}}}, \ and\ \bibinfo
  {author} {\bibfnamefont {Z.-H.}\ \bibnamefont {{Zhu}}},\ }\href {\doibase
  10.1088/1475-7516/2012/03/016} {\bibfield  {journal} {\bibinfo  {journal}
  {jcap}\ }\textbf {\bibinfo {volume} {3}},\ \bibinfo {eid} {016} (\bibinfo
  {year} {2012})},\ \Eprint {http://arxiv.org/abs/1105.6226} {arXiv:1105.6226}
  \BibitemShut {NoStop}%
\bibitem [{\citenamefont {Cao}\ \emph {et~al.}(2015{\natexlab{a}})\citenamefont
  {Cao}, \citenamefont {Biesiada}, \citenamefont {Gavazzi}, \citenamefont
  {Piórkowska},\ and\ \citenamefont {Zhu}}]{Cao:2015qja}%
  \BibitemOpen
  \bibfield  {author} {\bibinfo {author} {\bibfnamefont {S.}~\bibnamefont
  {Cao}}, \bibinfo {author} {\bibfnamefont {M.}~\bibnamefont {Biesiada}},
  \bibinfo {author} {\bibfnamefont {R.}~\bibnamefont {Gavazzi}}, \bibinfo
  {author} {\bibfnamefont {A.}~\bibnamefont {Piórkowska}}, \ and\ \bibinfo
  {author} {\bibfnamefont {Z.-H.}\ \bibnamefont {Zhu}},\ }\href {\doibase
  10.1088/0004-637X/806/2/185} {\bibfield  {journal} {\bibinfo  {journal}
  {Astrophys. J.}\ }\textbf {\bibinfo {volume} {806}},\ \bibinfo {pages} {185}
  (\bibinfo {year} {2015}{\natexlab{a}})},\ \Eprint
  {http://arxiv.org/abs/1509.07649} {arXiv:1509.07649 [astro-ph.CO]}
  \BibitemShut {NoStop}%
%%CITATION = ARXIV:1509.07649;%%
\bibitem [{\citenamefont {Amante}\ \emph {et~al.}(2019)\citenamefont {Amante},
  \citenamefont {Maga\~na}, \citenamefont {Motta}, \citenamefont
  {Garc\'ia-Aspeitia},\ and\ \citenamefont {Verdugo}}]{Amante:2019xao}%
  \BibitemOpen
  \bibfield  {author} {\bibinfo {author} {\bibfnamefont {M.~H.}\ \bibnamefont
  {Amante}}, \bibinfo {author} {\bibfnamefont {J.}~\bibnamefont {Maga\~na}},
  \bibinfo {author} {\bibfnamefont {V.}~\bibnamefont {Motta}}, \bibinfo
  {author} {\bibfnamefont {M.~A.}\ \bibnamefont {Garc\'ia-Aspeitia}}, \ and\
  \bibinfo {author} {\bibfnamefont {T.}~\bibnamefont {Verdugo}},\ }\href@noop
  {} {\  (\bibinfo {year} {2019})},\ \Eprint {http://arxiv.org/abs/1906.04107}
  {arXiv:1906.04107 [astro-ph.CO]} \BibitemShut {NoStop}%
%%CITATION = ARXIV:1906.04107;%%
\bibitem [{\citenamefont {{Ch{\'a}vez}}\ \emph {et~al.}(2012)\citenamefont
  {{Ch{\'a}vez}}, \citenamefont {{Terlevich}}, \citenamefont {{Terlevich}},
  \citenamefont {{Plionis}}, \citenamefont {{Bresolin}}, \citenamefont
  {{Basilakos}},\ and\ \citenamefont {{Melnick}}}]{Chavez2012}%
  \BibitemOpen
  \bibfield  {author} {\bibinfo {author} {\bibfnamefont {R.}~\bibnamefont
  {{Ch{\'a}vez}}}, \bibinfo {author} {\bibfnamefont {E.}~\bibnamefont
  {{Terlevich}}}, \bibinfo {author} {\bibfnamefont {R.}~\bibnamefont
  {{Terlevich}}}, \bibinfo {author} {\bibfnamefont {M.}~\bibnamefont
  {{Plionis}}}, \bibinfo {author} {\bibfnamefont {F.}~\bibnamefont
  {{Bresolin}}}, \bibinfo {author} {\bibfnamefont {S.}~\bibnamefont
  {{Basilakos}}}, \ and\ \bibinfo {author} {\bibfnamefont {J.}~\bibnamefont
  {{Melnick}}},\ }\href {\doibase 10.1111/j.1745-3933.2012.01299.x} {\bibfield
  {journal} {\bibinfo  {journal} {Monthly Notices of the Royal Astronomical
  Society}\ }\textbf {\bibinfo {volume} {425}},\ \bibinfo {pages} {L56}
  (\bibinfo {year} {2012})},\ \Eprint {http://arxiv.org/abs/1203.6222}
  {arXiv:1203.6222 [astro-ph.CO]} \BibitemShut {NoStop}%
\bibitem [{\citenamefont {{Ch{\'a}vez}}\ \emph {et~al.}(2014)\citenamefont
  {{Ch{\'a}vez}}, \citenamefont {{Terlevich}}, \citenamefont {{Terlevich}},
  \citenamefont {{Bresolin}}, \citenamefont {{Melnick}}, \citenamefont
  {{Plionis}},\ and\ \citenamefont {{Basilakos}}}]{Chavez2014}%
  \BibitemOpen
  \bibfield  {author} {\bibinfo {author} {\bibfnamefont {R.}~\bibnamefont
  {{Ch{\'a}vez}}}, \bibinfo {author} {\bibfnamefont {R.}~\bibnamefont
  {{Terlevich}}}, \bibinfo {author} {\bibfnamefont {E.}~\bibnamefont
  {{Terlevich}}}, \bibinfo {author} {\bibfnamefont {F.}~\bibnamefont
  {{Bresolin}}}, \bibinfo {author} {\bibfnamefont {J.}~\bibnamefont
  {{Melnick}}}, \bibinfo {author} {\bibfnamefont {M.}~\bibnamefont
  {{Plionis}}}, \ and\ \bibinfo {author} {\bibfnamefont {S.}~\bibnamefont
  {{Basilakos}}},\ }\href {\doibase 10.1093/mnras/stu987} {\bibfield  {journal}
  {\bibinfo  {journal} {Monthly Notices of the Royal Astronomical Society}\
  }\textbf {\bibinfo {volume} {442}},\ \bibinfo {pages} {3565} (\bibinfo {year}
  {2014})},\ \Eprint {http://arxiv.org/abs/1405.4010} {arXiv:1405.4010
  [astro-ph.GA]} \BibitemShut {NoStop}%
\bibitem [{\citenamefont {{Terlevich}}\ \emph {et~al.}(2015)\citenamefont
  {{Terlevich}}, \citenamefont {{Terlevich}}, \citenamefont {{Melnick}},
  \citenamefont {{Ch{\'a}vez}}, \citenamefont {{Plionis}}, \citenamefont
  {{Bresolin}},\ and\ \citenamefont {{Basilakos}}}]{Terlevich2015}%
  \BibitemOpen
  \bibfield  {author} {\bibinfo {author} {\bibfnamefont {R.}~\bibnamefont
  {{Terlevich}}}, \bibinfo {author} {\bibfnamefont {E.}~\bibnamefont
  {{Terlevich}}}, \bibinfo {author} {\bibfnamefont {J.}~\bibnamefont
  {{Melnick}}}, \bibinfo {author} {\bibfnamefont {R.}~\bibnamefont
  {{Ch{\'a}vez}}}, \bibinfo {author} {\bibfnamefont {M.}~\bibnamefont
  {{Plionis}}}, \bibinfo {author} {\bibfnamefont {F.}~\bibnamefont
  {{Bresolin}}}, \ and\ \bibinfo {author} {\bibfnamefont {S.}~\bibnamefont
  {{Basilakos}}},\ }\href {\doibase 10.1093/mnras/stv1128} {\bibfield
  {journal} {\bibinfo  {journal} {Monthly Notices of the Royal Astronomical
  Society}\ }\textbf {\bibinfo {volume} {451}},\ \bibinfo {pages} {3001}
  (\bibinfo {year} {2015})},\ \Eprint {http://arxiv.org/abs/1505.04376}
  {arXiv:1505.04376 [astro-ph.CO]} \BibitemShut {NoStop}%
\bibitem [{\citenamefont {{Ch{\'a}vez}}\ \emph {et~al.}(2016)\citenamefont
  {{Ch{\'a}vez}}, \citenamefont {{Plionis}}, \citenamefont {{Basilakos}},
  \citenamefont {{Terlevich}}, \citenamefont {{Terlevich}}, \citenamefont
  {{Melnick}}, \citenamefont {{Bresolin}},\ and\ \citenamefont
  {{Gonz{\'a}lez-Mor{\'a}n}}}]{Chavez2016}%
  \BibitemOpen
  \bibfield  {author} {\bibinfo {author} {\bibfnamefont {R.}~\bibnamefont
  {{Ch{\'a}vez}}}, \bibinfo {author} {\bibfnamefont {M.}~\bibnamefont
  {{Plionis}}}, \bibinfo {author} {\bibfnamefont {S.}~\bibnamefont
  {{Basilakos}}}, \bibinfo {author} {\bibfnamefont {R.}~\bibnamefont
  {{Terlevich}}}, \bibinfo {author} {\bibfnamefont {E.}~\bibnamefont
  {{Terlevich}}}, \bibinfo {author} {\bibfnamefont {J.}~\bibnamefont
  {{Melnick}}}, \bibinfo {author} {\bibfnamefont {F.}~\bibnamefont
  {{Bresolin}}}, \ and\ \bibinfo {author} {\bibfnamefont {A.~L.}\ \bibnamefont
  {{Gonz{\'a}lez-Mor{\'a}n}}},\ }\href {\doibase 10.1093/mnras/stw1813}
  {\bibfield  {journal} {\bibinfo  {journal} {Monthly Notices of the Royal
  Astronomical Society}\ }\textbf {\bibinfo {volume} {462}},\ \bibinfo {pages}
  {2431} (\bibinfo {year} {2016})},\ \Eprint {http://arxiv.org/abs/1607.06458}
  {arXiv:1607.06458 [astro-ph.CO]} \BibitemShut {NoStop}%
\bibitem [{\citenamefont {{Gonz{\'a}lez-Mor{\'a}n}}\ \emph
  {et~al.}(2019)\citenamefont {{Gonz{\'a}lez-Mor{\'a}n}}, \citenamefont
  {{Ch{\'a}vez}}, \citenamefont {{Terlevich}}, \citenamefont {{Terlevich}},
  \citenamefont {{Bresolin}}, \citenamefont {{Fern{\'a}ndez-Arenas}},
  \citenamefont {{Plionis}}, \citenamefont {{Basilakos}}, \citenamefont
  {{Melnick}},\ and\ \citenamefont {{Telles}}}]{GonzalezMoran2019}%
  \BibitemOpen
  \bibfield  {author} {\bibinfo {author} {\bibfnamefont {A.~L.}\ \bibnamefont
  {{Gonz{\'a}lez-Mor{\'a}n}}}, \bibinfo {author} {\bibfnamefont
  {R.}~\bibnamefont {{Ch{\'a}vez}}}, \bibinfo {author} {\bibfnamefont
  {R.}~\bibnamefont {{Terlevich}}}, \bibinfo {author} {\bibfnamefont
  {E.}~\bibnamefont {{Terlevich}}}, \bibinfo {author} {\bibfnamefont
  {F.}~\bibnamefont {{Bresolin}}}, \bibinfo {author} {\bibfnamefont
  {D.}~\bibnamefont {{Fern{\'a}ndez-Arenas}}}, \bibinfo {author} {\bibfnamefont
  {M.}~\bibnamefont {{Plionis}}}, \bibinfo {author} {\bibfnamefont
  {S.}~\bibnamefont {{Basilakos}}}, \bibinfo {author} {\bibfnamefont
  {J.}~\bibnamefont {{Melnick}}}, \ and\ \bibinfo {author} {\bibfnamefont
  {E.}~\bibnamefont {{Telles}}},\ }\href {\doibase 10.1093/mnras/stz1577}
  {\bibfield  {journal} {\bibinfo  {journal} {Monthly Notices of the Royal
  Astronomical Society}\ }\textbf {\bibinfo {volume} {487}},\ \bibinfo {pages}
  {4669} (\bibinfo {year} {2019})},\ \Eprint {http://arxiv.org/abs/1906.02195}
  {arXiv:1906.02195 [astro-ph.GA]} \BibitemShut {NoStop}%
\bibitem [{\citenamefont {Cao}\ \emph {et~al.}(2020)\citenamefont {Cao},
  \citenamefont {Ryan},\ and\ \citenamefont {Ratra}}]{Cao:2020jgu}%
  \BibitemOpen
  \bibfield  {author} {\bibinfo {author} {\bibfnamefont {S.}~\bibnamefont
  {Cao}}, \bibinfo {author} {\bibfnamefont {J.}~\bibnamefont {Ryan}}, \ and\
  \bibinfo {author} {\bibfnamefont {B.}~\bibnamefont {Ratra}},\ }\href@noop {}
  {\  (\bibinfo {year} {2020})},\ \Eprint {http://arxiv.org/abs/2005.12617}
  {arXiv:2005.12617 [astro-ph.CO]} \BibitemShut {NoStop}%
\bibitem [{\citenamefont {Jimenez}\ and\ \citenamefont
  {Loeb}(2002)}]{Jimenez:2001gg}%
  \BibitemOpen
  \bibfield  {author} {\bibinfo {author} {\bibfnamefont {R.}~\bibnamefont
  {Jimenez}}\ and\ \bibinfo {author} {\bibfnamefont {A.}~\bibnamefont {Loeb}},\
  }\href {\doibase 10.1086/340549} {\bibfield  {journal} {\bibinfo  {journal}
  {Astrophys. J.}\ }\textbf {\bibinfo {volume} {573}},\ \bibinfo {pages} {37}
  (\bibinfo {year} {2002})},\ \Eprint {http://arxiv.org/abs/astro-ph/0106145}
  {arXiv:astro-ph/0106145 [astro-ph]} \BibitemShut {NoStop}%
%%CITATION = ASTRO-PH/0106145;%%
\bibitem [{\citenamefont {Scolnic}\ \emph {et~al.}(2018)\citenamefont {Scolnic}
  \emph {et~al.}}]{Scolnic:2018}%
  \BibitemOpen
  \bibfield  {author} {\bibinfo {author} {\bibfnamefont {D.~M.}\ \bibnamefont
  {Scolnic}} \emph {et~al.},\ }\href {\doibase 10.3847/1538-4357/aab9bb}
  {\bibfield  {journal} {\bibinfo  {journal} {Astrophys. J.}\ }\textbf
  {\bibinfo {volume} {859}},\ \bibinfo {pages} {101} (\bibinfo {year}
  {2018})},\ \Eprint {http://arxiv.org/abs/1710.00845} {arXiv:1710.00845
  [astro-ph.CO]} \BibitemShut {NoStop}%
%%CITATION = ARXIV:1710.00845;%%
\bibitem [{\citenamefont {Hern\'andez-Almada}\ \emph
  {et~al.}(2020{\natexlab{b}})\citenamefont {Hern\'andez-Almada}, \citenamefont
  {Garc\'ia-Aspeitia}, \citenamefont {Maga\~na},\ and\ \citenamefont
  {Motta}}]{Hernandez-Almada:2020ulm}%
  \BibitemOpen
  \bibfield  {author} {\bibinfo {author} {\bibfnamefont {A.}~\bibnamefont
  {Hern\'andez-Almada}}, \bibinfo {author} {\bibfnamefont {M.~A.}\ \bibnamefont
  {Garc\'ia-Aspeitia}}, \bibinfo {author} {\bibfnamefont {J.}~\bibnamefont
  {Maga\~na}}, \ and\ \bibinfo {author} {\bibfnamefont {V.}~\bibnamefont
  {Motta}},\ }\href {\doibase 10.1103/PhysRevD.101.063516} {\bibfield
  {journal} {\bibinfo  {journal} {Phys. Rev. D}\ }\textbf {\bibinfo {volume}
  {101}},\ \bibinfo {pages} {063516} (\bibinfo {year} {2020}{\natexlab{b}})},\
  \Eprint {http://arxiv.org/abs/2001.08667} {arXiv:2001.08667 [astro-ph.CO]}
  \BibitemShut {NoStop}%
\bibitem [{\citenamefont {Maga\~na}\ \emph {et~al.}(2018)\citenamefont
  {Maga\~na}, \citenamefont {Amante}, \citenamefont {Garc\'ia-Aspeitia},\ and\
  \citenamefont {Motta}}]{Magana:2017nfs}%
  \BibitemOpen
  \bibfield  {author} {\bibinfo {author} {\bibfnamefont {J.}~\bibnamefont
  {Maga\~na}}, \bibinfo {author} {\bibfnamefont {M.~H.}\ \bibnamefont
  {Amante}}, \bibinfo {author} {\bibfnamefont {M.~A.}\ \bibnamefont
  {Garc\'ia-Aspeitia}}, \ and\ \bibinfo {author} {\bibfnamefont
  {V.}~\bibnamefont {Motta}},\ }\href {\doibase 10.1093/mnras/sty260}
  {\bibfield  {journal} {\bibinfo  {journal} {Mon. Not. Roy. Astron. Soc.}\
  }\textbf {\bibinfo {volume} {476}},\ \bibinfo {pages} {1036} (\bibinfo {year}
  {2018})},\ \Eprint {http://arxiv.org/abs/1706.09848} {arXiv:1706.09848
  [astro-ph.CO]} \BibitemShut {NoStop}%
%%CITATION = ARXIV:1706.09848;%%
\bibitem [{\citenamefont {Eisenstein}\ and\ \citenamefont
  {Hu}(1998)}]{Eisenstein:1997ik}%
  \BibitemOpen
  \bibfield  {author} {\bibinfo {author} {\bibfnamefont {D.~J.}\ \bibnamefont
  {Eisenstein}}\ and\ \bibinfo {author} {\bibfnamefont {W.}~\bibnamefont
  {Hu}},\ }\href {\doibase 10.1086/305424} {\bibfield  {journal} {\bibinfo
  {journal} {Astrophys. J.}\ }\textbf {\bibinfo {volume} {496}},\ \bibinfo
  {pages} {605} (\bibinfo {year} {1998})},\ \Eprint
  {http://arxiv.org/abs/astro-ph/9709112} {arXiv:astro-ph/9709112 [astro-ph]}
  \BibitemShut {NoStop}%
%%CITATION = ASTRO-PH/9709112;%%
\bibitem [{\citenamefont {Nunes}\ \emph {et~al.}(2020)\citenamefont {Nunes},
  \citenamefont {Yadav}, \citenamefont {Jesus},\ and\ \citenamefont
  {Bernui}}]{Nunes:2020hzy}%
  \BibitemOpen
  \bibfield  {author} {\bibinfo {author} {\bibfnamefont {R.~C.}\ \bibnamefont
  {Nunes}}, \bibinfo {author} {\bibfnamefont {S.~K.}\ \bibnamefont {Yadav}},
  \bibinfo {author} {\bibfnamefont {J.~F.}\ \bibnamefont {Jesus}}, \ and\
  \bibinfo {author} {\bibfnamefont {A.}~\bibnamefont {Bernui}},\ }\href
  {\doibase 10.1093/mnras/staa2036} {\bibfield  {journal} {\bibinfo  {journal}
  {Monthly Notices of the Royal Astronomical Society}\ }\textbf {\bibinfo
  {volume} {497}},\ \bibinfo {pages} {2133} (\bibinfo {year} {2020})},\ \Eprint
  {http://arxiv.org/abs/https://academic.oup.com/mnras/article-pdf/497/2/2133/33565883/staa2036.pdf}
  {https://academic.oup.com/mnras/article-pdf/497/2/2133/33565883/staa2036.pdf}
  \BibitemShut {NoStop}%
\bibitem [{\citenamefont {Carvalho}\ \emph {et~al.}(2016)\citenamefont
  {Carvalho}, \citenamefont {Bernui}, \citenamefont {Benetti}, \citenamefont
  {Carvalho},\ and\ \citenamefont {Alcaniz}}]{Carvalho:2016}%
  \BibitemOpen
  \bibfield  {author} {\bibinfo {author} {\bibfnamefont {G.~C.}\ \bibnamefont
  {Carvalho}}, \bibinfo {author} {\bibfnamefont {A.}~\bibnamefont {Bernui}},
  \bibinfo {author} {\bibfnamefont {M.}~\bibnamefont {Benetti}}, \bibinfo
  {author} {\bibfnamefont {J.~C.}\ \bibnamefont {Carvalho}}, \ and\ \bibinfo
  {author} {\bibfnamefont {J.~S.}\ \bibnamefont {Alcaniz}},\ }\href {\doibase
  10.1103/PhysRevD.93.023530} {\bibfield  {journal} {\bibinfo  {journal} {Phys.
  Rev. D}\ }\textbf {\bibinfo {volume} {93}},\ \bibinfo {pages} {023530}
  (\bibinfo {year} {2016})}\BibitemShut {NoStop}%
\bibitem [{\citenamefont {Alcaniz}\ \emph {et~al.}(2017)\citenamefont
  {Alcaniz}, \citenamefont {Carvalho}, \citenamefont {Bernui}, \citenamefont
  {Carvalho},\ and\ \citenamefont {Benetti}}]{Alcaniz2017}%
  \BibitemOpen
  \bibfield  {author} {\bibinfo {author} {\bibfnamefont {J.~S.}\ \bibnamefont
  {Alcaniz}}, \bibinfo {author} {\bibfnamefont {G.~C.}\ \bibnamefont
  {Carvalho}}, \bibinfo {author} {\bibfnamefont {A.}~\bibnamefont {Bernui}},
  \bibinfo {author} {\bibfnamefont {J.~C.}\ \bibnamefont {Carvalho}}, \ and\
  \bibinfo {author} {\bibfnamefont {M.}~\bibnamefont {Benetti}},\ }\enquote
  {\bibinfo {title} {Measuring baryon acoustic oscillations with angular
  two-point correlation function},}\ in\ \href {\doibase
  10.1007/978-3-319-51700-1_2} {\emph {\bibinfo {booktitle} {Gravity and the
  Quantum: Pedagogical Essays on Cosmology, Astrophysics, and Quantum
  Gravity}}},\ \bibinfo {editor} {edited by\ \bibinfo {editor} {\bibfnamefont
  {J.~S.}\ \bibnamefont {Bagla}}\ and\ \bibinfo {editor} {\bibfnamefont
  {S.}~\bibnamefont {Engineer}}}\ (\bibinfo  {publisher} {Springer
  International Publishing},\ \bibinfo {address} {Cham},\ \bibinfo {year}
  {2017})\ pp.\ \bibinfo {pages} {11--19}\BibitemShut {NoStop}%
\bibitem [{\citenamefont {Carvalho}\ \emph {et~al.}(2020)\citenamefont
  {Carvalho}, \citenamefont {Bernui}, \citenamefont {Benetti}, \citenamefont
  {Carvalho}, \citenamefont {{de Carvalho}},\ and\ \citenamefont
  {Alcaniz}}]{CARVALHO:2020}%
  \BibitemOpen
  \bibfield  {author} {\bibinfo {author} {\bibfnamefont {G.}~\bibnamefont
  {Carvalho}}, \bibinfo {author} {\bibfnamefont {A.}~\bibnamefont {Bernui}},
  \bibinfo {author} {\bibfnamefont {M.}~\bibnamefont {Benetti}}, \bibinfo
  {author} {\bibfnamefont {J.}~\bibnamefont {Carvalho}}, \bibinfo {author}
  {\bibfnamefont {E.}~\bibnamefont {{de Carvalho}}}, \ and\ \bibinfo {author}
  {\bibfnamefont {J.}~\bibnamefont {Alcaniz}},\ }\href {\doibase
  https://doi.org/10.1016/j.astropartphys.2020.102432} {\bibfield  {journal}
  {\bibinfo  {journal} {Astroparticle Physics}\ }\textbf {\bibinfo {volume}
  {119}},\ \bibinfo {pages} {102432} (\bibinfo {year} {2020})}\BibitemShut
  {NoStop}%
\bibitem [{\citenamefont {de~Carvalho}\ \emph {et~al.}(2018)\citenamefont
  {de~Carvalho}, \citenamefont {Bernui}, \citenamefont {Carvalho},
  \citenamefont {Novaes},\ and\ \citenamefont {Xavier}}]{de_Carvalho_2018}%
  \BibitemOpen
  \bibfield  {author} {\bibinfo {author} {\bibfnamefont {E.}~\bibnamefont
  {de~Carvalho}}, \bibinfo {author} {\bibfnamefont {A.}~\bibnamefont {Bernui}},
  \bibinfo {author} {\bibfnamefont {G.}~\bibnamefont {Carvalho}}, \bibinfo
  {author} {\bibfnamefont {C.}~\bibnamefont {Novaes}}, \ and\ \bibinfo {author}
  {\bibfnamefont {H.}~\bibnamefont {Xavier}},\ }\href {\doibase
  10.1088/1475-7516/2018/04/064} {\bibfield  {journal} {\bibinfo  {journal}
  {Journal of Cosmology and Astroparticle Physics}\ }\textbf {\bibinfo {volume}
  {2018}},\ \bibinfo {pages} {064} (\bibinfo {year} {2018})}\BibitemShut
  {NoStop}%
\bibitem [{\citenamefont {de~Carvalho}\ \emph {et~al.}(2020)\citenamefont
  {de~Carvalho}, \citenamefont {Bernui}, \citenamefont {Xavier},\ and\
  \citenamefont {Novaes}}]{de_Carvalho:2020mnras}%
  \BibitemOpen
  \bibfield  {author} {\bibinfo {author} {\bibfnamefont {E.}~\bibnamefont
  {de~Carvalho}}, \bibinfo {author} {\bibfnamefont {A.}~\bibnamefont {Bernui}},
  \bibinfo {author} {\bibfnamefont {H.~S.}\ \bibnamefont {Xavier}}, \ and\
  \bibinfo {author} {\bibfnamefont {C.~P.}\ \bibnamefont {Novaes}},\ }\href
  {\doibase 10.1093/mnras/staa119} {\bibfield  {journal} {\bibinfo  {journal}
  {Monthly Notices of the Royal Astronomical Society}\ }\textbf {\bibinfo
  {volume} {492}},\ \bibinfo {pages} {4469} (\bibinfo {year} {2020})},\ \Eprint
  {http://arxiv.org/abs/https://academic.oup.com/mnras/article-pdf/492/3/4469/32302539/staa119.pdf}
  {https://academic.oup.com/mnras/article-pdf/492/3/4469/32302539/staa119.pdf}
  \BibitemShut {NoStop}%
\bibitem [{\citenamefont {York}\ \emph {et~al.}(2000)\citenamefont {York},
  \citenamefont {Adelman}, \citenamefont {John E.~Anderson}, \citenamefont
  {Anderson}, \citenamefont {Annis}, \citenamefont {Bahcall}, \citenamefont
  {Bakken}, \citenamefont {Barkhouser}, \citenamefont {Bastian}, \citenamefont
  {Berman}, \citenamefont {Boroski}, \citenamefont {Bracker}, \citenamefont
  {Briegel}, \citenamefont {Briggs}, \citenamefont {Brinkmann}, \citenamefont
  {Brunner}, \citenamefont {Burles}, \citenamefont {Carey}, \citenamefont
  {Carr}, \citenamefont {Castander}, \citenamefont {Chen}, \citenamefont
  {Colestock}, \citenamefont {Connolly}, \citenamefont {Crocker}, \citenamefont
  {Csabai}, \citenamefont {Czarapata}, \citenamefont {Davis}, \citenamefont
  {Doi}, \citenamefont {Dombeck}, \citenamefont {Eisenstein}, \citenamefont
  {Ellman}, \citenamefont {Elms}, \citenamefont {Evans}, \citenamefont {Fan},
  \citenamefont {Federwitz}, \citenamefont {Fiscelli}, \citenamefont
  {Friedman}, \citenamefont {Frieman}, \citenamefont {Fukugita}, \citenamefont
  {Gillespie}, \citenamefont {Gunn}, \citenamefont {Gurbani}, \citenamefont
  {de~Haas}, \citenamefont {Haldeman}, \citenamefont {Harris}, \citenamefont
  {Hayes}, \citenamefont {Heckman}, \citenamefont {Hennessy}, \citenamefont
  {Hindsley}, \citenamefont {Holm}, \citenamefont {Holmgren}, \citenamefont
  {hao Huang}, \citenamefont {Hull}, \citenamefont {Husby}, \citenamefont
  {Ichikawa}, \citenamefont {Ichikawa}, \citenamefont {Ivezi{\'{c}}},
  \citenamefont {Kent}, \citenamefont {Kim}, \citenamefont {Kinney},
  \citenamefont {Klaene}, \citenamefont {Kleinman}, \citenamefont {Kleinman},
  \citenamefont {Knapp}, \citenamefont {Korienek}, \citenamefont {Kron},
  \citenamefont {Kunszt}, \citenamefont {Lamb}, \citenamefont {Lee},
  \citenamefont {Leger}, \citenamefont {Limmongkol}, \citenamefont
  {Lindenmeyer}, \citenamefont {Long}, \citenamefont {Loomis}, \citenamefont
  {Loveday}, \citenamefont {Lucinio}, \citenamefont {Lupton}, \citenamefont
  {MacKinnon}, \citenamefont {Mannery}, \citenamefont {Mantsch}, \citenamefont
  {Margon}, \citenamefont {McGehee}, \citenamefont {McKay}, \citenamefont
  {Meiksin}, \citenamefont {Merelli}, \citenamefont {Monet}, \citenamefont
  {Munn}, \citenamefont {Narayanan}, \citenamefont {Nash}, \citenamefont
  {Neilsen}, \citenamefont {Neswold}, \citenamefont {Newberg}, \citenamefont
  {Nichol}, \citenamefont {Nicinski}, \citenamefont {Nonino}, \citenamefont
  {Okada}, \citenamefont {Okamura}, \citenamefont {Ostriker}, \citenamefont
  {Owen}, \citenamefont {Pauls}, \citenamefont {Peoples}, \citenamefont
  {Peterson}, \citenamefont {Petravick}, \citenamefont {Pier}, \citenamefont
  {Pope}, \citenamefont {Pordes}, \citenamefont {Prosapio}, \citenamefont
  {Rechenmacher}, \citenamefont {Quinn}, \citenamefont {Richards},
  \citenamefont {Richmond}, \citenamefont {Rivetta}, \citenamefont {Rockosi},
  \citenamefont {Ruthmansdorfer}, \citenamefont {Sandford}, \citenamefont
  {Schlegel}, \citenamefont {Schneider}, \citenamefont {Sekiguchi},
  \citenamefont {Sergey}, \citenamefont {Shimasaku}, \citenamefont {Siegmund},
  \citenamefont {Smee}, \citenamefont {Smith}, \citenamefont {Snedden},
  \citenamefont {Stone}, \citenamefont {Stoughton}, \citenamefont {Strauss},
  \citenamefont {Stubbs}, \citenamefont {SubbaRao}, \citenamefont {Szalay},
  \citenamefont {Szapudi}, \citenamefont {Szokoly}, \citenamefont {Thakar},
  \citenamefont {Tremonti}, \citenamefont {Tucker}, \citenamefont {Uomoto},
  \citenamefont {Berk}, \citenamefont {Vogeley}, \citenamefont {Waddell},
  \citenamefont {i~Wang}, \citenamefont {Watanabe}, \citenamefont {Weinberg},
  \citenamefont {Yanny},\ and\ \citenamefont {Yasuda}}]{York_2000}%
  \BibitemOpen
  \bibfield  {author} {\bibinfo {author} {\bibfnamefont {D.~G.}\ \bibnamefont
  {York}}, \bibinfo {author} {\bibfnamefont {J.}~\bibnamefont {Adelman}},
  \bibinfo {author} {\bibfnamefont {J.}~\bibnamefont {John E.~Anderson}},
  \bibinfo {author} {\bibfnamefont {S.~F.}\ \bibnamefont {Anderson}}, \bibinfo
  {author} {\bibfnamefont {J.}~\bibnamefont {Annis}}, \bibinfo {author}
  {\bibfnamefont {N.~A.}\ \bibnamefont {Bahcall}}, \bibinfo {author}
  {\bibfnamefont {J.~A.}\ \bibnamefont {Bakken}}, \bibinfo {author}
  {\bibfnamefont {R.}~\bibnamefont {Barkhouser}}, \bibinfo {author}
  {\bibfnamefont {S.}~\bibnamefont {Bastian}}, \bibinfo {author} {\bibfnamefont
  {E.}~\bibnamefont {Berman}}, \bibinfo {author} {\bibfnamefont {W.~N.}\
  \bibnamefont {Boroski}}, \bibinfo {author} {\bibfnamefont {S.}~\bibnamefont
  {Bracker}}, \bibinfo {author} {\bibfnamefont {C.}~\bibnamefont {Briegel}},
  \bibinfo {author} {\bibfnamefont {J.~W.}\ \bibnamefont {Briggs}}, \bibinfo
  {author} {\bibfnamefont {J.}~\bibnamefont {Brinkmann}}, \bibinfo {author}
  {\bibfnamefont {R.}~\bibnamefont {Brunner}}, \bibinfo {author} {\bibfnamefont
  {S.}~\bibnamefont {Burles}}, \bibinfo {author} {\bibfnamefont
  {L.}~\bibnamefont {Carey}}, \bibinfo {author} {\bibfnamefont {M.~A.}\
  \bibnamefont {Carr}}, \bibinfo {author} {\bibfnamefont {F.~J.}\ \bibnamefont
  {Castander}}, \bibinfo {author} {\bibfnamefont {B.}~\bibnamefont {Chen}},
  \bibinfo {author} {\bibfnamefont {P.~L.}\ \bibnamefont {Colestock}}, \bibinfo
  {author} {\bibfnamefont {A.~J.}\ \bibnamefont {Connolly}}, \bibinfo {author}
  {\bibfnamefont {J.~H.}\ \bibnamefont {Crocker}}, \bibinfo {author}
  {\bibfnamefont {I.}~\bibnamefont {Csabai}}, \bibinfo {author} {\bibfnamefont
  {P.~C.}\ \bibnamefont {Czarapata}}, \bibinfo {author} {\bibfnamefont {J.~E.}\
  \bibnamefont {Davis}}, \bibinfo {author} {\bibfnamefont {M.}~\bibnamefont
  {Doi}}, \bibinfo {author} {\bibfnamefont {T.}~\bibnamefont {Dombeck}},
  \bibinfo {author} {\bibfnamefont {D.}~\bibnamefont {Eisenstein}}, \bibinfo
  {author} {\bibfnamefont {N.}~\bibnamefont {Ellman}}, \bibinfo {author}
  {\bibfnamefont {B.~R.}\ \bibnamefont {Elms}}, \bibinfo {author}
  {\bibfnamefont {M.~L.}\ \bibnamefont {Evans}}, \bibinfo {author}
  {\bibfnamefont {X.}~\bibnamefont {Fan}}, \bibinfo {author} {\bibfnamefont
  {G.~R.}\ \bibnamefont {Federwitz}}, \bibinfo {author} {\bibfnamefont
  {L.}~\bibnamefont {Fiscelli}}, \bibinfo {author} {\bibfnamefont
  {S.}~\bibnamefont {Friedman}}, \bibinfo {author} {\bibfnamefont {J.~A.}\
  \bibnamefont {Frieman}}, \bibinfo {author} {\bibfnamefont {M.}~\bibnamefont
  {Fukugita}}, \bibinfo {author} {\bibfnamefont {B.}~\bibnamefont {Gillespie}},
  \bibinfo {author} {\bibfnamefont {J.~E.}\ \bibnamefont {Gunn}}, \bibinfo
  {author} {\bibfnamefont {V.~K.}\ \bibnamefont {Gurbani}}, \bibinfo {author}
  {\bibfnamefont {E.}~\bibnamefont {de~Haas}}, \bibinfo {author} {\bibfnamefont
  {M.}~\bibnamefont {Haldeman}}, \bibinfo {author} {\bibfnamefont {F.~H.}\
  \bibnamefont {Harris}}, \bibinfo {author} {\bibfnamefont {J.}~\bibnamefont
  {Hayes}}, \bibinfo {author} {\bibfnamefont {T.~M.}\ \bibnamefont {Heckman}},
  \bibinfo {author} {\bibfnamefont {G.~S.}\ \bibnamefont {Hennessy}}, \bibinfo
  {author} {\bibfnamefont {R.~B.}\ \bibnamefont {Hindsley}}, \bibinfo {author}
  {\bibfnamefont {S.}~\bibnamefont {Holm}}, \bibinfo {author} {\bibfnamefont
  {D.~J.}\ \bibnamefont {Holmgren}}, \bibinfo {author} {\bibfnamefont
  {C.}~\bibnamefont {hao Huang}}, \bibinfo {author} {\bibfnamefont
  {C.}~\bibnamefont {Hull}}, \bibinfo {author} {\bibfnamefont {D.}~\bibnamefont
  {Husby}}, \bibinfo {author} {\bibfnamefont {S.-I.}\ \bibnamefont {Ichikawa}},
  \bibinfo {author} {\bibfnamefont {T.}~\bibnamefont {Ichikawa}}, \bibinfo
  {author} {\bibfnamefont {{\v{Z}}.}~\bibnamefont {Ivezi{\'{c}}}}, \bibinfo
  {author} {\bibfnamefont {S.}~\bibnamefont {Kent}}, \bibinfo {author}
  {\bibfnamefont {R.~S.~J.}\ \bibnamefont {Kim}}, \bibinfo {author}
  {\bibfnamefont {E.}~\bibnamefont {Kinney}}, \bibinfo {author} {\bibfnamefont
  {M.}~\bibnamefont {Klaene}}, \bibinfo {author} {\bibfnamefont {A.~N.}\
  \bibnamefont {Kleinman}}, \bibinfo {author} {\bibfnamefont {S.}~\bibnamefont
  {Kleinman}}, \bibinfo {author} {\bibfnamefont {G.~R.}\ \bibnamefont {Knapp}},
  \bibinfo {author} {\bibfnamefont {J.}~\bibnamefont {Korienek}}, \bibinfo
  {author} {\bibfnamefont {R.~G.}\ \bibnamefont {Kron}}, \bibinfo {author}
  {\bibfnamefont {P.~Z.}\ \bibnamefont {Kunszt}}, \bibinfo {author}
  {\bibfnamefont {D.~Q.}\ \bibnamefont {Lamb}}, \bibinfo {author}
  {\bibfnamefont {B.}~\bibnamefont {Lee}}, \bibinfo {author} {\bibfnamefont
  {R.~F.}\ \bibnamefont {Leger}}, \bibinfo {author} {\bibfnamefont
  {S.}~\bibnamefont {Limmongkol}}, \bibinfo {author} {\bibfnamefont
  {C.}~\bibnamefont {Lindenmeyer}}, \bibinfo {author} {\bibfnamefont {D.~C.}\
  \bibnamefont {Long}}, \bibinfo {author} {\bibfnamefont {C.}~\bibnamefont
  {Loomis}}, \bibinfo {author} {\bibfnamefont {J.}~\bibnamefont {Loveday}},
  \bibinfo {author} {\bibfnamefont {R.}~\bibnamefont {Lucinio}}, \bibinfo
  {author} {\bibfnamefont {R.~H.}\ \bibnamefont {Lupton}}, \bibinfo {author}
  {\bibfnamefont {B.}~\bibnamefont {MacKinnon}}, \bibinfo {author}
  {\bibfnamefont {E.~J.}\ \bibnamefont {Mannery}}, \bibinfo {author}
  {\bibfnamefont {P.~M.}\ \bibnamefont {Mantsch}}, \bibinfo {author}
  {\bibfnamefont {B.}~\bibnamefont {Margon}}, \bibinfo {author} {\bibfnamefont
  {P.}~\bibnamefont {McGehee}}, \bibinfo {author} {\bibfnamefont {T.~A.}\
  \bibnamefont {McKay}}, \bibinfo {author} {\bibfnamefont {A.}~\bibnamefont
  {Meiksin}}, \bibinfo {author} {\bibfnamefont {A.}~\bibnamefont {Merelli}},
  \bibinfo {author} {\bibfnamefont {D.~G.}\ \bibnamefont {Monet}}, \bibinfo
  {author} {\bibfnamefont {J.~A.}\ \bibnamefont {Munn}}, \bibinfo {author}
  {\bibfnamefont {V.~K.}\ \bibnamefont {Narayanan}}, \bibinfo {author}
  {\bibfnamefont {T.}~\bibnamefont {Nash}}, \bibinfo {author} {\bibfnamefont
  {E.}~\bibnamefont {Neilsen}}, \bibinfo {author} {\bibfnamefont
  {R.}~\bibnamefont {Neswold}}, \bibinfo {author} {\bibfnamefont {H.~J.}\
  \bibnamefont {Newberg}}, \bibinfo {author} {\bibfnamefont {R.~C.}\
  \bibnamefont {Nichol}}, \bibinfo {author} {\bibfnamefont {T.}~\bibnamefont
  {Nicinski}}, \bibinfo {author} {\bibfnamefont {M.}~\bibnamefont {Nonino}},
  \bibinfo {author} {\bibfnamefont {N.}~\bibnamefont {Okada}}, \bibinfo
  {author} {\bibfnamefont {S.}~\bibnamefont {Okamura}}, \bibinfo {author}
  {\bibfnamefont {J.~P.}\ \bibnamefont {Ostriker}}, \bibinfo {author}
  {\bibfnamefont {R.}~\bibnamefont {Owen}}, \bibinfo {author} {\bibfnamefont
  {A.~G.}\ \bibnamefont {Pauls}}, \bibinfo {author} {\bibfnamefont
  {J.}~\bibnamefont {Peoples}}, \bibinfo {author} {\bibfnamefont {R.~L.}\
  \bibnamefont {Peterson}}, \bibinfo {author} {\bibfnamefont {D.}~\bibnamefont
  {Petravick}}, \bibinfo {author} {\bibfnamefont {J.~R.}\ \bibnamefont {Pier}},
  \bibinfo {author} {\bibfnamefont {A.}~\bibnamefont {Pope}}, \bibinfo {author}
  {\bibfnamefont {R.}~\bibnamefont {Pordes}}, \bibinfo {author} {\bibfnamefont
  {A.}~\bibnamefont {Prosapio}}, \bibinfo {author} {\bibfnamefont
  {R.}~\bibnamefont {Rechenmacher}}, \bibinfo {author} {\bibfnamefont {T.~R.}\
  \bibnamefont {Quinn}}, \bibinfo {author} {\bibfnamefont {G.~T.}\ \bibnamefont
  {Richards}}, \bibinfo {author} {\bibfnamefont {M.~W.}\ \bibnamefont
  {Richmond}}, \bibinfo {author} {\bibfnamefont {C.~H.}\ \bibnamefont
  {Rivetta}}, \bibinfo {author} {\bibfnamefont {C.~M.}\ \bibnamefont
  {Rockosi}}, \bibinfo {author} {\bibfnamefont {K.}~\bibnamefont
  {Ruthmansdorfer}}, \bibinfo {author} {\bibfnamefont {D.}~\bibnamefont
  {Sandford}}, \bibinfo {author} {\bibfnamefont {D.~J.}\ \bibnamefont
  {Schlegel}}, \bibinfo {author} {\bibfnamefont {D.~P.}\ \bibnamefont
  {Schneider}}, \bibinfo {author} {\bibfnamefont {M.}~\bibnamefont
  {Sekiguchi}}, \bibinfo {author} {\bibfnamefont {G.}~\bibnamefont {Sergey}},
  \bibinfo {author} {\bibfnamefont {K.}~\bibnamefont {Shimasaku}}, \bibinfo
  {author} {\bibfnamefont {W.~A.}\ \bibnamefont {Siegmund}}, \bibinfo {author}
  {\bibfnamefont {S.}~\bibnamefont {Smee}}, \bibinfo {author} {\bibfnamefont
  {J.~A.}\ \bibnamefont {Smith}}, \bibinfo {author} {\bibfnamefont
  {S.}~\bibnamefont {Snedden}}, \bibinfo {author} {\bibfnamefont
  {R.}~\bibnamefont {Stone}}, \bibinfo {author} {\bibfnamefont
  {C.}~\bibnamefont {Stoughton}}, \bibinfo {author} {\bibfnamefont {M.~A.}\
  \bibnamefont {Strauss}}, \bibinfo {author} {\bibfnamefont {C.}~\bibnamefont
  {Stubbs}}, \bibinfo {author} {\bibfnamefont {M.}~\bibnamefont {SubbaRao}},
  \bibinfo {author} {\bibfnamefont {A.~S.}\ \bibnamefont {Szalay}}, \bibinfo
  {author} {\bibfnamefont {I.}~\bibnamefont {Szapudi}}, \bibinfo {author}
  {\bibfnamefont {G.~P.}\ \bibnamefont {Szokoly}}, \bibinfo {author}
  {\bibfnamefont {A.~R.}\ \bibnamefont {Thakar}}, \bibinfo {author}
  {\bibfnamefont {C.}~\bibnamefont {Tremonti}}, \bibinfo {author}
  {\bibfnamefont {D.~L.}\ \bibnamefont {Tucker}}, \bibinfo {author}
  {\bibfnamefont {A.}~\bibnamefont {Uomoto}}, \bibinfo {author} {\bibfnamefont
  {D.~V.}\ \bibnamefont {Berk}}, \bibinfo {author} {\bibfnamefont {M.~S.}\
  \bibnamefont {Vogeley}}, \bibinfo {author} {\bibfnamefont {P.}~\bibnamefont
  {Waddell}}, \bibinfo {author} {\bibfnamefont {S.}~\bibnamefont {i~Wang}},
  \bibinfo {author} {\bibfnamefont {M.}~\bibnamefont {Watanabe}}, \bibinfo
  {author} {\bibfnamefont {D.~H.}\ \bibnamefont {Weinberg}}, \bibinfo {author}
  {\bibfnamefont {B.}~\bibnamefont {Yanny}}, \ and\ \bibinfo {author}
  {\bibfnamefont {N.}~\bibnamefont {Yasuda}},\ }\href {\doibase 10.1086/301513}
  {\bibfield  {journal} {\bibinfo  {journal} {The Astronomical Journal}\
  }\textbf {\bibinfo {volume} {120}},\ \bibinfo {pages} {1579} (\bibinfo {year}
  {2000})}\BibitemShut {NoStop}%
\bibitem [{\citenamefont {Giostri}\ \emph {et~al.}(2012)\citenamefont
  {Giostri}, \citenamefont {dos Santos}, \citenamefont {Waga}, \citenamefont
  {Reis}, \citenamefont {Calv{\~{a}}o},\ and\ \citenamefont
  {Lago}}]{Giostri:2012}%
  \BibitemOpen
  \bibfield  {author} {\bibinfo {author} {\bibfnamefont {R.}~\bibnamefont
  {Giostri}}, \bibinfo {author} {\bibfnamefont {M.~V.}\ \bibnamefont {dos
  Santos}}, \bibinfo {author} {\bibfnamefont {I.}~\bibnamefont {Waga}},
  \bibinfo {author} {\bibfnamefont {R.}~\bibnamefont {Reis}}, \bibinfo {author}
  {\bibfnamefont {M.}~\bibnamefont {Calv{\~{a}}o}}, \ and\ \bibinfo {author}
  {\bibfnamefont {B.~L.}\ \bibnamefont {Lago}},\ }\href {\doibase
  10.1088/1475-7516/2012/03/027} {\bibfield  {journal} {\bibinfo  {journal}
  {Journal of Cosmology and Astroparticle Physics}\ }\textbf {\bibinfo {volume}
  {2012}},\ \bibinfo {pages} {027} (\bibinfo {year} {2012})}\BibitemShut
  {NoStop}%
\bibitem [{\citenamefont {Percival}\ \emph {et~al.}(2010)\citenamefont
  {Percival}, \citenamefont {Reid}, \citenamefont {Eisenstein}, \citenamefont
  {Bahcall}, \citenamefont {Budavari}, \citenamefont {Frieman}, \citenamefont
  {Fukugita}, \citenamefont {Gunn}, \citenamefont {Ivezic}, \citenamefont
  {Knapp}, \citenamefont {Kron}, \citenamefont {Loveday}, \citenamefont
  {Lupton}, \citenamefont {McKay}, \citenamefont {Meiksin}, \citenamefont
  {Nichol}, \citenamefont {Pope}, \citenamefont {Schlegel}, \citenamefont
  {Schneider}, \citenamefont {Spergel}, \citenamefont {Stoughton},
  \citenamefont {Strauss}, \citenamefont {Szalay}, \citenamefont {Tegmark},
  \citenamefont {Vogeley}, \citenamefont {Weinberg}, \citenamefont {York},\
  and\ \citenamefont {Zehavi}}]{Percival:2010}%
  \BibitemOpen
  \bibfield  {author} {\bibinfo {author} {\bibfnamefont {W.~J.}\ \bibnamefont
  {Percival}}, \bibinfo {author} {\bibfnamefont {B.~A.}\ \bibnamefont {Reid}},
  \bibinfo {author} {\bibfnamefont {D.~J.}\ \bibnamefont {Eisenstein}},
  \bibinfo {author} {\bibfnamefont {N.~A.}\ \bibnamefont {Bahcall}}, \bibinfo
  {author} {\bibfnamefont {T.}~\bibnamefont {Budavari}}, \bibinfo {author}
  {\bibfnamefont {J.~A.}\ \bibnamefont {Frieman}}, \bibinfo {author}
  {\bibfnamefont {M.}~\bibnamefont {Fukugita}}, \bibinfo {author}
  {\bibfnamefont {J.~E.}\ \bibnamefont {Gunn}}, \bibinfo {author}
  {\bibfnamefont {Z.}~\bibnamefont {Ivezic}}, \bibinfo {author} {\bibfnamefont
  {G.~R.}\ \bibnamefont {Knapp}}, \bibinfo {author} {\bibfnamefont {R.~G.}\
  \bibnamefont {Kron}}, \bibinfo {author} {\bibfnamefont {J.}~\bibnamefont
  {Loveday}}, \bibinfo {author} {\bibfnamefont {R.~H.}\ \bibnamefont {Lupton}},
  \bibinfo {author} {\bibfnamefont {T.~A.}\ \bibnamefont {McKay}}, \bibinfo
  {author} {\bibfnamefont {A.}~\bibnamefont {Meiksin}}, \bibinfo {author}
  {\bibfnamefont {R.~C.}\ \bibnamefont {Nichol}}, \bibinfo {author}
  {\bibfnamefont {A.~C.}\ \bibnamefont {Pope}}, \bibinfo {author}
  {\bibfnamefont {D.~J.}\ \bibnamefont {Schlegel}}, \bibinfo {author}
  {\bibfnamefont {D.~P.}\ \bibnamefont {Schneider}}, \bibinfo {author}
  {\bibfnamefont {D.~N.}\ \bibnamefont {Spergel}}, \bibinfo {author}
  {\bibfnamefont {C.}~\bibnamefont {Stoughton}}, \bibinfo {author}
  {\bibfnamefont {M.~A.}\ \bibnamefont {Strauss}}, \bibinfo {author}
  {\bibfnamefont {A.~S.}\ \bibnamefont {Szalay}}, \bibinfo {author}
  {\bibfnamefont {M.}~\bibnamefont {Tegmark}}, \bibinfo {author} {\bibfnamefont
  {M.~S.}\ \bibnamefont {Vogeley}}, \bibinfo {author} {\bibfnamefont {D.~H.}\
  \bibnamefont {Weinberg}}, \bibinfo {author} {\bibfnamefont {D.~G.}\
  \bibnamefont {York}}, \ and\ \bibinfo {author} {\bibfnamefont
  {I.}~\bibnamefont {Zehavi}},\ }\href {\doibase
  10.1111/j.1365-2966.2009.15812.x} {\bibfield  {journal} {\bibinfo  {journal}
  {Monthly Notices of the Royal Astronomical Society}\ }\textbf {\bibinfo
  {volume} {401}},\ \bibinfo {pages} {2148} (\bibinfo {year} {2010})},\ \Eprint
  {http://arxiv.org/abs/https://academic.oup.com/mnras/article-pdf/401/4/2148/3901461/mnras0401-2148.pdf}
  {https://academic.oup.com/mnras/article-pdf/401/4/2148/3901461/mnras0401-2148.pdf}
  \BibitemShut {NoStop}%
\bibitem [{\citenamefont {Blake}\ \emph {et~al.}(2011)\citenamefont {Blake},
  \citenamefont {Kazin}, \citenamefont {Beutler}, \citenamefont {Davis},
  \citenamefont {Parkinson}, \citenamefont {Brough}, \citenamefont {Colless},
  \citenamefont {Contreras}, \citenamefont {Couch}, \citenamefont {Croom},
  \citenamefont {Croton}, \citenamefont {Drinkwater}, \citenamefont {Forster},
  \citenamefont {Gilbank}, \citenamefont {Gladders}, \citenamefont
  {Glazebrook}, \citenamefont {Jelliffe}, \citenamefont {Jurek}, \citenamefont
  {Li}, \citenamefont {Madore}, \citenamefont {Martin}, \citenamefont
  {Pimbblet}, \citenamefont {Poole}, \citenamefont {Pracy}, \citenamefont
  {Sharp}, \citenamefont {Wisnioski}, \citenamefont {Woods}, \citenamefont
  {Wyder},\ and\ \citenamefont {Yee}}]{Blake:2011}%
  \BibitemOpen
  \bibfield  {author} {\bibinfo {author} {\bibfnamefont {C.}~\bibnamefont
  {Blake}}, \bibinfo {author} {\bibfnamefont {E.~A.}\ \bibnamefont {Kazin}},
  \bibinfo {author} {\bibfnamefont {F.}~\bibnamefont {Beutler}}, \bibinfo
  {author} {\bibfnamefont {T.~M.}\ \bibnamefont {Davis}}, \bibinfo {author}
  {\bibfnamefont {D.}~\bibnamefont {Parkinson}}, \bibinfo {author}
  {\bibfnamefont {S.}~\bibnamefont {Brough}}, \bibinfo {author} {\bibfnamefont
  {M.}~\bibnamefont {Colless}}, \bibinfo {author} {\bibfnamefont
  {C.}~\bibnamefont {Contreras}}, \bibinfo {author} {\bibfnamefont
  {W.}~\bibnamefont {Couch}}, \bibinfo {author} {\bibfnamefont
  {S.}~\bibnamefont {Croom}}, \bibinfo {author} {\bibfnamefont
  {D.}~\bibnamefont {Croton}}, \bibinfo {author} {\bibfnamefont {M.~J.}\
  \bibnamefont {Drinkwater}}, \bibinfo {author} {\bibfnamefont
  {K.}~\bibnamefont {Forster}}, \bibinfo {author} {\bibfnamefont
  {D.}~\bibnamefont {Gilbank}}, \bibinfo {author} {\bibfnamefont
  {M.}~\bibnamefont {Gladders}}, \bibinfo {author} {\bibfnamefont
  {K.}~\bibnamefont {Glazebrook}}, \bibinfo {author} {\bibfnamefont
  {B.}~\bibnamefont {Jelliffe}}, \bibinfo {author} {\bibfnamefont {R.~J.}\
  \bibnamefont {Jurek}}, \bibinfo {author} {\bibfnamefont {I.-h.}\ \bibnamefont
  {Li}}, \bibinfo {author} {\bibfnamefont {B.}~\bibnamefont {Madore}}, \bibinfo
  {author} {\bibfnamefont {D.~C.}\ \bibnamefont {Martin}}, \bibinfo {author}
  {\bibfnamefont {K.}~\bibnamefont {Pimbblet}}, \bibinfo {author}
  {\bibfnamefont {G.~B.}\ \bibnamefont {Poole}}, \bibinfo {author}
  {\bibfnamefont {M.}~\bibnamefont {Pracy}}, \bibinfo {author} {\bibfnamefont
  {R.}~\bibnamefont {Sharp}}, \bibinfo {author} {\bibfnamefont
  {E.}~\bibnamefont {Wisnioski}}, \bibinfo {author} {\bibfnamefont
  {D.}~\bibnamefont {Woods}}, \bibinfo {author} {\bibfnamefont {T.~K.}\
  \bibnamefont {Wyder}}, \ and\ \bibinfo {author} {\bibfnamefont {H.~K.~C.}\
  \bibnamefont {Yee}},\ }\href {\doibase 10.1111/j.1365-2966.2011.19592.x}
  {\bibfield  {journal} {\bibinfo  {journal} {Monthly Notices of the Royal
  Astronomical Society}\ }\textbf {\bibinfo {volume} {418}},\ \bibinfo {pages}
  {1707} (\bibinfo {year} {2011})},\ \Eprint
  {http://arxiv.org/abs/https://academic.oup.com/mnras/article-pdf/418/3/1707/18440857/mnras0418-1707.pdf}
  {https://academic.oup.com/mnras/article-pdf/418/3/1707/18440857/mnras0418-1707.pdf}
  \BibitemShut {NoStop}%
\bibitem [{\citenamefont {{Beutler}}\ \emph {et~al.}(2011)\citenamefont
  {{Beutler}}, \citenamefont {{Blake}}, \citenamefont {{Colless}},
  \citenamefont {{Jones}}, \citenamefont {{Staveley-Smith}}, \citenamefont
  {{Campbell}}, \citenamefont {{Parker}}, \citenamefont {{Saunders}},\ and\
  \citenamefont {{Watson}}}]{Beutler:2011hx}%
  \BibitemOpen
  \bibfield  {author} {\bibinfo {author} {\bibfnamefont {F.}~\bibnamefont
  {{Beutler}}}, \bibinfo {author} {\bibfnamefont {C.}~\bibnamefont {{Blake}}},
  \bibinfo {author} {\bibfnamefont {M.}~\bibnamefont {{Colless}}}, \bibinfo
  {author} {\bibfnamefont {D.~H.}\ \bibnamefont {{Jones}}}, \bibinfo {author}
  {\bibfnamefont {L.}~\bibnamefont {{Staveley-Smith}}}, \bibinfo {author}
  {\bibfnamefont {L.}~\bibnamefont {{Campbell}}}, \bibinfo {author}
  {\bibfnamefont {Q.}~\bibnamefont {{Parker}}}, \bibinfo {author}
  {\bibfnamefont {W.}~\bibnamefont {{Saunders}}}, \ and\ \bibinfo {author}
  {\bibfnamefont {F.}~\bibnamefont {{Watson}}},\ }\href {\doibase
  10.1111/j.1365-2966.2011.19250.x} {\bibfield  {journal} {\bibinfo  {journal}
  {mnras}\ }\textbf {\bibinfo {volume} {416}},\ \bibinfo {pages} {3017}
  (\bibinfo {year} {2011})},\ \Eprint {http://arxiv.org/abs/1106.3366}
  {arXiv:1106.3366 [astro-ph.CO]} \BibitemShut {NoStop}%
\bibitem [{\citenamefont {Eisenstein}\ \emph {et~al.}(2005)\citenamefont
  {Eisenstein} \emph {et~al.}}]{Wigglez:Eisenstein2005}%
  \BibitemOpen
  \bibfield  {author} {\bibinfo {author} {\bibfnamefont {D.~J.}\ \bibnamefont
  {Eisenstein}} \emph {et~al.} (\bibinfo {collaboration} {SDSS}),\ }\href
  {\doibase 10.1086/466512} {\bibfield  {journal} {\bibinfo  {journal}
  {Astrophys. J.}\ }\textbf {\bibinfo {volume} {633}},\ \bibinfo {pages} {560}
  (\bibinfo {year} {2005})}\BibitemShut {NoStop}%
%%CITATION = ASTRO-PH/0501171;%%
\bibitem [{\citenamefont {{Hinshaw}}\ \emph {et~al.}(2013)\citenamefont
  {{Hinshaw}}, \citenamefont {{Larson}}, \citenamefont {{Komatsu}},
  \citenamefont {{Spergel}}, \citenamefont {{Bennett}}, \citenamefont
  {{Dunkley}}, \citenamefont {{Nolta}}, \citenamefont {{Halpern}},
  \citenamefont {{Hill}}, \citenamefont {{Odegard}}, \citenamefont {{Page}},
  \citenamefont {{Smith}}, \citenamefont {{Weiland}}, \citenamefont {{Gold}},
  \citenamefont {{Jarosik}}, \citenamefont {{Kogut}}, \citenamefont {{Limon}},
  \citenamefont {{Meyer}}, \citenamefont {{Tucker}}, \citenamefont
  {{Wollack}},\ and\ \citenamefont {{Wright}}}]{Hinshaw:2013}%
  \BibitemOpen
  \bibfield  {author} {\bibinfo {author} {\bibfnamefont {G.}~\bibnamefont
  {{Hinshaw}}}, \bibinfo {author} {\bibfnamefont {D.}~\bibnamefont {{Larson}}},
  \bibinfo {author} {\bibfnamefont {E.}~\bibnamefont {{Komatsu}}}, \bibinfo
  {author} {\bibfnamefont {D.~N.}\ \bibnamefont {{Spergel}}}, \bibinfo {author}
  {\bibfnamefont {C.~L.}\ \bibnamefont {{Bennett}}}, \bibinfo {author}
  {\bibfnamefont {J.}~\bibnamefont {{Dunkley}}}, \bibinfo {author}
  {\bibfnamefont {M.~R.}\ \bibnamefont {{Nolta}}}, \bibinfo {author}
  {\bibfnamefont {M.}~\bibnamefont {{Halpern}}}, \bibinfo {author}
  {\bibfnamefont {R.~S.}\ \bibnamefont {{Hill}}}, \bibinfo {author}
  {\bibfnamefont {N.}~\bibnamefont {{Odegard}}}, \bibinfo {author}
  {\bibfnamefont {L.}~\bibnamefont {{Page}}}, \bibinfo {author} {\bibfnamefont
  {K.~M.}\ \bibnamefont {{Smith}}}, \bibinfo {author} {\bibfnamefont {J.~L.}\
  \bibnamefont {{Weiland}}}, \bibinfo {author} {\bibfnamefont {B.}~\bibnamefont
  {{Gold}}}, \bibinfo {author} {\bibfnamefont {N.}~\bibnamefont {{Jarosik}}},
  \bibinfo {author} {\bibfnamefont {A.}~\bibnamefont {{Kogut}}}, \bibinfo
  {author} {\bibfnamefont {M.}~\bibnamefont {{Limon}}}, \bibinfo {author}
  {\bibfnamefont {S.~S.}\ \bibnamefont {{Meyer}}}, \bibinfo {author}
  {\bibfnamefont {G.~S.}\ \bibnamefont {{Tucker}}}, \bibinfo {author}
  {\bibfnamefont {E.}~\bibnamefont {{Wollack}}}, \ and\ \bibinfo {author}
  {\bibfnamefont {E.~L.}\ \bibnamefont {{Wright}}},\ }\href {\doibase
  10.1088/0067-0049/208/2/19} {\bibfield  {journal} {\bibinfo  {journal}
  {ApJS}\ }\textbf {\bibinfo {volume} {208}},\ \bibinfo {eid} {19} (\bibinfo
  {year} {2013})},\ \Eprint {http://arxiv.org/abs/1212.5226} {arXiv:1212.5226
  [astro-ph.CO]} \BibitemShut {NoStop}%
\bibitem [{\citenamefont {Ade}\ \emph {et~al.}(2015{\natexlab{a}})\citenamefont
  {Ade} \emph {et~al.}}]{Planck:2015XIII}%
  \BibitemOpen
  \bibfield  {author} {\bibinfo {author} {\bibfnamefont {P.~A.~R.}\
  \bibnamefont {Ade}} \emph {et~al.} (\bibinfo {collaboration} {Planck}),\
  }\href@noop {} {\  (\bibinfo {year} {2015}{\natexlab{a}})},\ \Eprint
  {http://arxiv.org/abs/1502.01589} {arXiv:1502.01589 [astro-ph.CO]}
  \BibitemShut {NoStop}%
%%CITATION = ARXIV:1502.01589;%%
\bibitem [{\citenamefont {Aganim}\ \emph {et~al.}(2020)\citenamefont {Aganim}
  \emph {et~al.}}]{Planck:2020VI}%
  \BibitemOpen
  \bibfield  {author} {\bibinfo {author} {\bibfnamefont {N.}~\bibnamefont
  {Aganim}} \emph {et~al.},\ }\href {\doibase 10.1051/0004-6361/201833910}
  {\bibfield  {journal} {\bibinfo  {journal} {A\&A}\ }\textbf {\bibinfo
  {volume} {641}},\ \bibinfo {eid} {A6} (\bibinfo {year} {2020})},\ \Eprint
  {http://arxiv.org/abs/1807.06209} {arXiv:1807.06209 [astro-ph.CO]}
  \BibitemShut {NoStop}%
\bibitem [{\citenamefont {{Hu}}\ and\ \citenamefont
  {{Dodelson}}(2002)}]{Hu:2002}%
  \BibitemOpen
  \bibfield  {author} {\bibinfo {author} {\bibfnamefont {W.}~\bibnamefont
  {{Hu}}}\ and\ \bibinfo {author} {\bibfnamefont {S.}~\bibnamefont
  {{Dodelson}}},\ }\href {\doibase 10.1146/annurev.astro.40.060401.093926}
  {\bibfield  {journal} {\bibinfo  {journal} {Ann. Rev. Astron. Astrophys.}\
  }\textbf {\bibinfo {volume} {40}},\ \bibinfo {pages} {171} (\bibinfo {year}
  {2002})},\ \Eprint {http://arxiv.org/abs/astro-ph/0110414}
  {arXiv:astro-ph/0110414 [astro-ph]} \BibitemShut {NoStop}%
\bibitem [{\citenamefont {Komatsu}\ \emph
  {et~al.}(2011{\natexlab{a}})\citenamefont {Komatsu}, \citenamefont {Smith},
  \citenamefont {Dunkley}, \citenamefont {Bennett}, \citenamefont {Gold},
  \citenamefont {Hinshaw}, \citenamefont {Jarosik}, \citenamefont {Larson},
  \citenamefont {Nolta}, \citenamefont {Page}, \citenamefont {Spergel},
  \citenamefont {Halpern}, \citenamefont {Hill}, \citenamefont {Kogut},
  \citenamefont {Limon}, \citenamefont {Meyer}, \citenamefont {Odegard},
  \citenamefont {Tucker}, \citenamefont {Weiland}, \citenamefont {Wollack},\
  and\ \citenamefont {Wright}}]{Komatsu_2011}%
  \BibitemOpen
  \bibfield  {author} {\bibinfo {author} {\bibfnamefont {E.}~\bibnamefont
  {Komatsu}}, \bibinfo {author} {\bibfnamefont {K.~M.}\ \bibnamefont {Smith}},
  \bibinfo {author} {\bibfnamefont {J.}~\bibnamefont {Dunkley}}, \bibinfo
  {author} {\bibfnamefont {C.~L.}\ \bibnamefont {Bennett}}, \bibinfo {author}
  {\bibfnamefont {B.}~\bibnamefont {Gold}}, \bibinfo {author} {\bibfnamefont
  {G.}~\bibnamefont {Hinshaw}}, \bibinfo {author} {\bibfnamefont
  {N.}~\bibnamefont {Jarosik}}, \bibinfo {author} {\bibfnamefont
  {D.}~\bibnamefont {Larson}}, \bibinfo {author} {\bibfnamefont {M.~R.}\
  \bibnamefont {Nolta}}, \bibinfo {author} {\bibfnamefont {L.}~\bibnamefont
  {Page}}, \bibinfo {author} {\bibfnamefont {D.~N.}\ \bibnamefont {Spergel}},
  \bibinfo {author} {\bibfnamefont {M.}~\bibnamefont {Halpern}}, \bibinfo
  {author} {\bibfnamefont {R.~S.}\ \bibnamefont {Hill}}, \bibinfo {author}
  {\bibfnamefont {A.}~\bibnamefont {Kogut}}, \bibinfo {author} {\bibfnamefont
  {M.}~\bibnamefont {Limon}}, \bibinfo {author} {\bibfnamefont {S.~S.}\
  \bibnamefont {Meyer}}, \bibinfo {author} {\bibfnamefont {N.}~\bibnamefont
  {Odegard}}, \bibinfo {author} {\bibfnamefont {G.~S.}\ \bibnamefont {Tucker}},
  \bibinfo {author} {\bibfnamefont {J.~L.}\ \bibnamefont {Weiland}}, \bibinfo
  {author} {\bibfnamefont {E.}~\bibnamefont {Wollack}}, \ and\ \bibinfo
  {author} {\bibfnamefont {E.~L.}\ \bibnamefont {Wright}},\ }\href {\doibase
  10.1088/0067-0049/192/2/18} {\bibfield  {journal} {\bibinfo  {journal} {The
  Astrophysical Journal Supplement Series}\ }\textbf {\bibinfo {volume}
  {192}},\ \bibinfo {pages} {18} (\bibinfo {year}
  {2011}{\natexlab{a}})}\BibitemShut {NoStop}%
\bibitem [{\citenamefont {Wang}\ and\ \citenamefont
  {Mukherjee}(2006)}]{Wang:2006ts}%
  \BibitemOpen
  \bibfield  {author} {\bibinfo {author} {\bibfnamefont {Y.}~\bibnamefont
  {Wang}}\ and\ \bibinfo {author} {\bibfnamefont {P.}~\bibnamefont
  {Mukherjee}},\ }\href {\doibase 10.1086/507091} {\bibfield  {journal}
  {\bibinfo  {journal} {Astrophys. J.}\ }\textbf {\bibinfo {volume} {650}},\
  \bibinfo {pages} {1} (\bibinfo {year} {2006})},\ \Eprint
  {http://arxiv.org/abs/astro-ph/0604051} {arXiv:astro-ph/0604051} \BibitemShut
  {NoStop}%
\bibitem [{\citenamefont {Wang}\ \emph {et~al.}(2012)\citenamefont {Wang},
  \citenamefont {Chuang},\ and\ \citenamefont {Mukherjee}}]{Wang:2011sb}%
  \BibitemOpen
  \bibfield  {author} {\bibinfo {author} {\bibfnamefont {Y.}~\bibnamefont
  {Wang}}, \bibinfo {author} {\bibfnamefont {C.-H.}\ \bibnamefont {Chuang}}, \
  and\ \bibinfo {author} {\bibfnamefont {P.}~\bibnamefont {Mukherjee}},\ }\href
  {\doibase 10.1103/PhysRevD.85.023517} {\bibfield  {journal} {\bibinfo
  {journal} {Phys. Rev. D}\ }\textbf {\bibinfo {volume} {85}},\ \bibinfo
  {pages} {023517} (\bibinfo {year} {2012})},\ \Eprint
  {http://arxiv.org/abs/1109.3172} {arXiv:1109.3172 [astro-ph.CO]} \BibitemShut
  {NoStop}%
\bibitem [{\citenamefont {Hu}\ and\ \citenamefont
  {Sugiyama}(1996)}]{Hu:1995en}%
  \BibitemOpen
  \bibfield  {author} {\bibinfo {author} {\bibfnamefont {W.}~\bibnamefont
  {Hu}}\ and\ \bibinfo {author} {\bibfnamefont {N.}~\bibnamefont {Sugiyama}},\
  }\href {\doibase 10.1086/177989} {\bibfield  {journal} {\bibinfo  {journal}
  {Astrophys. J.}\ }\textbf {\bibinfo {volume} {471}},\ \bibinfo {pages} {542}
  (\bibinfo {year} {1996})},\ \Eprint {http://arxiv.org/abs/astro-ph/9510117}
  {arXiv:astro-ph/9510117 [astro-ph]} \BibitemShut {NoStop}%
%%CITATION = ASTRO-PH/9510117;%%
\bibitem [{\citenamefont {Bond}\ \emph {et~al.}(1997)\citenamefont {Bond},
  \citenamefont {Efstathiou},\ and\ \citenamefont {Tegmark}}]{Bond:1997}%
  \BibitemOpen
  \bibfield  {author} {\bibinfo {author} {\bibfnamefont {J.~R.}\ \bibnamefont
  {Bond}}, \bibinfo {author} {\bibfnamefont {G.}~\bibnamefont {Efstathiou}}, \
  and\ \bibinfo {author} {\bibfnamefont {M.}~\bibnamefont {Tegmark}},\ }\href
  {\doibase 10.1093/mnras/291.1.L33} {\bibfield  {journal} {\bibinfo  {journal}
  {Mon. Not. Roy. Astron. Soc.}\ }\textbf {\bibinfo {volume} {291}},\ \bibinfo
  {pages} {L33} (\bibinfo {year} {1997})},\ \Eprint
  {http://arxiv.org/abs/astro-ph/9702100} {arXiv:astro-ph/9702100} \BibitemShut
  {NoStop}%
\bibitem [{\citenamefont {Neveu}\ \emph {et~al.}(2017)\citenamefont {Neveu},
  \citenamefont {Ruhlmann-Kleider}, \citenamefont {Astier}, \citenamefont
  {Besan\c{c}on}, \citenamefont {Guy}, \citenamefont {M\"oller},\ and\
  \citenamefont {Babichev}}]{Neveu:2016gxp}%
  \BibitemOpen
  \bibfield  {author} {\bibinfo {author} {\bibfnamefont {J.}~\bibnamefont
  {Neveu}}, \bibinfo {author} {\bibfnamefont {V.}~\bibnamefont
  {Ruhlmann-Kleider}}, \bibinfo {author} {\bibfnamefont {P.}~\bibnamefont
  {Astier}}, \bibinfo {author} {\bibfnamefont {M.}~\bibnamefont
  {Besan\c{c}on}}, \bibinfo {author} {\bibfnamefont {J.}~\bibnamefont {Guy}},
  \bibinfo {author} {\bibfnamefont {A.}~\bibnamefont {M\"oller}}, \ and\
  \bibinfo {author} {\bibfnamefont {E.}~\bibnamefont {Babichev}},\ }\href
  {\doibase 10.1051/0004-6361/201628878} {\bibfield  {journal} {\bibinfo
  {journal} {Astron. Astrophys.}\ }\textbf {\bibinfo {volume} {600}},\ \bibinfo
  {pages} {A40} (\bibinfo {year} {2017})},\ \Eprint
  {http://arxiv.org/abs/1605.02627} {arXiv:1605.02627 [gr-qc]} \BibitemShut
  {NoStop}%
\bibitem [{\citenamefont {Moresco}\ \emph {et~al.}(2012)\citenamefont {Moresco}
  \emph {et~al.}}]{Moresco:2012}%
  \BibitemOpen
  \bibfield  {author} {\bibinfo {author} {\bibfnamefont {M.}~\bibnamefont
  {Moresco}} \emph {et~al.},\ }\href {\doibase 10.1088/1475-7516/2012/08/006}
  {\bibfield  {journal} {\bibinfo  {journal} {Journal of Cosmology and
  Astroparticle Physics}\ }\textbf {\bibinfo {volume} {2012}},\ \bibinfo
  {pages} {006} (\bibinfo {year} {2012})}\BibitemShut {NoStop}%
\bibitem [{\citenamefont {Chae}\ \emph {et~al.}(2002)\citenamefont {Chae} \emph
  {et~al.}}]{chae:2002}%
  \BibitemOpen
  \bibfield  {author} {\bibinfo {author} {\bibfnamefont {K.~H.}\ \bibnamefont
  {Chae}} \emph {et~al.},\ }\href {\doibase 10.1103/PhysRevLett.89.151301}
  {\bibfield  {journal} {\bibinfo  {journal} {Phys. Rev. Lett.}\ }\textbf
  {\bibinfo {volume} {89}},\ \bibinfo {pages} {151301} (\bibinfo {year}
  {2002})},\ \Eprint {http://arxiv.org/abs/astro-ph/0209602}
  {arXiv:astro-ph/0209602} \BibitemShut {NoStop}%
\bibitem [{\citenamefont {Biesiada}\ \emph {et~al.}(2010)\citenamefont
  {Biesiada}, \citenamefont {Piórkowska},\ and\ \citenamefont
  {Malec}}]{Biesiada}%
  \BibitemOpen
  \bibfield  {author} {\bibinfo {author} {\bibfnamefont {M.}~\bibnamefont
  {Biesiada}}, \bibinfo {author} {\bibfnamefont {A.}~\bibnamefont
  {Piórkowska}}, \ and\ \bibinfo {author} {\bibfnamefont {B.}~\bibnamefont
  {Malec}},\ }\href {\doibase 10.1111/j.1365-2966.2010.16725.x} {\bibfield
  {journal} {\bibinfo  {journal} {Monthly Notices of the Royal Astronomical
  Society}\ }\textbf {\bibinfo {volume} {406}},\ \bibinfo {pages} {1055}
  (\bibinfo {year} {2010})},\ \Eprint
  {http://arxiv.org/abs/https://academic.oup.com/mnras/article-pdf/406/2/1055/18711863/mnr0406-1055.pdf}
  {https://academic.oup.com/mnras/article-pdf/406/2/1055/18711863/mnr0406-1055.pdf}
  \BibitemShut {NoStop}%
\bibitem [{\citenamefont {Cao}\ \emph {et~al.}(2012{\natexlab{a}})\citenamefont
  {Cao}, \citenamefont {Pan}, \citenamefont {Biesiada}, \citenamefont
  {Godlowski},\ and\ \citenamefont {Zhu}}]{Cao_2012}%
  \BibitemOpen
  \bibfield  {author} {\bibinfo {author} {\bibfnamefont {S.}~\bibnamefont
  {Cao}}, \bibinfo {author} {\bibfnamefont {Y.}~\bibnamefont {Pan}}, \bibinfo
  {author} {\bibfnamefont {M.}~\bibnamefont {Biesiada}}, \bibinfo {author}
  {\bibfnamefont {W.}~\bibnamefont {Godlowski}}, \ and\ \bibinfo {author}
  {\bibfnamefont {Z.-H.}\ \bibnamefont {Zhu}},\ }\href {\doibase
  10.1088/1475-7516/2012/03/016} {\bibfield  {journal} {\bibinfo  {journal}
  {Journal of Cosmology and Astroparticle Physics}\ }\textbf {\bibinfo {volume}
  {2012}},\ \bibinfo {pages} {016} (\bibinfo {year}
  {2012}{\natexlab{a}})}\BibitemShut {NoStop}%
\bibitem [{\citenamefont {Cao}\ \emph {et~al.}(2015{\natexlab{b}})\citenamefont
  {Cao}, \citenamefont {Biesiada}, \citenamefont {Gavazzi}, \citenamefont
  {Pi{\'{o}}rkowska},\ and\ \citenamefont {Zhu}}]{Cao_2015}%
  \BibitemOpen
  \bibfield  {author} {\bibinfo {author} {\bibfnamefont {S.}~\bibnamefont
  {Cao}}, \bibinfo {author} {\bibfnamefont {M.}~\bibnamefont {Biesiada}},
  \bibinfo {author} {\bibfnamefont {R.}~\bibnamefont {Gavazzi}}, \bibinfo
  {author} {\bibfnamefont {A.}~\bibnamefont {Pi{\'{o}}rkowska}}, \ and\
  \bibinfo {author} {\bibfnamefont {Z.-H.}\ \bibnamefont {Zhu}},\ }\href
  {\doibase 10.1088/0004-637x/806/2/185} {\bibfield  {journal} {\bibinfo
  {journal} {The Astrophysical Journal}\ }\textbf {\bibinfo {volume} {806}},\
  \bibinfo {pages} {185} (\bibinfo {year} {2015}{\natexlab{b}})}\BibitemShut
  {NoStop}%
\bibitem [{\citenamefont {Maga{\~{n}}a}\ \emph {et~al.}(2015)\citenamefont
  {Maga{\~{n}}a}, \citenamefont {Motta}, \citenamefont {C{\'{a}}rdenas},
  \citenamefont {Verdugo},\ and\ \citenamefont {Jullo}}]{Maga_a_2015}%
  \BibitemOpen
  \bibfield  {author} {\bibinfo {author} {\bibfnamefont {J.}~\bibnamefont
  {Maga{\~{n}}a}}, \bibinfo {author} {\bibfnamefont {V.}~\bibnamefont {Motta}},
  \bibinfo {author} {\bibfnamefont {V.~H.}\ \bibnamefont {C{\'{a}}rdenas}},
  \bibinfo {author} {\bibfnamefont {T.}~\bibnamefont {Verdugo}}, \ and\
  \bibinfo {author} {\bibfnamefont {E.}~\bibnamefont {Jullo}},\ }\href
  {\doibase 10.1088/0004-637x/813/1/69} {\bibfield  {journal} {\bibinfo
  {journal} {The Astrophysical Journal}\ }\textbf {\bibinfo {volume} {813}},\
  \bibinfo {pages} {69} (\bibinfo {year} {2015})}\BibitemShut {NoStop}%
\bibitem [{\citenamefont {Maga{\~{n}}a}\ \emph {et~al.}(2018)\citenamefont
  {Maga{\~{n}}a}, \citenamefont {Acebr{\'{o}}n}, \citenamefont {Motta},
  \citenamefont {Verdugo}, \citenamefont {Jullo},\ and\ \citenamefont
  {Limousin}}]{Maga_a_2018}%
  \BibitemOpen
  \bibfield  {author} {\bibinfo {author} {\bibfnamefont {J.}~\bibnamefont
  {Maga{\~{n}}a}}, \bibinfo {author} {\bibfnamefont {A.}~\bibnamefont
  {Acebr{\'{o}}n}}, \bibinfo {author} {\bibfnamefont {V.}~\bibnamefont
  {Motta}}, \bibinfo {author} {\bibfnamefont {T.}~\bibnamefont {Verdugo}},
  \bibinfo {author} {\bibfnamefont {E.}~\bibnamefont {Jullo}}, \ and\ \bibinfo
  {author} {\bibfnamefont {M.}~\bibnamefont {Limousin}},\ }\href {\doibase
  10.3847/1538-4357/aada7d} {\bibfield  {journal} {\bibinfo  {journal} {The
  Astrophysical Journal}\ }\textbf {\bibinfo {volume} {865}},\ \bibinfo {pages}
  {122} (\bibinfo {year} {2018})}\BibitemShut {NoStop}%
\bibitem [{\citenamefont {Ofek}\ \emph {et~al.}(2003)\citenamefont {Ofek},
  \citenamefont {Rix},\ and\ \citenamefont {Maoz}}]{Ofek:2003}%
  \BibitemOpen
  \bibfield  {author} {\bibinfo {author} {\bibfnamefont {E.~O.}\ \bibnamefont
  {Ofek}}, \bibinfo {author} {\bibfnamefont {H.-W.}\ \bibnamefont {Rix}}, \
  and\ \bibinfo {author} {\bibfnamefont {D.}~\bibnamefont {Maoz}},\ }\href
  {\doibase 10.1046/j.1365-8711.2003.06707.x} {\bibfield  {journal} {\bibinfo
  {journal} {Mon. Not. Roy. Astron. Soc.}\ }\textbf {\bibinfo {volume} {343}},\
  \bibinfo {pages} {639} (\bibinfo {year} {2003})},\ \Eprint
  {http://arxiv.org/abs/astro-ph/0305201} {arXiv:astro-ph/0305201 [astro-ph]}
  \BibitemShut {NoStop}%
%%CITATION = ASTRO-PH/0305201;%%
\bibitem [{\citenamefont {Cao}\ \emph {et~al.}(2012{\natexlab{b}})\citenamefont
  {Cao}, \citenamefont {Pan}, \citenamefont {Biesiada}, \citenamefont
  {Godlowski},\ and\ \citenamefont {Zhu}}]{Cao:2011}%
  \BibitemOpen
  \bibfield  {author} {\bibinfo {author} {\bibfnamefont {S.}~\bibnamefont
  {Cao}}, \bibinfo {author} {\bibfnamefont {Y.}~\bibnamefont {Pan}}, \bibinfo
  {author} {\bibfnamefont {M.}~\bibnamefont {Biesiada}}, \bibinfo {author}
  {\bibfnamefont {W.}~\bibnamefont {Godlowski}}, \ and\ \bibinfo {author}
  {\bibfnamefont {Z.-H.}\ \bibnamefont {Zhu}},\ }\href {\doibase
  10.1088/1475-7516/2012/03/016} {\bibfield  {journal} {\bibinfo  {journal}
  {JCAP}\ }\textbf {\bibinfo {volume} {1203}},\ \bibinfo {pages} {016}
  (\bibinfo {year} {2012}{\natexlab{b}})},\ \Eprint
  {http://arxiv.org/abs/1105.6226} {arXiv:1105.6226 [astro-ph.CO]} \BibitemShut
  {NoStop}%
%%CITATION = ARXIV:1105.6226;%%
\bibitem [{\citenamefont {{Foreman-Mackey}}\ \emph {et~al.}(2013)\citenamefont
  {{Foreman-Mackey}}, \citenamefont {{Hogg}}, \citenamefont {{Lang}},\ and\
  \citenamefont {{Goodman}}}]{Foreman:2013}%
  \BibitemOpen
  \bibfield  {author} {\bibinfo {author} {\bibfnamefont {D.}~\bibnamefont
  {{Foreman-Mackey}}}, \bibinfo {author} {\bibfnamefont {D.~W.}\ \bibnamefont
  {{Hogg}}}, \bibinfo {author} {\bibfnamefont {D.}~\bibnamefont {{Lang}}}, \
  and\ \bibinfo {author} {\bibfnamefont {J.}~\bibnamefont {{Goodman}}},\ }\href
  {\doibase 10.1086/670067} {\bibfield  {journal} {\bibinfo  {journal} {PASP}\
  }\textbf {\bibinfo {volume} {125}},\ \bibinfo {pages} {306} (\bibinfo {year}
  {2013})},\ \Eprint {http://arxiv.org/abs/1202.3665} {arXiv:1202.3665
  [astro-ph.IM]} \BibitemShut {NoStop}%
\bibitem [{\citenamefont {Gelman}\ and\ \citenamefont
  {Rubin}(1992)}]{Gelman:1992}%
  \BibitemOpen
  \bibfield  {author} {\bibinfo {author} {\bibfnamefont {A.}~\bibnamefont
  {Gelman}}\ and\ \bibinfo {author} {\bibfnamefont {D.}~\bibnamefont {Rubin}},\
  }\href {\doibase 10.1103/PhysRevD.67.101301} {\bibfield  {journal} {\bibinfo
  {journal} {Statistical Science}\ }\textbf {\bibinfo {volume} {67}},\ \bibinfo
  {pages} {457} (\bibinfo {year} {1992})}\BibitemShut {NoStop}%
\bibitem [{\citenamefont {Garc\'\i{}a-Aspeitia}\ \emph
  {et~al.}(2019)\citenamefont {Garc\'\i{}a-Aspeitia}, \citenamefont
  {Hern\'andez-Almada}, \citenamefont {Maga\~na},\ and\ \citenamefont
  {Motta}}]{Garcia-Aspeitia:2019yod}%
  \BibitemOpen
  \bibfield  {author} {\bibinfo {author} {\bibfnamefont {M.~A.}\ \bibnamefont
  {Garc\'\i{}a-Aspeitia}}, \bibinfo {author} {\bibfnamefont {A.}~\bibnamefont
  {Hern\'andez-Almada}}, \bibinfo {author} {\bibfnamefont {J.}~\bibnamefont
  {Maga\~na}}, \ and\ \bibinfo {author} {\bibfnamefont {V.}~\bibnamefont
  {Motta}},\ }\href@noop {} {\  (\bibinfo {year} {2019})},\ \Eprint
  {http://arxiv.org/abs/1912.07500} {arXiv:1912.07500 [astro-ph.CO]}
  \BibitemShut {NoStop}%
\bibitem [{\citenamefont {Komatsu}\ \emph
  {et~al.}(2011{\natexlab{b}})\citenamefont {Komatsu}, \citenamefont {Smith},
  \citenamefont {Dunkley}, \citenamefont {Bennett}, \citenamefont {Gold},
  \citenamefont {Hinshaw}, \citenamefont {Jarosik}, \citenamefont {Larson},
  \citenamefont {Nolta}, \citenamefont {Page}, \citenamefont {Spergel},
  \citenamefont {Halpern}, \citenamefont {Hill}, \citenamefont {Kogut},
  \citenamefont {Limon}, \citenamefont {Meyer}, \citenamefont {Odegard},
  \citenamefont {Tucker}, \citenamefont {Weiland}, \citenamefont {Wollack},\
  and\ \citenamefont {Wright}}]{Komatsu:2011}%
  \BibitemOpen
  \bibfield  {author} {\bibinfo {author} {\bibfnamefont {E.}~\bibnamefont
  {Komatsu}}, \bibinfo {author} {\bibfnamefont {K.~M.}\ \bibnamefont {Smith}},
  \bibinfo {author} {\bibfnamefont {J.}~\bibnamefont {Dunkley}}, \bibinfo
  {author} {\bibfnamefont {C.~L.}\ \bibnamefont {Bennett}}, \bibinfo {author}
  {\bibfnamefont {B.}~\bibnamefont {Gold}}, \bibinfo {author} {\bibfnamefont
  {G.}~\bibnamefont {Hinshaw}}, \bibinfo {author} {\bibfnamefont
  {N.}~\bibnamefont {Jarosik}}, \bibinfo {author} {\bibfnamefont
  {D.}~\bibnamefont {Larson}}, \bibinfo {author} {\bibfnamefont {M.~R.}\
  \bibnamefont {Nolta}}, \bibinfo {author} {\bibfnamefont {L.}~\bibnamefont
  {Page}}, \bibinfo {author} {\bibfnamefont {D.~N.}\ \bibnamefont {Spergel}},
  \bibinfo {author} {\bibfnamefont {M.}~\bibnamefont {Halpern}}, \bibinfo
  {author} {\bibfnamefont {R.~S.}\ \bibnamefont {Hill}}, \bibinfo {author}
  {\bibfnamefont {A.}~\bibnamefont {Kogut}}, \bibinfo {author} {\bibfnamefont
  {M.}~\bibnamefont {Limon}}, \bibinfo {author} {\bibfnamefont {S.~S.}\
  \bibnamefont {Meyer}}, \bibinfo {author} {\bibfnamefont {N.}~\bibnamefont
  {Odegard}}, \bibinfo {author} {\bibfnamefont {G.~S.}\ \bibnamefont {Tucker}},
  \bibinfo {author} {\bibfnamefont {J.~L.}\ \bibnamefont {Weiland}}, \bibinfo
  {author} {\bibfnamefont {E.}~\bibnamefont {Wollack}}, \ and\ \bibinfo
  {author} {\bibfnamefont {E.~L.}\ \bibnamefont {Wright}},\ }\href
  {http://stacks.iop.org/0067-0049/192/i=2/a=18} {\bibfield  {journal}
  {\bibinfo  {journal} {The Astrophysical Journal Supplement Series}\ }\textbf
  {\bibinfo {volume} {192}},\ \bibinfo {pages} {18} (\bibinfo {year}
  {2011}{\natexlab{b}})}\BibitemShut {NoStop}%
\bibitem [{\citenamefont {Riess}\ \emph {et~al.}(2021)\citenamefont {Riess},
  \citenamefont {Casertano}, \citenamefont {Yuan}, \citenamefont {Bowers},
  \citenamefont {Macri}, \citenamefont {Zinn},\ and\ \citenamefont
  {Scolnic}}]{Riess:2020fzl}%
  \BibitemOpen
  \bibfield  {author} {\bibinfo {author} {\bibfnamefont {A.~G.}\ \bibnamefont
  {Riess}}, \bibinfo {author} {\bibfnamefont {S.}~\bibnamefont {Casertano}},
  \bibinfo {author} {\bibfnamefont {W.}~\bibnamefont {Yuan}}, \bibinfo {author}
  {\bibfnamefont {J.~B.}\ \bibnamefont {Bowers}}, \bibinfo {author}
  {\bibfnamefont {L.}~\bibnamefont {Macri}}, \bibinfo {author} {\bibfnamefont
  {J.~C.}\ \bibnamefont {Zinn}}, \ and\ \bibinfo {author} {\bibfnamefont
  {D.}~\bibnamefont {Scolnic}},\ }\href {\doibase 10.3847/2041-8213/abdbaf}
  {\bibfield  {journal} {\bibinfo  {journal} {Astrophys. J. Lett.}\ }\textbf
  {\bibinfo {volume} {908}},\ \bibinfo {pages} {L6} (\bibinfo {year} {2021})},\
  \Eprint {http://arxiv.org/abs/2012.08534} {arXiv:2012.08534 [astro-ph.CO]}
  \BibitemShut {NoStop}%
\bibitem [{\citenamefont {Chevallier}\ and\ \citenamefont
  {Polarski}(2001)}]{Chevallier:2000qy}%
  \BibitemOpen
  \bibfield  {author} {\bibinfo {author} {\bibfnamefont {M.}~\bibnamefont
  {Chevallier}}\ and\ \bibinfo {author} {\bibfnamefont {D.}~\bibnamefont
  {Polarski}},\ }\href {\doibase 10.1142/S0218271801000822} {\bibfield
  {journal} {\bibinfo  {journal} {Int. J. Mod. Phys.}\ }\textbf {\bibinfo
  {volume} {D10}},\ \bibinfo {pages} {213} (\bibinfo {year} {2001})},\ \Eprint
  {http://arxiv.org/abs/gr-qc/0009008} {arXiv:gr-qc/0009008 [gr-qc]}
  \BibitemShut {NoStop}%
%%CITATION = GR-QC/0009008;%%
\bibitem [{\citenamefont {Linder}(2003{\natexlab{a}})}]{Linder:2003}%
  \BibitemOpen
  \bibfield  {author} {\bibinfo {author} {\bibfnamefont {E.~V.}\ \bibnamefont
  {Linder}},\ }\href {\doibase 10.1103/PhysRevLett.90.091301} {\bibfield
  {journal} {\bibinfo  {journal} {Phys. Rev. Lett.}\ }\textbf {\bibinfo
  {volume} {90}},\ \bibinfo {pages} {091301} (\bibinfo {year}
  {2003}{\natexlab{a}})}\BibitemShut {NoStop}%
\bibitem [{\citenamefont {Linder}(2003{\natexlab{b}})}]{Linder:2002dt}%
  \BibitemOpen
  \bibfield  {author} {\bibinfo {author} {\bibfnamefont {E.~V.}\ \bibnamefont
  {Linder}},\ }\href {\doibase 10.1103/PhysRevD.68.083503} {\bibfield
  {journal} {\bibinfo  {journal} {Phys. Rev.}\ }\textbf {\bibinfo {volume}
  {D68}},\ \bibinfo {pages} {083503} (\bibinfo {year} {2003}{\natexlab{b}})},\
  \Eprint {http://arxiv.org/abs/astro-ph/0212301} {arXiv:astro-ph/0212301
  [astro-ph]} \BibitemShut {NoStop}%
%%CITATION = ASTRO-PH/0212301;%%
\bibitem [{\citenamefont {Maga\~na}\ \emph {et~al.}(2018)\citenamefont
  {Maga\~na}, \citenamefont {Amante}, \citenamefont {Garcia-Aspeitia},\ and\
  \citenamefont {Motta}}]{Magana:2017}%
  \BibitemOpen
  \bibfield  {author} {\bibinfo {author} {\bibfnamefont {J.}~\bibnamefont
  {Maga\~na}}, \bibinfo {author} {\bibfnamefont {M.~H.}\ \bibnamefont
  {Amante}}, \bibinfo {author} {\bibfnamefont {M.~A.}\ \bibnamefont
  {Garcia-Aspeitia}}, \ and\ \bibinfo {author} {\bibfnamefont {V.}~\bibnamefont
  {Motta}},\ }\href {\doibase 10.1093/mnras/sty260} {\bibfield  {journal}
  {\bibinfo  {journal} {Mon. Not. Roy. Astron. Soc.}\ }\textbf {\bibinfo
  {volume} {476}},\ \bibinfo {pages} {1036} (\bibinfo {year} {2018})},\ \Eprint
  {http://arxiv.org/abs/1706.09848} {arXiv:1706.09848 [astro-ph.CO]}
  \BibitemShut {NoStop}%
%%CITATION = ARXIV:1706.09848;%%
\bibitem [{\citenamefont {Jassal}\ \emph {et~al.}(2005)\citenamefont {Jassal},
  \citenamefont {Bagla},\ and\ \citenamefont {Padmanabhan}}]{Jassal}%
  \BibitemOpen
  \bibfield  {author} {\bibinfo {author} {\bibfnamefont {H.~K.}\ \bibnamefont
  {Jassal}}, \bibinfo {author} {\bibfnamefont {J.~S.}\ \bibnamefont {Bagla}}, \
  and\ \bibinfo {author} {\bibfnamefont {T.}~\bibnamefont {Padmanabhan}},\
  }\href {\doibase 10.1111/j.1745-3933.2005.08577.x} {\bibfield  {journal}
  {\bibinfo  {journal} {Mon. Not. Roy. Astron. Soc.}\ }\textbf {\bibinfo
  {volume} {356}},\ \bibinfo {pages} {L11} (\bibinfo {year} {2005})},\ \Eprint
  {http://arxiv.org/abs/astro-ph/0404378} {arXiv:astro-ph/0404378 [astro-ph]}
  \BibitemShut {NoStop}%
%%CITATION = ASTRO-PH/0404378;%%
\bibitem [{\citenamefont {Barboza}\ and\ \citenamefont
  {Alcaniz}(2008)}]{Barboza:2008rh}%
  \BibitemOpen
  \bibfield  {author} {\bibinfo {author} {\bibfnamefont {E.~M.}\ \bibnamefont
  {Barboza}, \bibfnamefont {Jr.}}\ and\ \bibinfo {author} {\bibfnamefont
  {J.~S.}\ \bibnamefont {Alcaniz}},\ }\href {\doibase
  10.1016/j.physletb.2008.08.012} {\bibfield  {journal} {\bibinfo  {journal}
  {Phys. Lett.}\ }\textbf {\bibinfo {volume} {B666}},\ \bibinfo {pages} {415}
  (\bibinfo {year} {2008})},\ \Eprint {http://arxiv.org/abs/0805.1713}
  {arXiv:0805.1713 [astro-ph]} \BibitemShut {NoStop}%
%%CITATION = ARXIV:0805.1713;%%
\bibitem [{\citenamefont {Feng}\ \emph {et~al.}(2012)\citenamefont {Feng},
  \citenamefont {Shen}, \citenamefont {Li},\ and\ \citenamefont
  {Li}}]{Feng:2012gf}%
  \BibitemOpen
  \bibfield  {author} {\bibinfo {author} {\bibfnamefont {C.-J.}\ \bibnamefont
  {Feng}}, \bibinfo {author} {\bibfnamefont {X.-Y.}\ \bibnamefont {Shen}},
  \bibinfo {author} {\bibfnamefont {P.}~\bibnamefont {Li}}, \ and\ \bibinfo
  {author} {\bibfnamefont {X.-Z.}\ \bibnamefont {Li}},\ }\href {\doibase
  10.1088/1475-7516/2012/09/023} {\bibfield  {journal} {\bibinfo  {journal}
  {JCAP}\ }\textbf {\bibinfo {volume} {1209}},\ \bibinfo {pages} {023}
  (\bibinfo {year} {2012})},\ \Eprint {http://arxiv.org/abs/1206.0063}
  {arXiv:1206.0063 [astro-ph.CO]} \BibitemShut {NoStop}%
%%CITATION = ARXIV:1206.0063;%%
\bibitem [{\citenamefont {{Sendra}}\ and\ \citenamefont
  {{Lazkoz}}(2012)}]{Sendra:2012}%
  \BibitemOpen
  \bibfield  {author} {\bibinfo {author} {\bibfnamefont {I.}~\bibnamefont
  {{Sendra}}}\ and\ \bibinfo {author} {\bibfnamefont {R.}~\bibnamefont
  {{Lazkoz}}},\ }\href {\doibase 10.1111/j.1365-2966.2012.20661.x} {\bibfield
  {journal} {\bibinfo  {journal} {Monthly Notices of the Royal Astronomical
  Society}\ }\textbf {\bibinfo {volume} {422}},\ \bibinfo {pages} {776}
  (\bibinfo {year} {2012})},\ \Eprint {http://arxiv.org/abs/1105.4943}
  {arXiv:1105.4943 [astro-ph.CO]} \BibitemShut {NoStop}%
\bibitem [{\citenamefont {{Luongo}}(2011)}]{Luongo:2011}%
  \BibitemOpen
  \bibfield  {author} {\bibinfo {author} {\bibfnamefont {O.}~\bibnamefont
  {{Luongo}}},\ }\href {\doibase 10.1142/S0217732311035894} {\bibfield
  {journal} {\bibinfo  {journal} {Modern Physics Letters A}\ }\textbf {\bibinfo
  {volume} {26}},\ \bibinfo {pages} {1459} (\bibinfo {year}
  {2011})}\BibitemShut {NoStop}%
\bibitem [{\citenamefont {{Aviles}}\ \emph {et~al.}(2012)\citenamefont
  {{Aviles}}, \citenamefont {{Gruber}}, \citenamefont {{Luongo}},\ and\
  \citenamefont {{Quevedo}}}]{Aviles:2012}%
  \BibitemOpen
  \bibfield  {author} {\bibinfo {author} {\bibfnamefont {A.}~\bibnamefont
  {{Aviles}}}, \bibinfo {author} {\bibfnamefont {C.}~\bibnamefont {{Gruber}}},
  \bibinfo {author} {\bibfnamefont {O.}~\bibnamefont {{Luongo}}}, \ and\
  \bibinfo {author} {\bibfnamefont {H.}~\bibnamefont {{Quevedo}}},\ }\href
  {\doibase 10.1103/PhysRevD.86.123516} {\bibfield  {journal} {\bibinfo
  {journal} {Phys. Rev. D}\ }\textbf {\bibinfo {volume} {86}},\ \bibinfo {eid}
  {123516} (\bibinfo {year} {2012})},\ \Eprint {http://arxiv.org/abs/1204.2007}
  {arXiv:1204.2007 [astro-ph.CO]} \BibitemShut {NoStop}%
\bibitem [{\citenamefont {Gruber}\ and\ \citenamefont
  {Luongo}(2014)}]{padeaprox}%
  \BibitemOpen
  \bibfield  {author} {\bibinfo {author} {\bibfnamefont {C.}~\bibnamefont
  {Gruber}}\ and\ \bibinfo {author} {\bibfnamefont {O.}~\bibnamefont
  {Luongo}},\ }\href {\doibase 10.1103/PhysRevD.89.103506} {\bibfield
  {journal} {\bibinfo  {journal} {Phys. Rev. D}\ }\textbf {\bibinfo {volume}
  {89}},\ \bibinfo {pages} {103506} (\bibinfo {year} {2014})}\BibitemShut
  {NoStop}%
\bibitem [{\citenamefont {Demianski}\ \emph {et~al.}(2012)\citenamefont
  {Demianski}, \citenamefont {Piedipalumbo}, \citenamefont {Rubano},\ and\
  \citenamefont {Scudellaro}}]{Demianski:2012}%
  \BibitemOpen
  \bibfield  {author} {\bibinfo {author} {\bibfnamefont {M.}~\bibnamefont
  {Demianski}}, \bibinfo {author} {\bibfnamefont {E.}~\bibnamefont
  {Piedipalumbo}}, \bibinfo {author} {\bibfnamefont {C.}~\bibnamefont
  {Rubano}}, \ and\ \bibinfo {author} {\bibfnamefont {P.}~\bibnamefont
  {Scudellaro}},\ }\href {\doibase 10.1111/j.1365-2966.2012.21568.x} {\bibfield
   {journal} {\bibinfo  {journal} {Monthly Notices of the Royal Astronomical
  Society}\ }\textbf {\bibinfo {volume} {426}},\ \bibinfo {pages} {1396}
  (\bibinfo {year} {2012})}\BibitemShut {NoStop}%
\bibitem [{\citenamefont {Zhang}\ \emph {et~al.}(2017)\citenamefont {Zhang},
  \citenamefont {Li},\ and\ \citenamefont {Xia}}]{Zhang:2016}%
  \BibitemOpen
  \bibfield  {author} {\bibinfo {author} {\bibfnamefont {M.-J.}\ \bibnamefont
  {Zhang}}, \bibinfo {author} {\bibfnamefont {H.}~\bibnamefont {Li}}, \ and\
  \bibinfo {author} {\bibfnamefont {J.-Q.}\ \bibnamefont {Xia}},\ }\href
  {\doibase 10.1140/epjc/s10052-017-5005-4} {\bibfield  {journal} {\bibinfo
  {journal} {Eur. Phys. J.}\ }\textbf {\bibinfo {volume} {C77}},\ \bibinfo
  {pages} {434} (\bibinfo {year} {2017})},\ \Eprint
  {http://arxiv.org/abs/1601.01758} {arXiv:1601.01758 [astro-ph.CO]}
  \BibitemShut {NoStop}%
%%CITATION = ARXIV:1601.01758;%%
\bibitem [{\citenamefont {{Al Mamon}}\ and\ \citenamefont
  {{Das}}(2016)}]{2016IJMPD..2550032A}%
  \BibitemOpen
  \bibfield  {author} {\bibinfo {author} {\bibfnamefont {A.}~\bibnamefont {{Al
  Mamon}}}\ and\ \bibinfo {author} {\bibfnamefont {S.}~\bibnamefont {{Das}}},\
  }\href {\doibase 10.1142/S0218271816500322} {\bibfield  {journal} {\bibinfo
  {journal} {International Journal of Modern Physics D}\ }\textbf {\bibinfo
  {volume} {25}},\ \bibinfo {eid} {1650032} (\bibinfo {year} {2016})},\ \Eprint
  {http://arxiv.org/abs/1507.00531} {arXiv:1507.00531 [gr-qc]} \BibitemShut
  {NoStop}%
\bibitem [{\citenamefont {Lizardo}\ \emph {et~al.}(2020)\citenamefont
  {Lizardo}, \citenamefont {Amante}, \citenamefont {Garc\'\i{}a-Aspeitia},
  \citenamefont {Maga\~na},\ and\ \citenamefont {Motta}}]{Lizardo:2020wxw}%
  \BibitemOpen
  \bibfield  {author} {\bibinfo {author} {\bibfnamefont {A.}~\bibnamefont
  {Lizardo}}, \bibinfo {author} {\bibfnamefont {M.~H.}\ \bibnamefont {Amante}},
  \bibinfo {author} {\bibfnamefont {M.~A.}\ \bibnamefont
  {Garc\'\i{}a-Aspeitia}}, \bibinfo {author} {\bibfnamefont {J.}~\bibnamefont
  {Maga\~na}}, \ and\ \bibinfo {author} {\bibfnamefont {V.}~\bibnamefont
  {Motta}},\ }\href@noop {} {\  (\bibinfo {year} {2020})},\ \Eprint
  {http://arxiv.org/abs/2008.10655} {arXiv:2008.10655 [astro-ph.CO]}
  \BibitemShut {NoStop}%
\bibitem [{\citenamefont {{Santos}}\ \emph {et~al.}(2011)\citenamefont
  {{Santos}}, \citenamefont {{Carvalho}},\ and\ \citenamefont
  {{Alcaniz}}}]{Santos:2011}%
  \BibitemOpen
  \bibfield  {author} {\bibinfo {author} {\bibfnamefont {B.}~\bibnamefont
  {{Santos}}}, \bibinfo {author} {\bibfnamefont {J.~C.}\ \bibnamefont
  {{Carvalho}}}, \ and\ \bibinfo {author} {\bibfnamefont {J.~S.}\ \bibnamefont
  {{Alcaniz}}},\ }\href {\doibase 10.1016/j.astropartphys.2011.04.002}
  {\bibfield  {journal} {\bibinfo  {journal} {Astroparticle Physics}\ }\textbf
  {\bibinfo {volume} {35}},\ \bibinfo {pages} {17} (\bibinfo {year} {2011})},\
  \Eprint {http://arxiv.org/abs/1009.2733} {arXiv:1009.2733 [astro-ph.CO]}
  \BibitemShut {NoStop}%
\bibitem [{\citenamefont {{del Campo}}\ \emph {et~al.}(2012)\citenamefont {{del
  Campo}}, \citenamefont {{Duran}}, \citenamefont {{Herrera}},\ and\
  \citenamefont {{Pav{\'o}n}}}]{delCampo:2012}%
  \BibitemOpen
  \bibfield  {author} {\bibinfo {author} {\bibfnamefont {S.}~\bibnamefont {{del
  Campo}}}, \bibinfo {author} {\bibfnamefont {I.}~\bibnamefont {{Duran}}},
  \bibinfo {author} {\bibfnamefont {R.}~\bibnamefont {{Herrera}}}, \ and\
  \bibinfo {author} {\bibfnamefont {D.}~\bibnamefont {{Pav{\'o}n}}},\ }\href
  {\doibase 10.1103/PhysRevD.86.083509} {\bibfield  {journal} {\bibinfo
  {journal} {Phys. Rev. D}\ }\textbf {\bibinfo {volume} {86}},\ \bibinfo {eid}
  {083509} (\bibinfo {year} {2012})},\ \Eprint {http://arxiv.org/abs/1209.3415}
  {arXiv:1209.3415 [gr-qc]} \BibitemShut {NoStop}%
\bibitem [{\citenamefont {{Nair}}\ \emph {et~al.}(2012)\citenamefont {{Nair}},
  \citenamefont {{Jhingan}},\ and\ \citenamefont {{Jain}}}]{Nair:2012}%
  \BibitemOpen
  \bibfield  {author} {\bibinfo {author} {\bibfnamefont {R.}~\bibnamefont
  {{Nair}}}, \bibinfo {author} {\bibfnamefont {S.}~\bibnamefont {{Jhingan}}}, \
  and\ \bibinfo {author} {\bibfnamefont {D.}~\bibnamefont {{Jain}}},\ }\href
  {\doibase 10.1088/1475-7516/2012/01/018} {\bibfield  {journal} {\bibinfo
  {journal} {Journal of Cosmology and Astroparticle Physics}\ }\textbf
  {\bibinfo {volume} {2012}},\ \bibinfo {eid} {018} (\bibinfo {year} {2012})},\
  \Eprint {http://arxiv.org/abs/1109.4574} {arXiv:1109.4574 [astro-ph.CO]}
  \BibitemShut {NoStop}%
\bibitem [{\citenamefont {{Mamon}}\ and\ \citenamefont
  {{Das}}(2017)}]{Mamon:2017}%
  \BibitemOpen
  \bibfield  {author} {\bibinfo {author} {\bibfnamefont {A.~A.}\ \bibnamefont
  {{Mamon}}}\ and\ \bibinfo {author} {\bibfnamefont {S.}~\bibnamefont
  {{Das}}},\ }\href {\doibase 10.1140/epjc/s10052-017-5066-4} {\bibfield
  {journal} {\bibinfo  {journal} {European Physical Journal C}\ }\textbf
  {\bibinfo {volume} {77}},\ \bibinfo {eid} {495} (\bibinfo {year} {2017})},\
  \Eprint {http://arxiv.org/abs/1610.07337} {arXiv:1610.07337 [gr-qc]}
  \BibitemShut {NoStop}%
\bibitem [{\citenamefont {{Rom{\'a}n-Garza}}\ \emph {et~al.}(2019)\citenamefont
  {{Rom{\'a}n-Garza}}, \citenamefont {{Verdugo}}, \citenamefont
  {{Maga{\~n}a}},\ and\ \citenamefont {{Motta}}}]{RomanGarza:2019}%
  \BibitemOpen
  \bibfield  {author} {\bibinfo {author} {\bibfnamefont {J.}~\bibnamefont
  {{Rom{\'a}n-Garza}}}, \bibinfo {author} {\bibfnamefont {T.}~\bibnamefont
  {{Verdugo}}}, \bibinfo {author} {\bibfnamefont {J.}~\bibnamefont
  {{Maga{\~n}a}}}, \ and\ \bibinfo {author} {\bibfnamefont {V.}~\bibnamefont
  {{Motta}}},\ }\href {\doibase 10.1140/epjc/s10052-019-7390-3} {\bibfield
  {journal} {\bibinfo  {journal} {European Physical Journal C}\ }\textbf
  {\bibinfo {volume} {79}},\ \bibinfo {eid} {890} (\bibinfo {year} {2019})},\
  \Eprint {http://arxiv.org/abs/1806.03538} {arXiv:1806.03538 [astro-ph.CO]}
  \BibitemShut {NoStop}%
\bibitem [{\citenamefont {Sharov}\ and\ \citenamefont
  {Vorontsova}(2014)}]{Sharov:2014voa}%
  \BibitemOpen
  \bibfield  {author} {\bibinfo {author} {\bibfnamefont {G.~S.}\ \bibnamefont
  {Sharov}}\ and\ \bibinfo {author} {\bibfnamefont {E.~G.}\ \bibnamefont
  {Vorontsova}},\ }\href {\doibase 10.1088/1475-7516/2014/10/057} {\bibfield
  {journal} {\bibinfo  {journal} {JCAP}\ }\textbf {\bibinfo {volume} {10}},\
  \bibinfo {pages} {057} (\bibinfo {year} {2014})},\ \Eprint
  {http://arxiv.org/abs/1407.5405} {arXiv:1407.5405 [gr-qc]} \BibitemShut
  {NoStop}%
\bibitem [{\citenamefont {Ade}\ \emph {et~al.}(2015{\natexlab{b}})\citenamefont
  {Ade} \emph {et~al.}}]{Planck:2015XIV}%
  \BibitemOpen
  \bibfield  {author} {\bibinfo {author} {\bibfnamefont {P.~A.~R.}\
  \bibnamefont {Ade}} \emph {et~al.} (\bibinfo {collaboration} {Planck}),\
  }\href@noop {} {\  (\bibinfo {year} {2015}{\natexlab{b}})},\ \Eprint
  {http://arxiv.org/abs/1502.01590} {arXiv:1502.01590 [astro-ph.CO]}
  \BibitemShut {NoStop}%
%%CITATION = ARXIV:1502.01590;%%
\bibitem [{\citenamefont {Maga\~na}\ \emph {et~al.}(2017)\citenamefont
  {Maga\~na}, \citenamefont {Motta}, \citenamefont {Cardenas},\ and\
  \citenamefont {Foex}}]{Magana:2017usz}%
  \BibitemOpen
  \bibfield  {author} {\bibinfo {author} {\bibfnamefont {J.}~\bibnamefont
  {Maga\~na}}, \bibinfo {author} {\bibfnamefont {V.}~\bibnamefont {Motta}},
  \bibinfo {author} {\bibfnamefont {V.~H.}\ \bibnamefont {Cardenas}}, \ and\
  \bibinfo {author} {\bibfnamefont {G.}~\bibnamefont {Foex}},\ }\href {\doibase
  10.1093/mnras/stx750} {\bibfield  {journal} {\bibinfo  {journal} {Mon. Not.
  Roy. Astron. Soc.}\ }\textbf {\bibinfo {volume} {469}},\ \bibinfo {pages}
  {47} (\bibinfo {year} {2017})},\ \Eprint {http://arxiv.org/abs/1703.08521}
  {arXiv:1703.08521 [astro-ph.CO]} \BibitemShut {NoStop}%
%%CITATION = ARXIV:1703.08521;%%
\bibitem [{\citenamefont {Shenavar}\ and\ \citenamefont
  {Javidan}(2020)}]{Shenavar:2020}%
  \BibitemOpen
  \bibfield  {author} {\bibinfo {author} {\bibfnamefont {H.}~\bibnamefont
  {Shenavar}}\ and\ \bibinfo {author} {\bibfnamefont {K.}~\bibnamefont
  {Javidan}},\ }\href {\doibase 10.3390/universe6010001} {\bibfield  {journal}
  {\bibinfo  {journal} {Universe}\ }\textbf {\bibinfo {volume} {6}} (\bibinfo
  {year} {2020}),\ 10.3390/universe6010001}\BibitemShut {NoStop}%
\bibitem [{\citenamefont {Dymnikova}\ \emph {et~al.}(2017)\citenamefont
  {Dymnikova}, \citenamefont {Dobosz},\ and\ \citenamefont
  {Sołtysek}}]{Dymnikova:2017}%
  \BibitemOpen
  \bibfield  {author} {\bibinfo {author} {\bibfnamefont {I.}~\bibnamefont
  {Dymnikova}}, \bibinfo {author} {\bibfnamefont {A.}~\bibnamefont {Dobosz}}, \
  and\ \bibinfo {author} {\bibfnamefont {B.}~\bibnamefont {Sołtysek}},\ }\href
  {\doibase 10.3390/universe3020039} {\bibfield  {journal} {\bibinfo  {journal}
  {Universe}\ }\textbf {\bibinfo {volume} {3}} (\bibinfo {year} {2017}),\
  10.3390/universe3020039}\BibitemShut {NoStop}%
\bibitem [{\citenamefont {{Deng}}(2011)}]{Deng:2011}%
  \BibitemOpen
  \bibfield  {author} {\bibinfo {author} {\bibfnamefont {X.-M.}\ \bibnamefont
  {{Deng}}},\ }\href {\doibase 10.1007/s13538-011-0044-z} {\bibfield  {journal}
  {\bibinfo  {journal} {Brazilian Journal of Physics}\ }\textbf {\bibinfo
  {volume} {41}},\ \bibinfo {pages} {333} (\bibinfo {year} {2011})}\BibitemShut
  {NoStop}%
\bibitem [{\citenamefont {{Bhadra}}\ and\ \citenamefont
  {{Debnath}}(2012)}]{Bhadra:2012}%
  \BibitemOpen
  \bibfield  {author} {\bibinfo {author} {\bibfnamefont {J.}~\bibnamefont
  {{Bhadra}}}\ and\ \bibinfo {author} {\bibfnamefont {U.}~\bibnamefont
  {{Debnath}}},\ }\href {\doibase 10.1140/epjp/i2012-12030-2} {\bibfield
  {journal} {\bibinfo  {journal} {European Physical Journal Plus}\ }\textbf
  {\bibinfo {volume} {127}},\ \bibinfo {eid} {30} (\bibinfo {year} {2012})},\
  \Eprint {http://arxiv.org/abs/1109.3578} {arXiv:1109.3578 [physics.gen-ph]}
  \BibitemShut {NoStop}%
\bibitem [{\citenamefont {{Pourhassan}}(2013)}]{Pourhassan:2013}%
  \BibitemOpen
  \bibfield  {author} {\bibinfo {author} {\bibfnamefont {B.}~\bibnamefont
  {{Pourhassan}}},\ }\href {\doibase 10.1142/S0218271813500612} {\bibfield
  {journal} {\bibinfo  {journal} {International Journal of Modern Physics D}\
  }\textbf {\bibinfo {volume} {22}},\ \bibinfo {eid} {1350061} (\bibinfo {year}
  {2013})},\ \Eprint {http://arxiv.org/abs/1301.2788} {arXiv:1301.2788 [gr-qc]}
  \BibitemShut {NoStop}%
\bibitem [{\citenamefont {Bento}\ \emph {et~al.}(2003)\citenamefont {Bento},
  \citenamefont {Bertolami},\ and\ \citenamefont {Sen}}]{Bento2003}%
  \BibitemOpen
  \bibfield  {author} {\bibinfo {author} {\bibfnamefont {M.~C.}\ \bibnamefont
  {Bento}}, \bibinfo {author} {\bibfnamefont {O.}~\bibnamefont {Bertolami}}, \
  and\ \bibinfo {author} {\bibfnamefont {A.~A.}\ \bibnamefont {Sen}},\ }\href
  {\doibase 10.1023/A:1026207312105} {\bibfield  {journal} {\bibinfo  {journal}
  {General Relativity and Gravitation}\ }\textbf {\bibinfo {volume} {35}},\
  \bibinfo {pages} {2063} (\bibinfo {year} {2003})}\BibitemShut {NoStop}%
\bibitem [{\citenamefont {Bento}\ \emph {et~al.}(2004)\citenamefont {Bento},
  \citenamefont {Bertolami},\ and\ \citenamefont {Sen}}]{Bento:2004}%
  \BibitemOpen
  \bibfield  {author} {\bibinfo {author} {\bibfnamefont {M.~C.}\ \bibnamefont
  {Bento}}, \bibinfo {author} {\bibfnamefont {O.}~\bibnamefont {Bertolami}}, \
  and\ \bibinfo {author} {\bibfnamefont {A.~A.}\ \bibnamefont {Sen}},\ }\href
  {\doibase 10.1103/PhysRevD.70.083519} {\bibfield  {journal} {\bibinfo
  {journal} {Phys. Rev. D}\ }\textbf {\bibinfo {volume} {70}},\ \bibinfo
  {pages} {083519} (\bibinfo {year} {2004})}\BibitemShut {NoStop}%
\bibitem [{\citenamefont {Chaplygin}(1904)}]{Chaplygin}%
  \BibitemOpen
  \bibfield  {author} {\bibinfo {author} {\bibfnamefont {S.~A.}\ \bibnamefont
  {Chaplygin}},\ }\href@noop {} {\bibfield  {journal} {\bibinfo  {journal}
  {Sci. Mem. Moscow Univ. Math. Phys.}\ }\textbf {\bibinfo {volume} {21}}
  (\bibinfo {year} {1904})}\BibitemShut {NoStop}%
\bibitem [{\citenamefont {Amendola}\ \emph {et~al.}(2003)\citenamefont
  {Amendola}, \citenamefont {Finelli}, \citenamefont {Burigana},\ and\
  \citenamefont {Carturan}}]{Amendola_2003}%
  \BibitemOpen
  \bibfield  {author} {\bibinfo {author} {\bibfnamefont {L.}~\bibnamefont
  {Amendola}}, \bibinfo {author} {\bibfnamefont {F.}~\bibnamefont {Finelli}},
  \bibinfo {author} {\bibfnamefont {C.}~\bibnamefont {Burigana}}, \ and\
  \bibinfo {author} {\bibfnamefont {D.}~\bibnamefont {Carturan}},\ }\href
  {\doibase 10.1088/1475-7516/2003/07/005} {\bibfield  {journal} {\bibinfo
  {journal} {Journal of Cosmology and Astroparticle Physics}\ }\textbf
  {\bibinfo {volume} {2003}},\ \bibinfo {pages} {005} (\bibinfo {year}
  {2003})}\BibitemShut {NoStop}%
\bibitem [{\citenamefont {Hova}\ and\ \citenamefont {Yang}(2017)}]{Hova2017}%
  \BibitemOpen
  \bibfield  {author} {\bibinfo {author} {\bibfnamefont {H.}~\bibnamefont
  {Hova}}\ and\ \bibinfo {author} {\bibfnamefont {H.}~\bibnamefont {Yang}},\
  }\href@noop {} {\bibfield  {journal} {\bibinfo  {journal} {International
  Journal of Modern Physics D}\ }\textbf {\bibinfo {volume} {26}},\ \bibinfo
  {pages} {1750178} (\bibinfo {year} {2017})}\BibitemShut {NoStop}%
\bibitem [{\citenamefont {Hernandez-Almada}\ \emph {et~al.}(2019)\citenamefont
  {Hernandez-Almada}, \citenamefont {Magana}, \citenamefont {Garcia-Aspeitia},\
  and\ \citenamefont {Motta}}]{Hernandez-Almada:2018osh}%
  \BibitemOpen
  \bibfield  {author} {\bibinfo {author} {\bibfnamefont {A.}~\bibnamefont
  {Hernandez-Almada}}, \bibinfo {author} {\bibfnamefont {J.}~\bibnamefont
  {Magana}}, \bibinfo {author} {\bibfnamefont {M.~A.}\ \bibnamefont
  {Garcia-Aspeitia}}, \ and\ \bibinfo {author} {\bibfnamefont {V.}~\bibnamefont
  {Motta}},\ }\href {\doibase 10.1140/epjc/s10052-018-6521-6} {\bibfield
  {journal} {\bibinfo  {journal} {Eur. Phys. J.}\ }\textbf {\bibinfo {volume}
  {C79}},\ \bibinfo {pages} {12} (\bibinfo {year} {2019})},\ \Eprint
  {http://arxiv.org/abs/1805.07895} {arXiv:1805.07895 [astro-ph.CO]}
  \BibitemShut {NoStop}%
%%CITATION = ARXIV:1805.07895;%%
\bibitem [{\citenamefont {Brevik}\ and\ \citenamefont
  {Gorbunova}(2005)}]{Brevik:2005bj}%
  \BibitemOpen
  \bibfield  {author} {\bibinfo {author} {\bibfnamefont {I.~H.}\ \bibnamefont
  {Brevik}}\ and\ \bibinfo {author} {\bibfnamefont {O.}~\bibnamefont
  {Gorbunova}},\ }\href {\doibase 10.1007/s10714-005-0178-9} {\bibfield
  {journal} {\bibinfo  {journal} {Gen. Rel. Grav.}\ }\textbf {\bibinfo {volume}
  {37}},\ \bibinfo {pages} {2039} (\bibinfo {year} {2005})},\ \Eprint
  {http://arxiv.org/abs/gr-qc/0504001} {arXiv:gr-qc/0504001} \BibitemShut
  {NoStop}%
\bibitem [{\citenamefont {Cruz}\ \emph {et~al.}(2017)\citenamefont {Cruz},
  \citenamefont {Cruz},\ and\ \citenamefont {Lepe}}]{Cruz:2017bcv}%
  \BibitemOpen
  \bibfield  {author} {\bibinfo {author} {\bibfnamefont {M.}~\bibnamefont
  {Cruz}}, \bibinfo {author} {\bibfnamefont {N.}~\bibnamefont {Cruz}}, \ and\
  \bibinfo {author} {\bibfnamefont {S.}~\bibnamefont {Lepe}},\ }\href {\doibase
  10.1103/PhysRevD.96.124020} {\bibfield  {journal} {\bibinfo  {journal} {Phys.
  Rev. D}\ }\textbf {\bibinfo {volume} {96}},\ \bibinfo {pages} {124020}
  (\bibinfo {year} {2017})},\ \Eprint {http://arxiv.org/abs/1710.02607}
  {arXiv:1710.02607 [gr-qc]} \BibitemShut {NoStop}%
\bibitem [{\citenamefont {Hern{\'a}ndez-Almada}(2019)}]{Almada:2019}%
  \BibitemOpen
  \bibfield  {author} {\bibinfo {author} {\bibfnamefont {A.}~\bibnamefont
  {Hern{\'a}ndez-Almada}},\ }\href {\doibase 10.1140/epjc/s10052-019-7264-8}
  {\bibfield  {journal} {\bibinfo  {journal} {The European Physical Journal C}\
  }\textbf {\bibinfo {volume} {79}},\ \bibinfo {pages} {751} (\bibinfo {year}
  {2019})}\BibitemShut {NoStop}%
\bibitem [{\citenamefont {Herrera-Zamorano}\ \emph {et~al.}(2020)\citenamefont
  {Herrera-Zamorano}, \citenamefont {Hern\'andez-Almada},\ and\ \citenamefont
  {Garc\'ia-Aspeitia}}]{Herrera:2020}%
  \BibitemOpen
  \bibfield  {author} {\bibinfo {author} {\bibfnamefont {L.}~\bibnamefont
  {Herrera-Zamorano}}, \bibinfo {author} {\bibfnamefont {A.}~\bibnamefont
  {Hern\'andez-Almada}}, \ and\ \bibinfo {author} {\bibfnamefont
  {M.}~\bibnamefont {Garc\'ia-Aspeitia}},\ }\href {\doibase
  10.1140/epjc/s10052-020-8225-y} {\bibfield  {journal} {\bibinfo  {journal}
  {Eur. Phys. J. C}\ }\textbf {\bibinfo {volume} {80}},\ \bibinfo {pages} {637}
  (\bibinfo {year} {2020})},\ \Eprint {http://arxiv.org/abs/2007.04507}
  {arXiv:2007.04507} \BibitemShut {NoStop}%
\bibitem [{\citenamefont {Zimdahl}\ and\ \citenamefont
  {Pav{\'o}n}(2003)}]{Zimdahl:2003}%
  \BibitemOpen
  \bibfield  {author} {\bibinfo {author} {\bibfnamefont {W.}~\bibnamefont
  {Zimdahl}}\ and\ \bibinfo {author} {\bibfnamefont {D.}~\bibnamefont
  {Pav{\'o}n}},\ }\href {\doibase 10.1023/A:1022369800053} {\bibfield
  {journal} {\bibinfo  {journal} {General Relativity and Gravitation}\ }\textbf
  {\bibinfo {volume} {35}},\ \bibinfo {pages} {413} (\bibinfo {year}
  {2003})}\BibitemShut {NoStop}%
\bibitem [{\citenamefont {Riess}\ \emph
  {et~al.}(2019{\natexlab{b}})\citenamefont {Riess}, \citenamefont {Casertano},
  \citenamefont {Yuan}, \citenamefont {Macri},\ and\ \citenamefont
  {Scolnic}}]{Riess_2019}%
  \BibitemOpen
  \bibfield  {author} {\bibinfo {author} {\bibfnamefont {A.~G.}\ \bibnamefont
  {Riess}}, \bibinfo {author} {\bibfnamefont {S.}~\bibnamefont {Casertano}},
  \bibinfo {author} {\bibfnamefont {W.}~\bibnamefont {Yuan}}, \bibinfo {author}
  {\bibfnamefont {L.~M.}\ \bibnamefont {Macri}}, \ and\ \bibinfo {author}
  {\bibfnamefont {D.}~\bibnamefont {Scolnic}},\ }\href {\doibase
  10.3847/1538-4357/ab1422} {\bibfield  {journal} {\bibinfo  {journal} {The
  Astrophysical Journal}\ }\textbf {\bibinfo {volume} {876}},\ \bibinfo {pages}
  {85} (\bibinfo {year} {2019}{\natexlab{b}})}\BibitemShut {NoStop}%
\bibitem [{\citenamefont {Randall}\ and\ \citenamefont
  {Sundrum}(1999{\natexlab{a}})}]{Randall-I}%
  \BibitemOpen
  \bibfield  {author} {\bibinfo {author} {\bibfnamefont {L.}~\bibnamefont
  {Randall}}\ and\ \bibinfo {author} {\bibfnamefont {R.}~\bibnamefont
  {Sundrum}},\ }\href {\doibase 10.1103/PhysRevLett.83.3370} {\bibfield
  {journal} {\bibinfo  {journal} {Phys. Rev. Lett.}\ }\textbf {\bibinfo
  {volume} {83}},\ \bibinfo {pages} {3370} (\bibinfo {year}
  {1999}{\natexlab{a}})}\BibitemShut {NoStop}%
%%CITATION = HEP-PH/9905221;%%
\bibitem [{\citenamefont {Randall}\ and\ \citenamefont
  {Sundrum}(1999{\natexlab{b}})}]{Randall-II}%
  \BibitemOpen
  \bibfield  {author} {\bibinfo {author} {\bibfnamefont {L.}~\bibnamefont
  {Randall}}\ and\ \bibinfo {author} {\bibfnamefont {R.}~\bibnamefont
  {Sundrum}},\ }\href {\doibase 10.1103/PhysRevLett.83.4690} {\bibfield
  {journal} {\bibinfo  {journal} {Phys. Rev. Lett.}\ }\textbf {\bibinfo
  {volume} {83}},\ \bibinfo {pages} {4690} (\bibinfo {year}
  {1999}{\natexlab{b}})},\ \Eprint {http://arxiv.org/abs/hep-th/9906064}
  {arXiv:hep-th/9906064 [hep-th]} \BibitemShut {NoStop}%
%%CITATION = HEP-TH/9906064;%%
\bibitem [{\citenamefont {Shiromizu}\ \emph {et~al.}(2000)\citenamefont
  {Shiromizu}, \citenamefont {Maeda},\ and\ \citenamefont {Sasaki}}]{sms}%
  \BibitemOpen
  \bibfield  {author} {\bibinfo {author} {\bibfnamefont {T.}~\bibnamefont
  {Shiromizu}}, \bibinfo {author} {\bibfnamefont {K.}~\bibnamefont {Maeda}}, \
  and\ \bibinfo {author} {\bibfnamefont {M.}~\bibnamefont {Sasaki}},\ }\href
  {http://link.aps.org/doi/10.1103/PhysRevD.62.024012} {\bibfield  {journal}
  {\bibinfo  {journal} {Phys. Rev. D}\ }\textbf {\bibinfo {volume} {62}},\
  \bibinfo {pages} {024012} (\bibinfo {year} {2000})}\BibitemShut {NoStop}%
\bibitem [{\citenamefont {Al~Mamon}\ and\ \citenamefont
  {Bamba}(2018)}]{Mamon:2018dxf}%
  \BibitemOpen
  \bibfield  {author} {\bibinfo {author} {\bibfnamefont {A.}~\bibnamefont
  {Al~Mamon}}\ and\ \bibinfo {author} {\bibfnamefont {K.}~\bibnamefont
  {Bamba}},\ }\href {\doibase 10.1140/epjc/s10052-018-6355-2} {\bibfield
  {journal} {\bibinfo  {journal} {Eur. Phys. J.}\ }\textbf {\bibinfo {volume}
  {C78}},\ \bibinfo {pages} {862} (\bibinfo {year} {2018})},\ \Eprint
  {http://arxiv.org/abs/1805.02854} {arXiv:1805.02854 [gr-qc]} \BibitemShut
  {NoStop}%
%%CITATION = ARXIV:1805.02854;%%
\bibitem [{\citenamefont {Ai}(2020)}]{Ai:2020peo}%
  \BibitemOpen
  \bibfield  {author} {\bibinfo {author} {\bibfnamefont {W.-Y.}\ \bibnamefont
  {Ai}},\ }\href {\doibase 10.1088/1572-9494/aba242} {\bibfield  {journal}
  {\bibinfo  {journal} {Commun. Theor. Phys.}\ }\textbf {\bibinfo {volume}
  {72}},\ \bibinfo {pages} {095402} (\bibinfo {year} {2020})},\ \Eprint
  {http://arxiv.org/abs/2004.02858} {arXiv:2004.02858 [gr-qc]} \BibitemShut
  {NoStop}%
\bibitem [{\citenamefont {G\"urses}\ \emph {et~al.}(2020)\citenamefont
  {G\"urses}, \citenamefont {\c{S}i\c{s}man},\ and\ \citenamefont
  {Tekin}}]{Gurses:2020ofy}%
  \BibitemOpen
  \bibfield  {author} {\bibinfo {author} {\bibfnamefont {M.}~\bibnamefont
  {G\"urses}}, \bibinfo {author} {\bibfnamefont {T.~c.}\ \bibnamefont
  {\c{S}i\c{s}man}}, \ and\ \bibinfo {author} {\bibfnamefont {B.}~\bibnamefont
  {Tekin}},\ }\href {\doibase 10.1140/epjc/s10052-020-8200-7} {\bibfield
  {journal} {\bibinfo  {journal} {Eur. Phys. J. C}\ }\textbf {\bibinfo {volume}
  {80}},\ \bibinfo {pages} {647} (\bibinfo {year} {2020})},\ \Eprint
  {http://arxiv.org/abs/2004.03390} {arXiv:2004.03390 [gr-qc]} \BibitemShut
  {NoStop}%
\bibitem [{\citenamefont {Lu}\ and\ \citenamefont {Pang}(2020)}]{Lu:2020iav}%
  \BibitemOpen
  \bibfield  {author} {\bibinfo {author} {\bibfnamefont {H.}~\bibnamefont
  {Lu}}\ and\ \bibinfo {author} {\bibfnamefont {Y.}~\bibnamefont {Pang}},\
  }\href {\doibase 10.1016/j.physletb.2020.135717} {\bibfield  {journal}
  {\bibinfo  {journal} {Phys. Lett. B}\ }\textbf {\bibinfo {volume} {809}},\
  \bibinfo {pages} {135717} (\bibinfo {year} {2020})},\ \Eprint
  {http://arxiv.org/abs/2003.11552} {arXiv:2003.11552 [gr-qc]} \BibitemShut
  {NoStop}%
\bibitem [{\citenamefont {Fernandes}\ \emph {et~al.}(2020)\citenamefont
  {Fernandes}, \citenamefont {Carrilho}, \citenamefont {Clifton},\ and\
  \citenamefont {Mulryne}}]{Fernandes:2020nbq}%
  \BibitemOpen
  \bibfield  {author} {\bibinfo {author} {\bibfnamefont {P.~G.~S.}\
  \bibnamefont {Fernandes}}, \bibinfo {author} {\bibfnamefont {P.}~\bibnamefont
  {Carrilho}}, \bibinfo {author} {\bibfnamefont {T.}~\bibnamefont {Clifton}}, \
  and\ \bibinfo {author} {\bibfnamefont {D.~J.}\ \bibnamefont {Mulryne}},\
  }\href {\doibase 10.1103/PhysRevD.102.024025} {\bibfield  {journal} {\bibinfo
   {journal} {Phys. Rev. D}\ }\textbf {\bibinfo {volume} {102}},\ \bibinfo
  {pages} {024025} (\bibinfo {year} {2020})},\ \Eprint
  {http://arxiv.org/abs/2004.08362} {arXiv:2004.08362 [gr-qc]} \BibitemShut
  {NoStop}%
\bibitem [{\citenamefont {Mahapatra}(2020)}]{Mahapatra:2020rds}%
  \BibitemOpen
  \bibfield  {author} {\bibinfo {author} {\bibfnamefont {S.}~\bibnamefont
  {Mahapatra}},\ }\href {\doibase 10.1140/epjc/s10052-020-08568-6} {\bibfield
  {journal} {\bibinfo  {journal} {Eur. Phys. J. C}\ }\textbf {\bibinfo {volume}
  {80}},\ \bibinfo {pages} {992} (\bibinfo {year} {2020})},\ \Eprint
  {http://arxiv.org/abs/2004.09214} {arXiv:2004.09214 [gr-qc]} \BibitemShut
  {NoStop}%
\bibitem [{\citenamefont {Clifton}\ \emph {et~al.}(2020)\citenamefont
  {Clifton}, \citenamefont {Carrilh}, \citenamefont {Fernandes},\ and\
  \citenamefont {Mulryne}}]{Clifton:2020xhc}%
  \BibitemOpen
  \bibfield  {author} {\bibinfo {author} {\bibfnamefont {T.}~\bibnamefont
  {Clifton}}, \bibinfo {author} {\bibfnamefont {P.}~\bibnamefont {Carrilh}},
  \bibinfo {author} {\bibfnamefont {P.~G.}\ \bibnamefont {Fernandes}}, \ and\
  \bibinfo {author} {\bibfnamefont {D.~J.}\ \bibnamefont {Mulryne}},\
  }\href@noop {} {\  (\bibinfo {year} {2020})},\ \Eprint
  {http://arxiv.org/abs/2006.15017} {arXiv:2006.15017 [gr-qc]} \BibitemShut
  {NoStop}%
\bibitem [{\citenamefont {{Efstathiou}}(2021)}]{Efstathiou:2021}%
  \BibitemOpen
  \bibfield  {author} {\bibinfo {author} {\bibfnamefont {G.}~\bibnamefont
  {{Efstathiou}}},\ }\href@noop {} {\bibfield  {journal} {\bibinfo  {journal}
  {arXiv e-prints}\ ,\ \bibinfo {eid} {arXiv:2103.08723}} (\bibinfo {year}
  {2021})},\ \Eprint {http://arxiv.org/abs/2103.08723} {arXiv:2103.08723
  [astro-ph.CO]} \BibitemShut {NoStop}%
\end{thebibliography}%



\end{document}